\documentclass[fleqn,usenatbib]{mnras}
\usepackage{mathptmx}
\usepackage{hyperref}
\usepackage[T1]{fontenc}
\usepackage{ae,aecompl}

\usepackage{graphicx}	
\usepackage{amsmath}	
\usepackage{amssymb}	
\usepackage{natbib}
\usepackage{graphicx}
\usepackage{epsfig}
\usepackage{txfonts}
\usepackage{hyperref}
\usepackage{float}
\usepackage{color}
\usepackage{lscape,graphicx}
\usepackage{journal_shortcuts}
\usepackage{amssymb}
\usepackage{multirow}
\usepackage{afterpage}
\usepackage{ulem}
\usepackage{threeparttable}
\usepackage{comment}
\usepackage{color}
\usepackage{morefloats}
\usepackage{longtable}
\usepackage{pdflscape}
\usepackage{mdwlist}
\usepackage{upgreek}
\newcommand{\kms}{km~s$^{-1}$}
\newcommand{\unitflux}{erg~cm$^{-2}$~s$^{-1}~$}
\newcommand{\unitlum}{erg~s$^{-1}~$}

\title[NIR Broad Lines detection in AGN2]{Detection of faint broad emission lines in type 2 AGN: I. Near infrared observations and spectral fitting}

\author[F. Onori et al. ]{F.~Onori,$^{1}$$^,$$^{2}$$^,$$^{3}$\thanks{E-mail:f.onori@sron.nl} 
F.~La~Franca,$^{3}$
F.~Ricci,$^{3}$
M.~Brusa,$^{4}$$^,$$^{5}$
E.~Sani,$^{6}$
 R.~Maiolino,$^{7}$
S.~Bianchi,$^{3}$ 
  \newauthor A.~Bongiorno,$^{8}$ F.~Fiore,$^{8}$ A.~Marconi$^{9,10}$ 
and C.~Vignali$^{4}$$^,$$^{5}$\\
$^{1}$SRON Netherlands Institute for Space Research, Sorbonnelaan 2, 3584 CA Utrecht, Netherlands\\
$^{2}$Department of Astrophysics/IMAPP, Radboud University, P.O. Box 9010, 6500 GL Nijmegen, the Netherlands\\
$^{3}$Dipartimento di Matematica e Fisica, Universit\`a Roma Tre,
   via della Vasca Navale 84, 00146 Roma, Italy\\
$^{4}$Dipartimento di Fisica e Astronomia, Universit\`a di Bologna, viale Berti Pichat 6/2, 40127 Bologna, Italy\\
$^{5}$INAF - Osservatorio Astronomico di Bologna, via Ranzani 1, 40127 Bologna, Italy\\
$^{6}$European Southern Observatory, Alonso de Cordova 3107, Casilla 19, Santiago 19001, Chile\\
$^{7}$Cavendish Laboratory, University of Cambridge, 19 J. J. Thomson Ave., Cambridge CB3 0HE, UK\\
$^{8}$INAF - Osservatorio Astronomico di Roma, via Frascati 33, 00044 Monte Porzio Catone, Italy\\
$^{9}$Dipartimento di Fisica e Astronomia, Universit\`a degli Studi di Firenze, Via G. Sansone 1, 50019 Sesto Fiorentino, Italy\\
$^{10}$INAF - Osservatorio Astronomico di Arcetri, Largo E. Fermi 5, 50125 Firenze, Italy}

\date{Accepted 2016 September 14. Received 2016 September 14; in original form 2016 May 24}
\pubyear{2016}

\begin{document}
\label{firstpage}
\pagerange{\pageref{firstpage}--\pageref{lastpage}}
\maketitle

\begin{abstract}


We present medium resolution near infrared spectroscopic observations of 41 obscured and intermediate class AGN (type 2, 1.9 and 1.8; AGN2) with redshift $z \lesssim$0.1, selected from the {\it Swift}/BAT 70-month catalogue. The observations have been carried out in the framework of a systematic study of the AGN2 near infrared spectral properties and have been executed using ISAAC/VLT, X-shooter/VLT and LUCI/LBT,
reaching an average S/N ratio of $\sim$30 per resolution element. For those objects observed with X-shooter we also obtained simultaneous optical and UV spectroscopy. We have identified  a component from the broad line region in 13 out of 41 AGN2, with FWHM ${\rm > 800 }$ \kms. We have verified that the detection of the broad line region components does not significantly depend on selection effects due to the quality of the spectra, the X-ray or near infrared fluxes, the orientation angle of the
host galaxy or the hydrogen column density measured in the X-ray band. The average broad line region components found in AGN2 has a significantly (a factor 2) smaller FWHM if  compared with a control sample of type 1 AGN.

\end{abstract}


\setcounter{figure}{0}

\section{Introduction}

 According to the original (``zero${\rm ^{th}}$ order'') standard unified model \citep{antonucci93},  the main different observational classes of Active Galactic Nuclei (AGN; AGN1 and AGN2) are believed to be the same kind of objects observed under different viewpoints (i.e. different orientations to the observer of an absorbing optically thick, probably clumpy, dusty torus). In particular, the broad lines ($> 1000$ \kms) observed in AGN1 are thought to be produced in the Broad Line Region (BLR),  a dense gas region near the central source (within a few thousands gravitational radii) and thus under the gravitational influence of the central supermassive Black Hole (BH). On parsec scales, the entire system is enshrouded in a dusty torus that is opaque to most of the electromagnetic radiation. The torus plays a key role in the framework of the Unified Model, since it enables the direct observation of the central region (including the BLR) only along particular directions. Narrow lines are generated in distant (on torus scale), rarefied gas regions, where the gravitational influence from the black hole is less intense, the Narrow Line Region (NLR). Therefore, an  AGN2 is observed in the torus plane (i.e. edge-on) and then the view of the innermost regions of the BLR is obstructed by the intercepting material of the torus. Only narrow emission lines are directly visible in this case. AGN1, instead, are believed to be observed with the torus nearly face-on and then they show in their rest-frame optical spectra both the BLR and the NLR emission lines.

The ``zero${\rm ^{th}}$ order'' AGN unified model implies that AGN with the same intrinsic (i.e. corrected for absorption) luminosity have the same properties (e.g. same BH masses, same accretion rates).   
Nevertheless, nowadays there is growing evidence that AGN1 and AGN2 are intrinsically different populations \citep[see e.g.][]{elitzur12},  having, on average, different  luminosities
\citep[smaller for AGN2;][]{lawrence82,ueda03,lafranca05,ueda14}, different accretion rates \citep[smaller for AGN2;][]{winter10},  different Eddington ratios \citep{lusso12}, different clustering, environment and halo mass properties \citep{allevato14,jiang16,dipompeo16}
 
One of the fundamental quantity describing an AGN is its BH mass, as it is of paramount importance in order to investigate all AGN-related science, such as  the evolution and phenomenology of AGN, the accretion physics and also the relations and interplay between supermassive BHs and their host galaxies through feedback processes. 

In the last decade, using virial based techniques in the optical band \citep[][]{blandford82,peterson93}, it has been possible to measure the BH mass on large AGN1 samples and therefore derive the supermassive BH mass function \citep{greene07b,kelly09,kelly10,merloni10,bongiorno14, schulze15}.  Many of these measurements are based on the Single Epoch (SE) black hole mass virial estimates.
By combining the velocity of the BLR clouds (assuming Keplerian orbits) along with their distance $R$ (which is proportional to the square root of the luminosity) it is possible to determine the total mass contained within the BLR \citep[which is dominated by the BH mass; e.g.][]{mclure01,vestergaard02,ho15}.
However these measurements cannot be applied to AGN2 as the  broad line component is not visible in the optical spectra.
In those few studies where AGN2 black hole masses have been derived  \citep[e.g.][from SDSS]{heckman04}, the authors used the BH-bulge scale relations which, however, 
were not verified to be valid also for AGN2 \citep[see][]{graham08,kormendy11}.

Since their discovery, there have been many attempts, using several methods (e.g. polarimetric and/or high S/N spectroscopy, both in the optical and near infrared bands) to find evidence of  the presence of the BLR also in AGN2. Several studies have shown that some AGN2 exhibit faint components of broad lines if observed with high S/N in the near infrared (NIR), where the dust absorption is less severe than in the optical (\citealt{Veilleux97}; \citealt{Riffel06}; \citealt{Cai10}). These studies have found evidence of faint BLR components only in few cases. However these works have not been carried out systematically on statistically well defined samples. 

This is the first in a series of papers dedicated to the study of the AGN2 population using NIR and optical spectra, together with multiwavelength photometric observations, of a subsample of AGN2 extracted from the  \textit{Swift}/BAT 70-month AGN X-ray catalogue \citep{Baumgartner13}. 
In this paper we describe the NIR observations and, through a detailed spectral analysis we investigate the presence 
of BLR components in the near infrared, and (if data are available) optical bands.  In the framework of this analysis, as nowadays used in many statistical studies on X-ray selected AGN samples, AGN2 refers to those X-ray selected AGN where no evident BLR emission component (or even no line at all, as in the case of the X-ray Bright Optically Normal Galaxies: XBONG) was identified in their optical spectra \citep[see e.g.][]{comastri02, civano07}.

\begin{figure}
\includegraphics[scale=0.17, angle = 0]{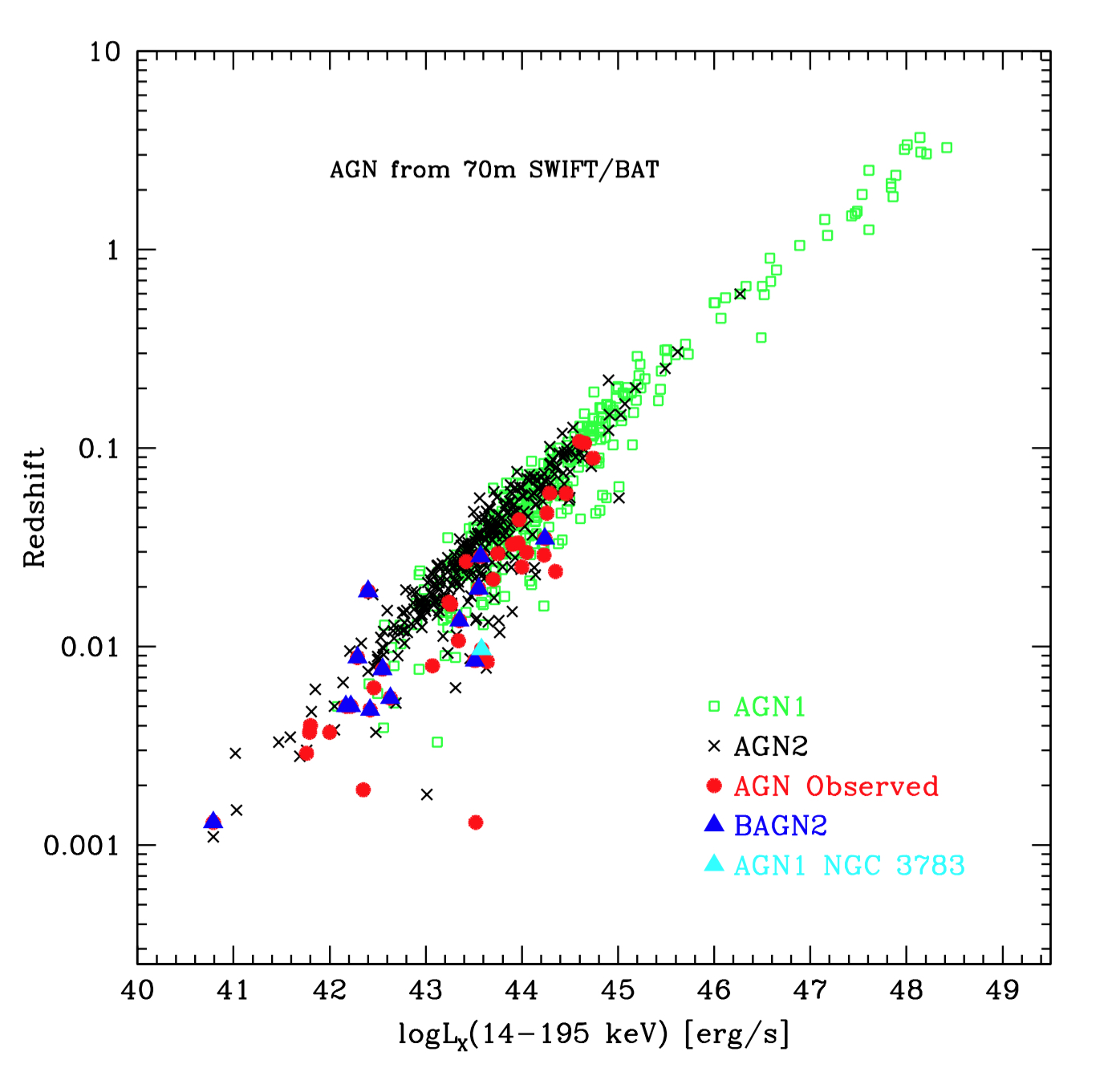}
\caption{Hubble diagram of the \textit{Swift}/BAT 70-month sample: AGN1 (green open squares), AGN2 (black crosses), AGN2 observed in the framework of our campaign (red filled dots) and those observed  `Broad AGN2', showing BLR components (BAGN2; blue filled triangles). The AGN1 NGC 3783, which has also been observed,
is shown by a cyan filled triangle.}
\label{fig:Lz}
\end{figure}

\begin{figure}
\includegraphics[scale=.17]{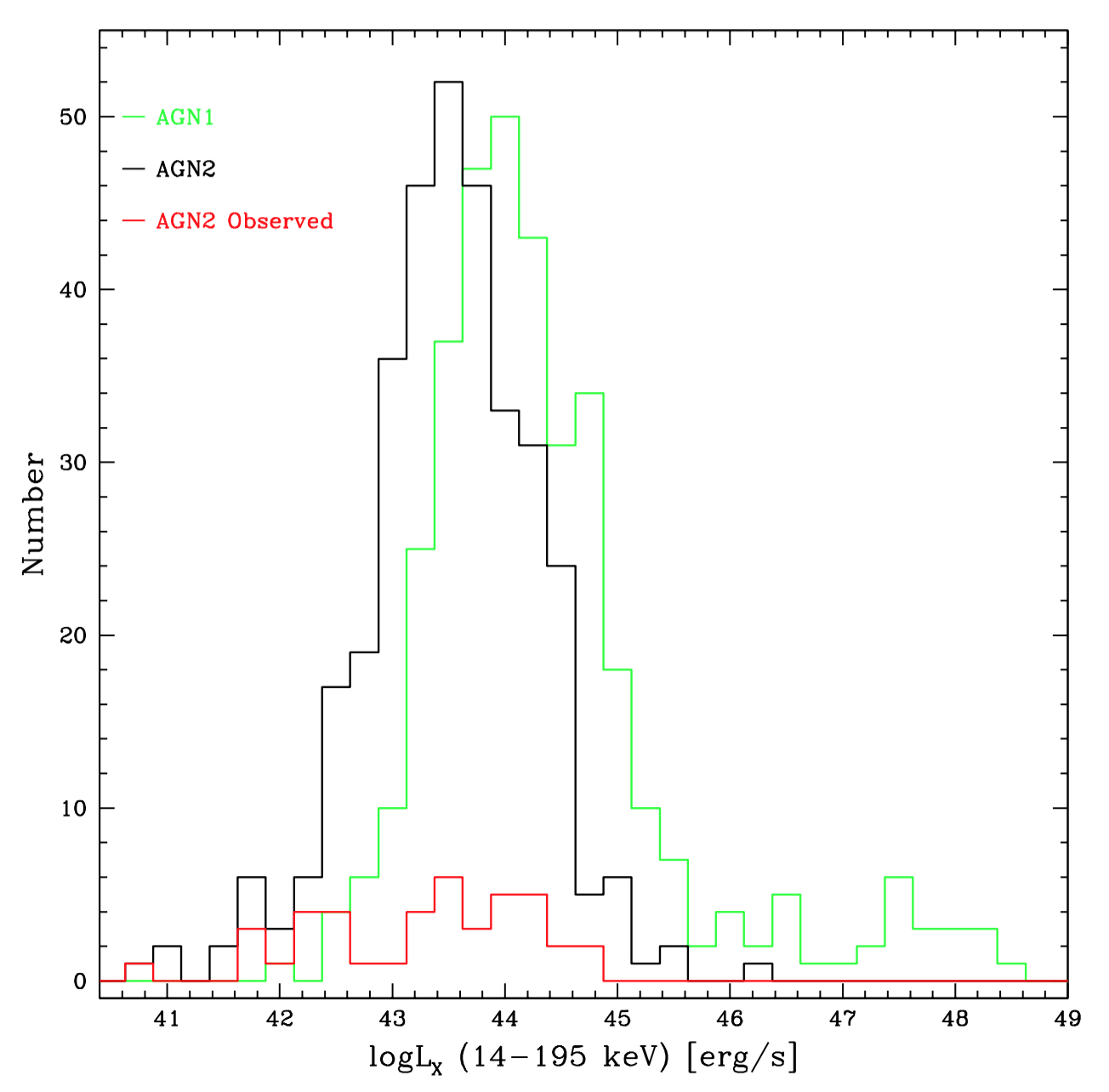}
\caption{Histogram of the 14--195 keV luminosity of the \textit{Swift}/BAT AGN1 (green line), \textit{Swift}/BAT AGN2 (black line) and our subsample of NIR spectroscopically observed  AGN2 (red line).}
\label{fig:Hist}
\end{figure}

The method and the results for NGC 4395, NCG 6221 and MCG $-$01$-$24$-$012 have already been partially presented in \citet[][]{lafranca15, lafranca16}. In section 2 and 3 the sample and the observations are described, while in section 4 the line fitting method and the results are shown. Section 5 is dedicated to a study of possible selection effects and in section 6 the conclusions are presented.
We adopt a $\Omega_{\rm m}$ = 0.3, $\Omega_\Lambda$ = 0.7 and H$_0$ = 70 \kms Mpc$^{-1}$ cosmology.

\section{The Sample}
\label{sec:obs}

The \textit{Swift} Gamma-ray burst observatory was launched in November 2004, and has been continually observing the hard X-ray sky in the 14--195 keV band with the Burst Alert Telescope (BAT), a large coded-mask telescope optimised to detect transient Gamma ray bursts. The wide field of view, the broad sky coverage and the improved sensitivity with respect to the previous \textit{Swift}/BAT surveys, allowed the construction of the 70-month catalogue, one of the most uniform and complete hard X-ray surveys, almost unbiased against Compton-thin ($N\rm _{H}\sim$10$^{22}$--10$^{24}$ cm$^{-2}$) X-ray absorbed AGN \citep{Baumgartner13}. The 70-month catalogue contains 1210 hard X-ray sources in the 14--195 keV band down to a significance level of 4.8 $\sigma$ and among these 711 are classified as AGN.

In order to look for the faint BLR components in the NIR, we selected a sample of 41 obscured and intermediate class AGN (type 2, 1.9 and 1.8, as classified in \citet{Baumgartner13}; in the following AGN2) with redshift $\lesssim$0.1 from the 312 AGN2 of the \textit{Swift}/BAT 70-month catalogue. The first 33 objects were observed with
LUCI/LBT (10) and ISAAC/VLT (23) and were randomly extracted from the parent sample according to the observability conditions at the telescopes, while 10 objects were observed with X-shooter/VLT and chosen in order to better cover the
X-ray luminosity range of the sample (40.79$\rm <$log$L\rm_{14-195}<$44.60$~erg~ s^{-1}$). Two sources (LEDA 093974 and MCG $-$05$-$23$-$016) were observed by both ISAAC/VLT and X-shooter/VLT. 
We have also observed one well known AGN1 with X-shooter (NGC 3783) in order to test our method.

\begin{table*}
\centering
 \begin{minipage}{140mm}
 \caption{General properties of the observed sample}
 \label{tbl:sample}
 \begin{center}
 \begin{tabular}{@{}lcclllcl}
 \hline
Object name & RA                & DEC             &  Redshift        & log$L\rm_{X}$   & log$N_{\rm H}$  & Cl                   & Instrument \\
(1)                 &  (2)                & (3)                &  (4)                 & (5)                & (6)                & (7)                  &      (8)        \\
\hline
2MASX J06411806+3249313               & 06:41:18.0           & +32:49:32           & 0.0470          & 44.26                   &   23.09  (P12)      &2               & LUCI\\
2MASX J09112999+4528060               &09:11:30.0            & +45:28:06           & 0.0268           &  43.42                  &  23.42 (T08)      & 2              & LUCI\\
2MASX J11271632+1909198               &11:27:16.3            & +19:09:20           & 0.1059           & 44.65                   &21.79                 & 1.8           & LUCI \\
 3C 403                                                  &  19:52:15.8           & +02:30:24          &0.059               & 44.46                  &23.60 (T08)       &2                &LUCI  \\
 Mrk 417                                                & 10:49:30.9            & +22:57:52          &0.0327             & 43.90                 & 23.60 (T08)      &2                & LUCI  \\
 NGC 3079                                             & 10:01:57.8            & +55:40:47           & 0.00372         & 42.00                 & 24.73 (A12)      &2                &LUCI \\
 NGC 4138                                             &  12:09:29.8           & +43:41:07           &0.0029             & 41.76                & 22.90 (T08)       &1.9            & LUCI  \\
 NGC 4388                                             &  12:25:46.7           & +12:39:44           &0.0084            & 43.64                 &23.63 (T08)        &2               & LUCI  \\
 NGC 4395                                             & 12:25:48.8            & +33:32:49           & 0.0013           & 40.79                & 22.30 (T08)        &1.9\footnote{Sy1.8 from \citealt{veron06}} &LUCI\\
 NGC 4686                                             &   12:46:39.9          & +54:32:03           &0.0167            & 43.24                &21.37 (V13)         &XBONG\footnote{X-Ray Bright Optically Normal Galaxy}   & LUCI \\
 2MASX J05054575$-$2351139               & 05:05:46.5            &  $-$23:51:22             & 0.0350          &44.24                & 23.50 (E09)        & 2                     & ISAAC \\
 3C 105                                                   & 04:07:16.4            &+03:42:26              & 0.089            & 44.74               &23.43 (T08)          & 2                     & ISAAC \\
 CGCG 420$-$015                                      & 04:53:25.7             & +04:03:42            & 0.0294          &43.75               &24.16 (A12)           & 2                     &ISAAC \\
 ESO 005$-$G004                                    & 06:05:44.0             & $-$86:37:57            &0.0062            & 42.46              &  23.88 (T08)          &2                     & ISAAC \\
 ESO 157$-$G023                                    & 04:22:24.2              & $-$56:13:33          &  0.0435           & 43.97              &22.80             & 2                     & ISAAC  \\
 ESO 297$-$G018              & 01:38:39.3             &$-$40:00:40           &0.0252              & 44.00              & 23.84 (T08)          & 2                     &ISAAC \\
 ESO  374$-$G044                                     & 10:13:20.4            & $-$35:59:07            & 0.0284            &43.57               & 23.71             & 2                     &ISAAC \\
 ESO 416$-$G002                                    & 02:35:14.1            & $-$29:36:26            & 0.0592            & 44.29              &$<$19.60 (T08)   &1.9                   & ISAAC  \\
 ESO 417$-$G006                                    & 02:56:21.6            & $-$32:11:26             & 0.0163           & 43.26              & 22.70-22.85 (HG15)                       & 2                      & ISAAC\\
 Fairall 272                                             & 08:23:01.1           & $-$04:56:05              & 0.0218          & 43.70               & 23.50 (B11)                        &  2                     & ISAAC  \\
 LEDA 093974                                       & 10:40:22.3            & $-$46:25:26             & 0.0239           & 43.35               & 22.96 (T08)       &2                       &ISAAC  \\
 MCG $-$01$-$24$-$012                                     & 09:20:46.2             & $-$08:03:22            & 0.0196           & 43.55                & 22.80 (T08)       & 2                      & ISAAC \\
 MCG $-$05$-$23$-$016                                     & 09:47:40.3             & $-$30:57:10           & 0.0085            & 43.51               & 22.47 (T08)        &2                       &ISAAC \\
 Mrk 1210                                              & 08:04:06.2              & +05:06:31         & 0.0135            & 43.35               & 23.34 (M12)       & 2                      & ISAAC \\
 NGC 612                                              & 01:33:59.3              & $-$36:29:41          &0.0298             & 44.05               & 23.70 (T08)       & 2                       & ISAAC  \\
 NGC 788                                             & 02:01:06.4              & $-$06:48:57          &0.0136           & 43.52              &23.48 (T08)         & 2                      & ISAAC \\
 NGC 1052                                           & 02:41:04.8              & $-$08:15:21         &0.005                &42.22               &23.30 (G00)        & 2                      & ISAAC \\
 NGC 1142                                           & 02:55:12.3              & $-$00:11:02         &0.0289              &44.23              &23.38 (T08)          &  2                      & ISAAC   \\
 NGC 1365                                          & 03:33:38.0              & $-$36:08:31         & 0.0055             & 42.63             &  23.60 (T08)        &1.8                     & ISAAC  \\
 NGC 2992                                           & 09:45:42.1              & $-$14:19:35         & 0.0077             & 42.55             & 22.00 (T08)         &2                        & ISAAC  \\
 NGC 3081                                           & 09:59:29.1              & $-$22:49:23         & 0.0080             & 43.07             &23.52 (T08)          & 2                       & ISAAC \\
 NGC 3281                                           & 10:31:52.1             & $-$34:51:13          & 0.0107              & 43.34            &24.30 (T08)          & 2                        & ISAAC \\
 PKS 0326$-$288                                    &  03:28:36.8             & $-$28:41:57          & 0.108                &44.60             & 20.49 (I12)                &1.9                    & ISAAC \\
 ESO 263$-$G013                                     &10:09:48.2               & $-$42:48:40         &0.0333               & 43.96            &  23.43 (M13)                         & 2                       &X-shooter\\
 LEDA 093974                                     & 10:40:22.3              & $-$46:25:26        & 0.0239                & 43.35           & 22.96 (T08)        & 2                        &X-shooter  \\
 MCG $-$05$-$23$-$016                                 & 09:47:40.3               & $-$30:57:10       &0.0085                  & 43.51           & 22.47 (T08)       & 2                        & X-shooter \\
  2MASX J18305065+0928414            & 18:30:50.6               &+09:28:41          & 0.0190             & 42.40          & 23.26                          &  2                        &X-shooter\\
 ESO 234$-$G050                                     & 20:35:57.8              & $-$50:11:32         & 0.0088              & 42.29          & 23.95                        & 2                          &X-shooter\\
 NGC 4941                                           &13:04:13.1                & $-$05:33:06          & 0.0037              & 41.79         & 21.38 (V13)       & 2                           & X-shooter\\
 NGC 4945                                           &13:05:27.3               & $-$49:28:04         &0.0019                & 42.35         & 24.60 (T08)       & 2                           &X-shooter\\
 NGC 5643                                           &14:32:40.8              & $-$44:10:29        & 0.0040                & 41.80         & 23.85 (G04) & 2                            &X-shooter\\
 NGC 6221                                              & 16:52:46.3             & $-$59:13:01        & 0.0050               &42.05         & 22.00 (B06)        &2                             &X-shooter \\
 NGC 7314                                           & 22:35:46.2              & $-$26:03:01       & 0.0048                & 42.42       & 21.79 (T08)         & 1.9                        &X-shooter\\
 \hline
  NGC 3783                                           &11:39:01.7                & $-$37:44:18        &0.0097               & 43.58          & 22.47 (T08)       & 1                           & X-shooter\\
\hline
\end{tabular}
\end{center}
Notes: (1) Source name; (2) and (3) R.A. and Dec. (J2000); (4) Redshift \citep{Baumgartner13}; (5)  14--195 keV luminosity (\unitlum); (6) Intrinsic hydrogen column density (cm$^{-2}$; A12 = \cite{Ajello12}; B06 = \cite{Beckmann06}; B11 = \cite{burlon11}; E09 = \cite{Eguchi09}; G00 = \cite{guainazzi00}; G04 = \cite{guainazzi04}; HG15 = \cite{hernandez15}; I12 = \cite{ichikawa12};  M12 = \cite{Marinucci12}; 
M13 = \cite{molina13}; P12 = \cite{parisi12} T08=\cite{Tueller08}; V13=\cite{Vasudevan13}); (7) SWIFT/BAT AGN classification;  (8) Instrument used.
\noindent
\end{minipage}
\end{table*}

\begin{table*}
\centering
\begin{minipage}{140mm}
\caption{Journal of observations}
\label{tbl:Log}
\begin{center}
\begin{tabular}{@{}lcclllcl}
\hline
Object name & Obs. Date  & UT         & Exposure & Seeing & Airmass &  Extraction &Scale  \\
                     &                    & [hh:mm] & [s]            & [$''$]    &               & [$''$]          & [pc]     \\
(1)                 &(2)               & (3)          & (4)           & (5)        &(6)          & (7)             & (8)      \\      
\hline  
 \multicolumn{8}{c}{ LUCI Observations} \\
 \multicolumn{8}{c}{ 1$''\times$2.8$'$ slit width; Filter: $zJ$spec; Grating: 200 $H$+$K$} \\
  \multicolumn{8}{c}{ } \\
\hline
2MASX J06411806+3249313    & 20 Oct 2012    & 08:15                        &8$\times$350             &1.47                         & 1.52            & 1.5 & 1402\\
2MASX J09112999+4528060    &20 Oct 2012        &10:57                         &8$\times$350             &0.75                      &1.39             & 1.5  & ~~801  \\
2MASX J11271632+1909198   & 06 Dec 2012        &11:06                       &8$\times$350             &0.53                       &1.10             & 1.5 &3146 \\
                3C 403                            &25 Oct 2012        &02:14                         &2$\times$350             &1.33                 &1.21              & 1.5 & 1778\\
                NGC 3079                      &24 Oct 2012        & 12:24                        &8$\times$350             &1.08                  &1.26              & 1.5  &~~117 \\
                NGC 4138                     & 06 Dec 2012       &10:55                        &8$\times$350             &0.54                   &1.37               & 1.5 & ~~~~93\\
                NGC 4388                     & 19 Feb 2013       &09:36                        &8$\times$350              &0.74                  &1.07               & 0.5 &~~~~82\\
                NGC 4395                      &05 Dec 2012      &12:09                        &8$\times$350             &0.54                    &1.21                & 1.5 &~~~~33\\
                NGC 4686                     &07 Dec 2012       &11:16                        &8$\times$350             &1.05                    &1.42                & 1.5 & ~~509\\
                Mrk 417                           &05 Dec 2012      &10:03                        &8$\times$350             &0.73                   &1.38                & 1.5 & ~~972\\
\hline
 \multicolumn{8}{c}{ ISAAC Observations} \\
  \multicolumn{8}{c}{ 0.8$''\times$120$''$ slit width; Filter: $J$; Grating: LR and MR} \\
  \multicolumn{8}{c}{ } \\
 \hline
 2MASX J05054575$-$2351139    & 03 Nov 2011              &03:41                         &6$\times$180               &0.90                          &1.41                        &0.6 & ~~413 \\
                3C 105                            & 21 Oct 2011                & 07:22                        &6$\times$180              &0.76                          & 1.15                        & 0.9 & 1593 \\
                CGCG 420$-$015            &19 Dec 2011                &06:04                       &6$\times$180               &0.71                          &1.45                         & 0.6  & ~~349 \\
                ESO 005$-$G004             & 04 Nov 2011             & 03:04                        &6$\times$180               &1.36                         & 2.30                        &0.9 & ~~100\\ 
                ESO 157$-$G023             & 19 Oct 2011               &07:41                         &6$\times$180              &1.03                          &1.17                         &0.9 & ~~159\\
                ESO 297$-$G018             & 13 Oct 2011                &04:34                         &6$\times$180              &1.22                       &1.04                            &0.6 & ~~~~62 \\
                ESO 374$-$G044             &22 Jan 2012                &01:59                        &6$\times$180               &0.98                           &1.82                       &0.9  & ~~102\\
                ESO 416$-$G002              & 08 Oct 2011                &02:37                         &6$\times$180              &1.05                           &1.48                        &0.6 & ~~703 \\
                ESO 417$-$G006              & 11 Oct 2011                &06:37                         &6$\times$180              &1.00                          &1.01                         &0.9 & ~~286\\
                Fairall 272                      &22 Dec 2011                 &04:49                      &6$\times$180               &0.96                          &1.26                          &0.6 & ~~253\\ 
                LEDA 093974                &08 Jan 2012                 &04:51                       &6$\times$180              &0.85                            &1.38                         &0.6 & ~~277 \\
                MCG $-$01$-$24$-$012           &07 Jan 2012                 &03:06                       &6$\times$180               &1.19                          &1.67                          &0.9 & ~~341\\ 
                MCG $-$05$-$23$-$016          &01 Jan 2012                  &04:39                       & 6$\times$180              &0.82                          &1.29                          & 0.6& ~~~~92 \\
                Mrk 1210                        &19 Dec 2011                 &07:56                       &6$\times$180               &0.69                          &1.24                          &0.6 & ~~156 \\
                NGC 612                         & 13 Oct 2011               &02:43                          &6$\times$180              &1.12                         &1.15                         &0.9 & ~~353 \\
                NGC 788                        & 07 Oct 2011                 &02:09                         &6$\times$180              &1.06                          &1.69                         &0.9 & ~~245 \\
                NGC 1052                      & 04 Nov 2011               &01:14                        &6$\times$180               &0.90                          &1.54                         &0.9  & ~~~~90\\
                NGC 1142                     &09 Oct 2011                   &04:53                         &6$\times$180              &0.90                         &1.20                         &0.9 & ~~517\\
                NGC 1365                       & 13 Oct 2011               & 05:56                        &6$\times$180              &0.87                          &1.03                         &0.6 & ~~~~60\\ 
                NGC 2992                      &07 Jan 2012                 &04:32                        &6$\times$180              &1.14                           &1.28                         &0.9 & ~~127\\ 
                NGC 3081                      &06 Jan 2012                 &03:55                       &6$\times$180               &1.16                          &1.59                          &0.4& ~~~~58\\ 
                NGC 3281                      & 14 Nov 2011               &07:26                       &6$\times$180                &1.09                          &1.64                         &0.9 & ~~178 \\
                PKS 0326$-$288               & 07 Oct 2011                & 03:01                        &6$\times$180              &0.88                          & 1.55                        &0.9& 1943\\
\hline  
\end{tabular}
\end{center}
\noindent
\end{minipage}
\end{table*}

\begin{table*}
\centering
\begin{minipage}{140mm}
\contcaption{}
\label{tbl:Log}
\begin{center}
\begin{tabular}{@{}lcclllllcr}
\hline
Object name & Obs. Date  & UT         & \multicolumn{3}{c}{ Exposure} & Seeing & Airmass &  Extraction &Scale  \\
                     &                    &               &UVB  & VIS &  NIR                   &            &                &                     &           \\                
                     &                    & [hh:mm] & [s]     &[s]    &[s]                       & [$''$]    &               & [$''$]           & [pc]     \\
(1)                 &(2)               & (3)          &         &(4)    &                           &(5)       &\phantom{Air}(6)           & (7)             & (8)      \\      
\hline  
\multicolumn{10}{c}{ X-shooter Observations} \\ 
  \multicolumn{10}{c}{ slit width: 1.0$''$$\times$11$''$ for UVB and a 0.9$''$$\times$11$''$ for VIS and NIR}\\
 \multicolumn{10}{c}{ } \\
  \hline
 ESO  263$-$G013     &11 Feb 2013  & 00:47   &\phantom{1}2$\times$50\phantom{0}   &\phantom{1}2$\times$50\phantom{0}  &\phantom{1}2$\times$150  &1.2   &1.93  &1.5  &980\\                  
 LEDA 093974             &11 Feb 2013  & 08:03  &\phantom{1}2$\times$225                      &\phantom{1}2$\times$250                   &\phantom{1}2$\times$300 &0.92  &1.28 &1.5  &693\\                 
 MCG $-$05$-$23$-$016  &11 Feb 2013 &01:10 &\phantom{1}2$\times$130                   &\phantom{1}2$\times$163                   &\phantom{1}2$\times$200 &0.85  &1.56 &1.5  &231\\
 2MASX J18305065+0928414  &05 Jun 2014 &04:15  &10$\times$225                          &10$\times$259                                      &10$\times$291                   &0.70 &1.42  &1.0 & 397\\          
 ESO  234-G050       &25 Jun 2014  &06:38                  &10$\times$225                         &10$\times$259                                      &10$\times$291                   & 0.78 &1.11 &1.0 & 172\\
 NGC 3783               &11 Feb 2013   &08:21     &\phantom{1}2$\times$230                   &\phantom{1}2$\times$263                      &\phantom{1}2$\times$300  &0.94 &1.09 &1.5 & 181\\
 NGC 4941               &21 Jul 2014    &22:54                  &10$\times$225                        &10$\times$259                                      &10$\times$291                     &0.84 &1.29 &1.0 & ~\,68\\                  
 NGC 4945              & 23 Apr 2014   &04:47                  &10$\times$225                       &10$\times$259                                       &10$\times$291                     &1.07 &1.14  &1.0 & ~\,26\\                  
 NGC 5643              &20 Jun 2014   &03:50                  &10$\times$225                       &10$\times$259                                       &10$\times$291                      &1.14 &1.24   &1.0 & ~\,71\\
 NGC 6221             &24 Apr  2014   &06:07                   &10$\times$225                      &10$\times$259                                       &10$\times$291                       &1.01 &1.25   &1.0 & ~\,92\\                  
 NGC 7314            &25 Jun 2014    &07:50                   &10$\times$225                      &10$\times$259                                        &10$\times$291                       & 0.67&1.01  &1.0 & ~\,98\\ 
 \hline
\end{tabular}
\end{center}
Notes: (1) Source name; (2) Date of observation; (3) Starting UT of acquisition; (4) Number of acquisitions and exposure time for each acquisition; (5) Seeing; (6) Airmass; (7) Width of the 1d spectra extraction; (8) Size of the nuclear region corresponding to the extraction width. The slit width, filter and grating of the instruments are also listed.
\noindent
\end{minipage}
\end{table*}

Our sample of 41 AGN2 and 1 AGN1 is listed in Table \ref{tbl:sample} along with the 14--195 keV luminosity (in the following hard X luminosity), the column density ($N\rm _H$),
the optical classification and the spectrograph used.
As shown in Table  \ref{tbl:sample}, for most of the sources the $N\rm _H$ values have been retrieved from the literature, while for 5 sources  publicly available X-ray data were analysed.
For 3XMM~J112716.3$+$190920 the {\it XMM}-{\it Newton} EPIC-pn data (exposure
time of $\approx$6.2~ks) plus {\it Swift}/BAT were used, while for ESO~157$-$G023,
ESO~374$-$G44, 2MASX1830$+$0928 and ESO~234$-$G50 the X-ray coverage was
provided by multiple {\it Swift}/XRT and BAT observations; the final
exposure time in XRT was $\approx$10.8, 18.5, 26.5 and 10.7~ks, respectively.
To derive the $N\rm _{H}$ and intrinsic (i.e., absorption-corrected) AGN luminosity of all
of these sources we adopted as a baseline model an absorbed
powerlaw plus a scattering component (for the soft X-ray emission);
only for 3XMM~J112716.3$+$190920 an iron K$\upalpha$ emission line
(with equivalent width of $\approx$80~eV) was detected. 
In Figure \ref{fig:Lz} the redshifts of the 41 observed AGN2 and the total 70-month {\it Swift}/BAT AGN1 and AGN2 samples, as a function of the hard X-ray luminosity, are shown, while
in Figure \ref{fig:Hist} we show the hard X-ray luminosity distribution.

\section{Observations and data reduction}
\label{sec:observations}
The observations have been carried out at ESO/VLT and at LBT in the period between October 2011 and June 2014. All the spectra were taken under a clear sky but with different seeing and airmass conditions (see Table \ref{tbl:Log}). Targets acquisition was carried out paying attention to the centring of the galaxy's nucleus at the best and, when possible, the slit has been rotated in order to include also a star for better OH telluric absorption correction. For each target the individual spectra were obtained using the nodding technique in the standard ABBA sequence, in order to obtain a good quality sky correction during the data reduction phase. We have also observed a bright star (O, B, A or Solar spectral type)  within 30 minutes to the target observations, which was used for the flux calibration and for correction of the OH absorptions every time the telluric star was not available. Flats and arcs were taken within one day of the observations.  

 In the following three sections we describe the spectrographs  and the observational setup used for the acquisition.

\subsection{LUCI/LBT observations}
\label{subsec:luci}

The LBT NIR Spectrograph Utility with Camera and Integral-Field Unit for Extragalactic Research \citep[LUCI;][]{seifert03}  is a NIR spectrograph and imager, mounted on the bent Gregorian focus of the SX mirror of the telescope at LBT observatory (LBTO) in Arizona. The instrument is equipped with Rockwell HAWAII-2 HdCdTe 2048$\times$2048 px$^{2}$ array and it works in the wavelength range from 0.85 $\upmu$m to 2.5 $\upmu$m, corresponding to the photometric $z$, $J$, $H$ and $K$ bands.  

We observed a total of 10 AGN2 of our sample in the period  October 2012 - February 2013, having 0$^h$$<\alpha<$22$^h$ and $\delta> -$10$^\circ$. All the objects have been acquired in the $zJ$ (0.92$-$1.5 $\upmu$m) band using the grating 200 $H$+$K$ in combination with the $zJ$spec filter. For each object 8 images were taken, with exposures of 350 s each.  A 1$''$$\times$2.8$'$  slit was used, corresponding to a resolution $R$ = ${\lambda/\Delta\lambda}$ = 1360 and to a velocity uncertainty of  $\sigma_v$ = 220 \kms\ for the $J$ band ($\lambda\rm_{c}$$\sim$11750 \AA) at redshift 0. 

\subsection{ISAAC/VLT observations} 
\label{subsec:isaac}

The VLT  Infrared Spectrometer And Array Camera \citep[ISAAC;][]{Moorwood98} is an IR (1$-$5 $\upmu$m) imager and spectrograph mounted at the Nasmyth A focus of the UT3 of the VLT in Chile. It has two arms, one equipped with the 1024$\times$1024 Hawaii Rockwell array, used for short wavelenght mode (SW; 1$-$1.5 $\upmu$m), and the other with a 1024$\times$1024 InSb Aladdin array, used for long wavelength mode (LW; 3$-$5 $\upmu$m). In spectroscopic mode ISAAC is equipped with two gratings, for Low and Medium resolution spectroscopy (LR and MR, respectively).

Twenty-three AGN2 of our sample, having 0$^h$$<\alpha<$12$^h$ and $\delta<$10$^\circ$, were observed in SW mode in the period October 2011 - January 2012.
All the targets have been observed in the $J$ band (1.1$-$1.4 $\upmu$m), using both LR and MR modes. In particular, in the MR mode the wavelength range was centred where either the Pa$\upbeta$ or the \ion{He}{I}10830 \AA\ emission lines were expected, according to the redshift of the source.
For each object we acquired 6 LR images with exposures of 180 s each and 4 MR images with exposures of 340 s each.  A  0.8$''$$\times$2$'$ slit was used, corresponding to a spectral resolution $R$ = 730 and $R$ = 4700 (at $\lambda\rm_{c}$$\sim$1.2 $\upmu$m) and to a velocity uncertainty of  $\sigma_v$ = 430 \kms and $\sigma_v$ = 60 \kms\  (at zero redshift) for LR and MR, respectively.  \\

\subsection{X-shooter/VLT observations}
\label{sec:xshooter}

X-shooter \citep{Vernet11} is a single slit spectrograph mounted to the Cassegrain focus of the VLT UT3, covering in a single exposure the spectral range from the UV to the $K$ band (300$-$2500 nm). The instrument is designed to maximize the sensitivity in the spectral range by splitting the incident light in three arms with optimized optics, coatings, dispersive elements and detectors. It operates at intermediate resolutions, $R$ = 4000--18000, depending on wavelength and slit width. 
The three arms are fixed format cross-dispersed \'echelle spectrographs and they operate in parallel.

We have observed 10 AGN2 and one AGN1 from our sample, with 10$^h$$<\alpha<$22$^h$and $\delta<$10$^\circ$. The first set of observations (4 sources) has been carried out in visitor mode on February 2013 and for each object we acquired 2 images with exposures, in the NIR band, in the range of 150--300 s each. The second set of  X-shooter observations, including 7 sources,  has been performed in service mode between April 2014 and June 2014. For each object we acquired 10 images with exposures, in the NIR band, of $\sim$290 s each.

For both set of observations, a 1.0$''$$\times$11$''$  slit for the UVB arm and a 0.9$''$$\times$11$''$ slit for the VIS and NIR arms were used,  corresponding to a spectral resolution $R$ = 4350 for the UVB arm,  $R$ = 7450 for the VIS arm and $R$ = 5300 for NIR arm, and to a velocity uncertainty  of $\sigma_v$$\sim$70/40/60 \kms\ at zero redshift,
in the UVB/VIS/NIR arms, respectively.

\subsection{Data reduction}

The data reduction was carried out using  standard IRAF  \citep{tody86} tasks and included flat field correction, cosmic ray cleaning, wavelength calibration, extraction of spectra from science frames using the optimized method by \citet{horne86},  telluric absorption correction and flux calibration. The wavelength calibrations made use of Xe and Ar arc lamps as a reference, and were eventually compared to the OH sky lines in order to apply, when necessary, small offsets due to instrument flexures during the observations. The telluric correction and flux calibration have been carried out by observing bright O, B, A or Solar spectral class type stars, just after or before the science observations (\citealt{maiolino96}; \citealt{vacca03}). In some cases, when possible, the calibration star has been put in the slit of the science observations, for a better telluric correction. As far as the X-shooter data are concerned, the above data reduction steps were carried out
using the REFLEX X-shooter pipeline \citep{freudling13}. 

\bigskip


\section{The emission line measurements }
\label{sec:lines}

\begin{table}
\centering
\caption{Observed optical and NIR emission lines}
\label{tbl:lines}
\begin{center}
\begin{tabular}{@{}lcclc}
\hline
\multicolumn{2}{c}{Optical}&        &\multicolumn{2}{c}{NIR}\\
\cline{1-2}\cline{4-5}\\
Element & Vacuum Wavelength &  & Element & Vacuum wavelength \\
              & [\AA]             &   &               &  [\AA]  \\
(1)         & (2)                     &   &  (3)         & (4) \\  
\hline
 $[\ion{Ne}{V}]$   & 3346.79    &  &  $[\ion{S}{III}]$   &\phantom{0}9069.0 \\
 $[\ion{Ne}{V}]$   & 3426.85    &  &  $[\ion{S}{III}]$   &\phantom{0}9531.0 \\
 $[\ion{O}{II}]$     & 3728.38    &   & Pa$\upepsilon$ &\phantom{0}9548.6 \\
 $[\ion{O}{II}]$     & 3729.86    &   & $[\ion{C}{I}]$     &\phantom{0}9853.0 \\
$[\ion{Fe}{VIII}]$  & 3759.69   &   &  $[\ion{S}{VIII}]$ & \phantom{0}9911.1 \\
$[\ion{Ne}{III}]$    & 3869.86      &  &Pa$\updelta$    & 10052.1 \\
$[\ion{[Fe}{V}]$    &3896.33      & & $\ion{He}{II}$               &10122.0 \\
$[\ion{Ne}{III}]$     &3968.59      & & $\ion{He}{I}$ &10830.0 \\
 H$\upgamma$ &4341.69 & & Pa$\upgamma$ & 10938.0\\
 $\ion{He}{II} $            & 4687.02  &  & $\ion{O}{I}$                 &11287.0 \\
H$\upbeta$ &4862.68      &  &  $[\ion{P}{II}]$               &11886.0 \\
$[\ion{O}{III}]$ & 4960.30    &  & $[\ion{S}{IX}]$              & 12523.0 \\
$[\ion{O}{III}]$ &5008.24     &  & $[\ion{Fe}{II}]$ &12570.0\\  
$[\ion{Fe}{VII}]$&5722.30     &  &Pa$\upbeta$ &12821.6\\
$\ion{He}{I}$         & 5877.25    &  &$[\ion{Fe}{II}]$        & 16436.0 \\
$[\ion{Fe}{VII}]$& 6087.98   &  &\\
$[\ion{O}{I}]$     &6302.05   &  &\\
$[\ion{O}{I}]$     &6365.54    &  &\\ 
$[\ion{N}{II}]$  & 6549.84  &  &  \\
H$\upalpha$& 6564.61  &  & \\
$[\ion{N}{II}]$ &6585.23      & & \\
$[\ion{S}{II}]$ &6718.32        & &\\
$[\ion{S}{II}]$ &6732.71        & & \\
$[\ion{O}{II}]$ &7322.01       &  & \\
$[\ion{Fe}{XI}]$& 7894.0\phantom{0}    &  &\\
\hline
\end{tabular}
\end{center}
\noindent
\end{table}

\begin{figure*}
\includegraphics[scale=0.38]{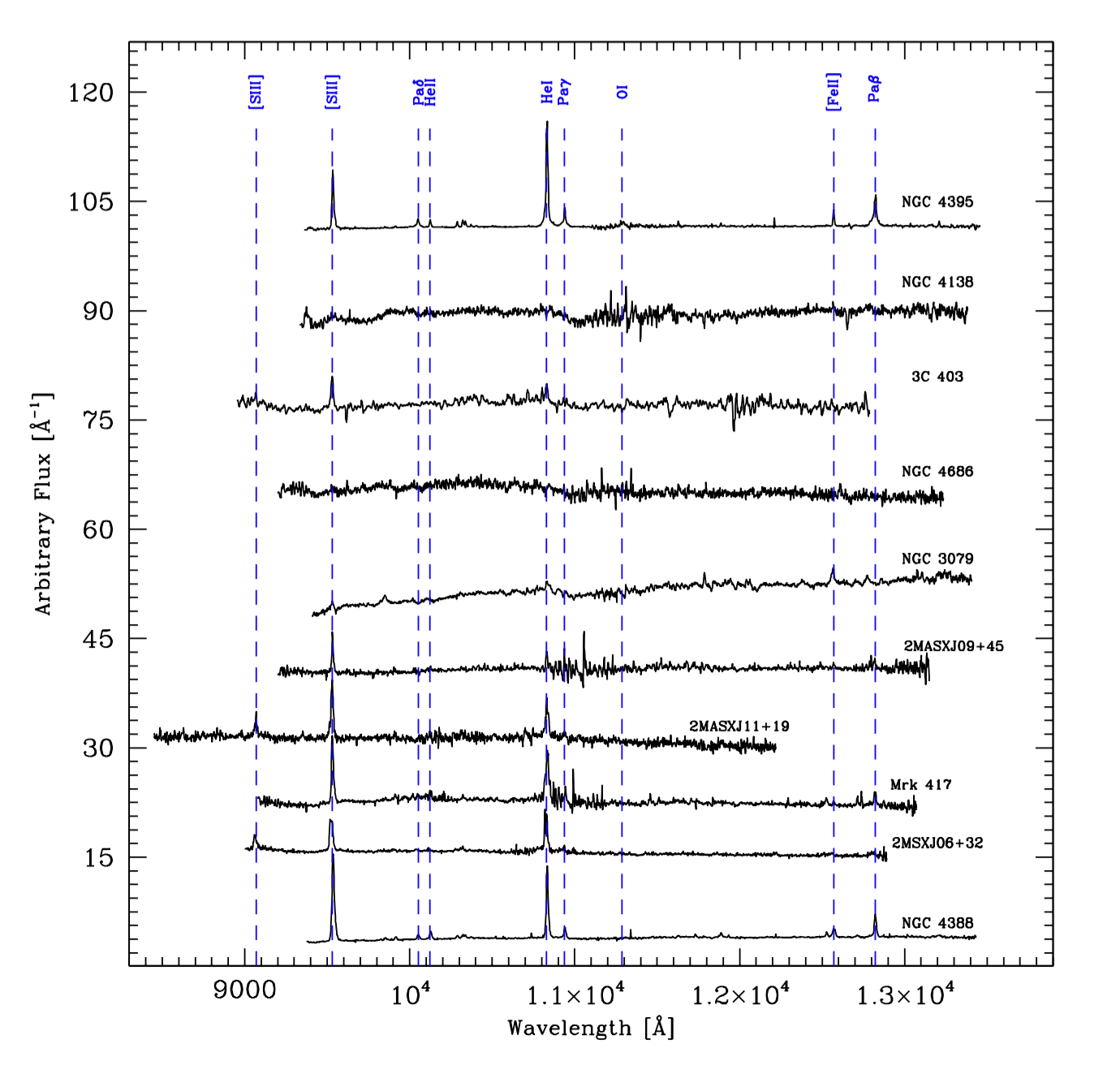}
\caption{Near infrared spectra, obtained with LUCI/LBT, corrected for redshift. The rest frame wavelength position of some of the most relevant emission lines are shown by dashed vertical lines. The spectra are flux calibrated, but are shown with arbitrary normalization.}
\label{fig:spettriluci}
\end{figure*}

In Figures 
\ref{fig:spettriluci}, 
\ref{fig:isaac1}, 
\ref{fig:XshNIR}, 
\ref{fig:OptXsh},  all the 42 spectra (41 AGN2 and 1 AGN1), divided according to the instrument used, are shown. 
We have identified and analysed in our spectra the most important NIR and optical emission lines, as listed in Table \ref{tbl:lines}.  The vacuum rest-frame wavelengths were used.
We have looked for faint broad components both in the Pa$\upbeta$ and the \ion{He}{I}10830 \AA\ emission lines as they are the most prominent permitted emission lines visible in the NIR $J$ band. For those objects observed with X-shooter the H$\upalpha$ and H$\upbeta$ spectral regions were also analysed. Moreover for those objects observed with either LUCI or ISAAC, where a BLR component was detected in the NIR, optical data taken from the literature were also studied.

The 1$\sigma$ uncertainties provided by the data reduction pipelines, or estimated  from featureless regions of the spectra, were used to carry out the fitting
using XSPEC 12.7.1 \citep{arnaud96}. 
The local
continuum was always modelled with a power-law and subtracted, then all significant (using the F-test) components were modelled with Gaussian profiles. All measurements were performed in the redshift corrected spectrum (i.e.
in the object rest frame). All the redshifts were taken from the {\it Swift}/BAT 70-month catalogue \citep{Baumgartner13}. We have identified as narrow (N) all the components having widths less than $\sim$500 \kms, well centered with the wavelength expected from the systemic redshift, and compatible with the forbidden lines widths. At variance, the largest components of the permitted \ion{H}{I} and \ion{He}{I} lines, significantly larger than the narrow components, have been classified as broad (B). In some cases other
intermediate (I) width components, blueshifted with respect to the narrow components, were also identified.
When possible, the narrow component of the permitted lines (\ion{H}{I} and \ion{He}{I}) has been modelled by imposing the same FWHM found for the narrow component of the forbidden lines in the same spectral band. In the optical we have imposed that the intensity ratios between the [\ion{O}{III}]4959 \AA\ and [\ion{O}{III}]5007  \AA\ and between [\ion{N}{II}]6548  \AA\ and [\ion{N}{II}]6583  \AA\ satisfied the 1:2.99 relation \citep{osterbrock06}. When intermediate components were found, we have fixed their FWHM and their blueshift $\Delta v$ to be equal to that measured in the corresponding intermediate components of the most intense forbidden line, observed in the same spectral region. 
We have found significant broad line components in 13 out of 41 observed AGN2. 
In Figures
\ref{fig:NGC4395},
\ref{fig:2MASXJ05-23_med},
\ref{fig:ESO374G44},
\ref{fig:MGC011224},
\ref{fig:MCG052316},
\ref{fig:mrk1210},
\ref{fig:NGC1052},
\ref{fig:NGC1365},
\ref{fig:NGC2992},
\ref{fig:2MASXJ18+09},
\ref{fig:ESO234G050},
\ref{fig:NGC6221},
\ref{fig:NGC7314}
the fitting models of the spectral regions including the Pa$\upbeta$, \ion{He}{I}, H$\upalpha$  and  H$\upbeta$ lines of these 13 AGN2  are shown. The corresponding fitting parameters (FWHM, not yet corrected for the instrumental resolution, $EW$ and $\Delta v$) are listed in Tables \ref{tbl_OptBAGN2lines} and
\ref{tbl:NIRBAGN2lines}, while the FWHM values, corrected for instrumental broadening, are listed in Table \ref{tbl_BLR}. In appendix \ref{sec:appendix1} more detailed descriptions of the line fittings are discussed, while in appendix \ref{sec:appendix2} the spectral fitting of those objects without (secure) evidence of a broad component in the region
of the Pa$\upbeta$ or \ion{He}{I} lines (when the Pa$\upbeta$ region was not observed) are shown. In appendix  \ref{sec:appendix3} we report  the main fitting parameters for all the analysed emission lines of all the 42 spectra: FWHM (not corrected for the instrumental resolution), $EW$ and line flux are listed (Tables
\ref{tbl_FWHMluciSpettri}, 
\ref{tbl_EWluciSpettri},
\ref{tbl_FLUXluciSpettri}
\ref{tbl_FWHMisaacSpettri},
\ref{tbl_EWisaacSpettri},
\ref{tbl_FLUXisaacSpettri},
\ref{tbl_FWHMxshNIRSpettri},
\ref{tbl_EWxshNIRSpettri},
\ref{tbl_FLUXxshNIRSpettri},
\ref{tbl_FWHMxshVISSpettri},
\ref{tbl_EWxshVISSpettri},
\ref{tbl_FLUXxshVISSpettri},
\ref{tbl_FWHMxshUVBSpettri},
\ref{tbl_EWxshUVBSpettri},
\ref{tbl_FLUXxshUVBSpettri}).

\clearpage

\clearpage

\begin{figure*}
\includegraphics[scale=0.38]{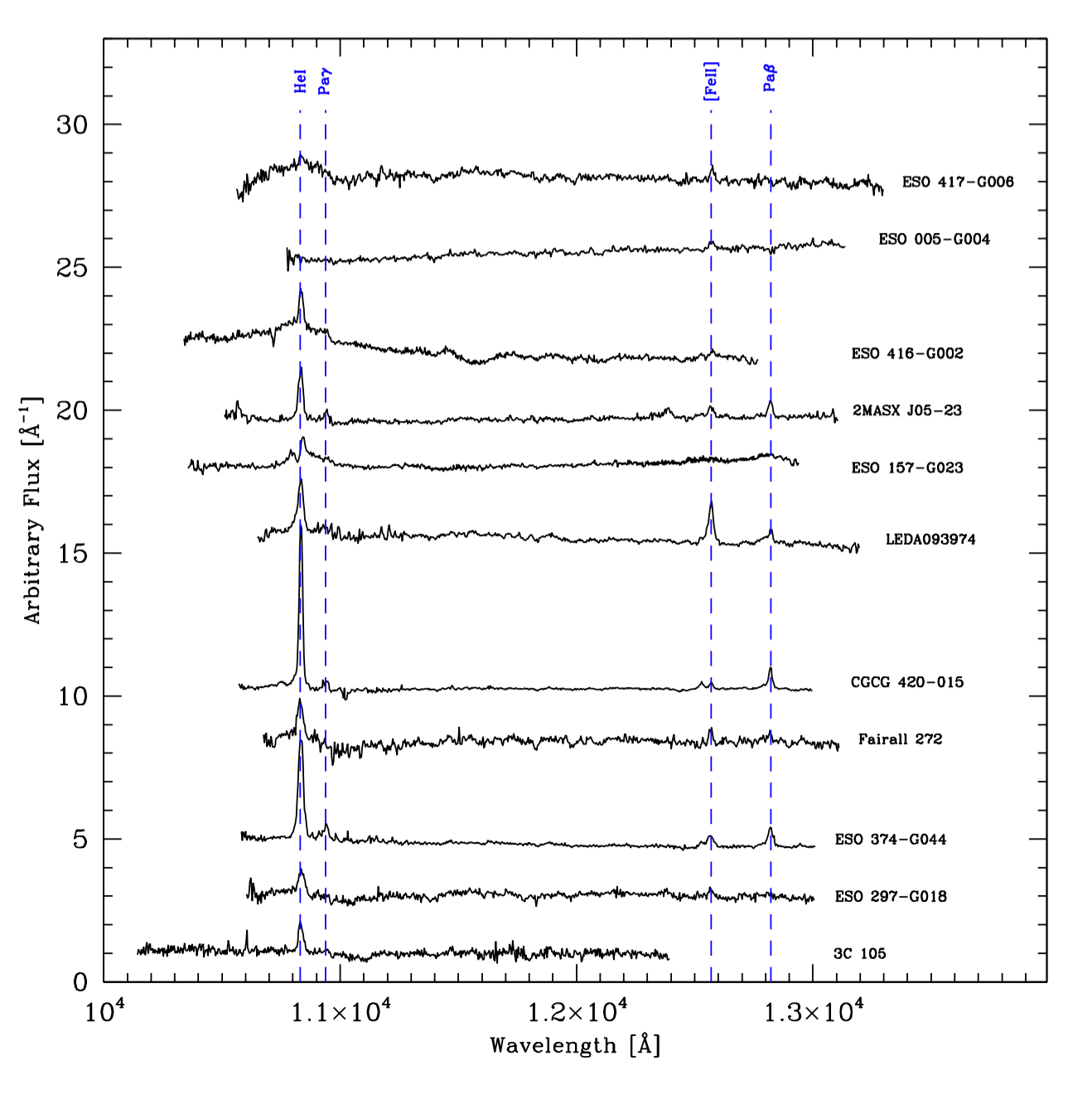}
\caption{Low resolution NIR spectra, obtained with ISAAC/VLT, corrected for redshift. The rest frame wavelength position of some of the most relevant emission lines are shown by dashed vertical lines. The spectra are flux calibrated, but are shown with arbitrary normalization.}
\label{fig:isaac1}
\end{figure*}
\clearpage

\setcounter{figure}{3}
\begin{figure*}
\includegraphics[scale=0.38]{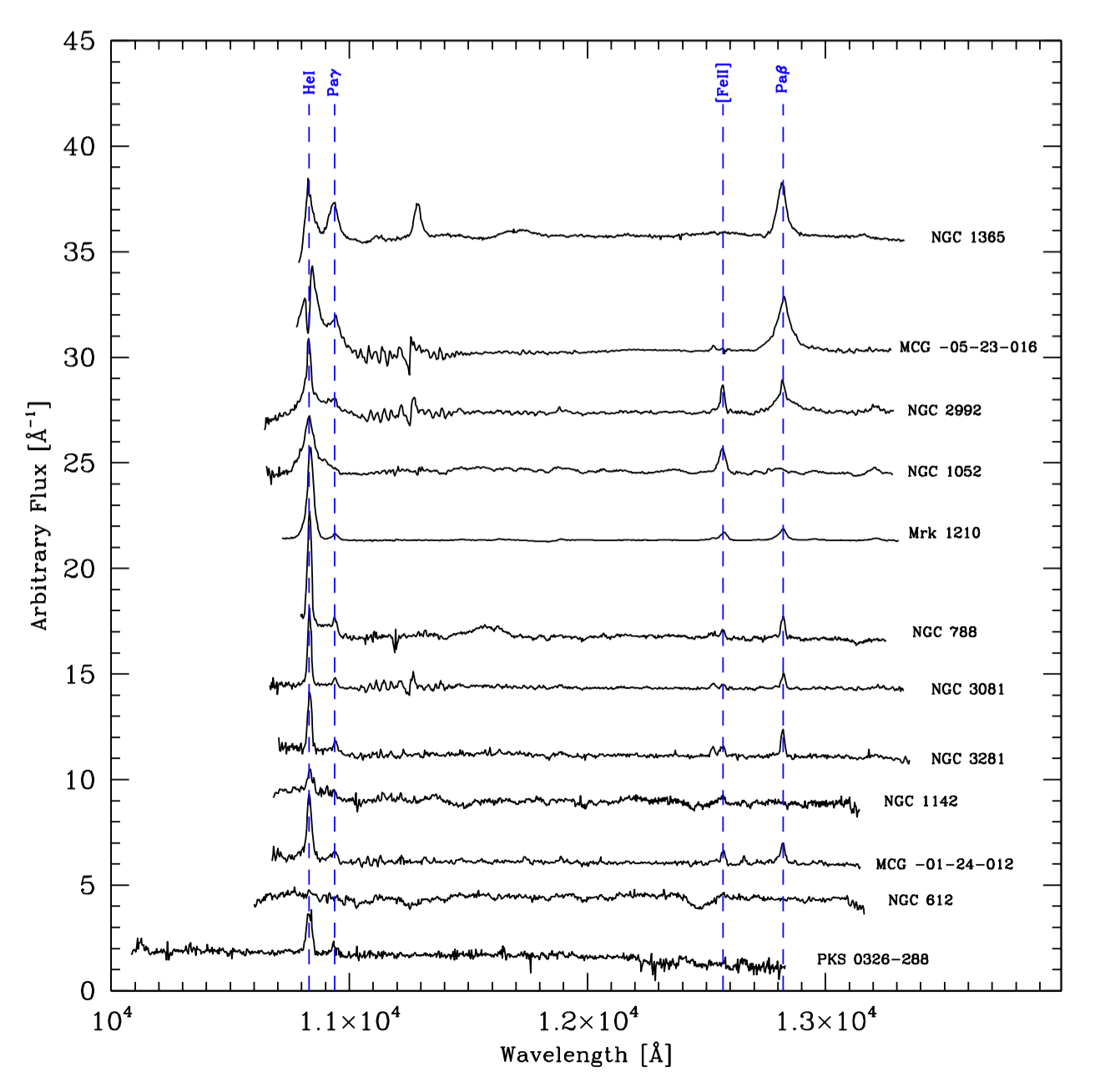}
\caption{Continued}
\end{figure*}
\clearpage

\begin{figure*}
\includegraphics[scale=0.38]{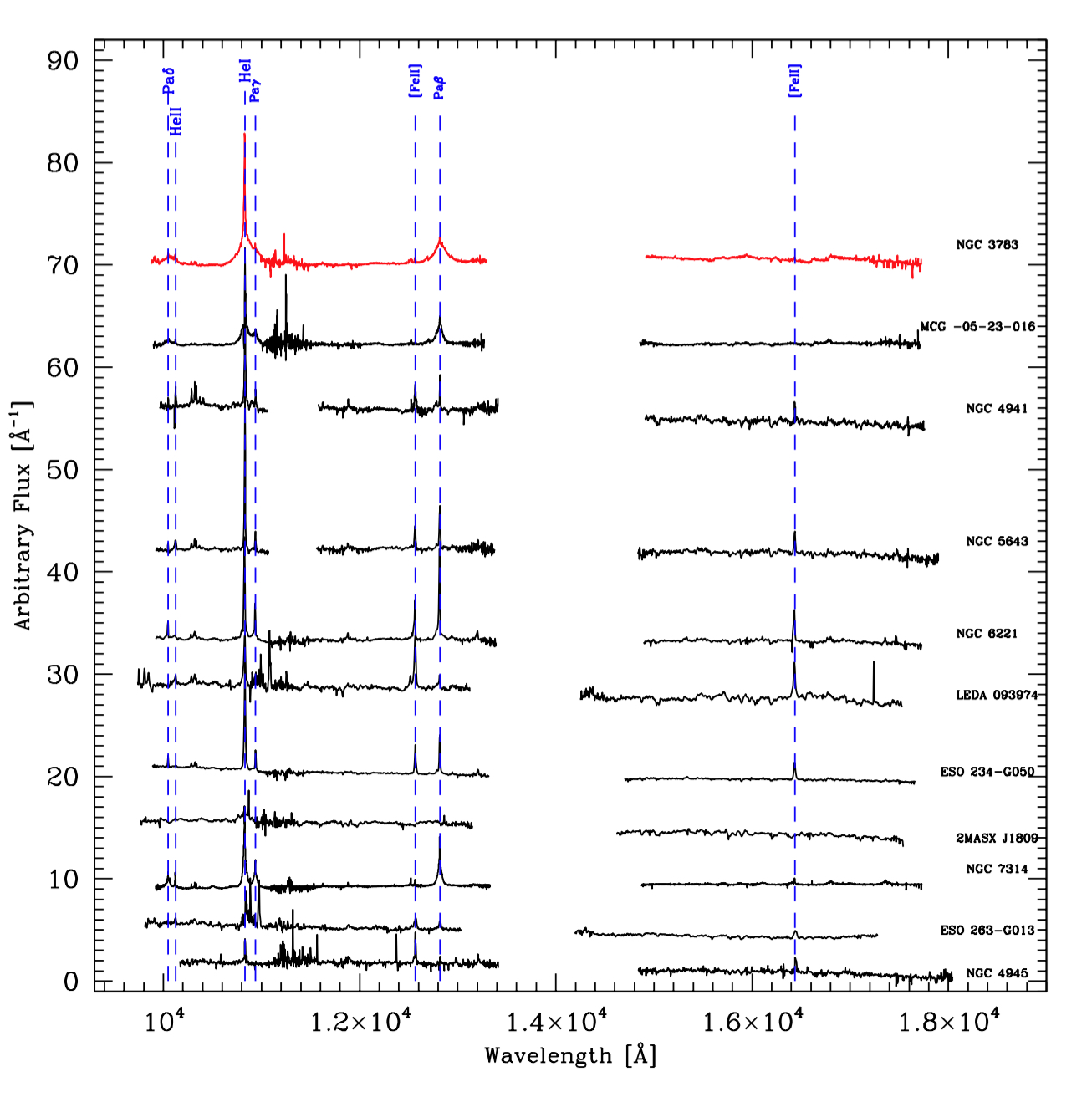}
\caption{Near infrared spectra obtained with X-shooter/VLT, corrected for redshift.  The spectrum of the AGN1 NGC 3783 is shown in red on the top. The rest frame wavelength position of some of the most relevant emission lines are shown by dashed vertical lines. Wavelength regions of bad atmospheric transmission have been masked out. The spectra are flux calibrated, but are shown with arbitrary normalization.}
\label{fig:XshNIR}
\end{figure*}
\clearpage

\begin{figure*}
\includegraphics[scale=0.38]{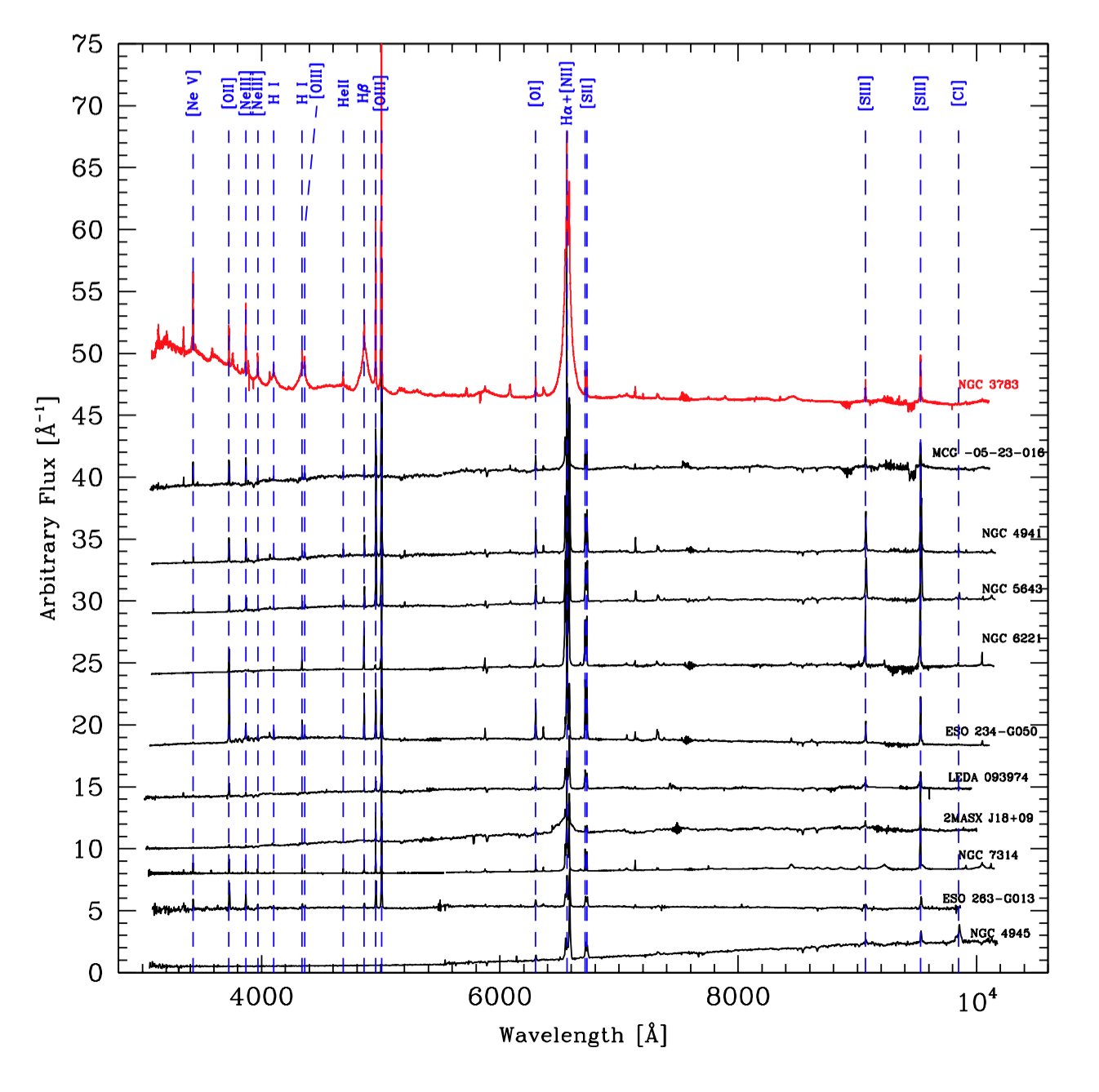}
\caption{UVB+VIS spectra obtained with X-shooter/VLT, corrected for redshift.  The spectrum of the AGN1 NGC 3783 is shown in red on the top. The rest frame wavelength position of some of the most relevant emission lines are shown by dashed vertical lines. The spectra are flux calibrated, but are shown with arbitrary normalization.}
\label{fig:OptXsh}
\end{figure*}
\clearpage


\begin{figure*}
\centering
\includegraphics[scale=.2, angle=-90]{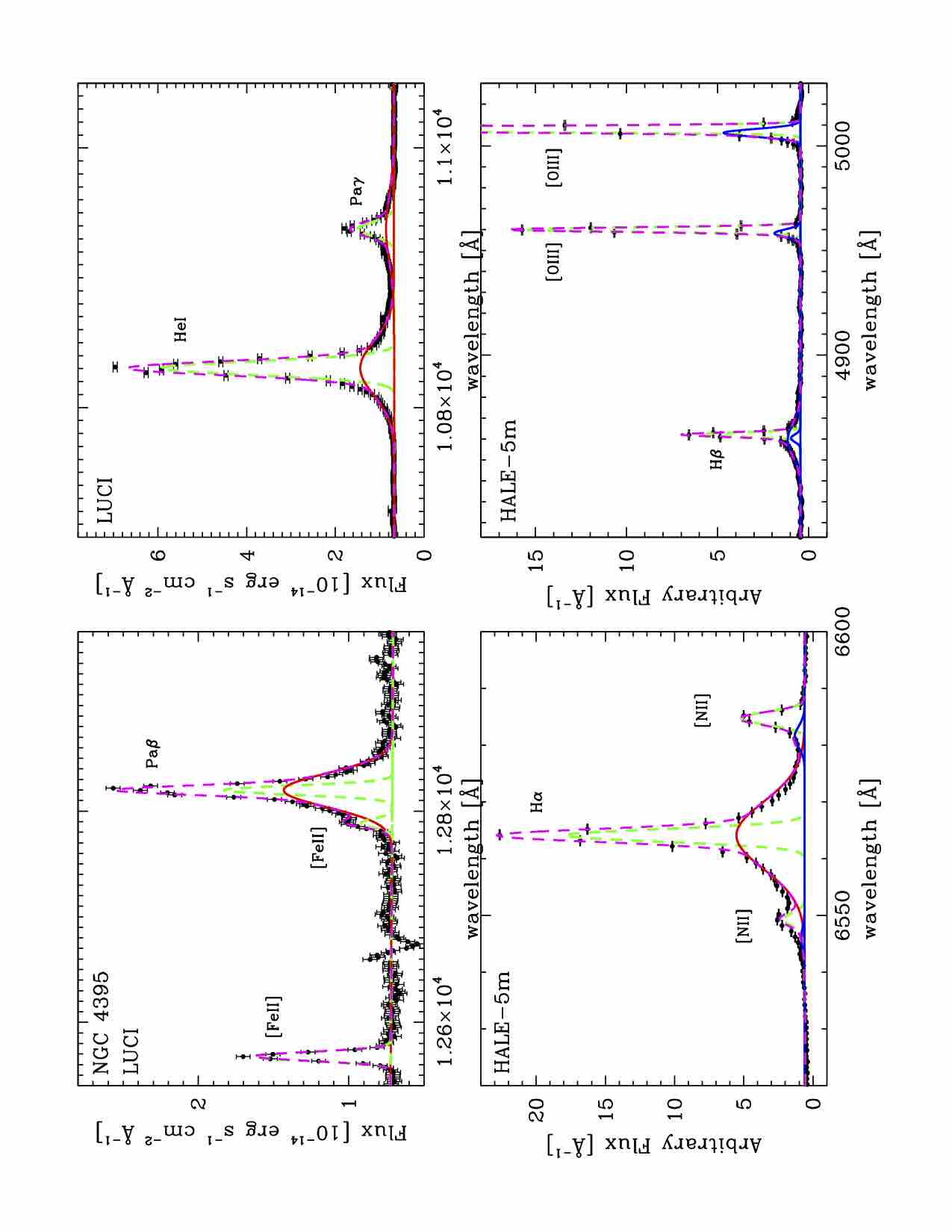}
\caption{Near infrared and optical spectra of NGC 4395, corrected for redshift \citep[see also][]{lafranca15}. {\it Top-left}: Pa$\upbeta$+[\ion{Fe}{II}] region.  {\it Top-right}: \ion{He}{I}+Pa$\upgamma$ region.
 {\it Bottom-left}: H$\upalpha$+[\ion{N}{II}] region.   {\it Bottom-right}: H$\upbeta$+[\ion{O}{III}] region. 
The narrow, intermediate and broad components are shown with green-dashed, blue and red lines, respectively. The magenta-dashed line shows the total fitting model.}
\label{fig:NGC4395}
\end{figure*}


\begin{figure*}
\centering
\includegraphics[scale=.2, angle= -90]{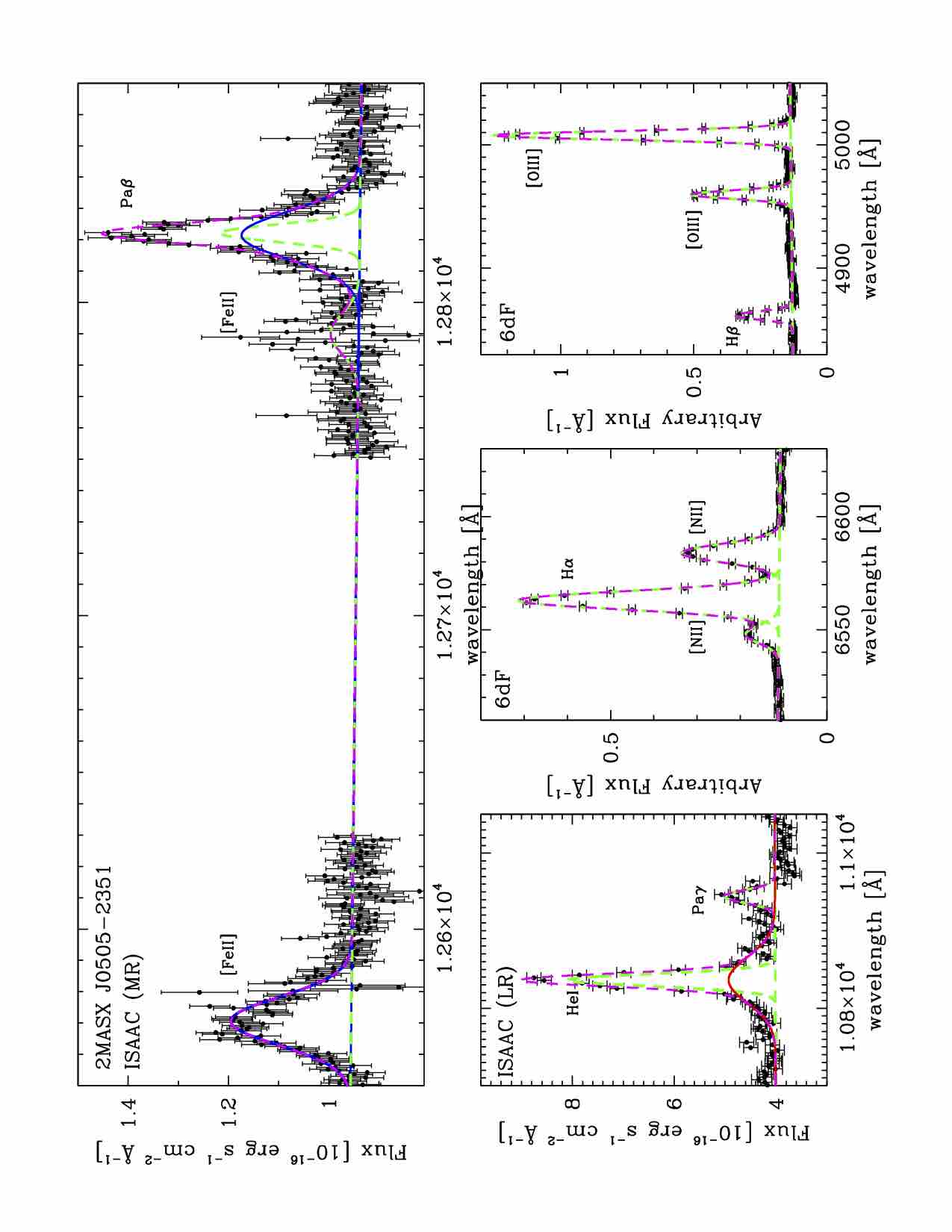}
\caption{Near infrared and optical spectra of 2MASX J05054575$-$2351139, corrected for redshift. {\it Top}: Pa$\upbeta$+[\ion{Fe}{II}] region.  {\it Bottom-left}:  \ion{He}{I}+Pa$\upgamma$ region. 
{\it Bottom-middle}: H$\upalpha$+[\ion{N}{II}] region. {\it Bottom-right}: H$\upbeta$+[\ion{O}{III}] region. 
The narrow, intermediate and broad components are shown with green-dashed, blue and red lines, respectively. The magenta-dashed line shows the total fitting model.
}
\label{fig:2MASXJ05-23_med}
\end{figure*}

\clearpage


\begin{figure*}
\centering
\includegraphics[scale=.2, angle= -90]{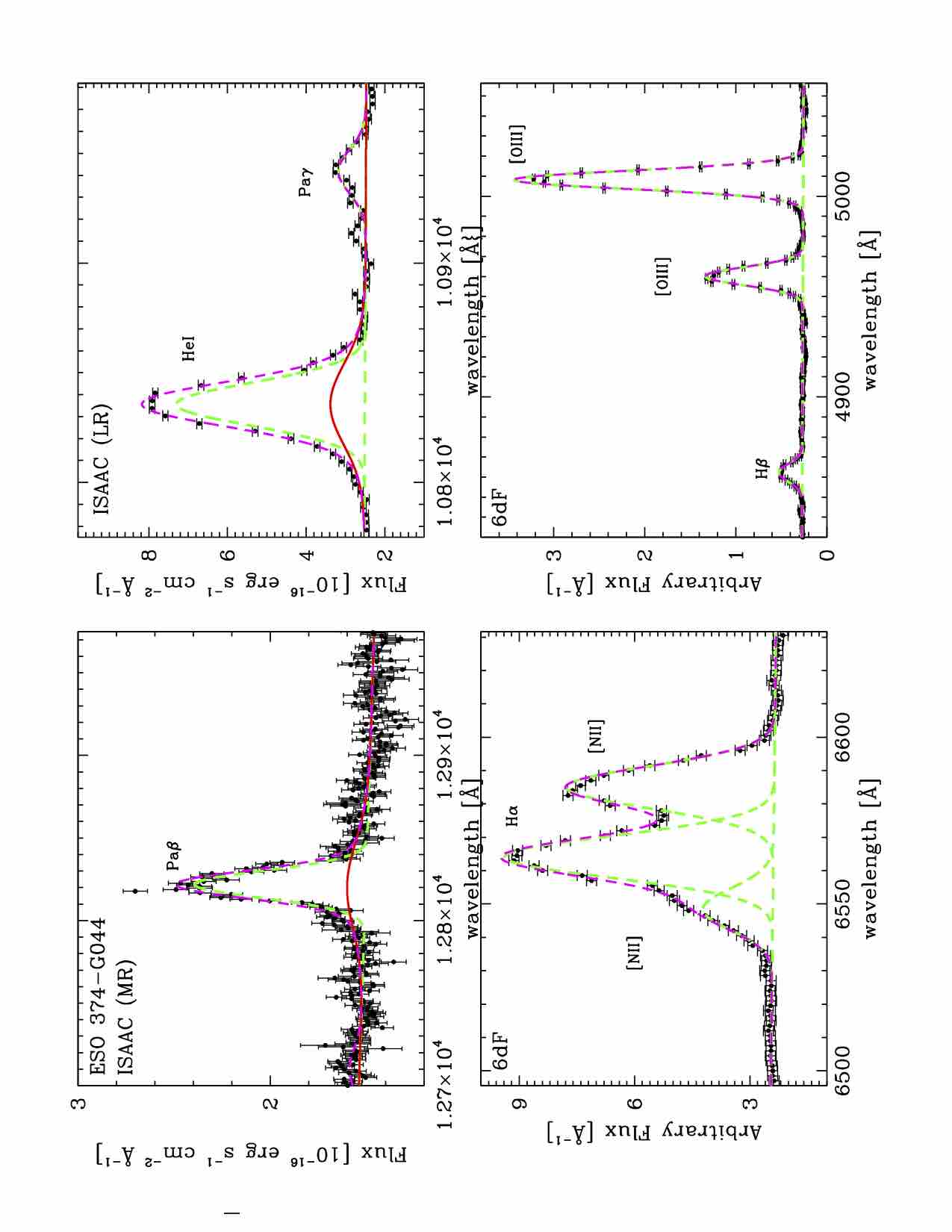}
\caption{Near infrared and optical spectra of ESO 374$-$G044, corrected for redshift. {\it Top-left}: Pa$\upbeta$+[\ion{Fe}{II}] region.  {\it Top-right}: \ion{He}{I}+Pa$\upgamma$ region.
 {\it Bottom-left}: H$\upalpha$+[\ion{N}{II}] region.   {\it Bottom-right}: H$\upbeta$+[\ion{O}{III}] region. 
The narrow, intermediate and broad components are shown with green-dashed, blue and red lines, respectively. The magenta-dashed line shows the total fitting model.
}
\label{fig:ESO374G44}
\end{figure*}


\begin{figure*}
\centering
\includegraphics[scale=.2, angle=-90]{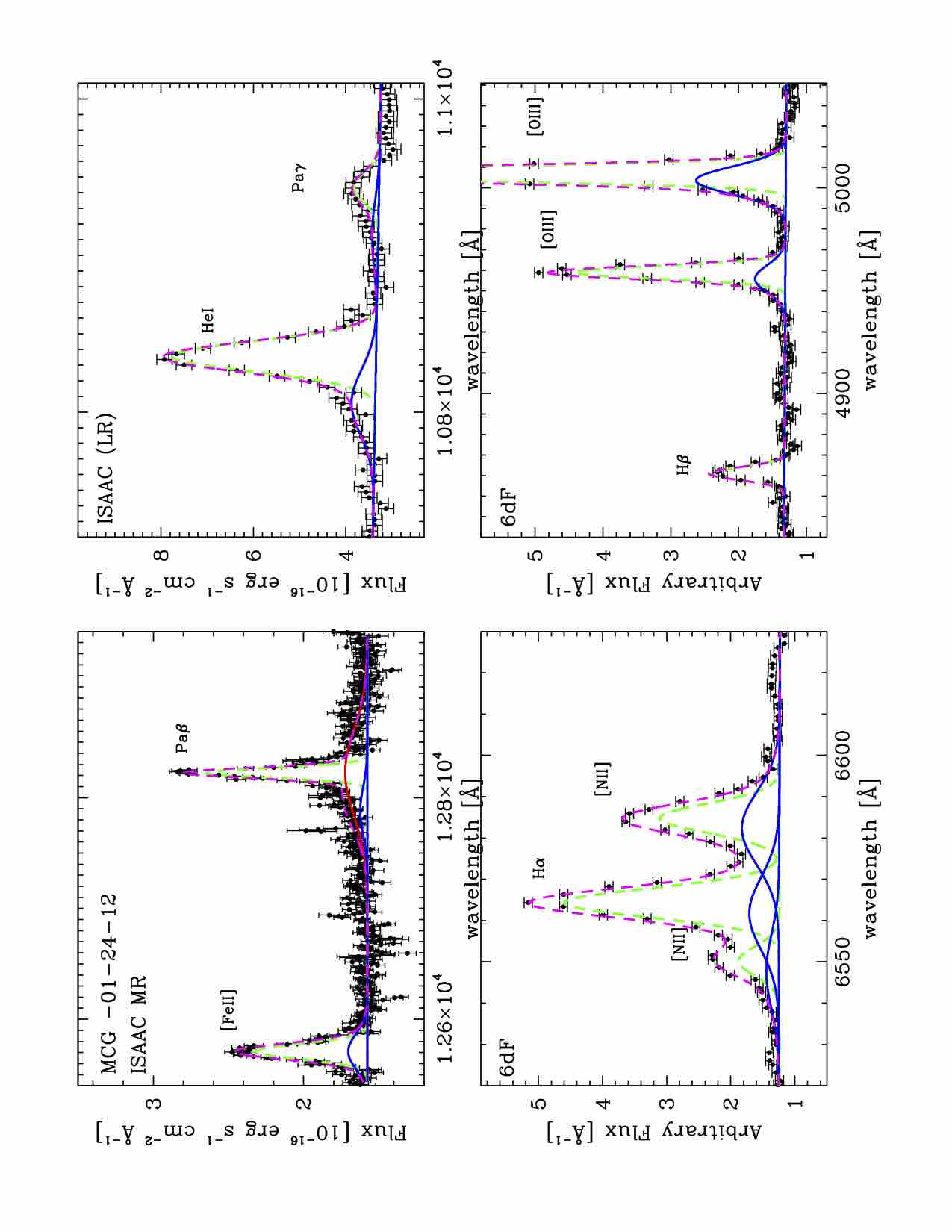}
\caption{Near infrared and optical spectra of MCG $-$01$-$24$-$012, corrected for redshift \citep[see also][]{lafranca15}. {\it Top-left}: Pa$\upbeta$+[\ion{Fe}{II}] region.  {\it Top-right}: \ion{He}{I}+Pa$\upgamma$ region.
 {\it Bottom-left}: H$\upalpha$+[\ion{N}{II}] region.   {\it Bottom-right}: H$\upbeta$+[\ion{O}{III}] region. 
The narrow, intermediate and broad components are shown with green-dashed, blue and red lines, respectively. The magenta-dashed line shows the total fitting model.
}
\label{fig:MGC011224}
\end{figure*}

\clearpage


\begin{figure*}
\centering
\includegraphics[scale=.2, angle=-90]{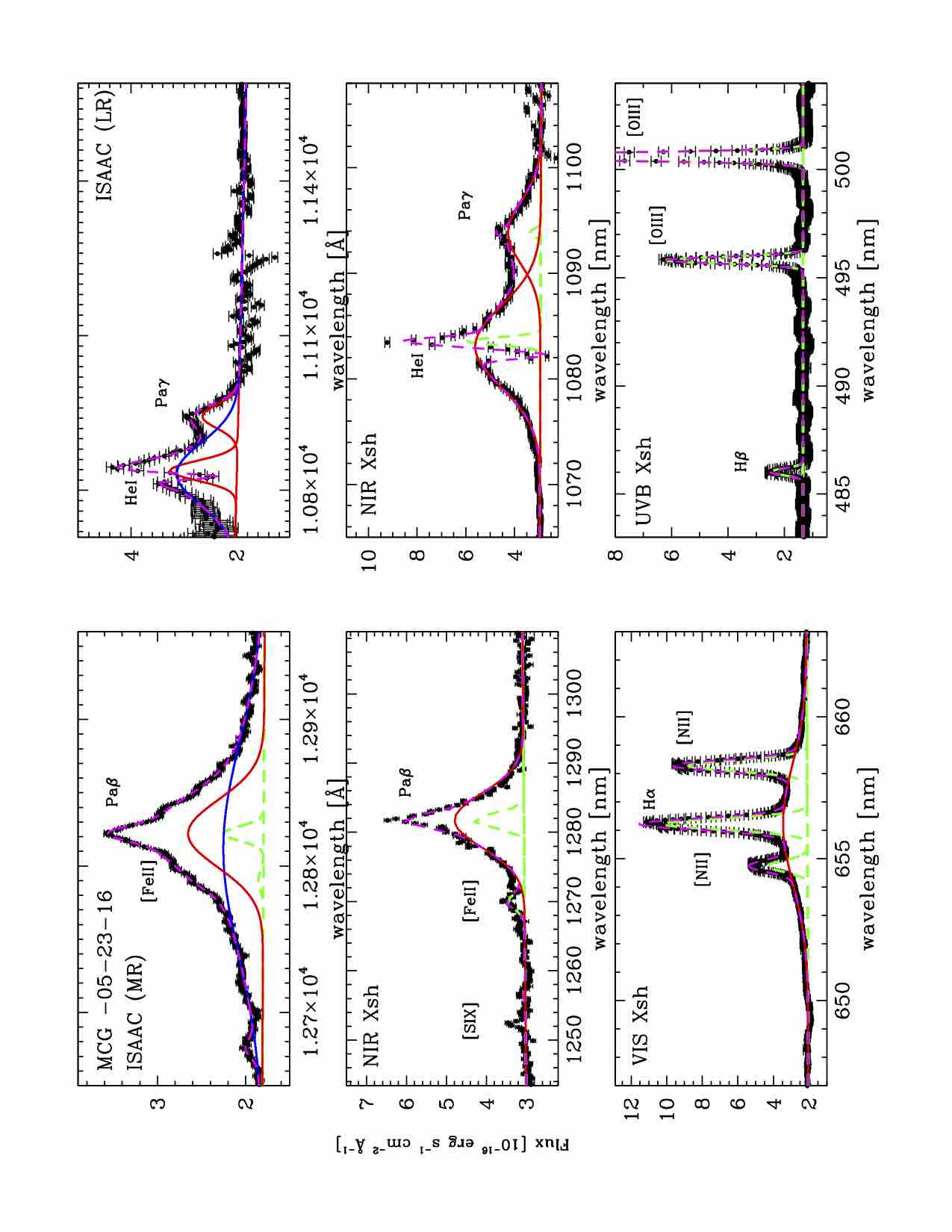}
\caption{Near infrared and optical spectra of MCG $-$05$-$23$-$016, corrected for redshift. {\it Top-left}: ISAAC MR Pa$\upbeta$+[\ion{Fe}{II}] region.  {\it Top-right}: ISAAC LR \ion{He}{I}+Pa$\upgamma$ region.
{\it Center-left}: X-shooter Pa$\upbeta$+[\ion{Fe}{II}] region.  {\it Center-right}: X-shooter \ion{He}{I}+Pa$\upgamma$ region.
 {\it Bottom-left}: X-shooter H$\upalpha$+[\ion{N}{II}] region.   {\it Bottom-right}: X-shooter H$\upbeta$+[\ion{O}{III}] region. 
The narrow, intermediate and broad components are shown with green-dashed, blue and red lines, respectively. The magenta-dashed line shows the total fitting model.}
\label{fig:MCG052316}
\end{figure*}


\begin{figure*}
\centering
\includegraphics[scale=.2, angle=-90]{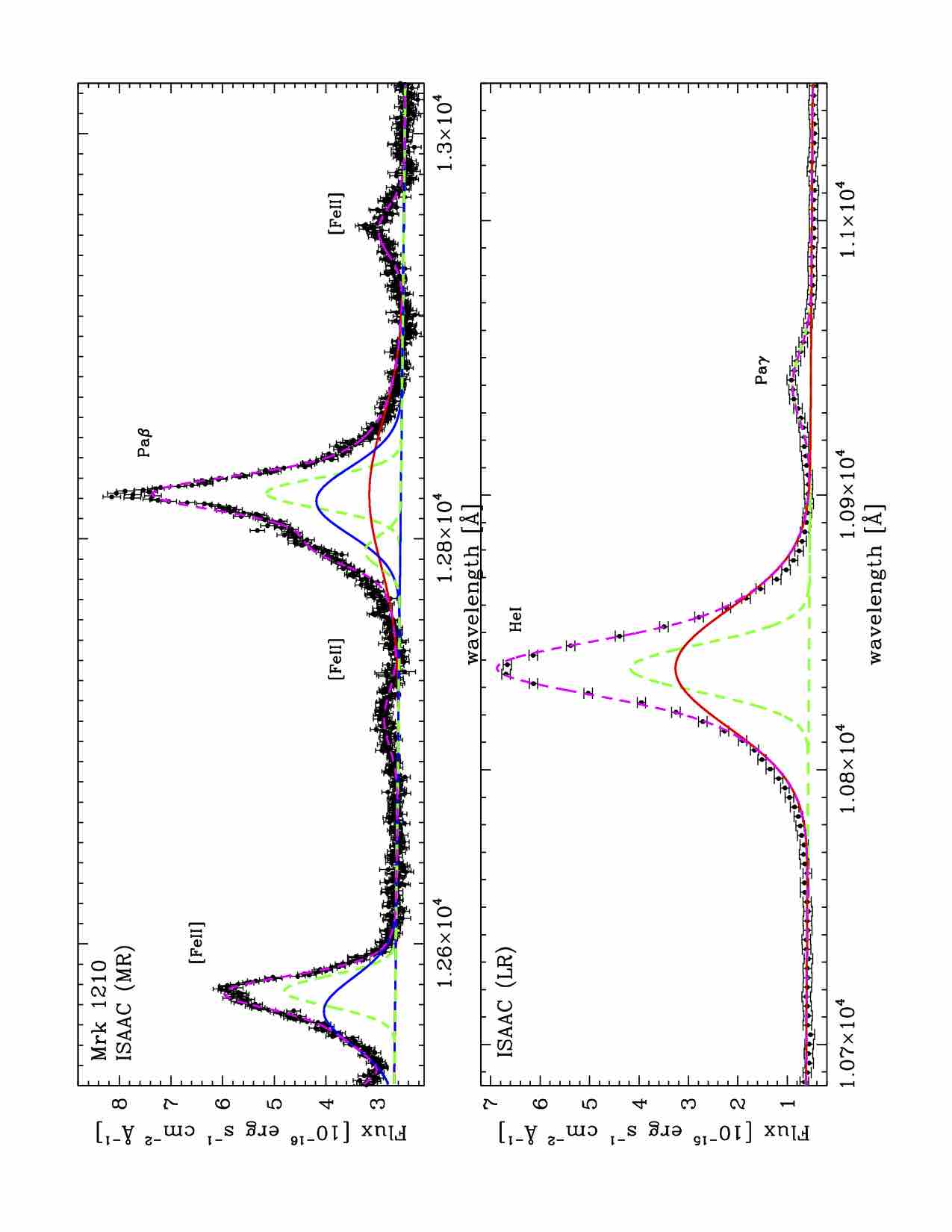}
\caption{Near infrared spectrum of Mrk 1210, corrected for redshift. {\it Top}: Pa$\upbeta$+[\ion{Fe}{II}] region.  {\it Bottom}: \ion{He}{I}+Pa$\upgamma$ region.
The narrow, intermediate and broad components are shown with green-dashed, blue and red lines, respectively. The magenta-dashed line shows the total fitting model.}
\label{fig:mrk1210}
\end{figure*}

\clearpage


\begin{figure*}
\centering
\includegraphics[scale=.2, angle=-90]{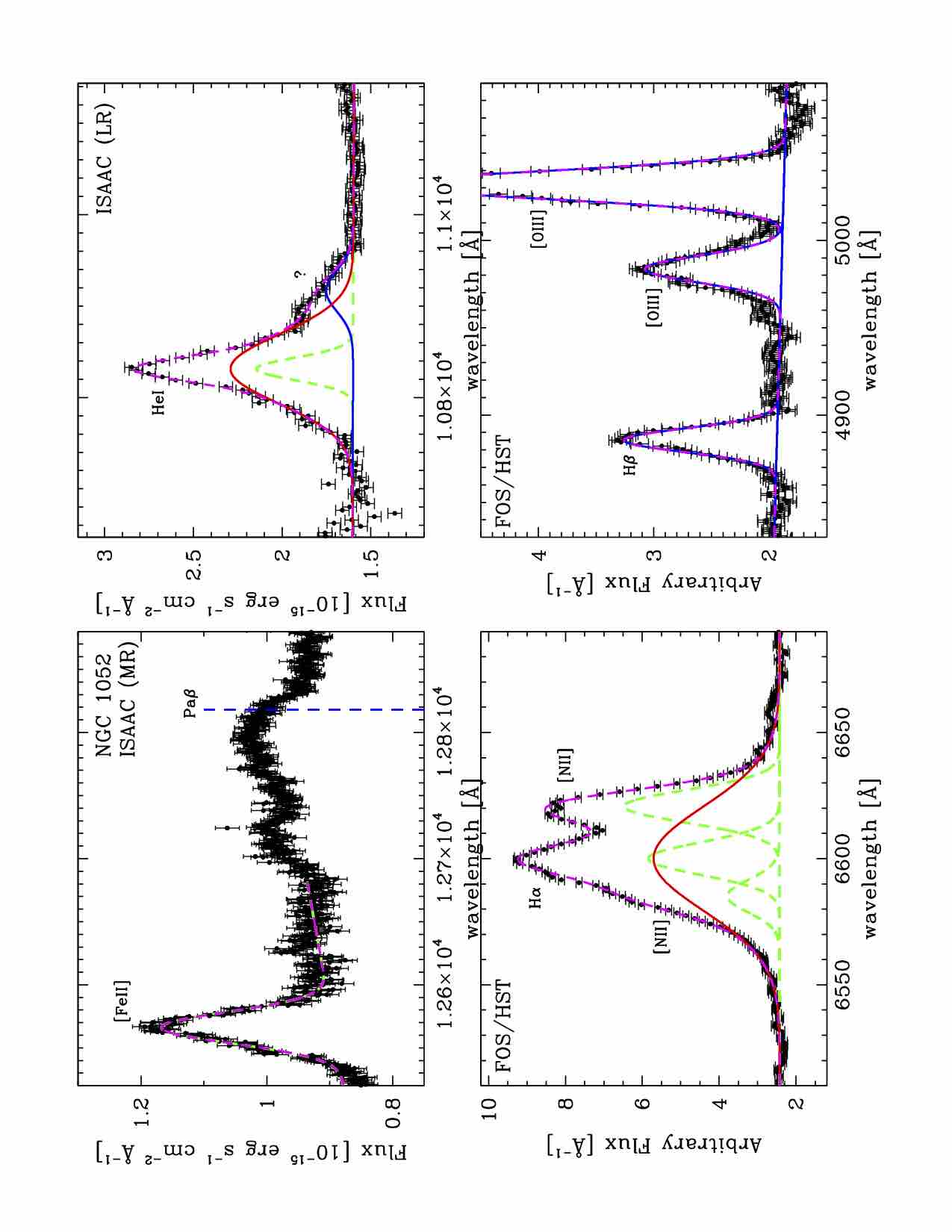}
\caption{Near infrared and optical spectra of NGC 1052, corrected for redshift. {\it Top-left}: Pa$\upbeta$+[\ion{Fe}{II}] region.  {\it Top-right}: \ion{He}{I}+Pa$\upgamma$ region.
 {\it Bottom-left}: H$\upalpha$+[\ion{N}{II}] region.   {\it Bottom-right}: H$\upbeta$+[\ion{O}{III}] region. 
The narrow, intermediate and broad components are shown with green-dashed, blue and red lines, respectively. The magenta-dashed line shows the total fitting model. The blue-dashed vertical line indicate the Pa$\upbeta$ position.}
\label{fig:NGC1052}
\end{figure*}


\begin{figure*}
\centering
\includegraphics[scale=.2, angle=-90]{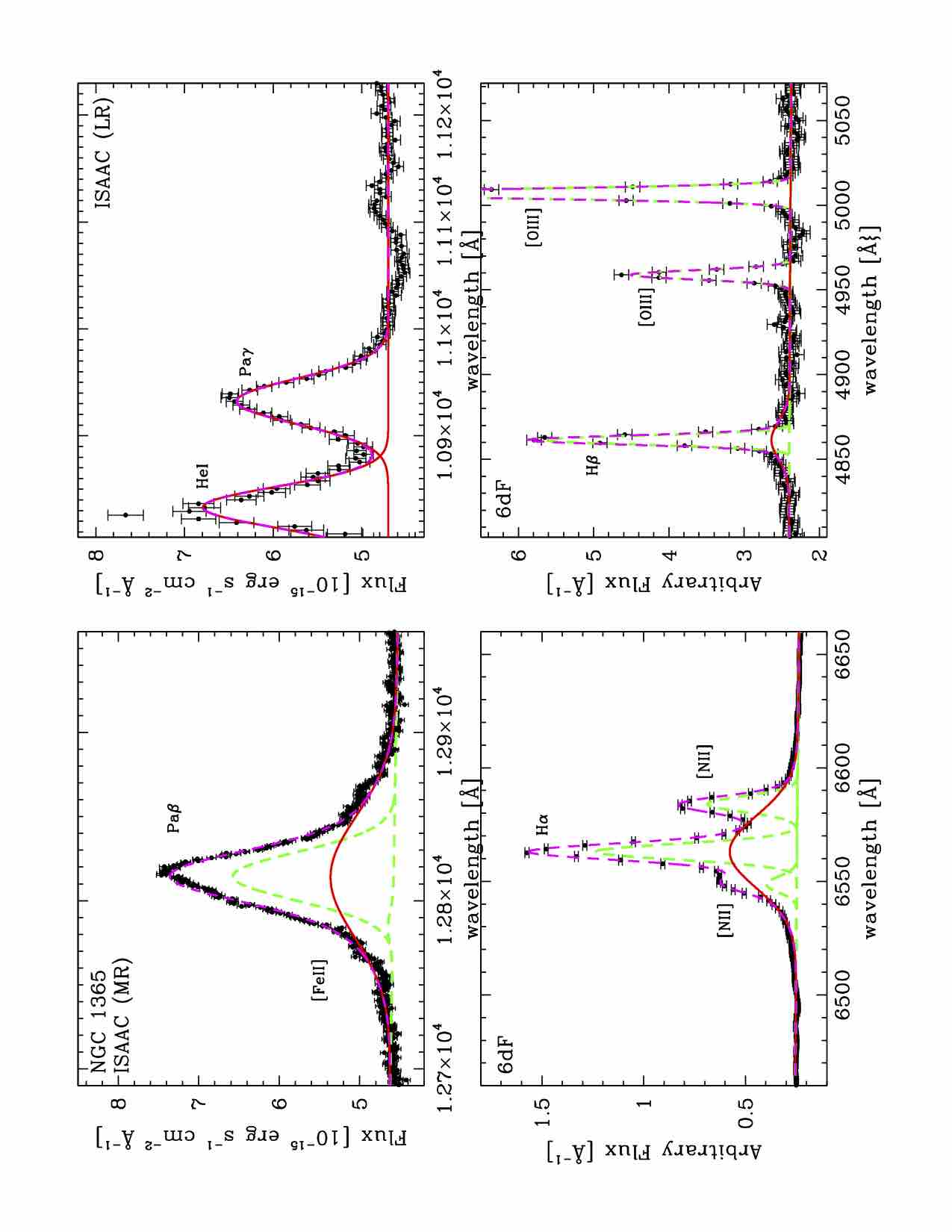}
\caption{Near infrared and optical spectra of NGC 1365, corrected for redshift. {\it Top-left}: Pa$\upbeta$ region.  {\it Top-right}: \ion{He}{I}+Pa$\upgamma$ region.
 {\it Bottom-left}: H$\upalpha$+[\ion{N}{II}] region.   {\it Bottom-right}: H$\upbeta$+[\ion{O}{III}] region. 
The narrow, intermediate and broad components are shown with green-dashed, blue and red lines, respectively. The magenta-dashed line shows the total fitting model.}
\label{fig:NGC1365}
\end{figure*}

\clearpage


\begin{figure*}
\centering
\includegraphics[scale=.2, angle=-90]{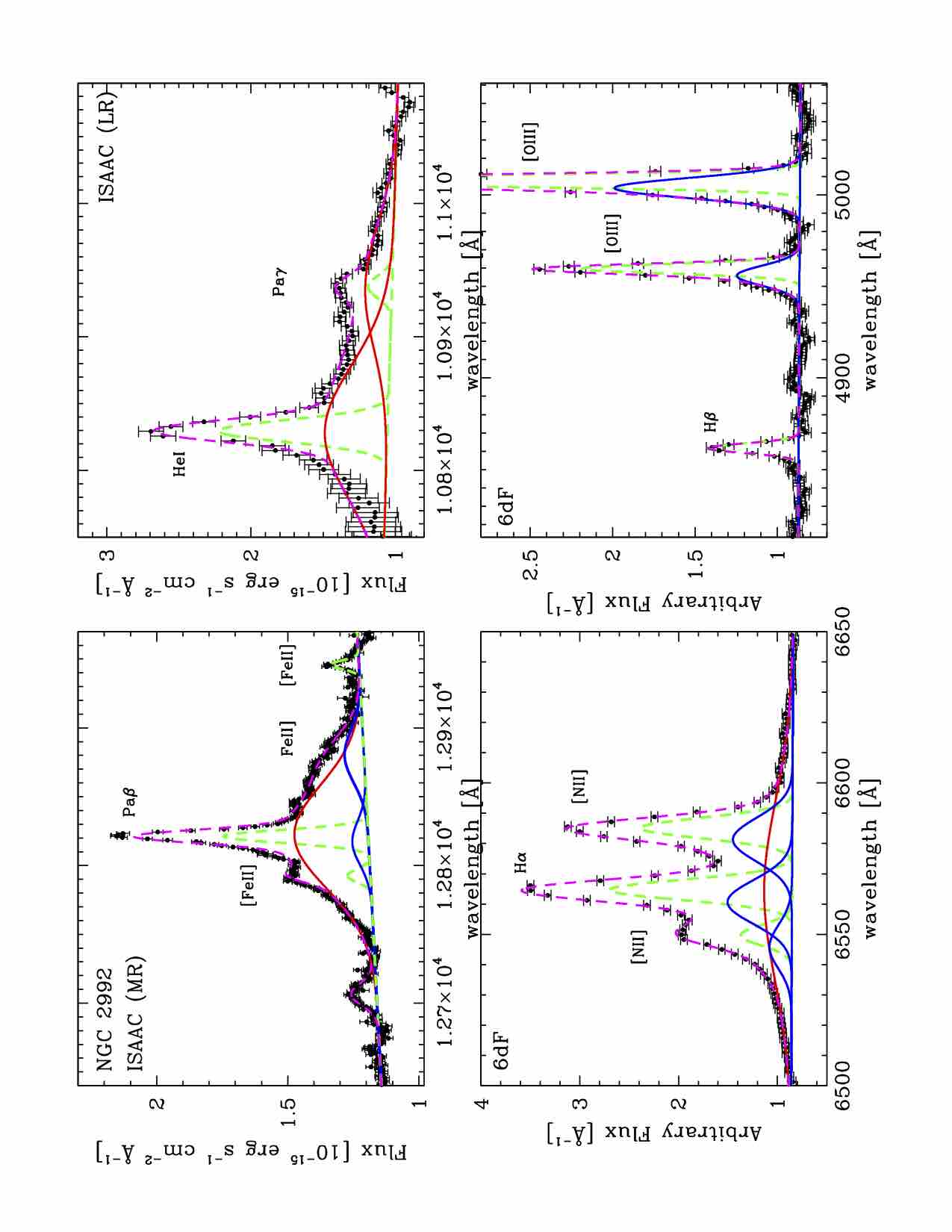}
\caption{Near infrared and optical spectra of NGC 2992, corrected for redshift. {\it Top-left}: Pa$\upbeta$+[\ion{Fe}{II}] region.  {\it Top-right}: \ion{He}{I}+Pa$\upgamma$ region.
 {\it Bottom-left}: H$\upalpha$+[\ion{N}{II}] region.   {\it Bottom-right}: H$\upbeta$+[\ion{O}{III}] region. 
The narrow, intermediate and broad components are shown with green-dashed, blue and red lines, respectively. The magenta-dashed line shows the total fitting model.}
\label{fig:NGC2992}
\end{figure*}

\begin{figure*}
\centering
\includegraphics[scale=.2, angle=-90]{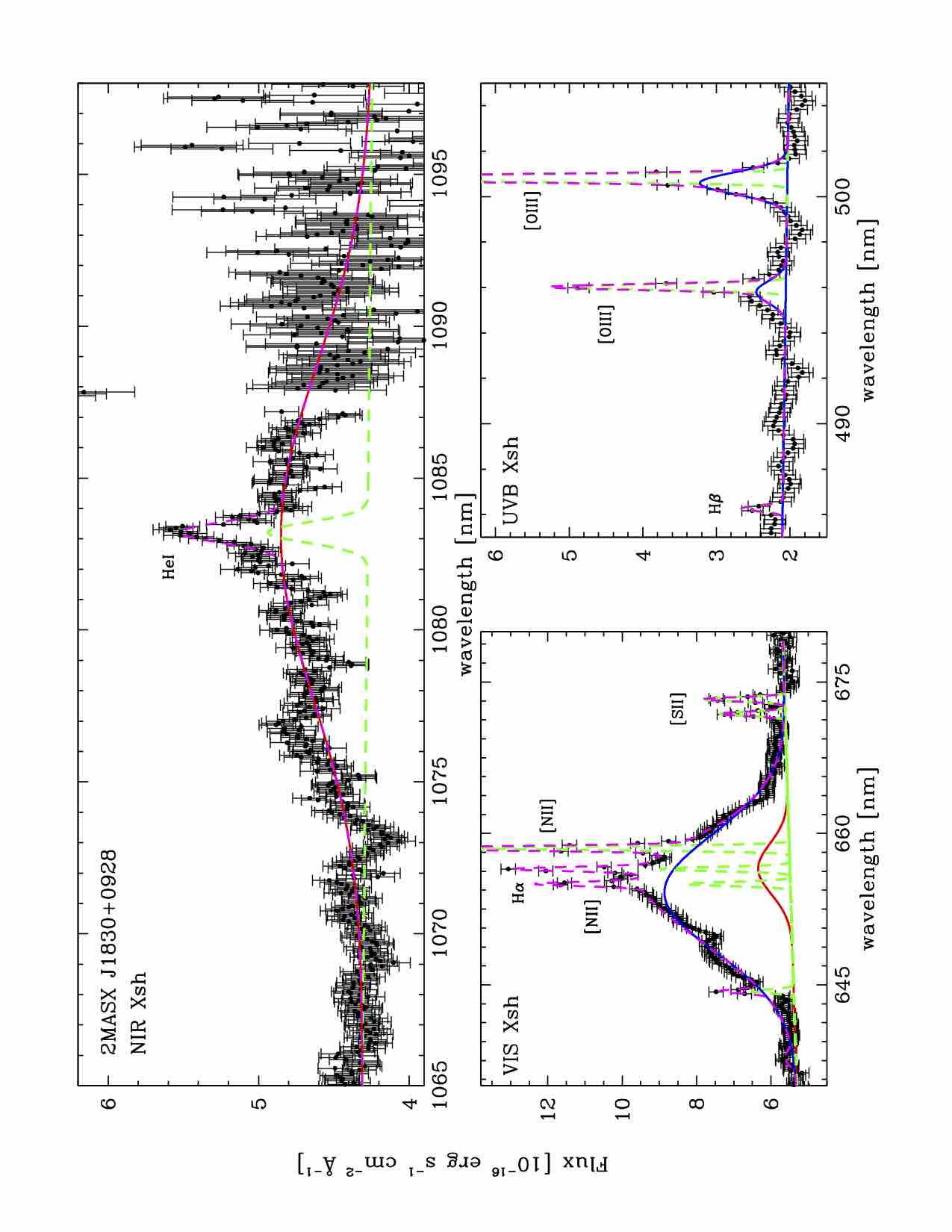}
\caption{Near infrared and optical spectra of 2MASX J18305065+0928414, corrected for redshift. {\it Top}: \ion{He}{I}  region.
 {\it Bottom-left}: H$\upalpha$+[\ion{N}{II}] region.   {\it Bottom-right}: H$\upbeta$+[\ion{O}{III}] region. 
The narrow, intermediate and broad components are shown with green-dashed, blue and red lines, respectively. The magenta-dashed line shows the total fitting model.}
\label{fig:2MASXJ18+09}
\end{figure*}

\clearpage


\begin{figure*}
\centering
\includegraphics[scale=.2, angle=-90]{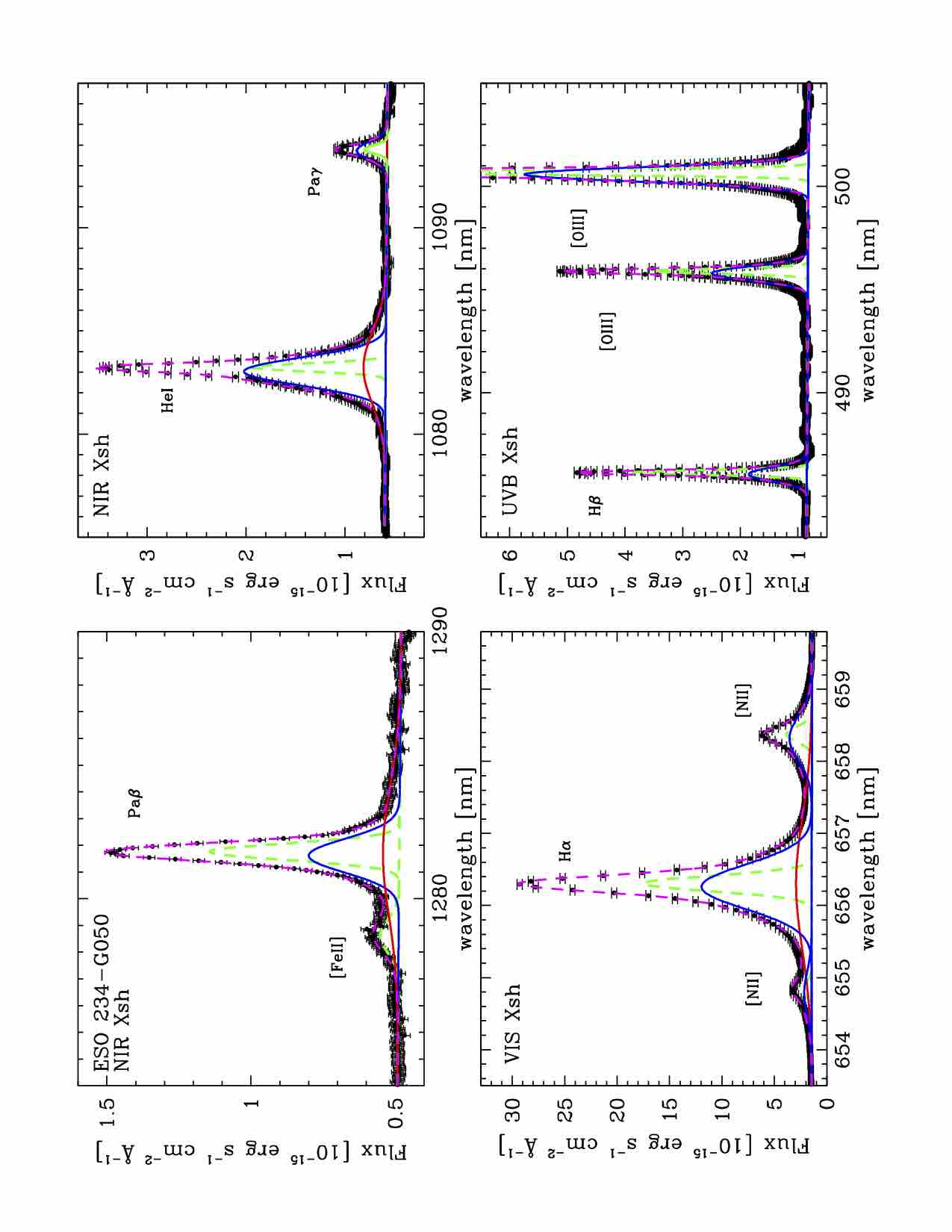}
\caption{Near infrared and optical spectra of ESO 234$-$G050, corrected for redshift. {\it Top-left}: Pa$\upbeta$+[\ion{Fe}{II}] region.  {\it Top-right}: \ion{He}{I}+Pa$\upgamma$ region.
{\it Bottom-left}: H$\upalpha$+[\ion{N}{II}] region.   {\it Bottom-right}: H$\upbeta$+[\ion{O}{III}] region. 
The narrow, intermediate and broad components are shown with green-dashed, blue and red lines, respectively. The magenta-dashed line shows the total fitting model.}
\label{fig:ESO234G050}
\end{figure*}

\begin{figure*}
\centering
\includegraphics[scale=.2, angle=-90]{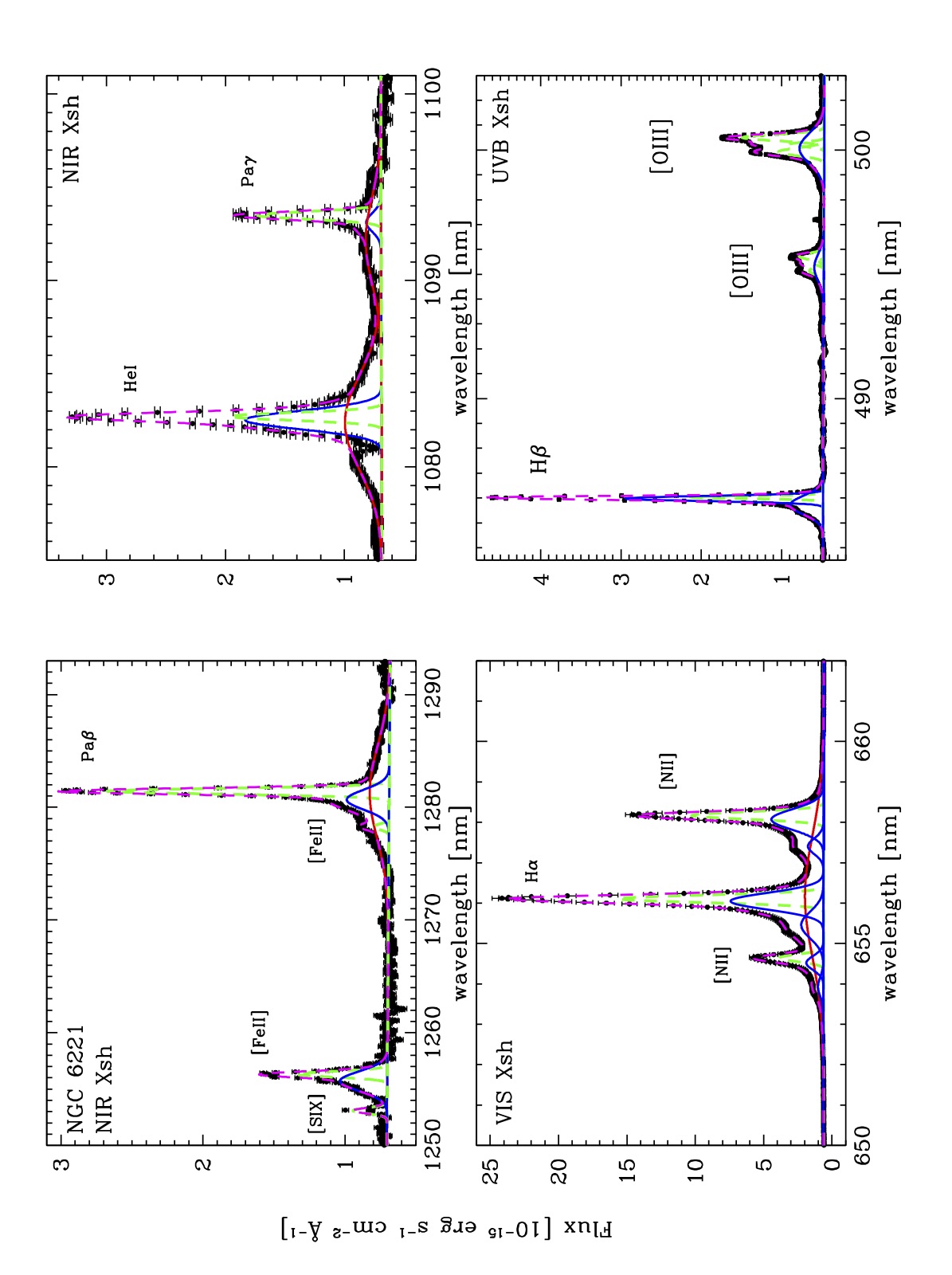}
\caption{Near infrared and optical spectra of NGC 6221, corrected for redshift \citep[see also][]{lafranca16}. {\it Top-left}: Pa$\upbeta$+[\ion{Fe}{II}] region.  {\it Top-right}: \ion{He}{I}+Pa$\upgamma$ region.
 {\it Bottom-left}: H$\upalpha$+[\ion{N}{II}] region.   {\it Bottom-right}: H$\upbeta$+[\ion{O}{III}] region. 
The narrow, intermediate and broad components are shown with green-dashed, blue and red lines, respectively. The magenta-dashed line shows the total fitting model.}
\label{fig:NGC6221}
\end{figure*}

\clearpage

\begin{figure*}
\centering
\includegraphics[scale=.2, angle = -90]{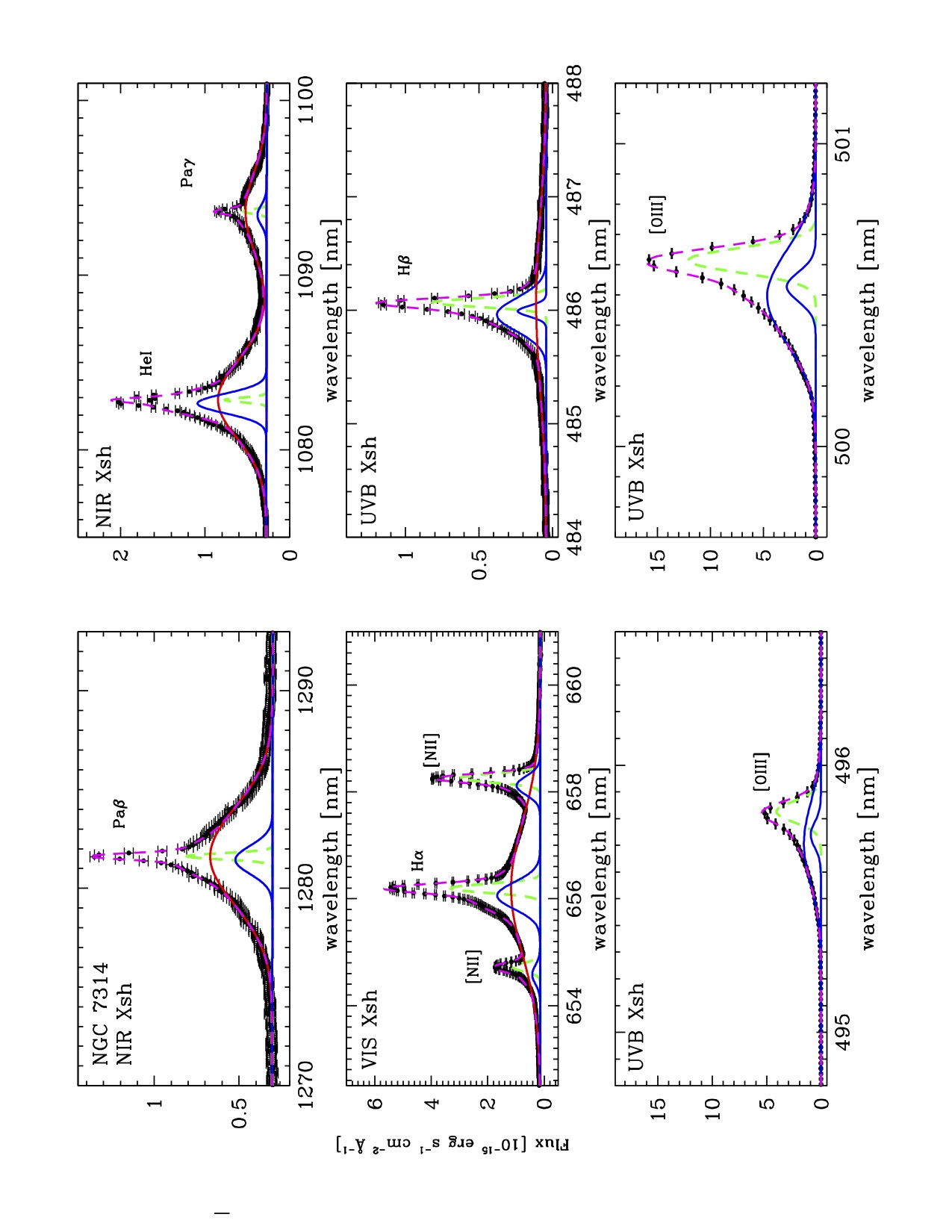}
\caption{Near infrared and optical spectra of  NGC 7314, corrected for redshift. {\it Top-left}: Pa$\upbeta$+[\ion{Fe}{II}] region.  {\it Top-right}: \ion{He}{I}+Pa$\upgamma$ region.
{\it Center-left}: H$\upalpha$+[\ion{N}{II}] region. {\it Center-right}: H$\upbeta$+[\ion{O}{III}] region.
 {\it Bottom-left}: [\ion{O}{III}]$\lambda$4959 region. {\it Bottom-right}: [\ion{O}{III}]$\lambda$5007 region.
The narrow, intermediate and broad components are shown with green-dashed, blue and red lines, respectively. The magenta-dashed line shows the total fitting model.}
\label{fig:NGC7314}
\end{figure*}

\clearpage
\newpage

\begin{table}
\centering
\begin{minipage}{140mm}
\caption{Line measurements of the optical emission lines of the  `broad' AGN2}
\label{tbl_OptBAGN2lines}
\resizebox{1.2\textwidth}{!}{
\begin{tabular}{@{}ccccccccccccccccccc}
\hline
\multicolumn{19}{c}{              } \\
\multicolumn{19}{c}{2MASX J05054575$-$2351139 (\citealt{jones09}); $R$ = 1000 ($\sigma_v\sim$300 \kms)} \\
\multicolumn{19}{c}{              } \\
\cline{1-19}\\
Comp &\multicolumn{3}{c}{{\bf H$\upbeta$}}&\multicolumn{3}{c}{{\bf[\ion{O}{III}]4959\AA}}&\multicolumn{3}{c}{{\bf[\ion{O}{III}]5007\AA\ }}&\multicolumn{3}{c}{{\bf[\ion{N}{II}]6548\AA}}&\multicolumn{3}{c}{{\bf H$\upalpha$}}&\multicolumn{3}{c}{{\bf[\ion{NII}]6583\AA}}\\
          &FWHM &$\Delta v$&$EW$  &FWHM &$\Delta v$ &$EW$  &FWHM  &$\Delta v$ &$EW$  &FWHM &$\Delta v$&$EW$ &FWHM&$\Delta v$&$EW$  &FWHM &$\Delta v$&$EW$\\
          (1)      & (2)      & (3)          & (4)   & (5)      & (6)           &(7)    & (8)        & (9)          &(10)  &(11)      &(12)         &(13)&(14)     &(15)         &(16) &(17)     &(18)          &(19)\\    
\cline{1-19}\\
N                     & 466$^{+4}_{-4}$ & - & 13.9 & 466$^{+4}_{-4}$ & - & 23.9  & 466$^{+4}_{-4}$ & - & 70.2 & 413$^{+6}_{-6}$ & - &6.8 & 413$^{+6}_{-6}$ & - &54.3 &413$^{+6}_{-6}$ & - &20.7  \\
\\
\cline{1-19}\\
\multicolumn{19}{c}{2MASX J18305065+0928414 X-shooter (UVB;VIS); $R$ = 4350/7350 ($\sigma_v\sim$70/40 \kms)} \\
\multicolumn{19}{c}{              } \\
\cline{1-19}\\
Comp &\multicolumn{3}{c}{{\bf H$\upbeta$}}&\multicolumn{3}{c}{{\bf[\ion{O}{III}]4959\AA}}&\multicolumn{3}{c}{{\bf[\ion{O}{III}]5007\AA\ }}&\multicolumn{3}{c}{{\bf[\ion{N}{II}]6548\AA}}&\multicolumn{3}{c}{{\bf H$\upalpha$}}&\multicolumn{3}{c}{{\bf[\ion{NII}]6583\AA}}\\
          &FWHM &$\Delta v$&$EW$  &FWHM &$\Delta v$ &$EW$  &FWHM  &$\Delta v$ &$EW$  &FWHM &$\Delta v$&EW &FWHM&$\Delta v$&$EW$  &FWHM &$\Delta v$&$EW$\\
\cline{1-19}\\
N                     & 176$^{+7}_{-8}$  & -   &0.8  &176$^{+7}_{-8}$ & -     &4.4      & 176$^{+7}_{-8}$   & -   &13.3        & 216$^{+12}_{-9}$     & - & 2.6     &  216$^{+12}_{-9}$ & -   &3.3         & 216$^{+12}_{-9}$ & - &  7.8    \\
I                      &                             & -  &   -  &663$^{+74}_{-65}$ & 134 &2.3    & 663$^{+74}_{-65}$ &134 &7.0      &                          -        &-  &   -   &     -                        &-   & -       & -                           & -   & -   \\
B                     & -                                   & - &-     &         -                     & -     &     -       & -                            & -   &  -       & -                         & - &-     &2660$^{+500}_{-460}$ &-    &  9.8       & -                           & -   &-     \\  
I                    & -                                    & - &-     &         -                     & -     &     -       & -                            & -   &  -       & -                         & - &-     &6194$^{+160}_{-160}$ &1103    & 90.9        & -                           & -   &-      \\ 
\\
\cline{1-19}\\
\multicolumn{19}{c}{ESO 234-G050 X-shooter (UVB;VIS); $R$ = 4350/7350 ($\sigma_v\sim$70/40 \kms)} \\
\multicolumn{19}{c}{              } \\
\cline{1-19}\\
Comp &\multicolumn{3}{c}{{\bf H$\upbeta$}}&\multicolumn{3}{c}{{\bf[\ion{O}{III}]4959\AA}}&\multicolumn{3}{c}{{\bf[\ion{O}{III}]5007\AA\ }}&\multicolumn{3}{c}{{\bf[\ion{N}{II}]6548\AA}}&\multicolumn{3}{c}{{\bf H$\upalpha$}}&\multicolumn{3}{c}{{\bf[\ion{NII}]6583\AA}}\\
          &FWHM &$\Delta v$&$EW$  &FWHM &$\Delta v$ &$EW$  &FWHM  &$\Delta v$ &$EW$  &FWHM &$\Delta v$&$EW$ &FWHM&$\Delta v$&$EW$  &FWHM &$\Delta v$&$EW$\\
\hline
N                     & 172$^{+2}_{-2}$ & -   &10.7  & 172$^{+2}_{-2}$  & -   &8.9    & 172$^{+2}_{-2}$  & -   &26.8  & 113$^{+3}_{-4}$     & - &1.5  &  113$^{+3}_{-4}$      & -   &29.7  & 113$^{+3}_{-4}$ & - &4.5  \\
I                      & 481$^{+2}_{-2}$  &50 & 10.0 & 481$^{+2}_{-2}$  &62 &15.4  & 481$^{+2}_{-2}$  &62 &46.3  &  327$^{+7}_{-10}$    & 25 &3.8  &    327$^{+7}_{-10}$    & 25 & 56.2& 327$^{+7}_{-10}$& 25&11.5 \\
B                     & -                                    & - &-     &         -                     & -     &     -       & -                            & -   &  -       & -                         & - &-             &972$^{+124}_{-104}$ &-  &20.7 & -                              & -   &-      \\
\\  
\cline{1-19}\\
\multicolumn{19}{c}{ESO 374-G044 (\citealt{jones09}); $R$ = 1000 ($\sigma_v\sim$300 \kms)} \\
\multicolumn{19}{c}{              } \\
\cline{1-19}\\
Comp &\multicolumn{3}{c}{{\bf H$\upbeta$}}&\multicolumn{3}{c}{{\bf[\ion{O}{III}]4959\AA}}&\multicolumn{3}{c}{{\bf[\ion{O}{III}]5007\AA\ }}&\multicolumn{3}{c}{{\bf[\ion{N}{II}]6548\AA}}&\multicolumn{3}{c}{{\bf H$\upalpha$}}&\multicolumn{3}{c}{{\bf[\ion{NII}]6583\AA}}\\
          &FWHM &$\Delta v$&$EW$  &FWHM &$\Delta v$ &$EW$  &FWHM  &$\Delta v$ &$EW$  &FWHM &$\Delta v$&$EW$ &FWHM&$\Delta v$&$EW$  &FWHM &$\Delta v$&$EW$\\
\cline{1-19}\\
N                     & 640$^{+5}_{-10}$ & - & 10.5 & 640$^{+5}_{-10}$ & - & 46.3  & 640$^{+5}_{-10}$ & - & 140.4& 669$^{+9}_{-9}$ & - &11.8 & 669$^{+9}_{-9}$ & - &45.9 &669$^{+9}_{-9}$ & - &35.9  \\
\\
\cline{1-19}\\
\multicolumn{19}{c}{MCG -01-24-012 (\citealt{jones09}); $R$ = 1000 ($\sigma_v\sim$300 \kms)} \\
\multicolumn{19}{c}{              } \\
\cline{1-19}\\
Comp &\multicolumn{3}{c}{{\bf H$\upbeta$}}&\multicolumn{3}{c}{{\bf[\ion{O}{III}]4959\AA}}&\multicolumn{3}{c}{{\bf[\ion{O}{III}]5007\AA\ }}&\multicolumn{3}{c}{{\bf[\ion{N}{II}]6548\AA}}&\multicolumn{3}{c}{{\bf H$\upalpha$}}&\multicolumn{3}{c}{{\bf[\ion{NII}]6583\AA}}\\
          &FWHM &$\Delta v$&$EW$  &FWHM &$\Delta v$ &$EW$  &FWHM  &$\Delta v$ &$EW$  &FWHM &$\Delta v$&$EW$ &FWHM&$\Delta v$&$EW$  &FWHM &$\Delta v$&$EW$\\
\cline{1-19}\\
N                     & 440$^{+8}_{-10}$ & - &6.5  &  440$^{+8}_{-10}$ & - &18.7   & 440$^{+8}_{-10}$ & - &56.2  & 371$^{+14}_{-13}$ & - &4.4 &  371$^{+14}_{-13}$ & - &23.7 & 371$^{+14}_{-13}$ & - &13.3  \\
I                      &                     -  & - &  &  926$^{+80}_{-70}$ & 189 &5.6   & 926$^{+80}_{-70}$ & 189 &16.8  & 926$^{+80}_{-70}$ & 111 &3.4 & 926$^{+80}_{-70}$ & 111 & 8.2& 926$^{+80}_{-70}$&111 &10.4  \\
\\
\cline{1-19}\\
\multicolumn{19}{c}{MCG -05-23-016 X-shooter (UVB;VIS); $R$ = 4350/7350 ($\sigma_v\sim$70/40 \kms)} \\
\multicolumn{19}{c}{              } \\
\cline{1-19}\\
Comp &\multicolumn{3}{c}{{\bf H$\upbeta$}}&\multicolumn{3}{c}{{\bf[\ion{O}{III}]4959\AA}}&\multicolumn{3}{c}{{\bf[\ion{O}{III}]5007\AA\ }}&\multicolumn{3}{c}{{\bf[\ion{N}{II}]6548\AA}}&\multicolumn{3}{c}{{\bf H$\upalpha$}}&\multicolumn{3}{c}{{\bf[\ion{NII}]6583\AA}}\\
          &FWHM &$\Delta v$&$EW$  &FWHM &$\Delta v$ &$EW$  &FWHM  &$\Delta v$ &$EW$  &FWHM &$\Delta v$&$EW$ &FWHM&$\Delta v$&$EW$  &FWHM &$\Delta v$&$EW$\\
\cline{1-19}\\
N                     & 221$^{+4}_{-3}$ & - &3.3  & 221$^{+4}_{-3}$ & - &14.1   & 221$^{+4}_{-3}$& - &42.2  & 215$^{+2}_{-2}$     & - &5.4 &  215$^{+2}_{-2}$      & - &19.4  & 215$^{+2}_{-2}$ & - &16.0  \\
B                     &                          - & - & -      & -                          & - & -        & -                        & - & -     &                             -  & - & -    &  2232$^{+36}_{-38}$& - & 34.2 & -                         & - & \\  
\\
\cline{1-19}\\
\multicolumn{19}{c}{NGC 1052 (\citealt{torrealba12}); $R$ = 2800 ($\sigma_v\sim$80 \kms)} \\
\multicolumn{19}{c}{              } \\
\cline{1-19}\\
Comp &\multicolumn{3}{c}{{\bf H$\upbeta$}}&\multicolumn{3}{c}{{\bf[\ion{O}{III}]4959\AA}}&\multicolumn{3}{c}{{\bf[\ion{O}{III}]5007\AA\ }}&\multicolumn{3}{c}{{\bf[\ion{N}{II}]6548\AA}}&\multicolumn{3}{c}{{\bf H$\upalpha$}}&\multicolumn{3}{c}{{\bf[\ion{NII}]6583\AA}}\\
          &FWHM &$\Delta v$&$EW$  &FWHM &$\Delta v$ &$EW$  &FWHM  &$\Delta v$ &$EW$  &FWHM &$\Delta v$&$EW$ &FWHM&$\Delta v$&$EW$  &FWHM &$\Delta v$&$EW$\\
\cline{1-19}\\
N                     &967$^{+49}_{-40}$ & - &11.6 &1091$^{+16}_{-16}$ & - &12.2 &1091$^{+16}_{-16}$& - &36.7 & 682$^{+7}_{-7}$     & -&9.0 &682$^{+7}_{-7}$      & - &22.6&682$^{+7}_{-7}$ & - &26.8  \\
B                     &-                             & - &-          &-                               & - &-         &-                               & - &-         & -                             & - & & 2193$^{+51}_{-45}$      & - &69.6&- & - &  \\
\\
\cline{1-19}\\
\end{tabular}
}
\noindent
\end{minipage}
\end{table} 

\clearpage

\begin{table}
\centering
\begin{minipage}{140mm}
\contcaption{}
\resizebox{1.2\textwidth}{!}{
\begin{tabular}{@{}ccccccccccccccccccc}
\hline
\multicolumn{19}{c}{NGC 1365 (\citealt{jones09}); $R$ = 1000 ($\sigma_v\sim$300 \kms)} \\
\multicolumn{19}{c}{              } \\
\cline{1-19}\\
Comp &\multicolumn{3}{c}{{\bf H$\upbeta$}}&\multicolumn{3}{c}{{\bf[\ion{O}{III}]4959\AA}}&\multicolumn{3}{c}{{\bf[\ion{O}{III}]5007\AA\ }}&\multicolumn{3}{c}{{\bf[\ion{N}{II}]6548\AA}}&\multicolumn{3}{c}{{\bf H$\upalpha$}}&\multicolumn{3}{c}{{\bf[\ion{NII}]6583\AA}}\\
          &FWHM &$\Delta v$&$EW$  &FWHM &$\Delta v$ &$EW$  &FWHM  &$\Delta v$ &$EW$  &FWHM &$\Delta v$&$EW$ &FWHM&$\Delta v$&$EW$  &FWHM &$\Delta v$&$EW$\\
                  (1)      & (2)      & (3)          & (4)   & (5)      & (6)           &(7)    & (8)        & (9)          &(10)  &(11)      &(12)         &(13)&(14)     &(15)         &(16) &(17)     &(18)          &(19)\\   
\cline{1-19}\\
N                     & 389$^{+7}_{-8}          $ & - &9.1 &  389$^{+7}_{-8}$ & - & 6.1     & 389$^{+7}_{-8}$ & - & 18.3  & 371$^{+3}_{-3}$ & - &5.2 &  371$^{+3}_{-3}$     & - &34.8 & 371$^{+3}_{-3}$ & - &15.8 \\
B                     & 1586$^{+517}_{-413}$ & - &2.7 &                             & -  &          & -                          & - &  -       & -                         & - &-     &1701$^{+31}_{-32}$ &-  & 53.4& -                         &-   &-       \\
\\
\cline{1-19}\\
\multicolumn{19}{c}{NGC 2992 (\citealt{jones09}); $R$ = 1000 ($\sigma_v\sim$300 \kms)} \\
\multicolumn{19}{c}{              } \\
\cline{1-19}\\
Comp &\multicolumn{3}{c}{{\bf H$\upbeta$}}&\multicolumn{3}{c}{{\bf[\ion{O}{III}]4959\AA}}&\multicolumn{3}{c}{{\bf[\ion{O}{III}]5007\AA\ }}&\multicolumn{3}{c}{{\bf[\ion{N}{II}]6548\AA}}&\multicolumn{3}{c}{{\bf H$\upalpha$}}&\multicolumn{3}{c}{{\bf[\ion{NII}]6583\AA}}\\
          &FWHM &$\Delta v$&$EW$  &FWHM &$\Delta v$ &$EW$  &FWHM  &$\Delta v$ &$EW$  &FWHM &$\Delta v$&$EW$ &FWHM&$\Delta v$&$EW$  &FWHM &$\Delta v$&$EW$\\
\cline{1-19}\\
N                     & 375$^{+10}_{-10}$       & - &4.1 &375$^{+10}_{-10}$& -       & 10.1     &375$^{+10}_{-10}$& -   & 30.4  & 358$^{+13}_{-13}$ & -     &5.1 &358$^{+13}_{-13}$  & -  & 18.2 &358$^{+13}_{-13}$  & - &15.2 \\
I                      & -                                    & - &-     &766$^{+35}_{-33}$ &230  &  5.6      &766$^{+35}_{-33}$&230&17.9   & 766$^{+35}_{-33}$ &174 &4.7 &766$^{+35}_{-33}$  &174&13.8 &766$^{+35}_{-33}$ &174&12.7  \\
B                     & -                                    & - &-     &         -                     & -     &     -       & -                            & -   &  -       & -                         & - &-             &3153$^{+436}_{-354}$ &-  &24.5 & -                              & -   &-       \\
\\

\cline{1-19}\\
\multicolumn{19}{c}{              } \\
\multicolumn{19}{c}{NGC 4395 (\citealt{Ho95}); $R$ = 2600 ($\sigma_v\sim$115 \kms)} \\
\multicolumn{19}{c}{              } \\
\cline{1-19}\\
Comp &\multicolumn{3}{c}{{\bf H$\upbeta$}}&\multicolumn{3}{c}{{\bf[\ion{O}{III}]4959\AA}}&\multicolumn{3}{c}{{\bf[\ion{O}{III}]5007\AA\ }}&\multicolumn{3}{c}{{\bf[\ion{N}{II}]6548\AA}}&\multicolumn{3}{c}{{\bf H$\upalpha$}}&\multicolumn{3}{c}{{\bf[\ion{NII}]6583\AA}}\\
          &FWHM &$\Delta v$&$EW$  &FWHM &$\Delta v$ &$EW$  &FWHM  &$\Delta v$ &$EW$  &FWHM &$\Delta v$&$EW$ &FWHM&$\Delta v$&$EW$  &FWHM &$\Delta v$&$EW$\\
\hline
N                        &151$^{+1}_{-1}$       & -      &32.3   &151$^{+1}_{-1}$     &  -     & 88.9 & 151$^{+1}_{-1}$      &  -   & 264.0 & 119$^{+2}_{-2}$      &  -    &7.1 & 119$^{+2}_{-2}$     & -    &79.4  & 119$^{+2}_{-2}$     &  -   &21.1     \\
 I                         & 258$^{+11}_{-12}$  & 257 &  5.2   &258$^{+11}_{-12}$ & 257 & 14.2  & 258$^{+11}_{-12}$  &257& 42.3   & 199$^{+81}_{-98}$  &148 & 1.8& -                              & -   &  -      & 199$^{+81}_{-98}$&148 & 5.5    \\
 I                        & 1145$^{+87}_{-90}$& 381  &30.7   & -                            &  -     &   -     & -                               & -    &           & -                              & -     &  -   & - &  - &-& -                             & -    &   -  \\
 B                       & -& -&-& -                            &  -     &   -     & -                               & -    &           & -                              & -     &  -   & 633$^{+20}_{-18}$  &  - &120.0 & -                             & -    &   -  \\
\\
\cline{1-19}\\
\multicolumn{19}{c}{NGC 6221 X-shooter (UVB;VIS); $R$ = 4350/7350 ($\sigma_v\sim$70/40 \kms)} \\
\multicolumn{19}{c}{              } \\
\cline{1-19}\\
Comp &\multicolumn{3}{c}{{\bf H$\upbeta$}}&\multicolumn{3}{c}{{\bf[\ion{O}{III}]4959\AA}}&\multicolumn{3}{c}{{\bf[\ion{O}{III}]5007\AA\ }}&\multicolumn{3}{c}{{\bf[\ion{N}{II}]6548\AA}}&\multicolumn{3}{c}{{\bf H$\upalpha$}}&\multicolumn{3}{c}{{\bf[\ion{NII}]6583\AA}}\\
          &FWHM &$\Delta v$&$EW$  &FWHM &$\Delta v$ &$EW$  &FWHM  &$\Delta v$ &$EW$  &FWHM &$\Delta v$&$EW$ &FWHM&$\Delta v$&$EW$  &FWHM &$\Delta v$&$EW$\\ 
\cline{1-19}\\
N                     & 79$^{+2}_{-1}$      & -     &   4.7  & 177$^{+4}_{-2}$  & -   &2.1   & 177$^{+4}_{-2}$    & -     &6.5   & 90$^{+4}_{-3}$     & - &9.9     &  90$^{+4}_{-3}$  & -    &48.0 & 90$^{+4}_{-3}$ & - &29.9  \\
I                      & 146$^{+2}_{-4}$     &31   &   12.9 & 125$^{+5}_{-5}$  &166& 0.6  & 125$^{+5}_{-5}$     &166  &2.0  & 192$^{+8}_{-13}$  & 21 &8.8  &192$^{+8}_{-13}$  & 21 & 50.9& 192$^{+8}_{-13}$& 21&26.6 \\
I                      & 541$^{+13}_{-13}$ &195 &   7.8 & 212$^{+3}_{-4}$  &346 &1.4  & 212$^{+3}_{-4}$  &346 & 4.3  & 500$^{+12}_{-7}$    &172 &11.3 & 500$^{+12}_{-7}$ &172 & 45.7& 500$^{+12}_{-7}$& 172&33.9 \\
I                      & -                            &-       &     -  &851$^{+16}_{-10}$  &261 & 3.0  &851$^{+16}_{-10}$   & 261 &  9.5        &   77$^{+16}_{-8}$   &375 &0.4  &77$^{+16}_{-8}$ &375 & 1.1& 77$^{+16}_{-8}$& 375&1.1 \\
B                     & -                                    & - &-     &         -                     & -     &     -       & -                            & -   &  -       & -                         & - &-             &1630$^{+12}_{-11}$ &-  &59.0 & -                              & -   &-      \\  
\\
\cline{1-19}\\
\multicolumn{19}{c}{NGC 7314 X-shooter (UVB;VIS); $R$ = 4350/7350 ($\sigma_v\sim$70/40 \kms)} \\
\multicolumn{19}{c}{              } \\
\cline{1-19}\\
Comp &\multicolumn{3}{c}{{\bf H$\upbeta$}}&\multicolumn{3}{c}{{\bf[\ion{O}{III}]4959\AA}}&\multicolumn{3}{c}{{\bf[\ion{O}{III}]5007\AA\ }}&\multicolumn{3}{c}{{\bf[\ion{N}{II}]6548\AA}}&\multicolumn{3}{c}{{\bf H$\upalpha$}}&\multicolumn{3}{c}{{\bf[\ion{NII}]6583\AA}}\\
          &FWHM &$\Delta v$&$EW$  &FWHM &$\Delta v$ &$EW$  &FWHM  &$\Delta v$ &$EW$  &FWHM &$\Delta v$&$EW$ &FWHM&$\Delta v$&$EW$  &FWHM &$\Delta v$&$EW$\\
\cline{1-19}\\
N                     & 56$^{+1}_{-1}$       & -   &  17.9  & 56$^{+1}_{-1}$    & -  &83.6   & 56$^{+1}_{-1}$   & -   &250.4  & 89$^{+1}_{-1}$     & - &13.5     &  89$^{+1}_{-1}$  & -    &45.2 & 89$^{+1}_{-1}$ & - &40.6  \\
I                      & 68$^{+1}_{-1}$        &47 &   5.5 & 68$^{+1}_{-1}$    &47 & 23.1   & 68$^{+1}_{-1}$   &47 &69.1    & 187$^{+2}_{-2}$  & 63 &8.5   &270$^{+2}_{-2}$  & 63 & 66.8& 187$^{+2}_{-2}$& 63&25.5 \\
I                      & 213$^{+1}_{-1}$      &67 &  29.5 & 213$^{+1}_{-1}$  &67 &120.5 & 213$^{+1}_{-1}$ &67 & 361.0 & -                           &-     &-       & -                         &  -   & -      & -                         &-   &-        \\
B                     & 1097$^{+57}_{-63}$& -  &  31.2  &    -                       & -    &    -    & -                         & -    &          & -                           & -     &-      &1330$^{+5}_{-4}$ &-     &223.0 & -                      & -   &-      \\  
\\
\hline
\end{tabular}
}
Notes: (1) Line components. B: BLR, N: NLR, I: intermediate (see sect \ref{sec:lines} for the classification criteria); (2) to (19) FWHM [\kms] (not deconvolved for the instrumental resolution), wavelength shift of the component's center relative to its narrow component [\kms], component's equivalent width [\AA ]. The object name, the reference of the optical spectrum, the instrumental resolution and the corresponding velocity resolution $\sigma_v$ [\kms ] are also reported.
\noindent
\end{minipage}
\end{table}

\clearpage

\begin{table*}
\centering
\begin{minipage}{140mm}
\caption{Line measurements of the NIR emission lines of the `broad' AGN2}
\label{tbl:NIRBAGN2lines}
\begin{center}
\begin{tabular}{@{}ccccccccccccc}
\hline
\multicolumn{13}{c}{              } \\
\multicolumn{13}{c}{2MASX J05054575-2351139, ISAAC/VLT (MR/LR), $R$ = 4700/$R$ = 730  ($\sigma_v\sim$60/430 \kms)} \\
\multicolumn{13}{c}{              } \\
\hline
Comp  &\multicolumn{3}{c}{{\bf \ion{He}{I}}}&\multicolumn{3}{c}{{\bf Pa$\upgamma$}}&\multicolumn{3}{c}{{\bf[\ion{Fe}{II}]12570\AA\  }}&\multicolumn{3}{c}{{\bf Pa$\upbeta$}} \\
            &  FWHM     &$\Delta v$& $EW$  & FWHM  &$\Delta v$& $EW$ &  FWHM& $\Delta v$   &   $EW$   &  FWHM  & $\Delta v$&$EW$  \\
            (1)        & (2)            & (3)          & (4)   &  (5)        & (6)          &  (7)  &   (8)     &   (9)            &   (10)  &   (11)       &  (12)        &(13) \\
\cline{1-13}\\
N                      &507$^{+49}_{-40}$       &   - & 20.5 & 507$^{+49}_{-40}$ &-  & 5.1 &  -                             &  - &   -   & 145$^{+30}_{-26}$ & -   &2.0  \\
I                        &-                                   & -   & -       & -                              & - & -     &413$^{+40}_{-38}$  &  - & 4.8 & 405$^{+67}_{-49}$ & -   &4.8  \\
B                      &1823$^{+419}_{-318}$ & -   &16.7 & -                              & - & -     &-                               & -   &-      &-                              &  - &  -     \\
\\
\cline{1-13}\\
\multicolumn{13}{c}{2MASX J18305065+0928414, X-shooter/VLT, $R$ = 5300  ($\sigma_v\sim$57 \kms)} \\
\multicolumn{13}{c}{              } \\
\hline
Comp  &\multicolumn{3}{c}{{\bf \ion{He}{I}}}&\multicolumn{3}{c}{{\bf Pa$\upgamma$}}&\multicolumn{3}{c}{{\bf[\ion{Fe}{II}]12570\AA\  }}&\multicolumn{3}{c}{{\bf Pa$\upbeta$}} \\
            &  FWHM     &$\Delta v$& $EW$  & FWHM  &$\Delta v$& $EW$ &  FWHM& $\Delta v$   &   $EW$   &  FWHM  & $\Delta v$&$EW$  \\
\cline{1-13}\\
N                      & 216$^{+12}_{-9}$        &   -   & 1.3        &-            &-     &  -     & -           &  -     & - & -     & -  &  -     \\
B                      & 3513$^{+232}_{-213}$ &  -    &18.6       & -           & -    & -      & -           & -      &-   & -     &-   & -   \\
\\
\cline{1-13}\\
\multicolumn{13}{c}{ESO 234-G050, X-shooter/VLT, $R$ = 5300  ($\sigma_v\sim$57 \kms)} \\
\multicolumn{13}{c}{              } \\
\hline
Comp  &\multicolumn{3}{c}{{\bf \ion{He}{I}}}&\multicolumn{3}{c}{{\bf Pa$\upgamma$}}&\multicolumn{3}{c}{{\bf[\ion{Fe}{II}]12570\AA\  }}&\multicolumn{3}{c}{{\bf Pa$\upbeta$}} \\
            &  FWHM     &$\Delta v$& $EW$  & FWHM  &$\Delta v$& $EW$ &  FWHM& $\Delta v$   &   $EW$   &  FWHM  & $\Delta v$&$EW$  \\
\cline{1-13}\\
N                      & 122$^{+12}_{-11}$      &   -   & 10.9  &122$^{+12}_{-11}$       &-       & 1.9  & 122$^{+12}_{-11}$       &  - &   3.9 & 167$^{+11}_{-13}$     & -       &  10.4     \\
I                        & 434$^{+7}_{-7}$          & 55  & 40.9 & 272$^{+26}_{-24}$      &19     &5.7   & 490$^{+26}_{-23}$      &  9 &14.8  &343$^{+63}_{-57}$     &35      &  10.2      \\
B                      & 1111$^{+63}_{-59}$     &  -    & 16.4 & -                                   & -       & -      & -                                  & -  &-      & 1305$^{+381}_{-322}$ &-        &   7.1   \\
\\
\cline{1-13}\\
\multicolumn{13}{c}{ESO 374-G044, ISAAC/VLT (MR/LR), $R$ = 4700/$R$ = 730  ($\sigma_v\sim$60/430 \kms)} \\
\multicolumn{13}{c}{              } \\
\hline
Comp  &\multicolumn{3}{c}{{\bf \ion{He}{I}}}&\multicolumn{3}{c}{{\bf Pa$\upgamma$}}&\multicolumn{3}{c}{{\bf[\ion{Fe}{II}]12570\AA\  }}&\multicolumn{3}{c}{{\bf Pa$\upbeta$}} \\
            &  FWHM     &$\Delta v$& $EW$  & FWHM  &$\Delta v$& $EW$ &  FWHM& $\Delta v$   &   $EW$   &  FWHM  & $\Delta v$&$EW$  \\
\cline{1-13}\\
N                      &632$^{+29}_{-36}$       &   - & 48.9 & 632$^{+29}_{-36}$ &-  &7.0 &  -                             &  - &   -     & 450$^{+27}_{-42}$     & -   & 12.2  \\
I                        &-                               & -   & -       & -                                 & - & -     &652$^{+74}_{-56}$  &  - & 10.2 &                                    & -   & -  \\
B                      &1202$^{+383}_{-221}$ & -   &16.3 & -                             & - & -     &-                               & -   &-       & 1413$^{+318}_{-294}$&  - &   4.2   \\
\\
\hline
\\
\multicolumn{13}{c}{MCG -01-24-012, ISAAC/VLT (MR/LR), $R$ = 4700/$R$ = 730  ($\sigma_v\sim$60/430 \kms)} \\
\multicolumn{13}{c}{              } \\
\hline
Comp  &\multicolumn{3}{c}{{\bf \ion{He}{I}}}&\multicolumn{3}{c}{{\bf Pa$\upgamma$}}&\multicolumn{3}{c}{{\bf[\ion{Fe}{II}]12570\AA\  }}&\multicolumn{3}{c}{{\bf Pa$\upbeta$}} \\
            &  FWHM     &$\Delta v$& $EW$  & FWHM  &$\Delta v$& $EW$ &  FWHM& $\Delta v$   &   $EW$   &  FWHM  & $\Delta v$&$EW$  \\
\cline{1-13}\\
N                      & 616$^{+28}_{-27}$      &   -   &31.9 & 616$^{+28}_{-27}$     &-       & 4.0  & 381$^{+78}_{-117}$      &  - & 8.3  & 245$^{+17}_{-16}$      & -   & 7.7  \\
I                        &1325$^{+347}_{-251}$ &806 & 7.9 &1325$^{+347}_{-251}$& 806 &2.2   & 857$^{+121}_{-273}$  &  - &  3.1  &1325$^{+347}_{-251}$ &806  &1.5  \\
B                      &                                     &  -    &       & -                                   & -      & -     & -                                   & -   &-      & 2070$^{+300}_{-280}$ &  - &8.9      \\
\\
\hline
\end{tabular}
\end{center}
\noindent
\end{minipage}
\end{table*} 

\clearpage

\begin{table*}
\centering
\begin{minipage}{140mm}
\contcaption{}
\begin{center}
\begin{tabular}{@{}ccccccccccccc}
\cline{1-13}\\
\multicolumn{13}{c}{MCG -05-23-016, ISAAC/VLT (MR/LR), $R$ = 4700/$R$ = 730  ($\sigma_v\sim$60/430 \kms)} \\
\multicolumn{13}{c}{              } \\
\hline
Comp  &\multicolumn{3}{c}{{\bf \ion{He}{I}}}&\multicolumn{3}{c}{{\bf Pa$\upgamma$}}&\multicolumn{3}{c}{{\bf[\ion{Fe}{II}]12570\AA\  }}&\multicolumn{3}{c}{{\bf Pa$\upbeta$}} \\
            &  FWHM     &$\Delta v$& $EW$  & FWHM  &$\Delta v$& $EW$ &  FWHM& $\Delta v$   &   $EW$   &  FWHM  & $\Delta v$&$EW$  \\
                                   (1)        & (2)            & (3)          & (4)   &  (5)        & (6)          &  (7)  &   (8)     &   (9)            &   (10)  &   (11)       &  (12)        &(13) \\
\cline{1-13}\\
N                      & -                                   &   -   & -       &-                                   &-       & -      & -                                    &  - &   -   & 247$^{+19}_{-18}$       & -       & 3.1 \\
B                       &1223$^{+90}_{-80}$     &  -    & 29.8 &1451$^{+134}_{-149}$& -       &19.5 & -                                   &  - &  -    &1165$^{+27}_{-18}$       &-       &27.4  \\
I                      & 3855$^{+172}_{-158}$ &  -    & 84.7 & -                                  & -      & -      & -                                   & -  &-      & 3695$^{+335}_{-196}$  & 227 &46.1   \\
\\
\cline{1-13}\\
\multicolumn{13}{c}{MCG -05-23-016, X-shooter/VLT, $R$ = 5300  ($\sigma_v\sim$57 \kms)} \\
\multicolumn{13}{c}{              } \\
\hline
Comp  &\multicolumn{3}{c}{{\bf \ion{He}{I}}}&\multicolumn{3}{c}{{\bf Pa$\upgamma$}}&\multicolumn{3}{c}{{\bf[\ion{Fe}{II}]12570\AA\  }}&\multicolumn{3}{c}{{\bf Pa$\upbeta$}} \\
            &  FWHM     &$\Delta v$& $EW$  & FWHM  &$\Delta v$& $EW$ &  FWHM& $\Delta v$   &   $EW$   &  FWHM  & $\Delta v$&$EW$  \\
\cline{1-13}\\
N                      & 233$^{+15}_{-15}$      &   -   & 9.2  &233$^{+15}_{-15}$       &-       & 1.4      & 560$^{+80}_{-77}$       &  - & 0.5 & 560$^{+80}_{-77}$       & -       &  10.7     \\
B                      &2474$^{+67}_{-64}$     &  -    & 87.4 &1911$^{+105}_{-79}$    & -       &34.3& -                                   &  - &  -    &2134$^{+93}_{-89}$       &-        &  55.7      \\
\\
\cline{1-13}\\
\multicolumn{13}{c}{Mrk 1210, ISAAC/VLT (MR/LR), $R$ = 4700/$R$ = 730  ($\sigma_v\sim$60/430 \kms)} \\
\multicolumn{13}{c}{              } \\
\hline
Comp  &\multicolumn{3}{c}{{\bf \ion{He}{I}}}&\multicolumn{3}{c}{{\bf Pa$\upgamma$}}&\multicolumn{3}{c}{{\bf[\ion{Fe}{II}]12570\AA\  }}&\multicolumn{3}{c}{{\bf Pa$\upbeta$}} \\
            &  FWHM     &$\Delta v$& $EW$  & FWHM  &$\Delta v$& $EW$ &  FWHM& $\Delta v$   &   $EW$   &  FWHM  & $\Delta v$&$EW$  \\
\cline{1-13}\\
N                      & 564$^{+8}_{-8}$          &   -   &141.4  &-                                   &-       & -      & 414$^{+18}_{-18}$       &  - & 15.1  & 414$^{+18}_{-18}$       & -       & 19.5     \\
I                       &  -                                  &   -    &      -   & 791$^{+122}_{-116}$ & -       &22.0 & 902$^{+114}_{-61}$  & 136 &21.1 &902$^{+114}_{-61}$    & 136 & 26.7       \\
B                      & 1374$^{+73}_{-32}$    &  -     & 258.0& -                                  & -      & -      & -                                   & -    &-      & 1937$^{+118}_{-225}$       &    -     &21.14      \\
\\
\cline{1-13}\\
\multicolumn{13}{c}{NGC 1052, ISAAC/VLT (MR/LR), $R$ = 4700/$R$ = 730  ($\sigma_v\sim$60/430 \kms)} \\
\multicolumn{13}{c}{              } \\
\hline
Comp  &\multicolumn{3}{c}{{\bf \ion{He}{I}}}&\multicolumn{3}{c}{{\bf Pa$\upgamma$}}&\multicolumn{3}{c}{{\bf[\ion{Fe}{II}]12570\AA\  }}&\multicolumn{3}{c}{{\bf Pa$\upbeta$}} \\
            &  FWHM     &$\Delta v$& $EW$  & FWHM  &$\Delta v$& $EW$ &  FWHM& $\Delta v$   &   $EW$   &  FWHM  & $\Delta v$&$EW$  \\
\cline{1-13}\\
N                      & 792$^{+61}_{-51}$   &   -   &10.4 &-                                   &-       & -      & 693$^{+20}_{-20}$&  - &9.5 &  -   & -       &      \\
B                      & 2455$^{+143}_{-128}$ &  -     &40.9 & -                                  & -      & -      & -                            & -  &-      &-      &-        &       \\
\\
\cline{1-13}\\
\multicolumn{13}{c}{NGC 1365, ISAAC/VLT (MR/LR), $R$ = 4700/$R$ = 730  ($\sigma_v\sim$60/430 \kms)} \\
\multicolumn{13}{c}{              } \\
\hline
Comp  &\multicolumn{3}{c}{{\bf \ion{He}{I}}}&\multicolumn{3}{c}{{\bf Pa$\upgamma$}}&\multicolumn{3}{c}{{\bf[\ion{Fe}{II}]12570\AA\  }}&\multicolumn{3}{c}{{\bf Pa$\upbeta$}} \\
            &  FWHM     &$\Delta v$& $EW$  & FWHM  &$\Delta v$& $EW$ &  FWHM& $\Delta v$   &   $EW$   &  FWHM  & $\Delta v$&$EW$  \\
\cline{1-13}\\
N                       & -                                      &   -   & -       &-                               &-       & -       & -                                    &  - &   -   &   787$^{+8}_{-8}$           & -       &15.7      \\
B                       & 1243$^{+100  }_{-92   }$&   -    &21.3  &1363$^{+67}_{-54}$& -      & 19.3 & -                                   &  - &  -    & 1972$^{+85}_{-75}$       &-        &15.4       \\
\\
\hline
\end{tabular}
\end{center}
\noindent
\end{minipage}
\end{table*} 

\clearpage

\begin{table*}
\centering
\begin{minipage}{140mm}
\contcaption{}
\begin{center}
\begin{tabular}{@{}ccccccccccccc}
\cline{1-13}\\
\multicolumn{13}{c}{NGC 2992, ISAAC/VLT (MR/LR), $R$ = 4700/$R$ = 730  ($\sigma_v\sim$60/430 \kms)} \\
\multicolumn{13}{c}{              } \\
\hline
Comp  &\multicolumn{3}{c}{{\bf \ion{He}{I}}}&\multicolumn{3}{c}{{\bf Pa$\upgamma$}}&\multicolumn{3}{c}{{\bf[\ion{Fe}{II}]12570\AA\  }}&\multicolumn{3}{c}{{\bf Pa$\upbeta$}} \\
            &  FWHM     &$\Delta v$& $EW$  & FWHM  &$\Delta v$& $EW$ &  FWHM& $\Delta v$   &   $EW$   &  FWHM  & $\Delta v$&$EW$  \\
                (1)        & (2)            & (3)          & (4)   &  (5)        & (6)          &  (7)  &   (8)     &   (9)            &   (10)  &   (11)       &  (12)        &(13) \\
\cline{1-13}\\
N                       & 528$^{+56}_{-58}$       &   -   & 22.8  &528$^{+56}_{-58}$    &-  & 3.4   & 254$^{+3}_{-8}$   &  - & 7.7   &254$^{+3}_{-8}$  & -       &5.4     \\
I                       & -                                     &   -    & -       &-                                 &  -&  -      &869$^{+34}_{-56}$&111 & 5.4 & 869$^{+34}_{-56}$&111 &2.1       \\
B                       & 3186$^{+586}_{-400}$ &   -    &50.9  &2989$^{+660}_{-488}$& -  & 21.2 & -                          &  -  &  -    & 2056$^{+29}_{-30}$&-     &22.2      \\
\\
\cline{1-13}\\
\multicolumn{13}{c}{              } \\
\multicolumn{13}{c}{NGC 4395, LUCI/LBT; $R$ = 1360  ($\sigma_v\sim$220 \kms)} \\
\multicolumn{13}{c}{              } \\
\cline{1-13}\\
Comp  &\multicolumn{3}{c}{{\bf \ion{He}{I}}}&\multicolumn{3}{c}{{\bf Pa$\upgamma$}}&\multicolumn{3}{c}{{\bf[\ion{Fe}{II}]12570\AA\  }}&\multicolumn{3}{c}{{\bf Pa$\upbeta$}} \\
            &  FWHM     &$\Delta v$& $EW$  & FWHM  &$\Delta v$& $EW$ &  FWHM& $\Delta v$   &   $EW$   &  FWHM  & $\Delta v$&$EW$  \\
\cline{1-13}\\
N                        &348$^{+5}_{-6}$       & -       &104.1   &348$^{+5}_{-6}$     &  -     & 16.1 & 235$^{+7}_{-6}$ &  -   & 13.2   & 235$^{+7}_{-6}$      &  -    & 16.9     \\
 I                         & -                              &         &  -         & -                                 &  -     & -       & -                         &-     &           &                                 & -     &             \\
 B                       & 1350$^{+93}_{-70}$& -       &58.7    & 1591$^{+310}_{-245}$& -      & 16.3 & -                         & -    &           & 879$^{+29}_{-34}$  & -     &40.6       \\
 \\
\cline{1-13}\\
\multicolumn{13}{c}{NGC 6221, X-shooter/VLT, $R$ = 5300  ($\sigma_v\sim$57 \kms)} \\
\multicolumn{13}{c}{              } \\
\hline
Comp  &\multicolumn{3}{c}{{\bf \ion{He}{I}}}&\multicolumn{3}{c}{{\bf Pa$\upgamma$}}&\multicolumn{3}{c}{{\bf[\ion{Fe}{II}]12570\AA\  }}&\multicolumn{3}{c}{{\bf Pa$\upbeta$}} \\
            &  FWHM     &$\Delta v$& $EW$  & FWHM  &$\Delta v$& $EW$ &  FWHM& $\Delta v$   &   $EW$   &  FWHM  & $\Delta v$&$EW$  \\
\cline{1-13}\\
N                      & 141$^{+1}_{-1}$      &   -   & 10.7  &141$^{+1}_{-1}$        &-       & 8.6  & 141$^{+1}_{-1}$      &  -     &   5.9 & 141$^{+1}_{-1}$     & -       &  18.6     \\
I                        & 343$^{+9}_{-9}$      & 52  & 21.7  & 343$^{+9}_{-9}$       &213   &2.0   & 483$^{+12}_{-12}$ & 176 &10.3  & 483$^{+12}_{-12}$ &176    &   9.5      \\
B                      & 2142$^{+110}_{-141}$ &  -    & 29.6 & 1433$^{+70}_{-70}$ & -       & 9.2  & -                              & -      &-      & 2257$^{+99}_{-93 }$&-        &   20.5   \\
\\
\cline{1-13}\\
\multicolumn{13}{c}{NGC 7314, X-shooter/VLT, $R$ = 5300  ($\sigma_v\sim$57 \kms)} \\
\multicolumn{13}{c}{              } \\
\hline
Comp  &\multicolumn{3}{c}{{\bf \ion{He}{I}}}&\multicolumn{3}{c}{{\bf Pa$\upgamma$}}&\multicolumn{3}{c}{{\bf[\ion{Fe}{II}]12570\AA\  }}&\multicolumn{3}{c}{{\bf Pa$\upbeta$}} \\
            &  FWHM     &$\Delta v$& $EW$  & FWHM  &$\Delta v$& $EW$ &  FWHM& $\Delta v$   &   $EW$   &  FWHM  & $\Delta v$&$EW$  \\
\cline{1-13}\\
N                      & 84$^{+2}_{-2}$         &   -   & 11.7  &84$^{+2}_{-2}$        &-     & 3.3  & 84$^{+2}_{-2}$       &  -     & 1.4 & 84$^{+2}_{-2}$     & -       &  6.4     \\
I                        & 361$^{+10}_{-9}$     & 55  & 41.0  & 361$^{+10}_{-9}$   &55  & 5.6  & 392$^{+13}_{-13}$ & 42 &1.4  & 392$^{+13}_{-13}$ &42    &  13.0      \\
B                      & 1428$^{+46}_{-38}$ &  -    &114.0 & 1430$^{+78}_{-99}$ & -   &50.9 & -                              & -      &-      & 1348$^{+15}_{-15 }$&-   & 74.9   \\
\\
\hline
\end{tabular}
\end{center}
Notes: (1) Line components. B: BLR, N: NLR, I: intermediate (see sect \ref{sec:lines} for the classification criteria); (2) to (13) FWHM [\kms] (not deconvolved for the instrumental resolution), wavelength shift of the component's center relative to its narrow component [\kms], component's equivalent width [\AA ]. The object name, the instrument used for the observations, the instrumental resolution and the corresponding velocity resolution $\sigma_v$ [\kms] are also reported.

\end{minipage}
\end{table*}

\clearpage

\subsection{{\bf AGN2 classification}}
\label{sec:classAGN}

\begin{figure*}
\includegraphics[scale=0.3]{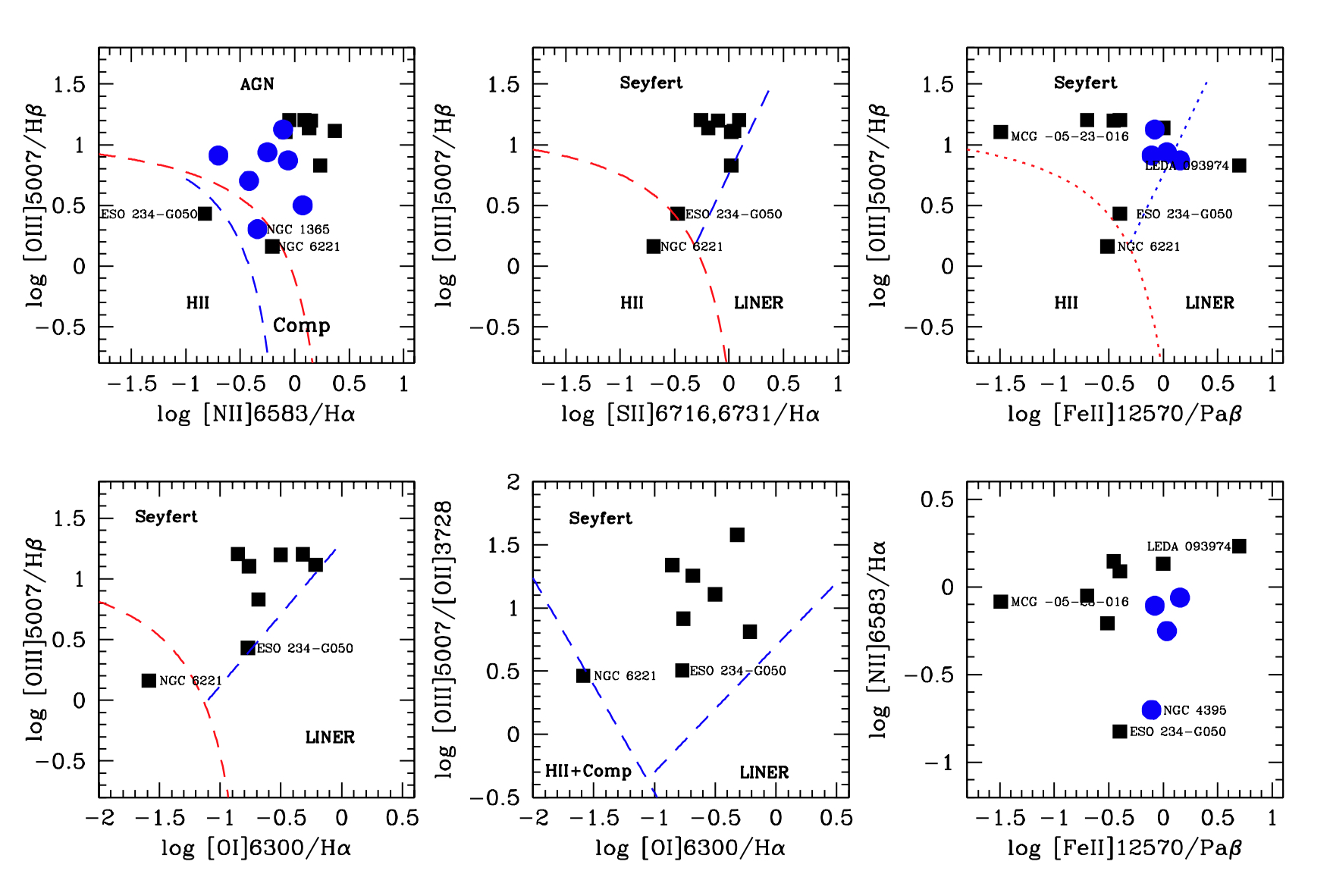}
\caption {Line ratio diagnostic diagrams of the activity classification. The X-shooter objects are shown by black filled squares. For the ISAAC and LUCI targets (shown by blue filled circles), we show only the results for the broad AGN2 sample, which were derived using the optical spectra taken from literature. 
The lines used to define the activity regions are the following: extreme starbursts \citep[red dashed;][]{kewley01}, pure star formation \citep[blue dashed;][]{kauffmann03}, Seyfert-LINER \citep[purple solid;][]{kewley06},  Seyfert-HII+Composite galaxies \citep[red dot dashed;][]{kewley06}.
 In the [\ion{O}{III}]5007/H$\upbeta$ versus [\ion{Fe}{II}]/Pa$\upbeta$  diagram, as no activity region definition was available, we used as a reference those computed for the  [\ion{O}{III}]5007/H$\upbeta$ versus  [\ion{S}{II}]6716,6731/H$\upalpha$ diagram.}
\label{fig:classAGN}
\end{figure*}

As the activity classification of our AGN sample has been taken from the inhomogeneous compilation of \citet{Baumgartner13}, we decided to use our line measures to  independently re-determine the AGN classes trough the  line ratio diagnostic diagrams \citep[e.g. BPT diagrams:][]{baldwin81}. The narrow line ratios [\ion{O}{III}]5007/H$\upbeta$, [\ion{O}{I}]6003/H$\upalpha$, [\ion{O}{III}]5007/[\ion{O}{II}]3728, [\ion{N}{II}]6583/H$\upalpha$ and [\ion{S}{II}]6716,6731/H$\upalpha$ (in the optical)  and [\ion{Fe}{II}]/Pa$\upbeta$ (in the NIR; see \citealt{Veilleux97}) were used. The line ratios were computed using the fluxes reported in Tables  \ref{tbl_FLUXxshVISSpettri} and \ref{tbl_FLUXxshUVBSpettri}, for the X-shooter AGNs, while we used the $EW$ reported in Table \ref{tbl_OptBAGN2lines} and \ref{tbl:NIRBAGN2lines}, for the non-X-shooter broad AGN2 (for which optical spectra were taken from the literature). The diagnostic diagrams are shown in Figure \ref{fig:classAGN}. The pure star formation (blue dashed line) and extreme starburst (red dashed line) regions were defined according to  \cite{kauffmann03} and \cite{kewley01}, respectively. The separation between AGN2 and LINERS was carried out according to \citet[][purple solid line]{kewley06}. Moreover, in the diagram showing  the [\ion{O}{III}]5007/[\ion{O}{II}]3728 ratio, the starburst region as defined by  \citet[][dot-dashed red line]{kewley06} was used.
The line ratio diagnostic analysis confirms in general  the AGN activity nature of the sources. However, 
it results that both NGC 1365 and NGC 6221 fall in the high excitation part of the HII-AGN composite galaxy region, confirming the presence of a strong starburst component in their spectra, as already suggested by other authors (for NGC 6221 see \cite{lafranca16}, and references therein, and for NGC 1365 see \cite{trippe10}).
Interestingly, also ESO 234$-$G050 is placed in a intermediate zone between the Seyfert and the HII region. The activity classification results, in addition to our new spectroscopic redshift estimates
(which agree, within the uncertainties, with the values listed by \cite{Baumgartner13}), are summarized in Table \ref{tbl:classAGN}.

\clearpage

\begin{table*}
\centering
\begin{minipage}{140mm}
\caption{Activity classification and new spectroscopic redshifts}
\label{tbl:classAGN}
\begin{center}
\begin{tabular}{@{}llcc}
\hline
Object                                           & Redshift                 &Swift/BAT Classification   & Classification \\ 
(1)                                                 &(2)             &    (3)                       & (4)                  \\      
\hline                     
 2MASX J05054575$-$2351139       & 0.035        & 2                   & AGN2\\
 2MASX J06411806+3249313      & 0.048        &2                    &-  \\
 2MASX J09112999+4528060      & 0.0269       & 2                   &-  \\
 2MASX J11271632+1909198      & 0.1057       & 1.8                &-  \\
 2MASX J18305065+0928414      & 0.0193       &  2                  & AGN2\\
 3C 105                                         & 0.0884        & 2                    &- \\
 3C 403                                         & 0.0589        & 2                    &-  \\
 CGCG 420$-$015                            & 0.029        & 2                    &-\\
 ESO 005$-$G004                            & 0.006        & 2                     &-  \\
 ESO 157$-$G023                            & 0.044       & 2                     &-  \\
 ESO 234$-$G050                            & 0.0088       & 2                     &AGN2/Starburst\\
 ESO 263$-$G013                            & 0.0334        & 2                     & AGN2\\
 ESO 297$-$G018                            & 0.0253        & 2                     &- \\
 ESO  374$-$G044                           & 0.0284       & 2                     &AGN2 \\
 ESO 416$-$G002                            & 0.059       &1.9                   & -\\
 ESO 417$-$G006                            & 0.0163       & 2                      &- \\
 Fairall 272                                    & 0.022        &  2                     & - \\
 LEDA 093974                               & 0.0239      & 2                      & AGN2 \\
 MCG $-$01$-$24$-$012                          & 0.0197      & 2                      & AGN2\\
 MCG $-$05$-$23$-$016                          & 0.0084       & 2                      & AGN2  \\
 Mrk 417                                        &0.0327        & 2                      & - \\
 Mrk 1210                                      & 0.014        & 2                      & - \\
 NGC 612                                      &-                   & 2                      & -  \\
 NGC 788                                      &0.0135        & 2                      & - \\
 NGC 1052                                    &0.0047         & 2                      & AGN2 \\
 NGC 1142                                    &0.0294        &  2                      &-    \\
 NGC 1365                                    & 0.005        & 1.8                     &AGN2/Starburst   \\
 NGC 2992                                    & 0.0077       & 2                        &AGN2   \\
 NGC 3079                                    & 0.0036      & 2                        &-   \\
 NGC 3081                                    & 0.0077       & 2                       & -\\
  NGC 3281                                   & 0.0113       & 2                        &-  \\
 NGC 4138                                    & -                & 1.9                    & -  \\
 NGC 4388                                    & 0.0089       & 2                       & - \\
 NGC 4395                                    & 0.0014       & 1.9                    &AGN2\\
 NGC 4686                                    & -                 &XBONG             & - \\
 NGC 4941                                     & 0.0038       & 2                       &AGN2\\
 NGC 4945                                    & 0.0017       & 2                       &-\\
 NGC 5643                                    & 0.0040       & 2                       &AGN2\\
 NGC 6221                                    & 0.0045       & 2                        &AGN2/Starburst\\
 NGC 7314                                    & 0.0047       & 1.9                     &AGN2\\
 PKS 0326$-$288                              & 0.1096         & 1.9                    & - \\
\hline
 \end{tabular}
\end{center}
Notes: (1) Source name; (2) spectroscopic redshift (this work); (2) {\it Swift}/BAT Classification \citep{Baumgartner13}; (4) line ratio diagnostic classification (see sect. \ref{sec:classAGN}). 
\noindent
\end{minipage}
\end{table*}

\clearpage

\begin{table*}
\centering
\begin{minipage}{140mm}
\caption{FWHM of the BLR Components}
\label{tbl_BLR}
\begin{center}
\begin{tabular}{@{}llcccc}
\hline
             &              & &\multicolumn{3}{c}{FWHM [\kms]}\\
 \cline{4-6}\\
Object (Instr.)  & Redshift\phantom{00} & Cl &H$\upalpha$ & \ion{He}{I} &Pa$\upbeta$\\ 
(1)                   & ~~~(2)          & (3)             & (4)  &(5)  & (6)          \\      
\hline                     
 2MASX J05054575$-$2351139 (I)     & 0.035 & 2 &-                         &1772$^{+419}_{-318}$ & - \\ 
 2MASX J18305065+0928414 (X) & 0.0193 & 2 &2660$^{+500}_{-460}$ & 3513$^{+232}_{-213}$ & -  \\ 
  ESO 234$-$G050 (X)                      &0.0088   &2 &\phantom{0}971$^{+124}_{-104}$ &1110$^{+63\phantom{0}}_{-59}$ &1304$^{+381}_{-322}$\\ 
  ESO 374$-$G044 (I)                          & 0.0284 & 2 &-                           &1123$^{+383}_{-221}$ &1412$^{+318}_{-294}$ \\
  MCG $-$01$-$24$-$012 (I)                        & 0.0197 & 2 &-                           & -                                           &2069$^{+300}_{-280}$  \\
  MCG $-$05$-$23$-$016 (X)                        & 0.0084   & 2 &2232$^{+36\phantom{0}}_{-38}$ &2474$^{+67\phantom{0}}_{-64}$ &2133$^{+93\phantom{0}}_{-89}$  \\ 
  Mrk 1210              (I)                     &0.014  & 2 &NA                &1305$^{+73\phantom{0}}_{-32}$ &1936$^{+118}_{-225}$ \\
  NGC 1052            (I)                     &0.0047  &2 &2192$^{+51\phantom{0}}_{-45}$ &2417$^{+143}_{-128}$ & - \\
  NGC 1365            (I)                     & 0.005  &1.8 &1674$^{+47\phantom{0}}_{-46}$ & -                             &1971$^{+85\phantom{0}}_{-75}$ \\
 NGC 2992  (I)                                 & 0.0077 & 2 &3139$^{+436}_{-354}$ &3157$^{+586}_{-400}$ & 2055$^{+29\phantom{0}}_{-30}$\\ 
 NGC 4395   (L)                                & 0.0014 & 1.9 &\phantom{0}622$^{+20\phantom{0}}_{-18}$ & 1332$^{+93\phantom{0}}_{-70}$ & \phantom{0}851$^{+29\phantom{0}}_{-34}$ \\
 NGC 6221    (X)                            &0.0045  & 2 &1630$^{+12\phantom{0}}_{-11}$  &2141$^{+110}_{-141}$ & 2256$^{+99\phantom{0}}_{-82}$ \\
 NGC 7314     (X)                            & 0.0047 &1.9 &1328$^{+4\phantom{00}}_{-5}$ & 1427$^{+46\phantom{0}}_{-38}$ & 1347$^{+46\phantom{0}}_{-39}$ \\
 \hline
 \end{tabular}
\end{center}
Notes: (1) Object (Instrument used. X: X-shooter, I: ISAAC, L: LUCI); (2) Redshift as reported in Table \ref{tbl:classAGN}; (3) AGN spectral classification according to \citet{Baumgartner13}; (4) to (6) FWHM of the BLR component deconvolved for the instrumental spectral resolution. The FWHM of the optical lines of the objects not observed with X-shooter have been obtained by analysis of spectra retrieved using the NED\footnote{NASA/IPAC Extragalactic Database: \url{https://ned.ipac.caltech.edu}} database.
\noindent
\end{minipage}
\end{table*}

\subsection{The intrinsic FWHM of the BLR components}
\label{sec:broadAGN2}

\begin{figure}
\centering
\includegraphics[scale=.15, angle=-90 ]{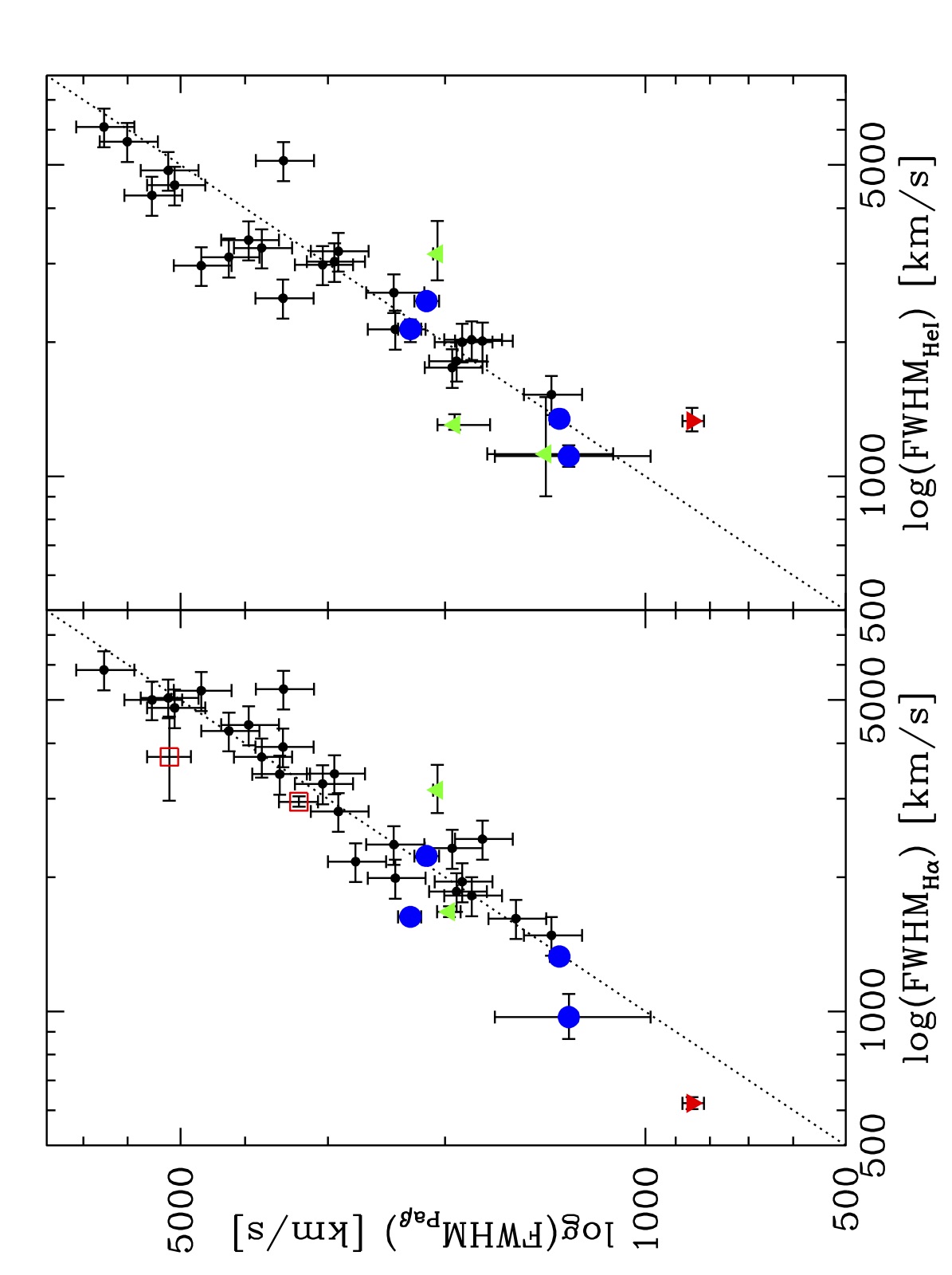}
\caption{{\it Left.} Best fit FWHM of the broad components of the Pa$\upbeta$ line as a function of the FWHM of the H$\upalpha$ line. The open red squares  shows those
two objects lacking a Pa$\upbeta$ measurement and whose FWHM value has been replaced by the \ion{He}{I} measurement (see text).
{\it Right.} Best fit FWHM of the broad components of the Pa$\upbeta$ line as a function of the FWHM of the \ion{He}{I} line. Blue circles: X-shooter; Green triangles: ISAAC; Red, up side down triangles: LUCI. In both panels the black dotted line shows the 1:1 locus, while the black dots show the  FWHM (corrected for instrumental broadening) distribution of a sample of AGN1 from \citet{landt08}.
}

\label{fig:HaHeIPab}
\end{figure}

In summary, we have found significant evidence of the presence of BLR components  in 13 out of 41 AGN2 spectra (in the following called broad AGN2). However, in order to derive the real width of the lines
it is necessary to deconvolve the FWHM measurements with the broadening due to the spectral resolution of the instruments used.
In Table \ref{tbl_BLR} we list the intrinsic FWHM of the broad components of the H$\upalpha$, \ion{He}{I} and Pa$\upbeta$ lines, once corrected for the instrumental broadening. 
The measured FWHM of the Pa$\upbeta$ ranges from $\sim$800 \kms (NGC 4395) to $\sim$2250 \kms (NGC 6221).

According to \citet{Baumgartner13}, three out of our 13 classified broad AGN2 had been previously classified as intermediate (NGC 1365, NGC 4395 and NGC 7314) in the optical, and indeed we detected a faint BLR H$\upalpha$ component in all of them. Moreover, four other optically intermediate class AGN have not shown a significant BLR component in their NIR spectra (2MASX J11271632+1909198, NGC 4138,
ESO 416$-$G002 and PKS 0326$-$288).
It should also be noted that a BLR component of the H$\upalpha$ was detected in 10 out of the 13 broad AGN2.
For 8 (7) sources we found a BLR detection on both the Pa$\upbeta$ and the \ion{He}{I} (H$\upalpha$) lines.
In Figure \ref{fig:HaHeIPab}  we show the comparison between the FWHM of the  H$\upalpha$, \ion{He}{I} and Pa$\upbeta$ lines when measured on the same object. As already found in other studies \citep[see the data by][in Figure \ref{fig:HaHeIPab}]{landt08} a fair agreement is found (if the $\sim$10\% uncertainties are taken into account) among all these measurements.
 
\section{Analysis of possible selection effects}

As we have found evidence of BLR components in 13 out of 41 AGN2, it is relevant to investigate whether some selection effect could
affect the sample of AGN2 where the BLR was found, or if there are some physical differences between the samples of AGN2 showing (or not) faint BLR components. 
We have therefore analysed the distributions of the spectral S/N ratio, $J$-band magnitude, X-ray flux, X-ray luminosity, hydrogen column density $N_{\rm H}$,  and the orientation ($a/b$ = major axis/minor axis) of the hosting galaxy \citep[as reported in Hyperleda\footnote{\url{http://leda.univ-lyon1.fr/}};][see Table \ref{tab:diagnostici}]{makarov14}. 

\begin{figure}
\centering
\includegraphics[scale=.14, angle= -90]{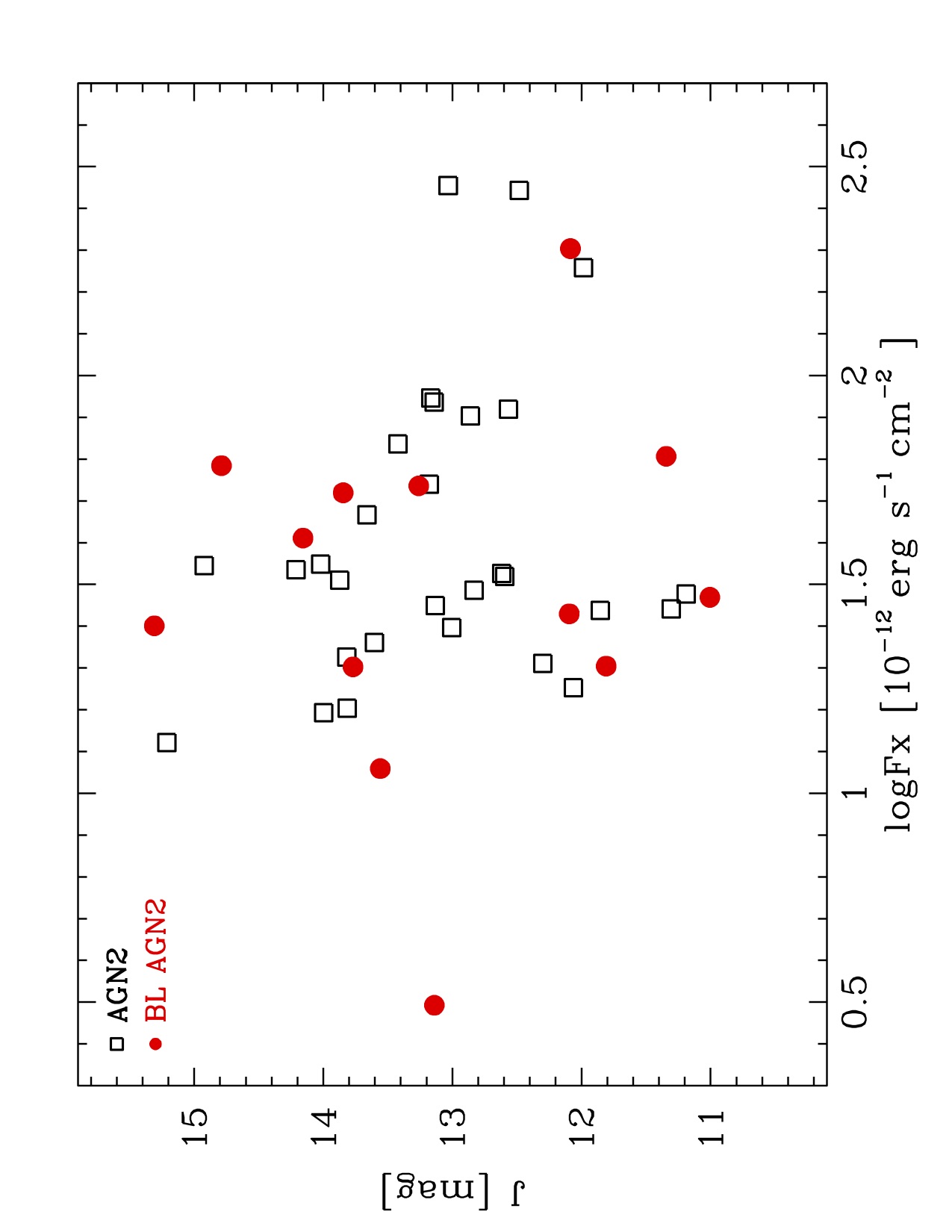}
\caption{$J$-band magnitude as a function of the 14--195 keV flux  of the `non-broad' (black open squares) and the `broad' (red filled circles) AGN2 samples .}
\label{fig:mJ_logFx}
\end{figure}

In Figure \ref{fig:mJ_logFx} the 14--195 keV  flux as a function of the 2MASS $J$-band magnitude for the `broad' and `non-broad' AGN2 samples is shown: no significant difference  between the two samples is found. The `broad' AGN2 sample has an average $J$-band magnitude value (13.1$\pm$0.4) which is undistinguishable  from that of the `non-broad' AGN2 sample (13.1$\pm$0.2; see Table \ref{tab:diagnostici}). Moreover,  also the average X-ray flux (log$F_X$ = 1.5$\pm$0.1 10$^{-12}$ \unitflux) is  almost identical to that of the `non-broad' AGN2 sample (log$F_X$ = 1.6$\pm$0.1 10$^{-12}$ \unitflux).

\begin{figure}
\centering
\includegraphics[scale=.14, angle= -90]{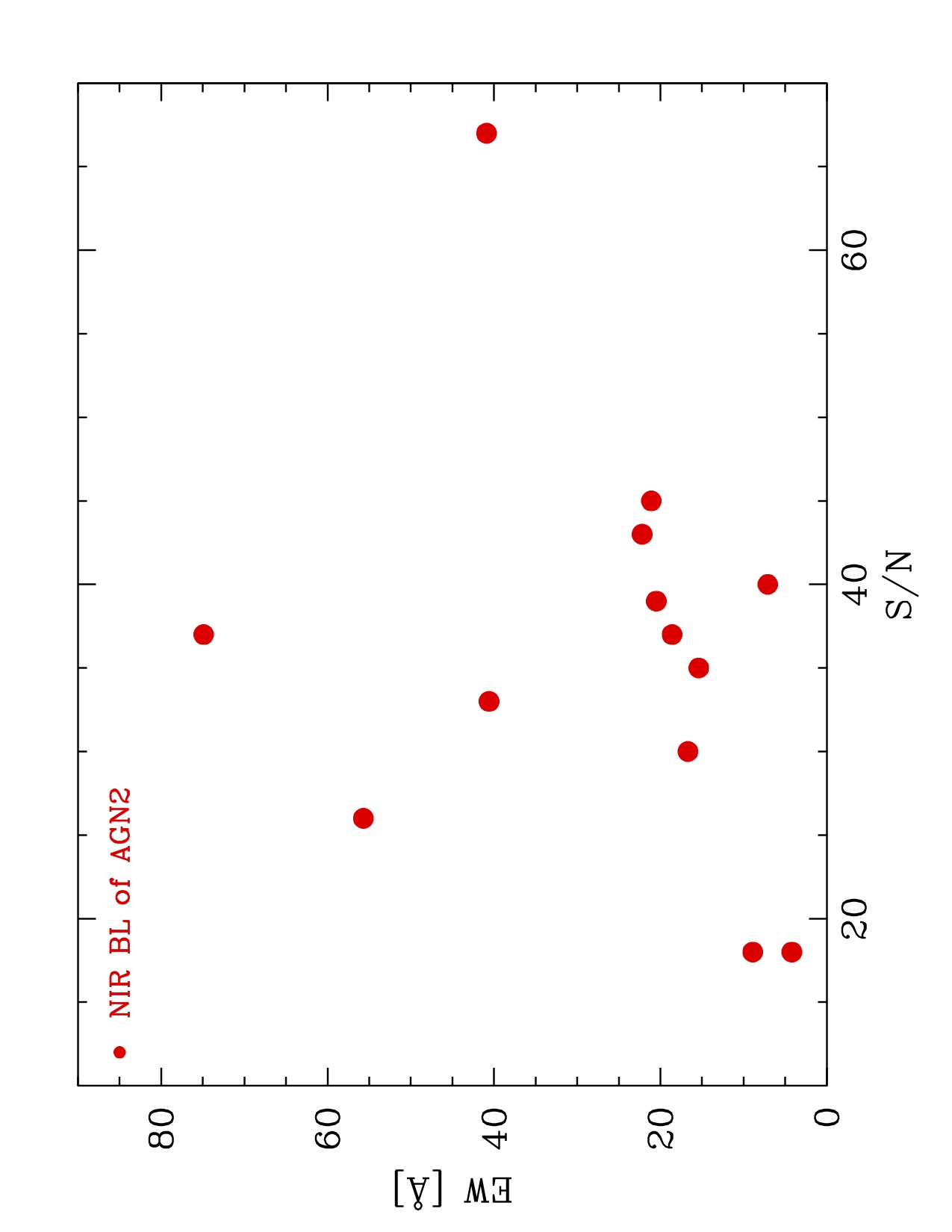}
\caption{Equivalent widths  of the broad components as a function of the spectral S/N ratio.}
\label{fig:EW_SN}
\end{figure}

\begin{figure}
\centering
\includegraphics[scale=.14, angle= -90]{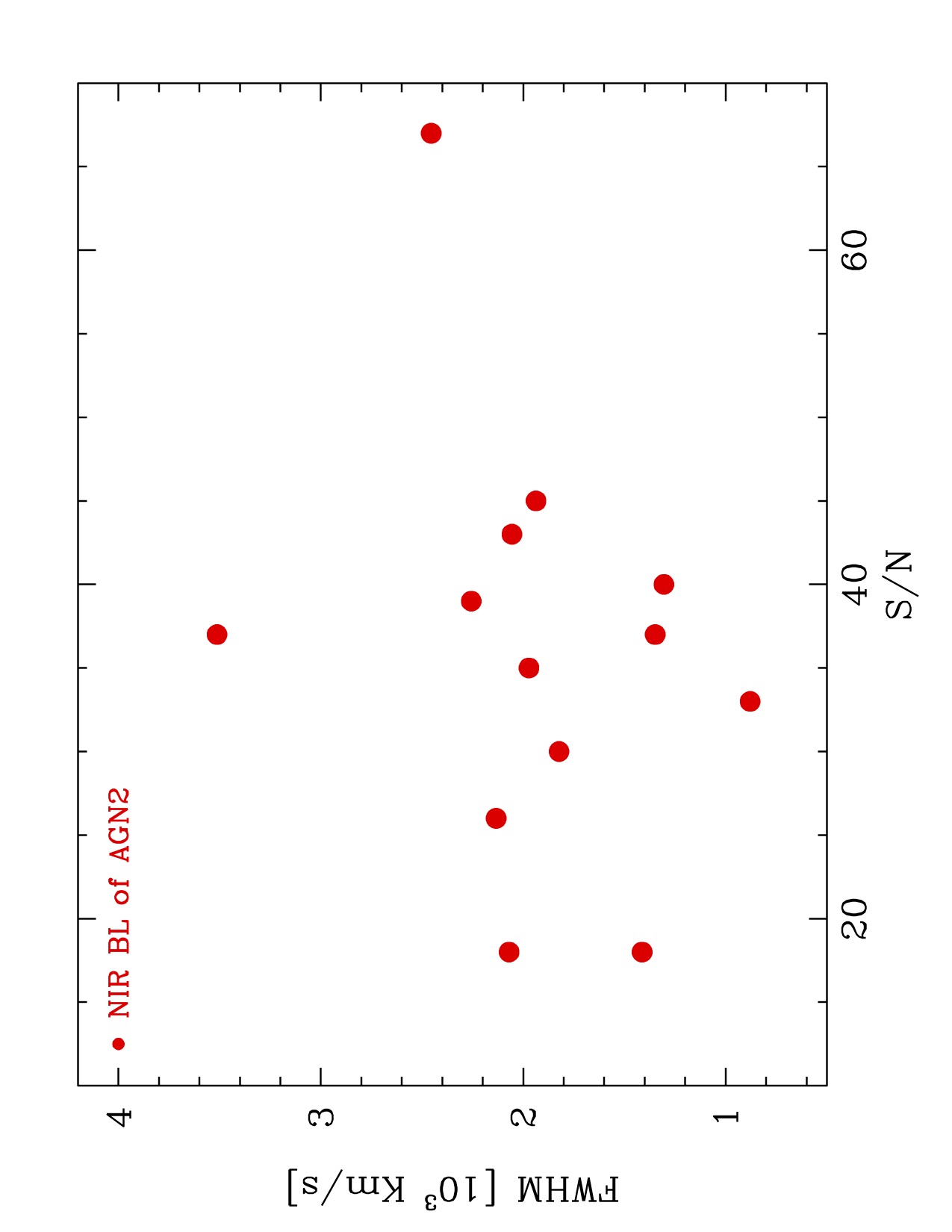}
\caption{FWHM of the broad components as a function of the spectral S/N ratio. }
\label{fig:FWHM_SN}
\end{figure}

In Figures \ref{fig:EW_SN} and \ref{fig:FWHM_SN} we show the dependence of the $EW$ and FWHM of the BLR components as a function of the spectral S/N ratio. For each of the 13 `broad' AGN2 the values of the
Pa$\upbeta$ line, when found (10 sources), are shown, otherwise the values obtained for the \ion{He}{I} line (3 sources) are shown. Indeed, as shown in Figure \ref{fig:HaHeIPab} and according to \citet{landt08}, no significant difference is observed among the FWHM measurements of the Pa$\upbeta$ and \ion{He}{I} lines. 
No evidence is found of a dependence of the $EW$ and FWHM measurements from the S/N ratio. More precisely, 
on the contrary of what it could be expected, it is not observed that the smallest $EW$ were measured in the highest S/N spectra. 

\begin{figure}
\centering
\includegraphics[scale=.14, angle= -90]{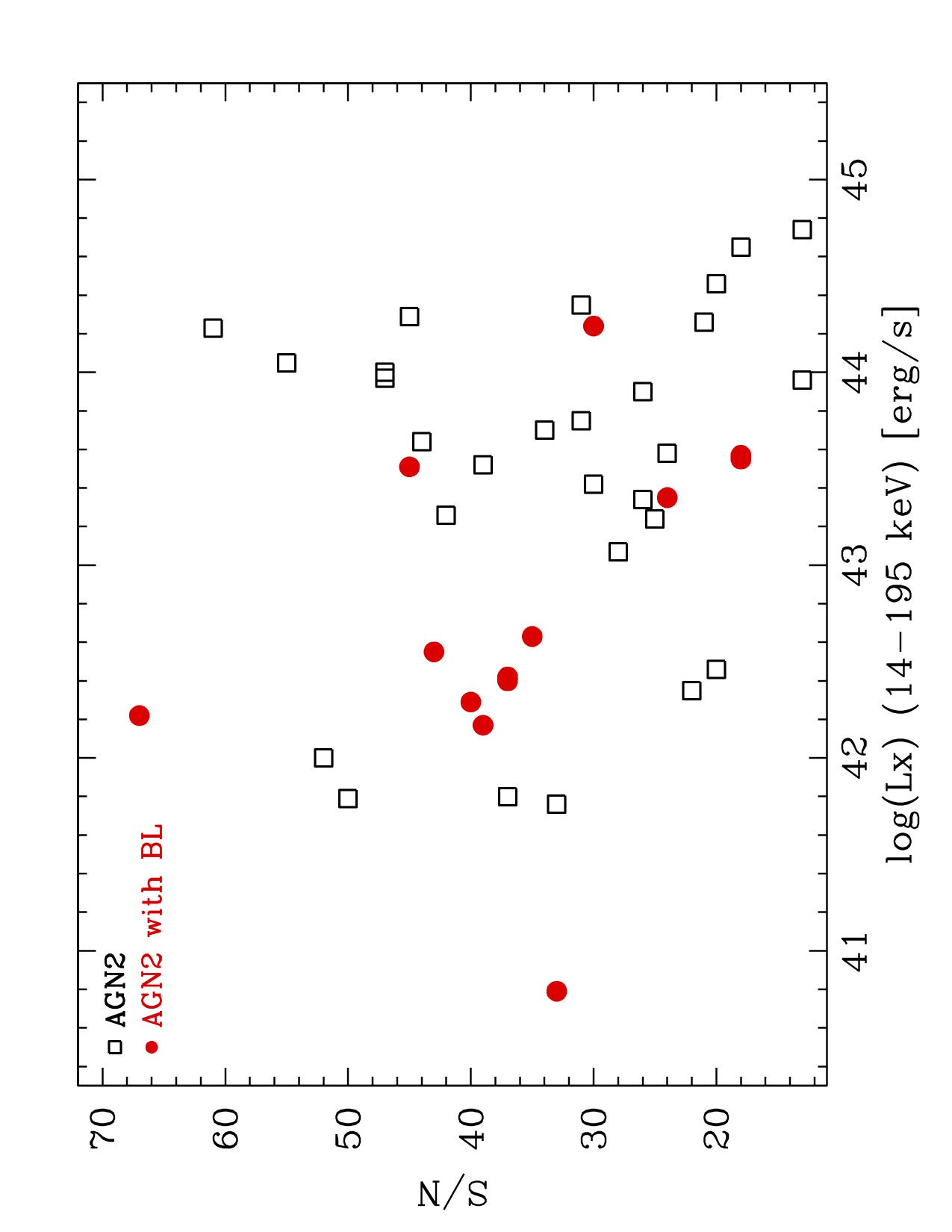}
\caption{Spectral S/N ratio as a function of the 14--195 keV luminosity of the `non-broad' (black open squares) and the `broad' (red filled circles) AGN2 samples.}
\label{fig:SN_Lx}
\end{figure}

In Figure \ref{fig:SN_Lx} the S/N ratio as a function of the 14--195 keV luminosity, of the `non broad' and the `broad' AGN2 samples, is shown.
The `broad' AGN2 sample has an average S/N ratio (35.8$\pm$3.5) which is undistinguishable  from that of the `non broad' AGN2 sample (32.8$\pm$2.5).
Nonetheless, it should be noticed that there is marginal (2$\sigma$) evidence for the `broad' AGN2 sample having, on average, lower X-ray luminosities (log$L_X$ = 42.7$\pm$0.2 \unitlum) than the `non-broad' AGN2 sample (log$L_X$ = 43.5$\pm$0.2 \unitlum).  This difference could be ascribed to the intrinsic differences among the different observed samples. The 11 objects observed with X-shooter (the most efficient  instrument for this project: 5 `broad' AGN2 found) were chosen to be among the less luminous of the {\it Swift}/BAT sample.

\begin{figure}
\centering
\includegraphics[scale=.14, angle= -90]{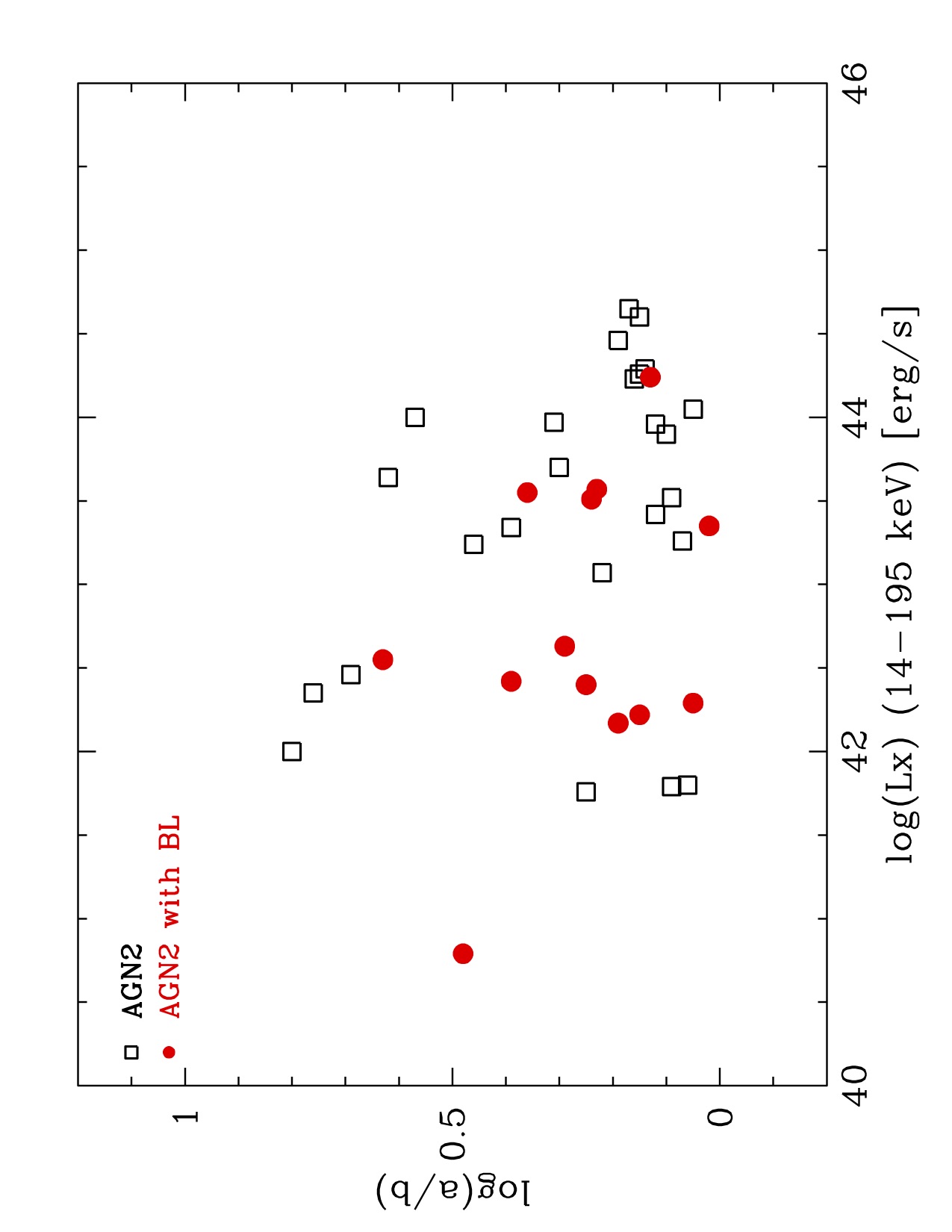}
\caption{The axis ratio ($a/b$ = major axis/minor axis) of the AGN2 host galaxies as a function of the 14--195 keV luminosity for the `non-broad' AGN2 sample (black open squares) and for the `broad' AGN2 sample (red filled circles). The
axis ratio has been evaluated at the isophote 25 mag/arcsec$^2$ in the  $B$-band \citep[from Hyperleda;][]{makarov14}.}
\label{fig:Lx_logr25}
\end{figure}

In Figure \ref{fig:Lx_logr25} the host galaxy axis ratio ($a/b$ = major-axis/minor-axis)  as a function of the 14--195 keV luminosity is shown.  The axis ratio has been evaluated at the isophote 25 mag/arcsec$^2$ in the  $B$-band.
No significant difference between the two samples was found.
The `broad' AGN2 sample shows an average axis ratio log($a/b$) = 0.26$\pm$0.05 while the `non-broad' AGN2 sample has an average axis ratio log($a/b$) = 0.27$\pm$0.05.

\begin{figure}
\centering
\includegraphics[scale=.14, angle= -90]{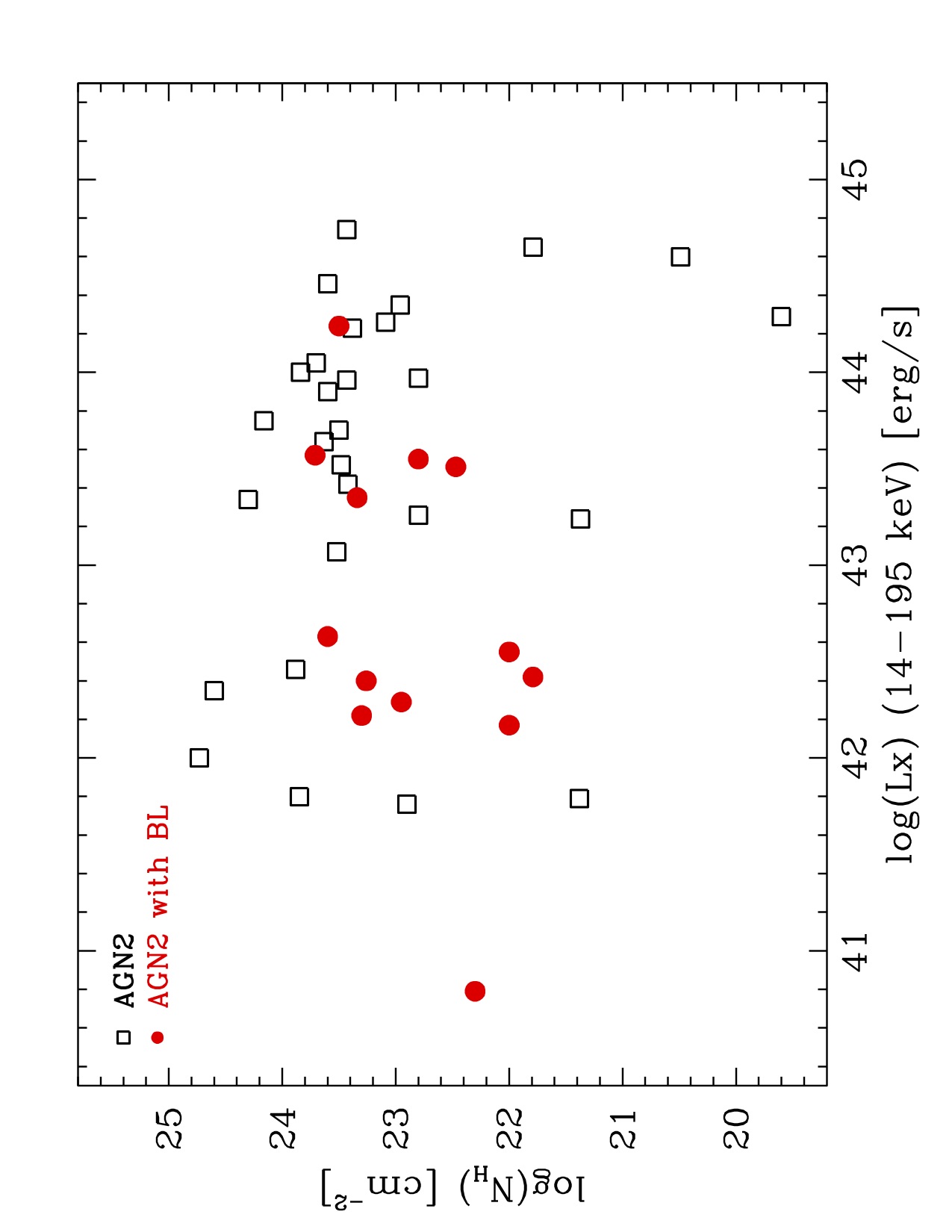}
\caption{Hydrogen column density, $N\rm_{H}$,  as a function of the 14--195 keV luminosity of the `non-broad' (black open squares) and the `broad' (red filled circles) AGN2 samples.}
\label{fig:Lx_Nh}
\end{figure}

In Figure \ref{fig:Lx_Nh} the hydrogen column density, $N_{\rm H}$,  as a function of the 14--195 keV  luminosity is shown. There is marginal (2$\upsigma$) evidence
that the `broad' AGN2 have, on average, lower
$N_{\rm H}$ (log$N_{\rm H}$ = 22.8$\pm$0.2 cm$^{-2}$) than measured in the `non-broad' AGN2 (log$N_{\rm H}$ = 23.1$\pm$0.2 cm$^{-2}$). The largest column density measured in the `broad' AGN2 sample is log$N_{\rm H}$ = 23.7 cm$^{-2}$, while  there are 7 `non-broad' AGN2 with  log$N_{\rm H}$ $>$23.7 cm$^{-2}$. 

\begin{figure*}
\centering
\includegraphics[scale=.3, angle=-90]{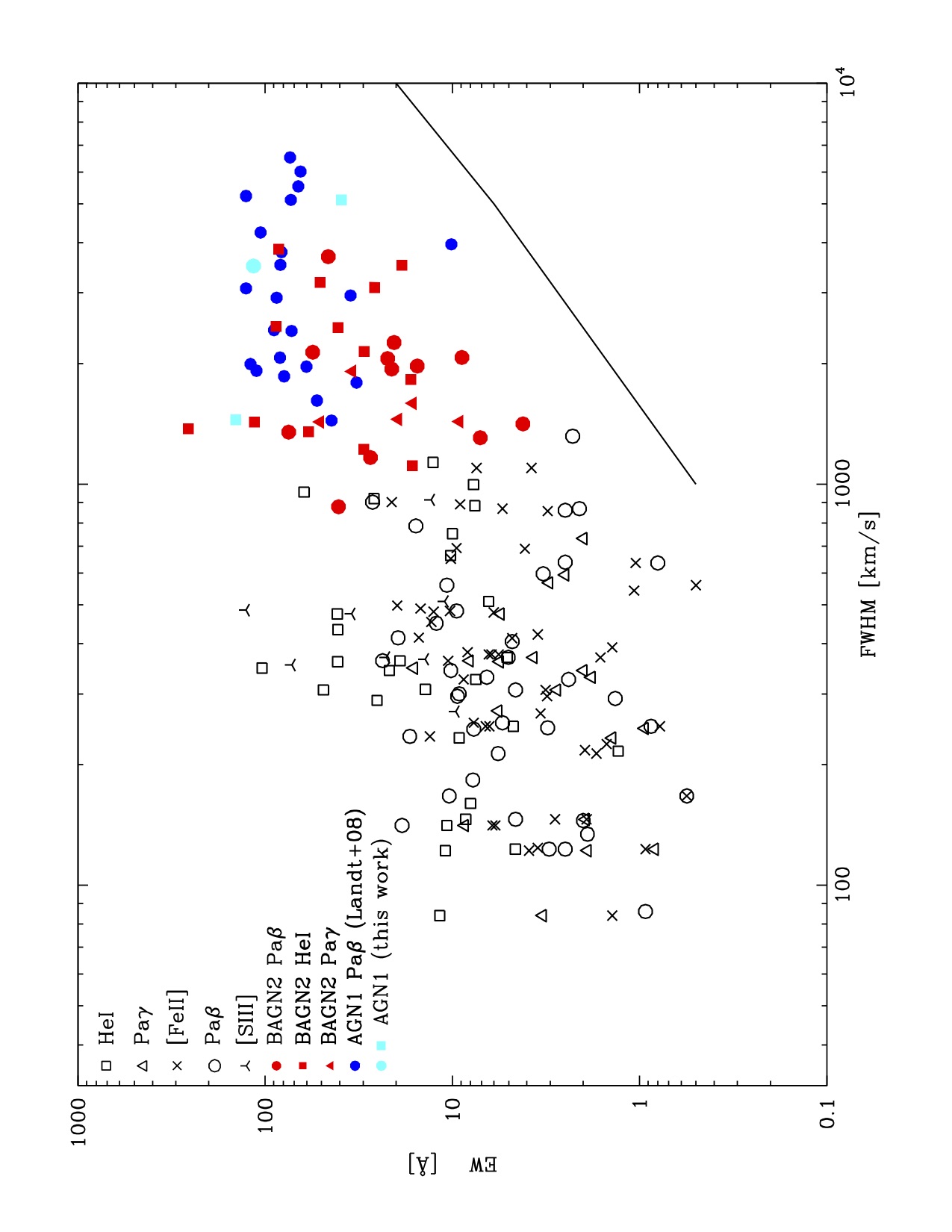}
\caption{FWHM of the most relevant line components of all 42 AGN analized in this work as a function of their $EW$. 
The broad AGN2 components are shown by red filled circles, squares and triangles for the Pa$\upbeta$, \ion{He}{I} and Pa$\upgamma$ lines, respectively.
Broad Pa$\upbeta$ measurements  of a control sub-sample of AGN1 \citep{landt08}, from the {\it Swift}/BAT 70-month catalogue,  are shown by filled blue circles. Other narrow components are in black as explained in the legend.
The black solid line shows the lower detection limit for the Pa$\upbeta$ emission line, according to the average S/N ratio and spectral resolution of the observations.}
\label{fig:FWHM_EW}
\end{figure*}

Finally, Figure \ref{fig:FWHM_EW} shows the FWHM as a function of their $EW$, of all line components (both broad and narrow) detected in the whole sample of the 41 studied AGN2.
For comparison, the Pa$\upbeta$ emission line fitting values of a sub-sample of 24, low redshift, AGN1 \citep{landt08}, from the {\it Swift}/BAT 70-month catalogue, and our measurement of the AGN1 NGC 3783, are also shown. As is seen, AGN1 show, on average, larger FWHM and $EW$ than the AGN2. In particular, AGN1 have an average FWHM = 3350$\pm$310 \kms and an average $EW$ = 80$\pm$7 \AA , while AGN2 have an average FWHM = 1680$\pm$610 \kms and an average $EW$ = 39$\pm$8 \AA . It should be noticed that, as expected from the AGN X-ray Luminosity Function (see introduction; the samples come from an hard X-ray flux limited survey), AGN1 are, on average, about one dex more luminous than AGN2: log$L\rm_X$ = 44.0$\pm$0.9 \unitlum and log$L\rm_X$ = 42.9$\pm$1.0 \unitlum, respectively.
The detection limits for our observations in the $EW$-FWHM plane (black solid line in Figure \ref{fig:FWHM_EW}) has been derived through simulations with our spectra on the detection of emission lines having different $EW$s and FWHMs.
According to this analysis, the observed difference on the FWHM and $EW$ measurements among the AGN1 and AGN2 samples is not due to selection biases, as FWHM and $EW$ values
as large as measured in the AGN1 population could have been easily detected also in the AGN2 sample, if present.

We have also tested whether the results of the above described analyses would change if those AGN2 having relevant starburst or LINER component in their spectra (see sect. \ref{sec:classAGN}) were excluded. As is shown in Table \ref{tab:diagnostici} no difference was found.

\begin{table*}
\centering
\begin{minipage}{140mm}
\caption{Average diagnostic values of the samples}
\label{tab:diagnostici}
\begin{center}
\begin{tabular}{@{}lllllll}
\hline
Sample& $J$       &log$F\rm_{X}$                                  & log$N\rm_{H}$ & log$L\rm_{X}$ &log($a/b$) &S/N \\
            &[mag] &[10$^{-12}$ erg s$^{-1}$ cm$^{-2}$]  & [cm$^{-2}$] & erg s$^{-1}$  &             &      \\
(1)        &(2)     &   (3)                                    &  (4)             &  (5)     & (6)        & (7) \\ 
 \hline
{\it Non Broad} AGN2              & 13.1$\pm$0.2 & 1.6$\pm$0.1 & 23.1$\pm$0.2 & 43.5$\pm$0.2&0.27$\pm$0.05& 32.8$\pm$2.5\\
N. Objects                               &  29                 & 29                 &  29                  &   29                        &   26                   &  29 \\
\\
{\it Broad} AGN2 & 13.1$\pm$0.4     & 1.5$\pm$0.1 & 22.8$\pm$0.2 & 42.7$\pm$0.2& 0.26$\pm$0.05& 35.8$\pm$3.5\\
N. Objects          & 13                       & 13                 & 13                    &  13                     & 13                &      13 \\           
\\
{\it Broad} AGN2\footnote{Excluding those objects showing a substantial starburst component (NGC 1365, NGC 6221, ESO 234$-$G050) and the LINER galaxy NGC 1052 (see sec. \ref{sec:classAGN}).} & 13.6$\pm$0.4     & 1.5$\pm$0.2 & 22.8$\pm$0.2 & 42.9$\pm$0.3& 0.30$\pm$0.06& 31.7$\pm$3.4\\
N. Objects          & \phantom{1}9       & \phantom{1}9  &\phantom{1}9   &  \phantom{1}9  & \phantom{1}9   &      \phantom{1}9 \\           
\hline
 \end{tabular}
\end{center}
Notes: (1) Sample; (2) average J-band magnitude (from 2MASS); (3) average 14--195 keV flux (from SWIFT70M); (4) average hydrogen column density; (5) average 14--195 keV luminosity; (6) average axis ratio ($a/b$ = major-axis/minor-axis) at the $B$-band isophote 25 mag/arcsec$^2$ \citep[from Hyperleda;][]{makarov14}; (7) average spectral signal to noise ratio. The number of objects is also reported. 
\noindent
\end{minipage}
\end{table*}

\section{Conclusions}

Thanks to NIR spectroscopy with LUCI/LBT, ISAAC/VLT and X-shooter/VLT we have been able to detect faint BLR components in 13 out of 41 AGN2
drawn from the {\it Swift}/BAT 70-month catalogue. 

The sample of AGN2 where the BLR components have been found does not show significant differences in spectral S/N ratio, X-ray fluxes and galaxy orientation, from the rest of the AGN2.
The possibility can therefore be excluded that relevant selection effects affect the AGN2 sample where the BLR lines have been detected.

The only significant differences are instead physical: a) no BLR was found in the most (heavily, log($N_{\rm H}) > $23.7 cm$^{-2}$) X-ray obscured sources and b)
AGN2 show smaller FWHMs and $EW$s if compared to AGN1.
The first result most likely implies that, as expected by the AGN unification models, the absence of the BLR in the optical and NIR spectra is linked to the presence of strong obscuration of the central parts of the AGN, where the BLR is located.
As far as the second result is concerned, according to the virial methods, where the black hole mass depends on the square of the BLR FWHM (and on the square root of the luminosity), smaller broad line region FWHM (and luminosity) values of the AGN2, if compared to the AGN1 population, imply that AGN2 have smaller black hole masses than AGN1. 
The study of the differences on the black hole mass distributions, Eddington ratios, and host galaxy properties of the AGN1 and AGN2 populations is beyond the scope of this work and will be discussed in following papers (Onori et al. in prep; Ricci et al. in prep.).

\section*{Acknowledgments}

We thank the anonymous referee for her/his helpful comments which have improved the paper. 
Based on observations collected at the European Organisation for Astronomical Research in the Southern Hemisphere 
under ESO programme(s) 088.A-0839(A), 090.A-0830(A)  and 093.A-0766(A).
We acknowledge the support from the LBT-Italian Coordination Facility for the
execution of observations and data distribution.
The LBT is an international collaboration among institutions in the United States,
Italy and Germany. LBT Corporation partners are: The University of Arizona on 
behalf of the Arizona university system; Istituto Nazionale di Astrofisica, Italy;
LBT Beteiligungsgesellschaft, Germany, representing the Max-Planck Society,
the Astrophysical Institute Potsdam, and Heidelberg University; The Ohio State
University, and The Research Corporation, on behalf of The University of Notre Dame,
University of Minnesota, and University of Virginia.
We acknowledge the usage of the HyperLeda (\url{http://leda.univ-lyon1.fr}) databases.
This publication makes use of data products from the Two Micron All Sky Survey, which is a joint project of the University of Massachusetts and the Infrared Processing and Analysis Center/California Institute of Technology, funded by the National Aeronautics and Space Administration and the National Science Foundation.
This research has made use of the NASA/IPAC Extragalactic Database (NED) which is operated by the Jet Propulsion Laboratory,
California Institute of Technology, under contract with the National Aeronautics and Space Administration. We acknowledge funding from PRIN/MIUR and PRIN/INAF.

\bibliographystyle{mnras}
\bibliography{mybib} 

\label{lastpage}

\appendix

\section{Comments on the BLR best fit models}
\label{sec:appendix1}

{\bf NGC 4395}

\noindent 
NGC 4395 is a nearby, small bulgeless (Sd) galaxy, known to host one of the smallest super massive BH (10$^4$-10$^5$ M$_{\odot}$) ever found. \cite{filippenko89} classified the nucleus of NGC 4395 as a type 1.8 or type 1.9 Seyfert using the relative intensities of the narrow and broad components. On the basis of the appearence of the Balmer lines, \cite{veron06} classified it as a type 1.8 Seyfert. 

\noindent The best fit of the most relevant emission lines of  NGC 4395 are shown in Figure \ref{fig:NGC4395}.
In the optical spectrum,  besides the narrow and broad components, evidence was found of intermediate
components in the [\ion{O}{III}] and [\ion{N}{II}] lines which are blueshifted of 259 \kms and 291 \kms, respectively. In the $J$-band spectrum it was not necessary or possible to
include a similar intermediate component due to the lower velocity resolution of 
LUCI ($\Delta v$ $\sim$ 220 \kms).\\

\begin{figure*}
\centering
\includegraphics[scale=.25, angle= -90]{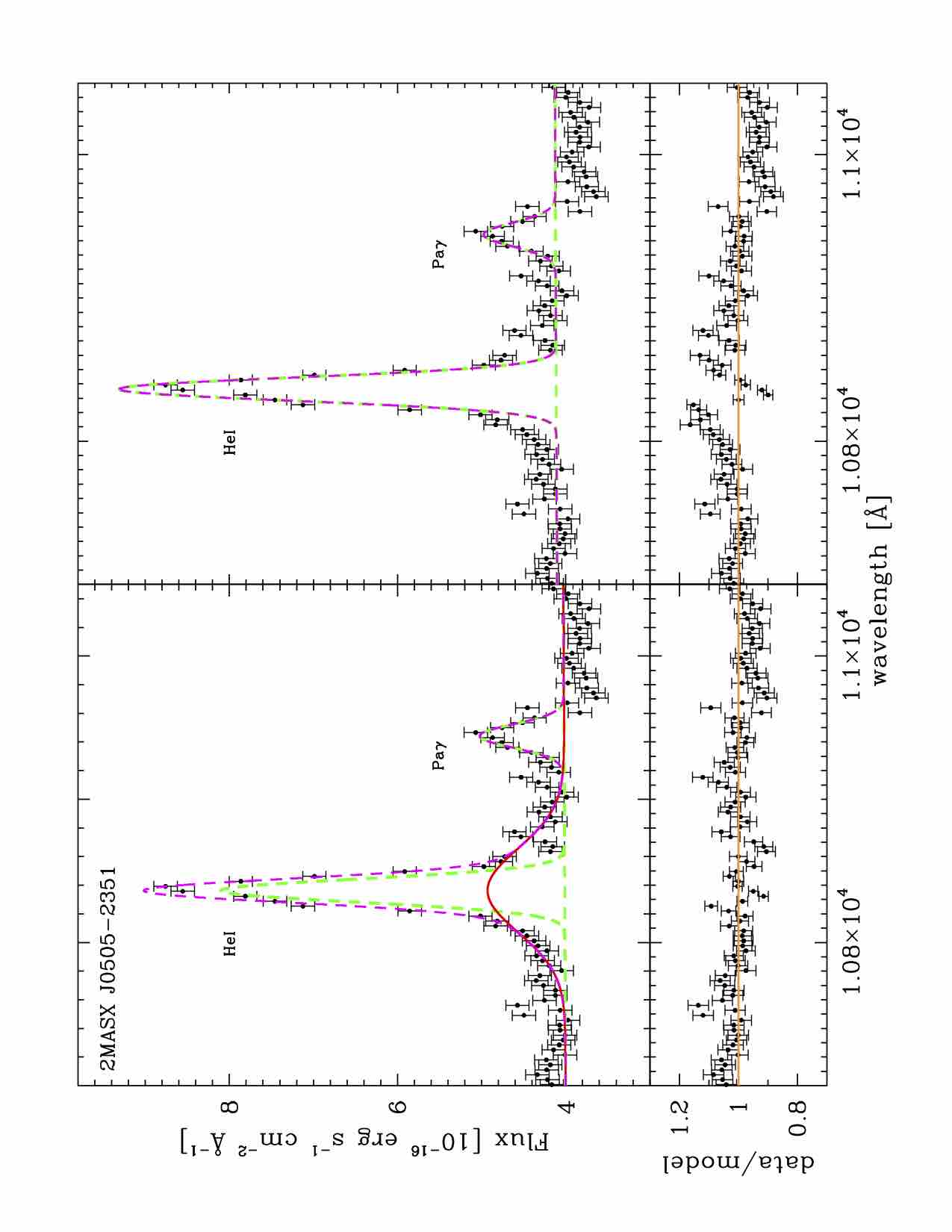}
\caption{{\it Left}: Best fit of the \ion{He}{I} and Pa$\upgamma$  lines of 2MASX J0505$-$2351 including the broad (FWHM = 1736 \kms) \ion{He}{I} component. {\it Right}: Same as before without including the broad \ion{He}{I} component. Lower panels show the data to model ratio.}
\label{fig:Res2MASXJ0523}
\end{figure*}

\noindent{\bf 2MASX J05054575$-$2351139}

\noindent 2MASX J05054575$-$2351139 is a nearby galaxy hosting a Seyfert 2 nucleus. The optical spectrum analysed by \cite{parisi09} shows very narrow H$\upalpha$ and H$\upbeta$ emission lines, confirming the Seyfert 2 classification. In the X-ray band, using {\it XMM} data, \cite{vasudevan13b} found a complex X-ray spectrum indicating a partially covered absorption. The authors also  suggested only moderate variations in the intrinsic luminosity and in the column density. Moreover, \cite{riccic14} reported the detection of Fe K$\upalpha$ emission in {\it Suzaku} data. 

The best fit of the most relevant emission lines of   2MASX J05054575$-$2351139 are shown in Figure \ref{fig:2MASXJ05-23_med}.
The ISAAC $J$-band LR spectrum shows an intense \ion{He}{I} line, well separated from a faint  Pa$\upgamma$ line. The Pa$\upbeta$ and [\ion{Fe}{II}]12570\AA\   lines have been also observed in MR mode. 
The Pa$\upbeta$ line shows a narrow component with FWHM = 145$^{+30}_{-26}$ \kms and a wider intermediate component with FWHM = 405$^{+67}_{-49}$ \kms. In contrast, the [\ion{Fe}{II}]12570\AA\ line can be modelled only by a single intermediate component  with FWHM = 413$^{+40}_{-38}$ \kms (compatible with the width of the intermediate  component of the Pa$\upbeta$ line). 
A narrow component of FWHM = 337 \kms was also added to take into account for the [\ion{Fe}{II}]12791\AA\ blending to the Pa$\upbeta$.
The \ion{He}{I} line shows a narrow component having  FWHM = 507$^{+49}_{-40}$ \kms (FWHM$\sim$269 \kms if the instrumental resolution is deconvolved) and a broad component having FWHM = 1823$^{+419}_{-318}$ \kms (FWHM $\sim$ 1772 \kms if the instrumental resolution is deconvolved).  The need of adding a broad \ion{He}{I} component can be seen in figure \ref{fig:Res2MASXJ0523} where the fits with residuals, with or without the broad \ion{He}{I} component, are shown.
The Pa$\upgamma$ line shows only a single narrow component.
We have also analysed the 6dF optical spectrum of 2MASX J05054575$-$2351139, having a spectral resolution $R$ = 1000 \citep{jones09}. 
All the lines show only a narrow component ($<$ 400 \kms if the instrumental resolution is subctracted).

In summary, we found a broad component only in the LR \ion{He}{I} line, having FWHM$\sim$1820 \kms, while, although the Pa$\upbeta$ line has been observed with a higher resolution (in the MR mode), it shows no signs of a broad component. Nevertheless both the H$\upalpha$+[\ion{N}{II}] and H$\upbeta$+[\ion{O}{III}] are well represented only by single components, all having FWHM in the range 413--466 \kms , in agreement with what was found in the optical by \cite{parisi09}. As the broad \ion{He}{I} component is significantly larger than the other components, it can be attributed to the BLR.\\

\begin{figure*}
\centering
\includegraphics[scale=.25, angle= -90]{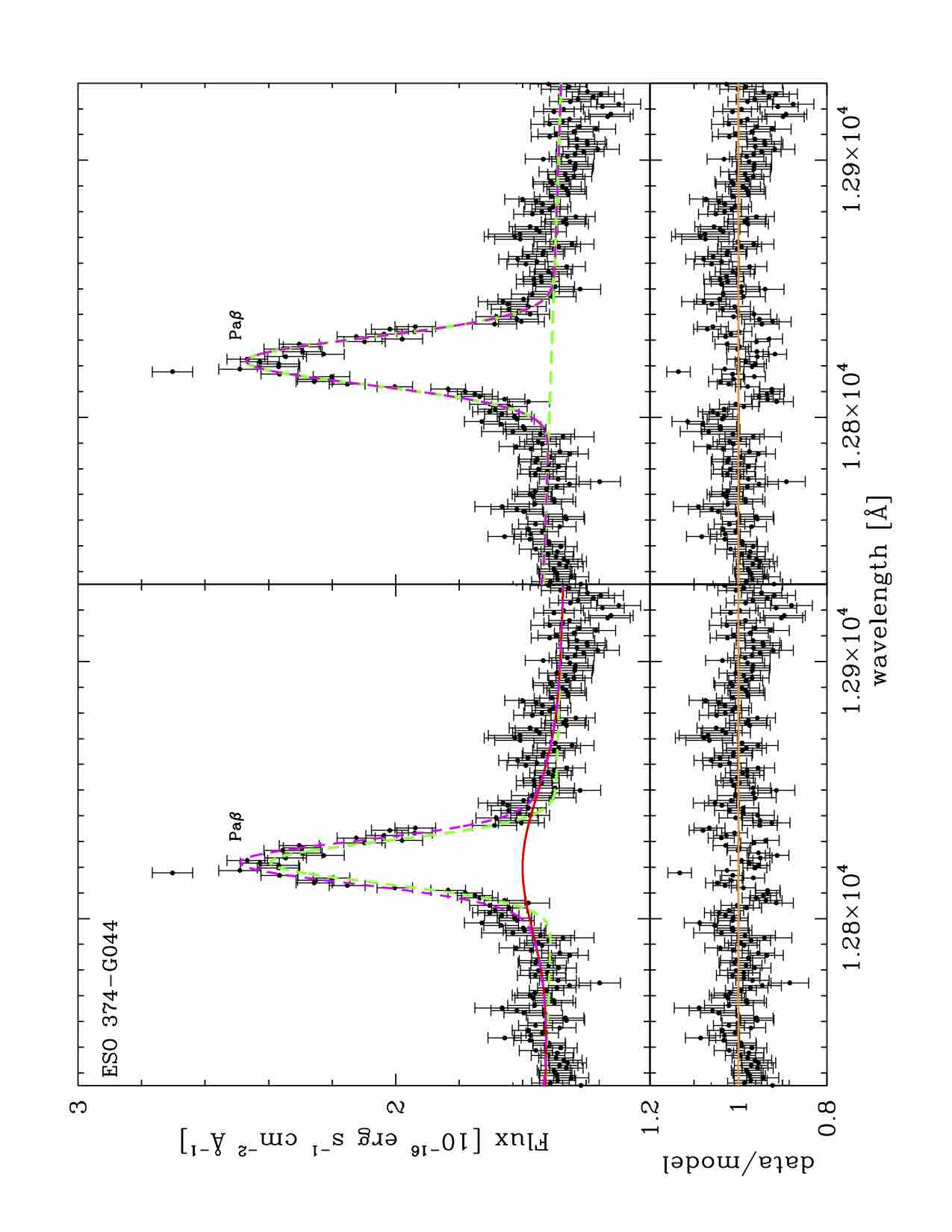}
\caption{{\it Left}: Best fit of the Pa$\upbeta$  lines of ESO 374$-$G044 including a broad (FWHM = 1413 \kms) Pa$\upbeta$ component. {\it Right}: Same as before without including the broad Pa$\upbeta$ component. Lower panels show the data to model ratio.}
\label{fig:ResESO374G44}
\end{figure*}
 
 \noindent {\bf ESO 374$-$G044}
 
 \noindent ESO 374$-$G044 is a nearby SBa galaxy in the Antlia galaxy cluster hosting a Seyfert 2 nucleus \citep{veron10}.
 
The best fit of the most relevant emission lines of   ESO 374$-$G044 are shown in Figure \ref{fig:ESO374G44}.
The MR spectrum in the region of the Pa$\upbeta$ line has been fitted by three components: the narrow (FWHM = 450 \kms) Pa$\upbeta$, the broad (1413 \kms) Pa$\upbeta$ and an intermediate (FWHM = 652 \kms) [\ion{Fe}{II}]12570\AA\ . The LR spectrum of the \ion{He}{I} region has been fitted by three components: the narrow (FWHM = 632 \kms; 463 \kms if deconvolved for the instrumental resolution) and broad (1202 \kms;  1122 \kms if deconvolved for the instrumental resolution) \ion{He}{I} and the narrow Pa$\upgamma$.
The optical spectrum does not show evidence of broad (or intermediate) line components, confirming the Seyfert 2 optical classification of \cite{veron10}. 
The need for the broad Pa$\upbeta$ component can be seen in Figure \ref{fig:ResESO374G44}, where the fits with residuals, including or not the broad Pa$\upbeta$ component, are shown.
The F test gives a probability of 0.06/0.01 that the improvement obtained with the inclusion of a broad Pa$\upbeta$ component is due to statistical fluctuations. We therefore conclude that the presence of these faint broad components in both \ion{He}{I} and Pa$\upbeta$ lines should not be ignored. \\

\noindent {\bf MCG $-$01$-$24$-$012}


\begin{figure*}
\centering
\includegraphics[scale=.25, angle= -90]{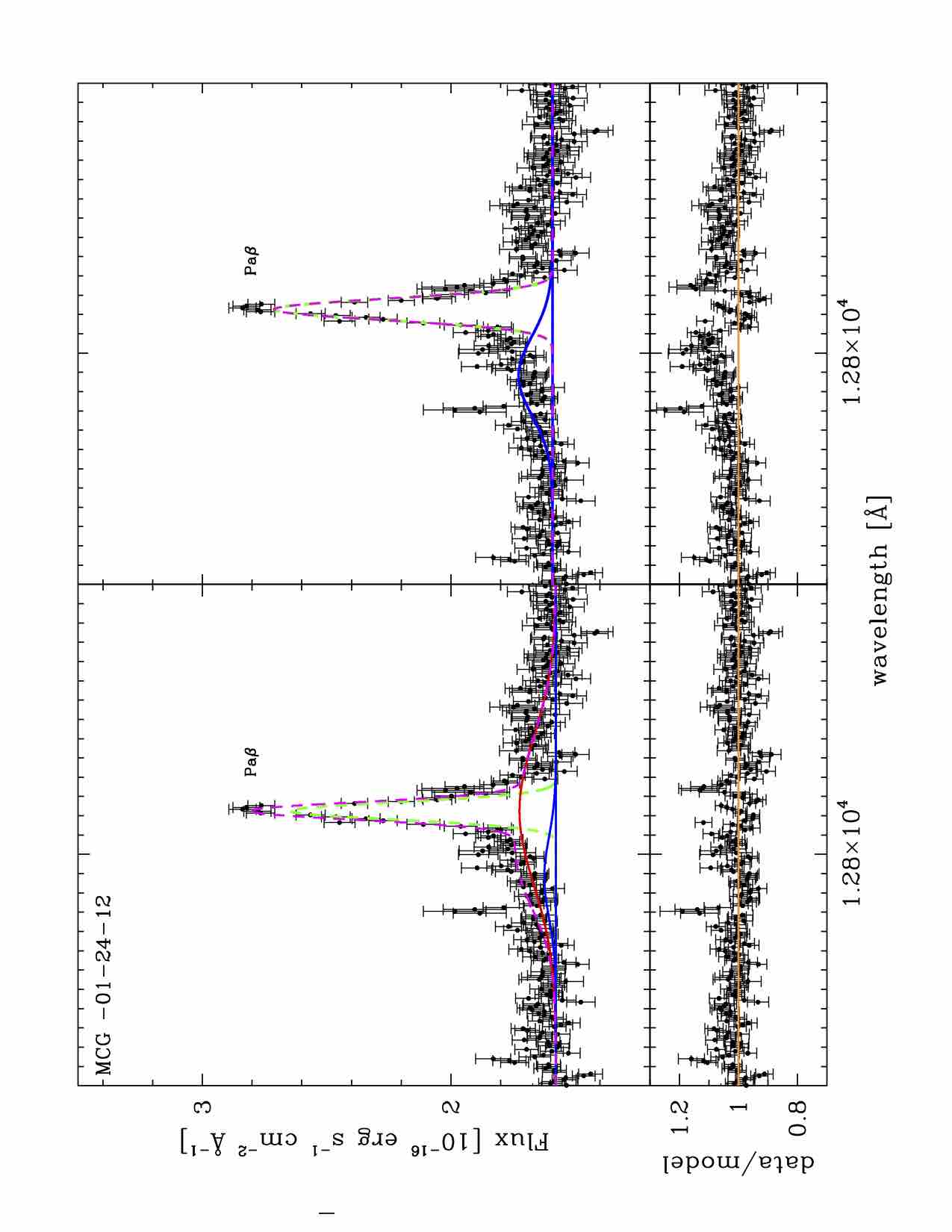}
\caption{{\it Left}: Best fit of the Pa$\upbeta$  lines of MCG $-$01$-$24$-$012 including a broad (FWHM = 2070 \kms) Pa$\upbeta$ component. {\it Right}: Same as before without including the broad Pa$\upbeta$ component. Lower panels show the data to model ratio.}
\label{fig:ResMCG012412}
\end{figure*}

\noindent MCG $-$01$-$24$-$012 is a nearby spiral galaxy hosting a Compton-thin ($N\rm_{H} \sim$ 7$\times$10$^{22}$ cm$^{-2}$) Seyfert 2 nucleus \citep{veron06}. It was detected in the hard X-rays by \cite{malizia02} with {\it Beppo}SAX/PDS observations, and it was identified as the counterpart of the X-ray source H0917$-$074, detected by \cite{piccinotti82}. Interestingly, the X-ray spectrum shows the presence of a Fe K$\upalpha$ emission line and an absorption feature at $\sim$8.7 keV which cannot be explained with the presence of a warm absorber. 
From {\it HST} images, \cite{schmitt03} found a resolved emission in the [\ion{O}{III}] which was explained as an extended (1.15 arcsec $\times$ 2.3 arcsec; 460 pc $\times$ 910 pc) NLR with the major axis along PA = 75$^\circ$. From WiFeS observation \cite{dopita15} traced further the extended NLR out to 3.7$\times$2.5 kpc and found a faint ring of HII regions of about 3 kpc in radius surrounding the NLR.
 
\noindent The best fit of the most relevant emission lines of  MCG $-$01$-$24$-$012 is shown in Figure \ref{fig:MGC011224}.
The 6dF optical spectrum \citep[R = 1000;][]{jones09} shows evidence of  intermediate components having a blueshift of $\Delta v$ = 189 \kms\ in the [\ion{O}{III}] lines and of $\Delta v$ = 111 \kms\ in the [\ion{N}{II}]+H$\upalpha$ lines with respect to the NLR components.
Similarly, in the NIR spectrum an intermediate component having a blueshift of $\Delta v$ = 806 \kms\ is present in the LR \ion{He}{I} and Pa$\upgamma$ lines and in the MR Pa$\upbeta$ line.
Moreover the inclusion of a broad (FWHM = 2070 \kms) Pa$\upbeta$  component is necessary. In Figure \ref{fig:ResMCG012412} the fit with and without the inclusion of this  broad Pa$\upbeta$ component is shown. The F test gives a probability of 2$\times$10$^{-15}$ that the improvement of the fit obtained with the inclusion of a broad Pa$\upbeta$ component is due to statistical fluctuations. \\

\noindent  {\bf MCG $-$05$-$23$-$016}

\noindent MCG $-$05$-$23$-$016 is a nearby X-ray bright S0 galaxy, optically classified as a Seyfert 1.9 \citep{veron80}. Moreover, clear evidence was found for a broad H$\upalpha$ in the polarized flux \citep{lumsden04}. Its X-ray spectrum resembles a classical Compton-thin Seyfert 2 galaxy and it shows both narrow and broad components in the iron K$\upalpha$ line \citep{reeves07}.

The best fit of the most relevant emission lines of MCG $-$05$-$23$-$016 is shown in Figure \ref{fig:MCG052316}. In the X-shooter optical spectrum, beside the narrow component, a clear evidence was found for a H$\upalpha$ BLR component with FWHM = 2232 \kms, while no evidence for a BLR component in H$\upbeta$ was found. Furthermore, both the ISAAC and X-shooter NIR spectra show clear evidence for a BLR component in the Pa$\upbeta$ and in the \ion{He}{I} (FWHM = 2134 \kms and FWHM = 2474 \kms, respectively).\\

\noindent {\bf Mrk 1210}


\begin{figure*}
\centering
\includegraphics[scale=.25, angle= -90]{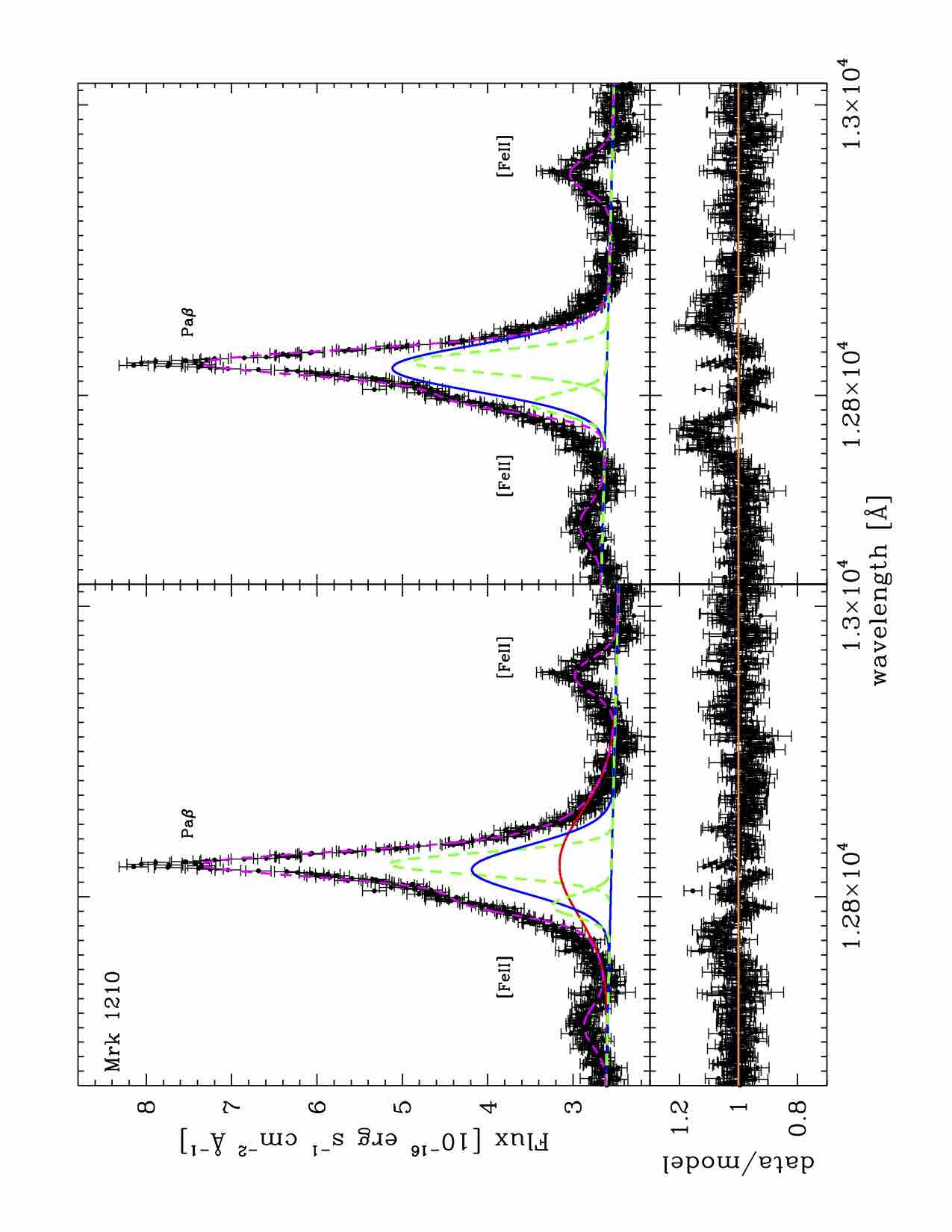}
\caption{{\it Left}: Best fit of the Pa$\upbeta$  lines of Mrk 1210 including a broad (FWHM = 2019 \kms) Pa$\upbeta$ component. {\it Right}: Same as before without including a broad Pa$\upbeta$ component. Lower panels show the data to model ratio.}
\label{fig:ResMrk1210}
\end{figure*}
 
\noindent  Mrk 1210 is a nearby Sa galaxy, hosting a Seyfert 2 nucleus \citep{veron10}. The spectral energy distribution peaking near 60 $\upmu$m made it a member of the sample of the so-called 60 $\upmu$m peakers galaxies (60PKs, \citealt{heisler99}). In the optical, \cite{storchibergmann98}  found  Wolf-Rayet features in the central 200 pc of the galaxy, indicating the presence of a circumnuclear starburst. Moreover, broad H$\upalpha$ and H$\upbeta$ components (FWHM$\sim$2400 \kms)  were detected in the polarized light (\citealt{tran92}; \citealt{tran95}). Near infrared spectra reported by \cite{Veilleux97} show a Pa$\upbeta$ profile characterized by a strong narrow component on top of a broad base with FWHM$\sim$1600 \kms, suggesting the presence of a hidden broad line region. 
However, \cite{mazzalay07} found broad components of similar shape both in permitted (H$\upbeta$ and Pa$\upbeta$) and in forbidden ([\ion{O}{III}] and [\ion{Fe}{II}]) emission lines, suggesting that the broad permitted component is not produced in a genuine high density BLR. In the X-ray band, Mrk 1210 is one of the very few cases of an AGN in transition between a Compton-thick, reflection dominated state, and a Compton-thin state. This transition could be attributed to a clumpy structure in the torus \citep{guainazzi02}. 

\noindent The ISAAC LR spectrum of Mrk 1210 shows an intense \ion{He}{I} line, well separated from the Pa$\upgamma$ line.
The region of the Pa$\upbeta$+[\ion{Fe}{II}]12570\AA\   lines has been observed in MR mode. In this region we have found six components: the Pa$\upbeta$ narrow, broad and intermediate  components, the [\ion{Fe}{II}]12570\AA\    narrow and  intermediate components and the [\ion{Fe}{II}]12791\AA\ narrow component in blend with the Pa$\upbeta$ line. The narrow components have a  FWHM = 414$^{+18}_{-18}$ \kms, 
while the intermediate components have a  FWHM = 902$^{+114}_{-61}$ \kms\ and show a blueshift of 136 \kms\ with respect to its narrow line components. The broad Pa$\upbeta$ line
has a FWHM = 1937$^{+118}_{-225}$ \kms. In Figure \ref{fig:ResMrk1210} the fits with and without the inclusion of the broad Pa$\upbeta$ component are shown. The F test gives a probability of 1$\times$10$^{-24}$ that the improvement of the fit is due to statistical fluctuations.

In summary, we found evidence for a broad component in the Pa$\upbeta$, having FWHM$\sim$1900 \kms, that can be attributed to the BLR emission. This finding is also supported by the detection of broad H$\upalpha$ and H$\upbeta$ in polarized light, indicating the presence of a hidden broad line region. Furthermore, the detection of a blueshifted intermediate component both in the [\ion{Fe}{II}] and in the Pa$\upbeta$ could be attributed to the presence of outflows in the NLR which are compatible to the observations of  the [\ion{O}{III}] and [\ion{Fe}{II}] line profiles by \cite{mazzalay07}.\\

\noindent {\bf NGC 1052}

\noindent NGC1052 is morphologically classified as an elliptical (E4) galaxy  and it is the brightest member of the Cetus I cluster. It has long been considered one of the prototypical LINERs and, after the detection of a faint broad component in the H$\upalpha$, it was classified as a LINER1.9 by \cite{ho97}. Moreover, a broad component in the H$\upalpha$ emission line (FWHM$\sim$2100 \kms) was detected also in polarization by \cite{barth99a}. Furthermore, \cite{mould12} presented  a  NIR  spectrum  of  NGC 1052,  with prominent Pa$\upbeta$ and [Fe II] emission lines. 
NGC 1052 shows H$_{2}$O megamaser emission \citep{claussen98} and variability in radio and X-rays  (\citealt{vermeulen03}; \citealt{hernandezgarcia13}). Radio observations of NGC 1052 revealed a double sided jet emerging from the nucleus \citep{kellermann98}. The X-ray spectrum is flat with a high absorbing column density \citep{guainazzi00} and there is a clear indication for the presence of an unresolved nuclear source in the hard bands (\citealt{satyapal05}). \cite{gonzalezmartin14} suggested that this object might be more similar to a Seyfert than to a LINER galaxy, from the X-ray point of view. 

\noindent The optical spectrum of NGC 1052 was taken with FOS/{\it HST}, and has a spectral resolution $R$ = 2800, corresponding to a  $\sigma_v\sim$ 80 \kms\  \citep{torrealba12}. All the typical narrow [\ion{O}{III}] and [\ion{N}{II}] lines are fairly well modelled by single components having a FWHM of 1091 \kms\ and 682 \kms\, respectively. These FWHM values are larger than observed in the typical narrow lines (see sect. 3), possibly due to the presence of unresolved outflow components.  The H$\upalpha$ shows a broad component of FWHM = 2193 \kms , in agreement with previous observations. The ISAAC LR spectrum shows an intense \ion{He}{I} with evidence of a broad component having
a FWHM of 2455 \kms\  (FWHM = 2417 \kms\ if the instrumental resolution is subtracted). We attributed this detection to the BLR emission. Finally we notice that, although the [\ion{Fe}{II}]12570\AA\ is quite intense, there is no evidence for the Pa$\upgamma$  and Pa$\upbeta$  emission lines (even in the MR observations), in contrast with what was found by \cite{mould12}. \\

\noindent {\bf NGC 1365}

\begin{figure*}
\centering
\includegraphics[scale=.25, angle= -90]{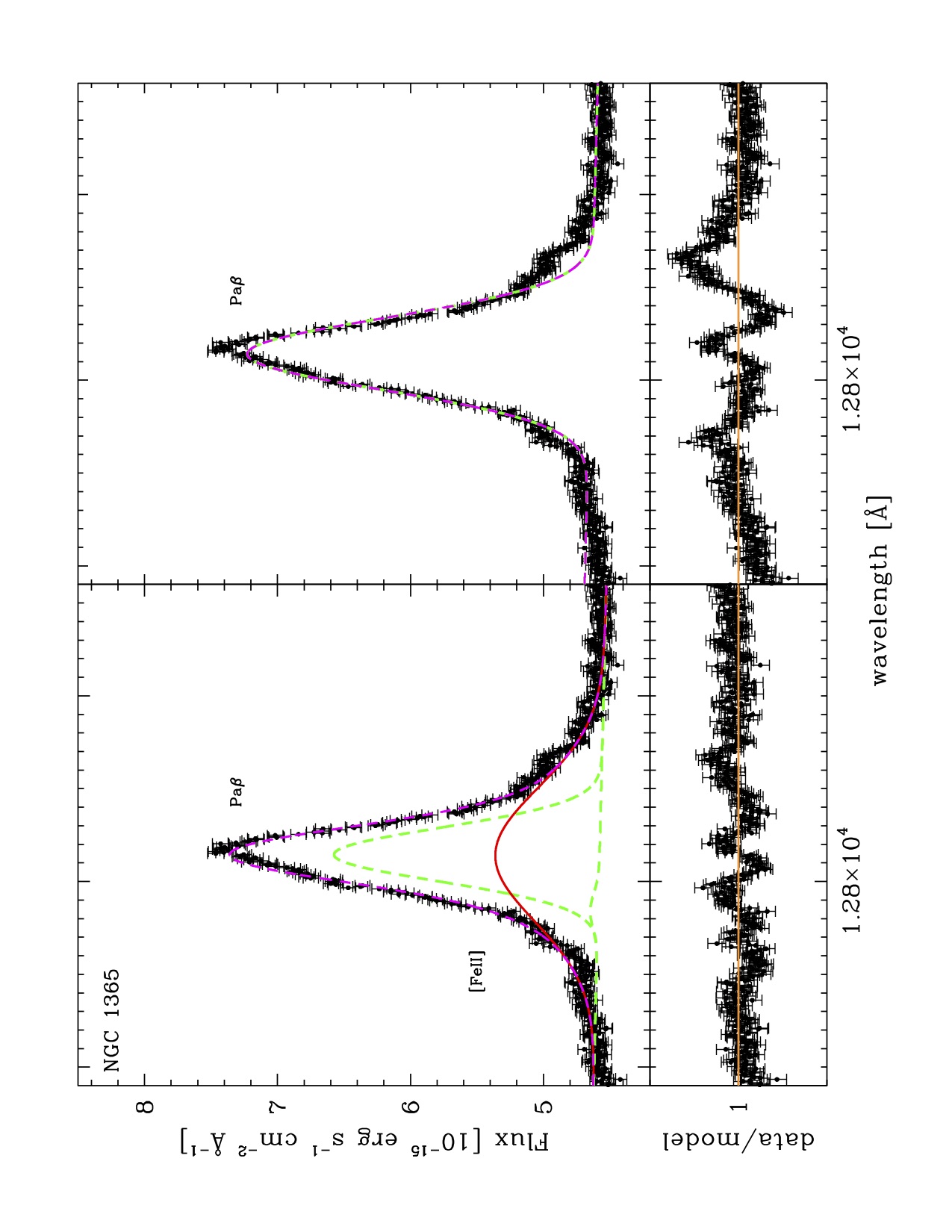}
\caption{{\it Left}: Best fit of the Pa$\upbeta$  lines of NGC 1365  including a broad (FWHM = 1791 \kms) Pa$\upbeta$ component. {\it Right}: Same as before without including a broad Pa$\upbeta$ component. Lower panels show the data to model ratio.}
\label{fig:ResNGC1365}
\end{figure*}

\noindent
NGC 1365 is a SB galaxy hosting a Seyfert 1.9 nucleus \citep{trippe10}. It displays strong X-ray variability (on time scales of hours to years) and evidence for a relativistically broadened iron line, indicative of a a rapidly rotating black hole (\citealt{risaliti09}; \citealt{risaliti13}). Furthermore, it displayed a complex and variable absorption, with rapid variations in column density, attributed to an occultation event originating in the BLR clouds (\citealt{risaliti05}; \citealt{risaliti07}). In the optical spectra of NGC 1365 there is some evidence of variability. \cite{veron80} classified it as a Seyfert 1.5 on the basis of a strong H$\upalpha$ detected in the nucleus. Moreover, \cite{schulz99} detected a broad component in H$\upbeta$ with a FWHM$\sim$1900 \kms. However, \cite{trippe10} found an extremely faint component in H$\upalpha$ and a purely narrow H$\upbeta$ that appears to be enhanced with respect to the [\ion{O}{III}] lines, indicating a very strong starburst emission component.

In the optical spectrum of \cite{jones09} we have found clear indication of a BLR component both in H$\upbeta$ and in H$\upalpha$ (FWHM$\sim$1586 \kms and FWHM$\sim$1700 \kms, respectively), in agreement with the results of \cite{schulz99}.
The ISAAC MR NIR spectrum of NGC 1365 shows evidence of a broad Pa$\upbeta$ component having a FWHM = 1972 \kms.  In Figure \ref{fig:ResNGC1365} the fit with and without the inclusion of this broad Pa$\upbeta$ component is shown. The F test gives a probability of 1$\times$10$^{-89}$ that the improvement of the fit is due to statistical fluctuations. We attribute this component to the BLR emission. The \ion{He}{I} line, was observed with ISAAC in LR mode and has been fitted by a single large component having a FWHM = 1243 \kms (1166 \kms if the instrumental resolution is subtracted). However, as the line is placed at the lower wavelength limit of the spectrum and it is not fully covered, we preferred not to include its best fit measurements in the dataset of the (secure) BLR detections (see Table \ref{tbl_BLR}). Finally, using line ratio diagnostic diagrams (see Section \ref{sec:classAGN}), we found some indication of a starburst component beside the AGN emission, in agreement with \cite{trippe10}.
\\

\noindent {\bf NGC 2992} 

\noindent NGC 2992 is a nearby Sa galaxy, highly inclined to our line of sight (about 70$^{\circ}$), showing a broad disturbed lane of dust in the equatorial plane. It is an interacting system, linked to NGC 2993 by a tidal tail with a projected length of 2.9 arcmin. Such an interaction could have induced a starburst activity \citep{glass97}. It is optically classified as Seyfert 1.9 \citep{veron80}, but its classification type has been observed to vary conspicuously in the past, leading to classifications ranging from Seyfert 2 to Seyfert 1.9 on the basis of a broad H$\upalpha$ with no corresponding H$\upbeta$ component in its nuclear spectrum \citep{ward80}, suggesting the existence of an obscured BLR. This was also confirmed later in the infrared by the detection of a broad component (FWHM$\sim$2900 \kms) in the Pa$\upbeta$ (\citealt{goodrich94}; \citealt{Veilleux97}). \cite{gilli00} correlated the presence (or the absence) of the broad component of H$\upalpha$ with the nuclear X-ray flux, suggesting that the observed optical variations were due to different phases of rebuilding of the central accretion disk (see also \citealt{Trippe08}). The X-ray flux and spectrum were observed to vary along 16 years of observations. These spectral variations were interpreted as evidence for radiation reprocessed by the molecular, obscuring torus and as an indication of a re-emergence of the AGN nuclear emission. \citep{gilli00}.
 
\noindent The optical spectrum of \cite{jones09} shows a broad component in H$\upalpha$, having FWHM = 3153 \kms, and a purely narrow H$\upbeta$. The ISAAC MR  spectrum of NGC 2992 shows intense  [\ion{Fe}{II}]12570\AA\ and Pa$\upbeta$ lines with evidence of a strong broad Pa$\upbeta$
component having FWHM = 2056 \kms. We therefore attribute it to the BLR emission. In order to properly reproduce the line profiles, it was necessary to include (likewise observed for the [\ion{O}{III}] and [\ion{N}{II}] optical lines) two intermediate [\ion{Fe}{II}]
components, tied together both in velocity and FWHM, having a velocity offset of 111 \kms\ with respect to the narrow component. Some other [\ion{Fe}{II}] lines were also included (see Figure \ref{fig:NGC2992}).
\\

\noindent {\bf 2MASX J18305065+0928414}

\noindent 2MASX J18305065+0928414 is a nearby galaxy, classified as a Seyfert 2 on the basis of the detection of pure narrow H$\upalpha$ and H$\upbeta$ in the optical spectra by \cite{masetti10}.

\begin{figure*}
\centering
\includegraphics[scale=.25, angle= -90]{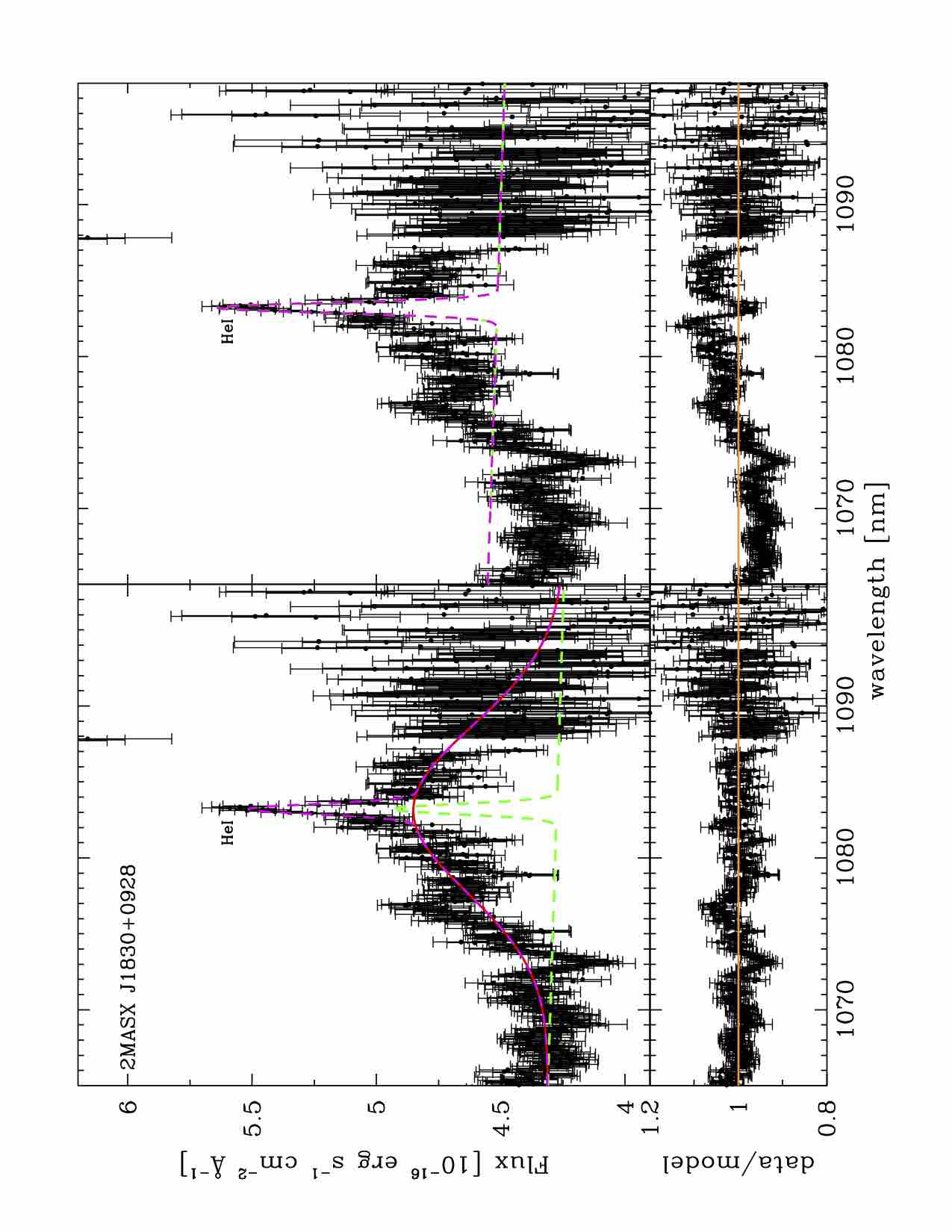}
\caption{{\it Left}: Best fit of the \ion{He}{I} line of 2MASX J18305065+0928414 including a broad (FWHM = 3513  \kms) \ion{He}{I} component. {\it Right}: Same as before without including a broad \ion{He}{I} component. Lower panels show the data to model ratio.}
\label{fig:Res2masxj1809}
\end{figure*}

\begin{figure*}
\centering
\includegraphics[scale=.25, angle= -90]{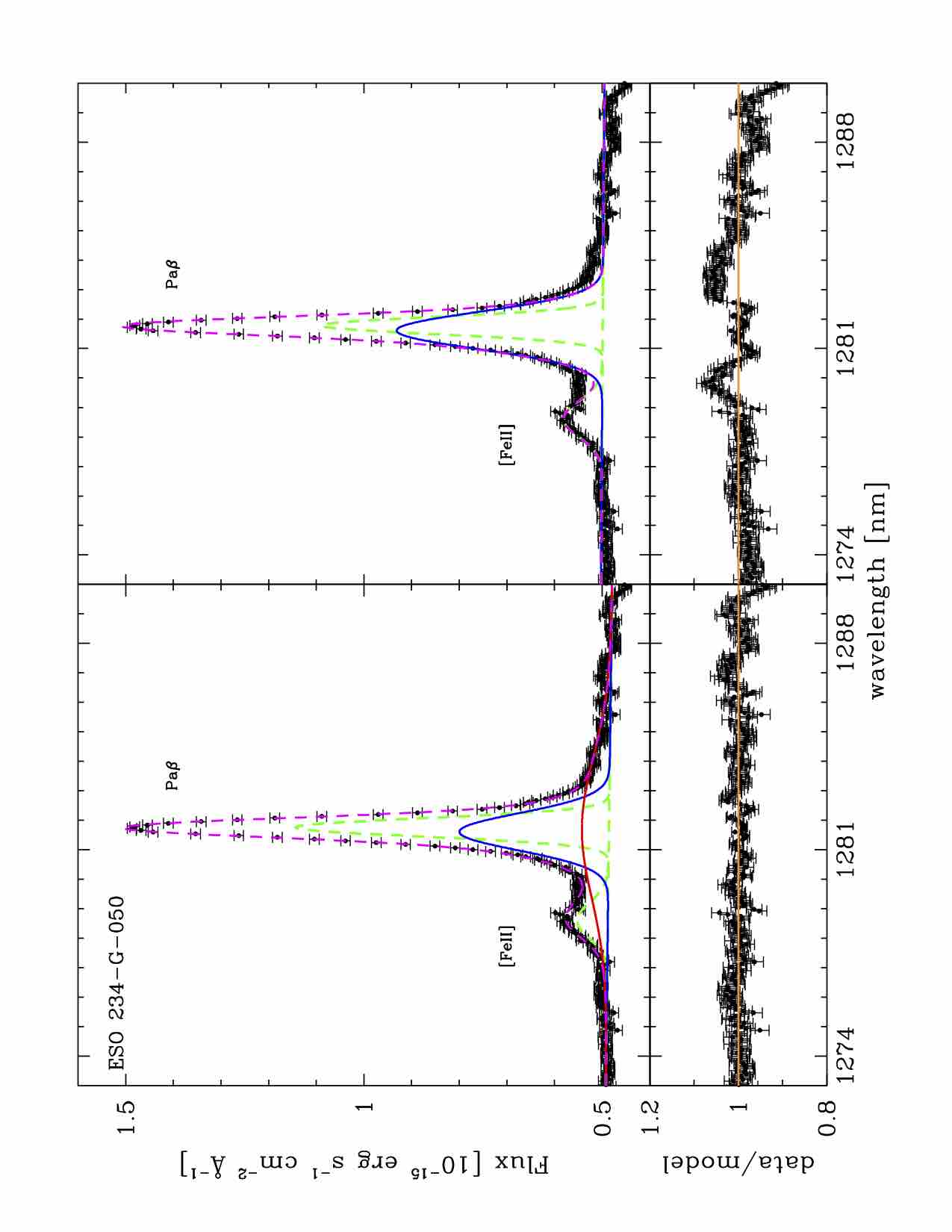}
\caption{{\it Left}: Best fit of the Pa$\upbeta$ line of ESO 234$-$G050 including a broad (FWHM = 1305 \kms) Pa$\upbeta$ component. {\it Right}: Same as before without including the broad Pa$\upbeta$ component. Lower panels show the data to model ratio.}
\label{fig:ResESO234G050}
\end{figure*}

\begin{figure*}
\centering
\includegraphics[angle=-90,scale=.25]{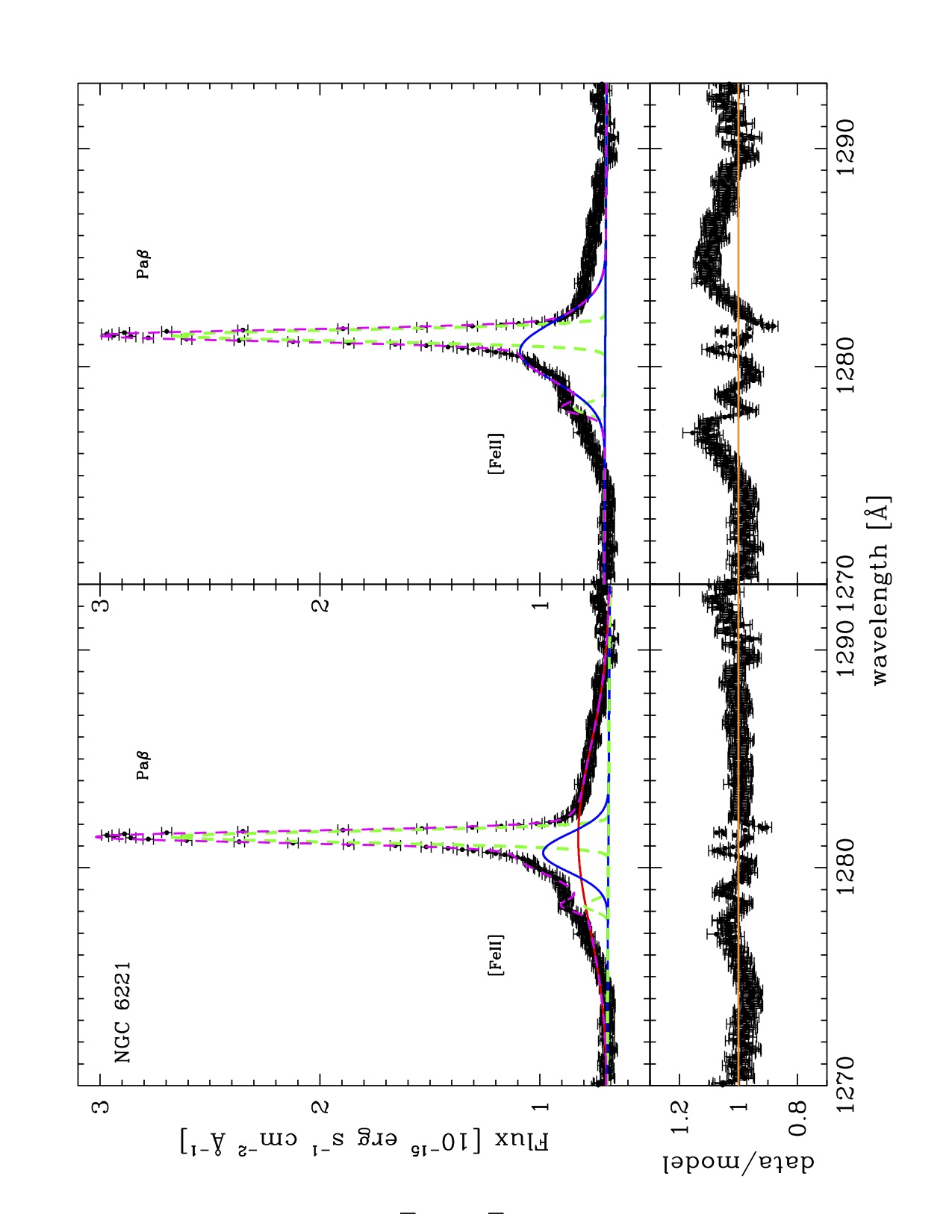}
\caption{{\it Left}: Best fit of the Pa$\upbeta$ line of NGC 6221 including a broad (FWHM = 2257 \kms) component. {\it Right}: Same as before without including a broad Pa$\upbeta$ component. Lower panels show the data to model ratio.}
\label{fig:ResNGC6221}
\end{figure*}

\noindent The X-shooter NIR spectrum of 2MASX J18305065+0928414 shows a faint \ion{He}{I} with the signature of a broad component which falls near a telluric absorption band (see Figure \ref{fig:2MASXJ18+09}). Moreover there is no evidence of the [\ion{Fe}{II}]12570\AA\ and Pa$\upbeta$ emission lines.
The \ion{He}{I} broad component has a FWHM of 3513 \kms\ while the narrow component is associated with the FWHM = 216  \kms\ component found in the
optical [\ion{N}{II}] lines.  In Figure \ref{fig:Res2masxj1809} the fit with and without the inclusion of this broad \ion{He}{I} component is shown. The F test gives a probability of
1$\times$10$^{-121}$ that the improvement of the fit is due to statistical fluctuations. 
In the H$\upalpha$ region of the optical spectrum, beside the standard NLR components, there is a very broad component
having a FWHM of 6194 \kms\ and blueshifted by 1103 \kms\ with respect to the narrow component. However, in order to
properly fit the spectrum, it is necessary to include also a broad H$\upalpha$ component having FWHM = 2660 \kms, with the same center of the
NLR component, which we attribute to the BLR. The F test gives a probability of
6$\times$10$^{-8}$ that the improvement of the fit obtained including this last BLR H$\upalpha$ line is due to statistical fluctuations. \\

\noindent {\bf ESO 234$-$G050 }

\noindent ESO 234$-$G050 is a blue compact dwarf (BCD) elliptical galaxy hosting a Seyfert 2 nucleus \citep{aguero93}. No detection of broad components in the permitted lines was reported so far.  In our optical X-shooter spectra we have found no evidence for broad components in the H$\upbeta$, while a faint broad H$\upalpha$ having FWHM$\sim$970 \kms\ has been detected.

\noindent The Pa$\upbeta$ has been fitted by three components: the NLR component having FHWM = 167 \kms, an intermediate component having 
FWHM = 343 \kms and blueshifted by 35 \kms\ with respect to the NLR (similarly as observed in the optical band) and a 
BLR component having FWHM = 1305 \kms\ , with the same center of the
NLR component. In Figure \ref{fig:ResESO234G050} the fit with and without the inclusion of the broad Pa$\upbeta$ component is shown. The F test gives a probability of 1$\times$10$^{-60}$ that the improvement of the fit is due to statistical fluctuations. We therefore attribute it to the BLR emission. Finally, using line ratio diagnostic diagrams (see Section \ref{sec:classAGN}), we found some indication of a starburst component beside the AGN emission.   
\\

\noindent {\bf NGC 6221}

\noindent NGC 6221 is a nearby, spiral galaxy classified as SBc(s) by \cite{devaucouleurs91}. The bar is clearly visible in the optical and in the infrared and lies at a PA of 118$^{\circ}$ with a length of $\sim$6 kpc. NGC 6221 forms an apparent physical pair with the spiral galaxy NGC 6215, which is $\sim$110 kpc distant, and is also possibly interacting with two nearby galaxies \citep{koribalski04}. While its optical spectrum resembles that of a typical reddened (A$_{V}$=3) starburst galaxy \citep{storchibergmann95}, in the X-ray band NGC 6221 shows a typical type 2 AGN spectrum, variable on timescales of days and years, with a 2--10 keV intrinsic luminosity of $L_{2-10}$ = 6.6$\times$10$^{41}$ \unitlum (\cite{levenson01}, Bianchi et al. in prep.). The presence of a sign of non-stellar activity in the optical band (the [\ion{O}{III}] shows a component broader and blueshifted with respect to the H$\upbeta$) and the early detection of NGC 6221 as an X-ray source \citep{marshall79}, motivated \cite{veron81} to propose a composite Seyfert 2/starburst classification for this object. 

No indications of broad (FWHM$>$1000 \kms) permitted emission lines were found both in the optical and in the NIR (see \cite{levenson01} and reference within). The optical X-shooter spectrum of NGC 6221 is quite complex. Beside the NLR components both in the H$\upbeta$ and H$\upalpha$ regions, three
intermediate components having a blueshift in the range 20--380 \kms\ have been fitted (see Figure \ref{fig:NGC6221}). It was also necessary to add
a BLR H$\upalpha$ component having FWHM = 1630 \kms. The F test gives a probability of 1$\times$10$^{-256}$ that the improvement of the fit is due to statistical fluctuations. Beside the presence of the NLR and one intermediate component (FWHM = 483 \kms) blueshifted by 176 \kms, the
Pa$\upbeta$ line has been fitted with a BLR component having FWHM = 2257 \kms. In Figure \ref{fig:ResNGC6221} the fit with and without the inclusion of the broad Pa$\upbeta$ component is shown. The F test gives a probability of 1$\times$10$^{-118}$ that the improvement of the fit is due to statistical fluctuations \citep[see][for a more detailed discussion]{lafranca16}. Moreover, using line ratio diagnostic diagrams (see Section \ref{sec:classAGN}), we confirm the presence of a starburst component beside the AGN emission.\\

\noindent {\bf NGC 7314}

\noindent  NGC 7314 is a barred spiral galaxy (SABb) hosting a Seyfert 1.9 active nucleus \citep{zoghbi13}, confirmed by the the detection of a broad H$\upalpha$ in the {\it HST} spectrum. Moreover, a clear evidence was found for a broad H$\upalpha$ in the polarized flux \citep{lumsden04}. The X-ray behaviour of the source is extreme, with a strong variability on all observed timescales and it is thought to be a type 2 counterpart to the NLS1 class \citep{dewangan05}. Furthermore, X-ray observations (from {\it ASCA}, {\it Chandra} and {\it XMM}) revealed a broad component in the Fe K$\upalpha$ at 6.4 keV. Interestingly, the line response to the rapid variations of the continuum is different for the narrow and the broad components, suggesting different origins of the corresponding gas \citep{ebrero11}.   

The best fit of the most relevant emission lines of NGC 7314 are shown in Figure \ref{fig:NGC7314}. In the X-shooter optical spectrum, beside the narrow and broad components of H$\upbeta$ and H$\upalpha$  (FWHM = 1097 \kms and FWHM = 1330 \kms, respectively) evidence was found of intermediate components. In particular, in the H$\upbeta$+[\ion{O}{III}] region two intermediate components with a blueshift of 47 \kms and 67 \kms respectively were found. In the H$\upalpha$+[\ion{N}{II}] region one intermediate component, with a blueshift of  63 \kms was found. In the NIR spectrum, beside the presence of the narrow component and one intermediate component having a blueshift of 42 \kms, the Pa$\upbeta$ line clearly shows a BLR component having FWHM = 1348 \kms.

\vfill\eject
\section{Best fits figures of the AGN2 without secure evidence of BLR components}
\label{sec:appendix2}


\begin{figure*}
\includegraphics[scale=0.4]{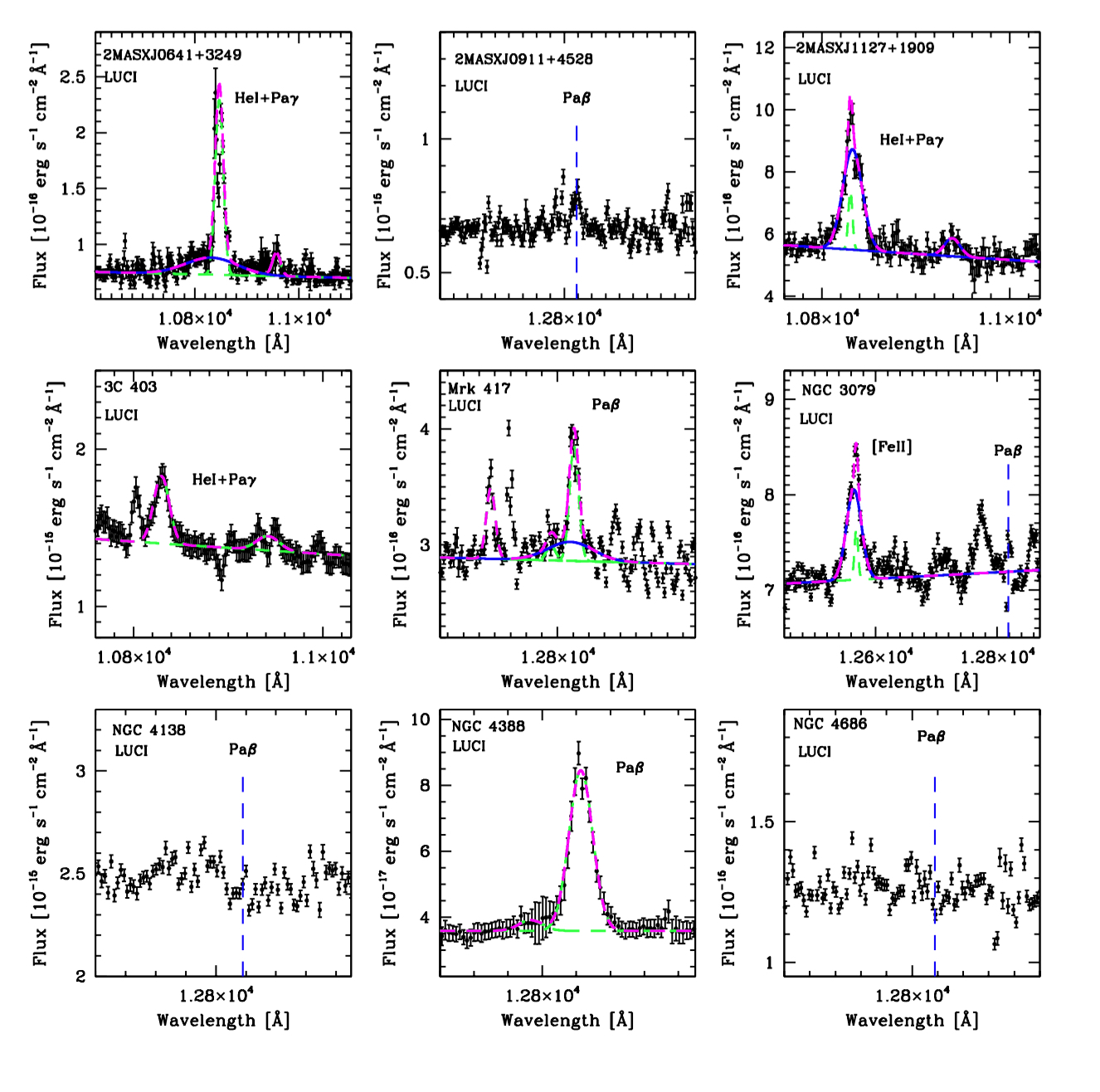}
\caption{Lines fit of LUCI `non-broad' AGN2}
\label{fig:narrowluci}
\end{figure*}


\begin{figure*}
\includegraphics[scale=0.4]{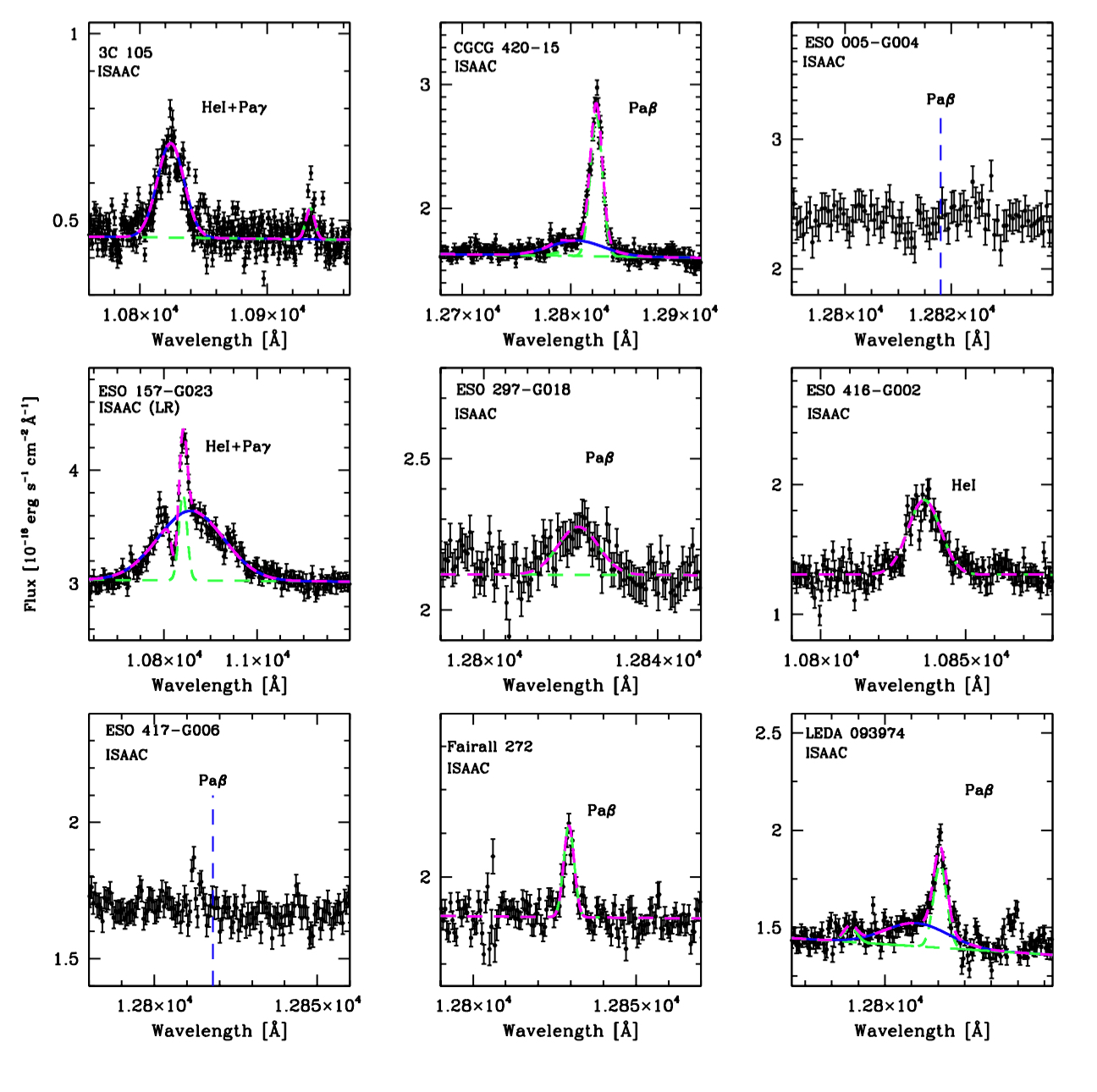}
\caption{Lines fit of ISAAC `non-broad' AGN2}
\label{fig:spettriIsaac}
\end{figure*}

\clearpage

\begin{figure*}
\includegraphics[scale=0.4]{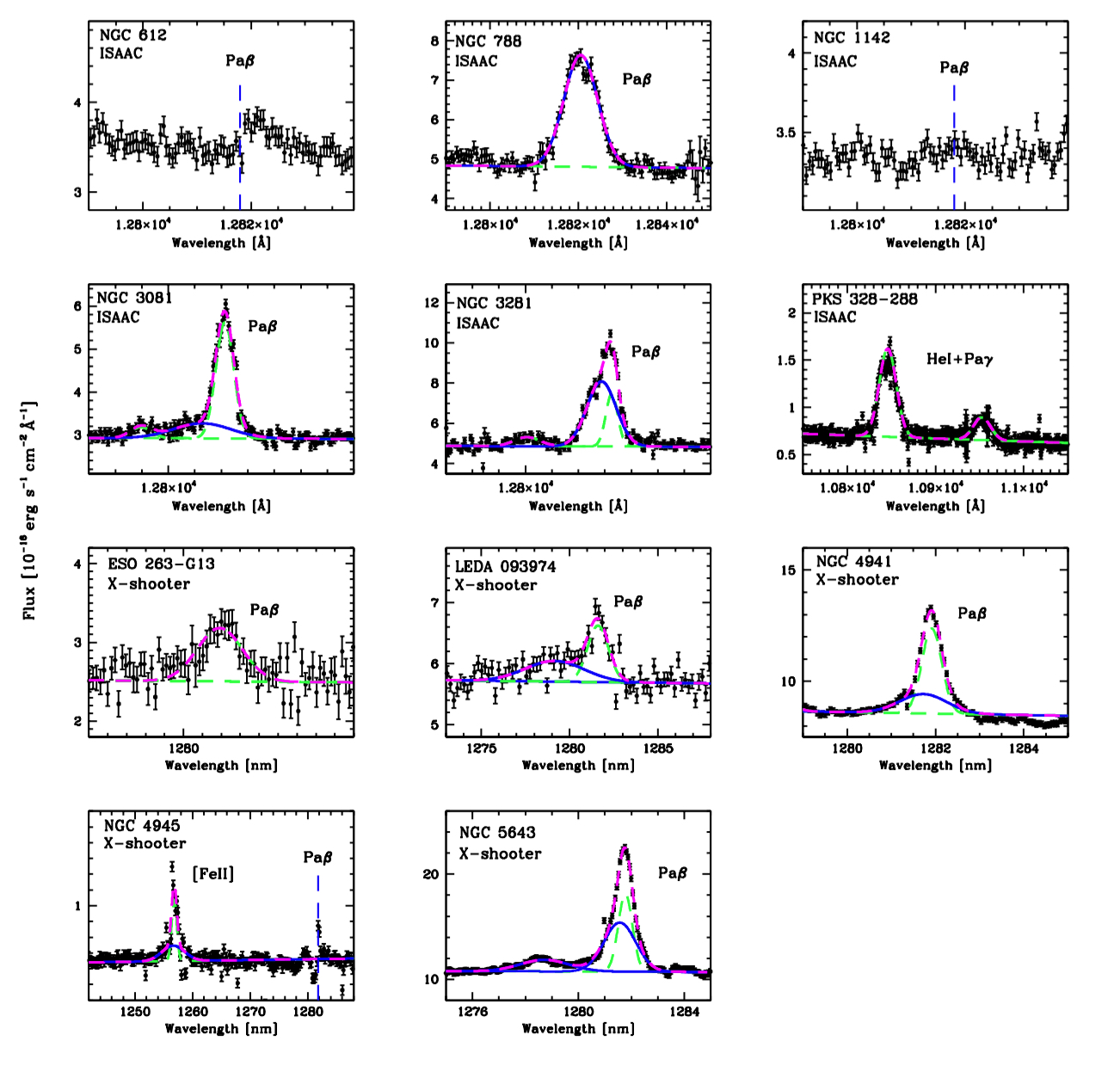}
\caption{Lines fit of ISAAC and X-shooter `non-broad' AGN2}
\label{fig:spettriIsaacXsh}
\end{figure*}
\clearpage

\section{Best fits values of the optical and NIR emission lines of the whole sample}
\label{sec:appendix3}

\clearpage
\begin{landscape}
\begin{table}
\begin{minipage}{140mm}
\caption{FWHM measurements of the emission lines in LUCI spectra.}
\label{tbl_FWHMluciSpettri}
\resizebox{1.7\textwidth}{!}{
\begin{tabular}{@{}llcccccccccccccccc}
\hline
Object & Redshift & Cl & S/N & Comp.  & [\ion{S}{III}]   & [\ion{S}{III}]   & Pa$\upepsilon$ & [\ion{C}{I}] & [\ion{S}{VIII}]  & Pa$\updelta$ & \ion{He}{II} & \ion{He}{I}  & Pa$\upgamma$ & [\ion{P}{II}]       & [\ion{S}{IX}] & [\ion{Fe}{II}] & Pa$\upbeta$ \\
           &               &       &        &            &$\lambda$9069\AA\ & $\lambda$9531\AA\ &  &$\lambda$9853\AA\  &$\lambda$9911\AA\ &    &$\lambda$10122\AA\ &$\lambda$10830\AA\ &  &$\lambda$11886\AA\ &$\lambda$12523\AA\  &$\lambda$12570\AA\    &     \\
(1)      & (2)          & (3)  & (4)  &  (5)      &  (6)         &  (7)     &  (8)    &  (9)    &  (10)    &  (11) &  (12)   & (13)   &  (14)  &  (15)       &  (16)                     &  (17)                         &  (18)   \\
\hline
2MASX J06411806+3249313\footnote{The intermediate component shows a $\Delta v$ = 423 \kms} 
 & 0.048 & 2 & 21 & N& 475$^{+86}_{-52}$& 475$^{+86}_{-52}$ &  475$^{+86}_{-52}$      & - & - & -  &-  &475$^{+86}_{-52}$        &475$^{+86}_{-52}$  &- & -  & - & -  \\
&          &    &      & I & -                            &                                & - & -  &-   & -   &-   &3080$^{+695}_{-414}$ &-                                &-  & -   &  -  &-    \\
2MASX J09112999+4528060 &0.0269 & 2 & 30 &N  & - & 271$^{+16}_{-15}$     &  181$^{+65}_{-48}$   & -     &   -       & -  & -  & -  & -  &   -        &-   & -  &- \\
2MASX J11271632+1909198  &0.1057  &1.8& 18&N &511$^{+15}_{-27}$ &511$^{+15}_{-27}$  &  -     &   - & -  &  -  & -    &511$^{+15}_{-27}$     &511$^{+15}_{-27}$ & -   & - &   -  &  -   \\
3C 403                                     &0.0589  &2  &  20 &N &                  -          &510$^{+57}_{-51}$  &  -      & -  & - & -   &  - & 510$^{+57}_{-51}$     & -& -&-& - & -\\
                                                 &           &     &       & I  & -                          &                            -   & -       & -  & -  &  -  &  - & -                                 &732$^{+362}_{-277}$ &-&- &- &-   \\
Mrk 417                                    &0.0327 &2   & 26 &N  & -   &370$^{+8}_{-15}$&370$^{+8}_{-15}$& - &399$^{+150}_{-114}$ &399$^{+109}_{-84}$&-&370$^{+8}_{-15}$&370$^{+8}_{-15}$ & - &370$^{+8}_{-15}$&370$^{+8}_{-15}$&370$^{+8}_{-15}$ \\
 &           &     &      & B   & - &-                           & -                          & - & - &  - &-& 921$^{+270}_{-174}$ \footnote{This component is centered on its narrow component, but it is on the edge of telluric absorption, thus we did not include it in the `broad' AGN2 sample.} & -& -& -&- &- \\
NGC 3079                                &0.0036&  2 & 52 &N & - &554$^{+50}_{-44}$& - &467$^{+44}_{-42}$ & - & - & - & - & - & - & - &167$^{-26}_{+30}$& -  \\
                                                 &               &     &      & I  & - & -                            &- &1434$^{+82}_{-60}$& - & - & - & - & - & - & - &690$^{+30}_{-30}$ & -  \\ 
NGC 4138                                & -              &1.9& 33 &N & - & -                            & - &    -                          & -  & - & - & - & - & - & - & -                            & - \\ 
NGC 4388                                & 0.0089         &  2 & 44 &N & - &487$^{+6}_{-6}$&487$^{+6}_{-6}$& - & 318$^{+58}_{-48}$&363$^{+17}_{-16}$&363$^{+17}_{-16}$&363$^{+17}_{-16}$ &363$^{+17}_{-16}$    & 464$^{+102}_{-77}$&363$^{+17}_{-16}$ &363$^{+17}_{-16}$ &363$^{+17}_{-16}$ \\
                                                 &               &     &      &I   & - &-                        & -                        & - &-                               & -                           & -                            & 956$^{+69}_{-61}$ \footnote{This broad component, although well centered with the systemic redshift frame, was not included among the BLR detections as the Pa$\upbeta$ emission line is not detected.}&568$^{+152}_{-131}$ &-   & -& - &- \\
NGC 4395                                & 0.0014  &  2 & 33 &N  & - &389$^{+7}_{-6}$&389$^{+7}_{-6}$&251$^{+130}_{-78}$&363$^{+138}_{-88}$&211$^{+6}_{-6}$&211$^{+6}_{-6}$&348$^{+5}_{-6}$&348$^{+5}_{-6}$&172$^{+87}_{-60}$ & - &235$^{+7}_{-6}$&235$^{+7}_{-6}$\\
                                                 &               &     &      &B & -  &                  -     & -                          & -                             &                           -    &975$^{+76}_{-72}$&589$^{+263}_{-140}$&1350$^{+93}_{-70}$&1591$^{+310}_{-245}$& - & - & -& 879$^{+29}_{-34}$ \\   
NGC 4686                                &  -     & XB &25  &N &-&-&-&-&-&-& -& - &- &-& -&- &-\\                                                           
\hline
 \end{tabular}
 }
Notes: (1) Source name; (2) Redshift as reported in Table \ref{tbl:classAGN}; (3) SWIFT70M optical spectral classification; (4) S/N near the Pa$\upbeta$; (5) line components; N: NLR, B: BLR, I: intermediate (see sect \ref{sec:lines} for the classification criteria); (6) to (18) FWHM (\kms) not deconvolved for instrumental resolution.
\noindent
\end{minipage}
\end{table} 
\end{landscape}

\clearpage
\begin{landscape}
\begin{table}
\begin{minipage}{140mm}
\caption{Equivalent width measurements of the emission lines in LUCI spectra.}
\label{tbl_EWluciSpettri}
\resizebox{1.7\textwidth}{!}{
\begin{tabular}{@{}llcccccccccccccccc}
\hline
Object & Redshift & Cl & S/N & Comp.  & [\ion{S}{III}]   & [\ion{S}{III}]   & Pa$\upepsilon$ & [\ion{C}{I}] & [\ion{S}{VIII}] & Pa$\updelta$ & \ion{He}{II} & \ion{He}{I}  & Pa$\upgamma$ & [\ion{P}{II}]       & [\ion{S}{IX}] & [\ion{Fe}{II}] & Pa$\upbeta$ \\
           &               &       &        &            &$\lambda$9069\AA\  &$\lambda$9531\AA\ &  &$\lambda$9853\AA\  &$\lambda$9911\AA\ &    &$\lambda$10122\AA\ &$\lambda$10830\AA\ &  &$\lambda$11886\AA\ &$\lambda$12523\AA\  &$\lambda$12570\AA\  &  \\
(1)      & (2)          & (3)  & (4)  &  (5)      &  (6)         &  (7)     &  (8)    &  (9)    &  (10)    &  (11) &  (12)   & (13)   &  (14)  &  (15)       &  (16)                     &  (17)                         &  (18)   \\
\hline
2MASX J06411806+3249313 &  0.048   & 2 & 21 & N& 11.0& 34.5  & 9.5   & - & - & -  &-     &40.0  &5.6  &-  & -  &- & -   \\
                                                &             &     &      & I & -      &    -      & -     & -  &-   & -   &-   &25.1  &-      &-   & - & -&-    \\
2MASX J09112999+4528060 & 0.0269   & 2 & 30 &N  &    -   & 9.6  &  1.4   & -     &   -  & -  & -  & -  &  &-      &-   & - & -  \\
2MASX J11271632+1909198  &0.1057 &1.8& 18 &N & 7.3   &24.6   &  -     &  -  & -   &  -  & -    &18.0& 2.5 & -   & - &   -  &  -    \\
3C 403                                     &0.0589   &2  &  20 &N &   -     &11.0   &   -     & -  & - & -   &  - & 6.4  & -   & -&-& - & -\\
                                                 &           &     &       & I  & -      &   -     & -       & -  & -  &  -  &  - & -      &2.0&-&- &- &-   \\
Mrk 417                                    &0.0327 &2   & 26 &N  & -      &22.5  & 5.2    & - &1.7 &1.6& - &4.4   &3.7  & - &2.7 &1.3&6.1\\
                                                 &           &     &      & B   & - &-                           & -                          & - & - &  - &-&27.2& -& -& -&- & -  \\
NGC 3079                                &0.0036&  2 & 52 &N & - &2.4 & - &1.1 & - & - & - & F & - & - & - & 0.56 & -  \\
                                                 &               &     &      & I  & - & -    &- &3.5 & - & - & - & - & - & - & - &4.1& -  \\  
NGC 4138                                & -              &1.9& 33 &N & - & -                        & - &    -                          & -  & - & - & - & - & - & - & -   & - \\ 
NGC 4388                                & 0.0089        &  2 & 44 &N & - &127.2 &20.8 & F & 2.4 &4.9 &8.8 &19.04 &  8.11   & 5.6 &4.1 &10.5 &23.3 \\
                                                 &               &     &      &I   & - &-        & -      & -  &-     & -     & -    & 62.08 &3.05&-   & -& - &- \\
NGC 4395                                &  0.0014    &  2 & 33 &N  & - &79.0&10.9&1.0 &1.9 &3.8 &3.7 &104.1&16.1&2.0 & - &13.2 &16.9\\
                                                 &               &     &      &B & -  &     - & -     & -   & -    &12.9&4.8 &58.7&16.3 & -    & - & -& 40.9 \\   
NGC 4686                                &  -    & XB &25  &N &-&-&-&-&-&-& -& - &- &-& -&- &-\\
 \hline
 \end{tabular}
 }
Notes: (1) Source name; (2) Redshift as reported in Table \ref{tbl:classAGN}; (3) SWIFT70M optical spectral classification; (4) S/N near the Pa$\upbeta$; (5) line components; N: NLR, B: BLR, I: intermediate (see sect \ref{sec:lines} for the classification criteria); (6) to (18) $EW$ in \AA\. 
\noindent
\end{minipage}
\end{table} 
\end{landscape}

\clearpage
\begin{landscape}
\begin{table}
\begin{minipage}{140mm}
\caption{Flux measurements of the emission lines in LUCI spectra.}
\label{tbl_FLUXluciSpettri}
\resizebox{1.7\textwidth}{!}{
\begin{tabular}{@{}llcccccccccccccccc}
\hline
Object & Redshift & Cl & S/N & Comp.  & [\ion{S}{III}]   & [\ion{S}{III}]   & Pa$\upepsilon$ & [\ion{C}{I}] & [\ion{S}{VIII}] & Pa$\updelta$ & \ion{He}{II} & \ion{He}{I}  & Pa$\upgamma$ & [\ion{P}{II}]       & [\ion{S}{IX}] & [\ion{Fe}{II}] & Pa$\upbeta$ \\
           &               &       &        &            &$\lambda$9069\AA\  &$\lambda$9531\AA\ &  &$\lambda$9853\AA\  &$\lambda$9911\AA\ &    &$\lambda$10122\AA\ &$\lambda$10830\AA\ &  &$\lambda$11886\AA\ &$\lambda$12523\AA\  &$\lambda$12570\AA\  &  \\
(1)      & (2)          & (3)  & (4)  &  (5)      &  (6)         &  (7)     &  (8)    &  (9)    &  (10)    &  (11) &  (12)   & (13)   &  (14)  &  (15)       &  (16)                     &  (17)                         &  (18)  \\
\hline
2MASX J06411806+3249313 &  0.048    & 2 & 21 & N& 0.9E-15&2.5E-15 & 0.7E-15 & - & -& -  &-  &2.8E-15 &0.4E-15 &-  & -  & -& -   \\
                                                &             &     &      & I & -         &       -       & -  & -  &-   & -   &-   &1.7E-15 &-      &-   & -   &  - &-    \\
2MASX J09112999+4528060  &0.0269    & 2 & 30 &N  &   -         &5.6E-15&0.8E-15  & -     &   -    & -  & -  & -  & -  &-  &-   &  -&  -   \\
2MASX J11271632+1909198  &0.1057   &1.8& 18 &N & 3.6E-15&11.7E-15 &  -     &  -  & -  &  -  & -    &8.9E-15&1.2E-15& -   & - &   -  &  -     \\
3C 403                                     &0.0589   &2  &  20 &N &   -     &13.2E-15   &   -     & -  & - & -   &  - & 8.5E-15  & -   & -&-& - & -\\
                                                 &           &     &       & I  & -      &   -     & -       & -  & -  &  -  &  - & -      &2.6E-15&-&- &- &-  \\
Mrk 417                                    &0.0327  &2   & 26 &N  & -      &6.8E-15 & 1.6E-15 & - &0.6E-15 &0.5E-15& - &13.8E-15  &1.1E-15 & - &0.8E-15&0.4E-15&1.2E-15 \\
                                                 &           &     &      & B  & -     &-             & -            & - & -            &  -         &-  &8.5E-15 & -& -& -&- &- \\                                                 
NGC 3079                                &0.0036 &  2 & 52 &N & - &1.2E-15& - &0.6E-15 & - & - & - & F & - & - & - & 0.4E-15 & - \\
                                                 &               &     &      & I  & - & -           &- &1.2E-15 & - & - & - & - & - & - & - &2.9E-15& -  \\  
 NGC 4138                                & -     &1.9& 33 &N & - & -                            & - &    -                          & -  & - & - & - & - & - & - & -                            & - \\                                             
 NGC 4388                                & 0.0089 &  2 & 44 &N & - &3.3E-15 &0.5E-15 & - &0.07E-15&0.1E-15 &0.3E-15 &0.6E-15 &0.3E-15 &0.2E-15 & 0.1E-15 &0.4E-15 &0.8E-15 \\
                                                 &               &     &      &I   & - &-        & -      & -  &-     & -     & -    &2.0E-15 &0.1E-15&-   & -& - &- \\
 NGC 4395                                & 0.0014  &  2 & 33 &N  & - &4.5E-13&6.1E-14&6.0E-15&1.2E-14&23.8E-15 &23.4E-15&70.2E-14&10.9E-14&13.8E-15& - &95.1E-15 &12.0E-14\\
                                                 &               &     &      &B   & -  &     -      & -          & -          & -           &80.7E-15&30.4E-15 &39.6E-14&11.0E-14& -    & - & -& 29.0E-14 \\   
NGC 4686                                &  -     & XB &25  &N &-&-&-&-&-&-& -& - &- &-& -&- &-\\                                            
\hline
 \end{tabular}
 }
Notes: (1) Source name; (2) Redshift as reported in Table \ref{tbl:classAGN}; (3) SWIFT70M optical spectral classification; (4) S/N near the Pa$\upbeta$; (5) line components; N: NLR, B: BLR, I: intermediate (see sect \ref{sec:lines} for the classification criteria); (6) to (18) emission line Flux in erg s$^{-1}$ cm$^{-2}$. 
\noindent
\end{minipage}
\end{table} 
\end{landscape}

\clearpage
\begin{table}
\begin{minipage}{140mm}
\caption{FWHM measurements of the emission lines in ISAAC spectra.}
\label{tbl_FWHMisaacSpettri}
\begin{tabular}{@{}llcccccccccc}
\hline
Object & Redshift & Cl & S/N & Comp.  & \ion{He}{I}  & Pa$\upgamma$ & \ion{O}{II}      & [\ion{P}{II}]                       & [\ion{S}{IX}]                     & [\ion{Fe}{II}]                     & Pa$\upbeta$\\
           &               &       & LR/MR       &            &$\lambda$10830\AA\         &           &$\lambda$11287\AA\    &$\lambda$11886\AA\ &$\lambda$12523\AA\ &$\lambda$12570\AA\ &    \\
(1)      & (2)          & (3)  & (4)  &  (5)      &  (6)         &  (7)     &  (8)    &  (9)    &  (10)    &  (11)  & (12)  \\
\hline
2MASX J05054575$-$2351139& 0.035  & 2  &  30/24  & N &  507$^{+49}_{-40}$ &  507$^{+49}_{-40}$ & - &  - & - & - &145$^{+30}_{-26}$ \\
                                              &              &    &              & I  & -                                 & -                             &-   & -  & - &413$^{+40}_{-38}$& 405$^{+67}_{-49}$\\
                                              &              &    &             & B &1823$^{+419}_{-318}$& -                             &-  & -  & - &-   & -  \\
3C 105                                  & 0.0884    & 2  & 27/13  & N & -                                 &200$^{+112}_{-72}$   & - & - & - & -  & -  \\
                                              &              &    &             & I  &   627$^{+55}_{-54}$   &   -    & - & - & - & -  & -    \\ 
 CGCG 420$-$015 \footnote{Intermediate components of FWHM$\sim$1100 \kms were found also in the [\ion{O}{III}] lines in the 6dF optical spectra.} 
                                             &0.029  &  2    & 64/ 31 & N &456$^{+14}_{-12}$       &  - &  - &  -   &634$^{+350}_{-284}$ &296$^{+10}_{-10}$  & 296$^{+10}_{-10}$      \\
                                             &            &        &             & I  &1331$^{+272}_{-134}$ & -   &-   & -  &                     -         &           -                    & 1317$^{+177}_{-186}$\\
  ESO 005$-$G004                  &0.006& 2    & 48/20    &N & -                                    & -   &  - & -    &          -                     & 225$^{+39}_{-49}$&  -                                 \\
ESO 157$-$G023\footnote{ Since we have found no signs of Pa$\upbeta$ and no optical spectra to validate our measurement of this \ion{He}{I} wide component, we decided to not include it in the broad lines  AGN2 sample.}   
                                            & 0.044 &   2   &47/37    & N  & 545$^{+85}_{-75}$ & -  &- & - & - &  -  & -  \\
                                           &              &       &              &  I &4043$^{+293}_{-273}$ & - & - & - &-  & -   &   - \\  
 ESO 297$-$G018                 &0.0253 & 2     & 47/38     & N  & -                                 & -  & - &- &     -  & 249$^{+43}_{-39}$& 249$^{+43}_{-39}$\\     
                                          &             &       &                 & I  &  929$^{+77}_{-78}$ \footnote{This broad component was not conservatively included among the BLR detections as  its narrow component was not resolved and no other broad components are observed in the NIR spectrum.}  & -  &  - &- &     -   &        -                     & -                        \\   
 ESO  374$-$G044                &0.0284 & 2    & 45/18       & N & 632$^{+29}_{-36}$   & 632$^{+29}_{-36}$   & - & 763$^{+474}_{-347}$&429$^{+153}_{-64}$&  - & 450$^{+27}_{-42}$ \\
                                           &           &       &                  & I  &      -                           &  -                              & - &   -  &  -   & 652$^{+74}_{-56}$ &  -   \\
                                           &           &       &                 & B   &   1202$^{+383}_{-221}$ & -                        &  - &-    &   -   &     -     & 1413$^{+318}_{-294}$   \\                                                                                       
ESO 416$-$G002                  &0.059 & 1.9 &45/17      & N  & 347$^{+34}_{-30}$           & -                        & - &-    &  -    & -         &   - \\
ESO 417$-$G006                 &0.0163    & 2   &60/42      & N  &   -                                    & -                       &-  &   -    & -    & 376$^{+45}_{-40}$ & -          \\
                                          &              &       &             & B   &   -                                    & -                       & -  &-       & -    &      -                         & -    \\ 
Fairall 272                         &0.022   & 2    &60/34    & N  & 793$^{+52}_{-101}$        & -                        & - &  -     &  -    & 422$^{+15}_{-30}$& 86$^{+15}_{-10}$   \\
LEDA 093974                    & 0.0239  &2    & 59/35  &N &555$^{+45}_{-42}$            & -                         & -  &  -        &555$^{+45}_{-42}$  &453$^{+12}_{-12}$ & 134$^{+20}_{-21}$    \\
                                         &               &      &            & I & 1058$^{+135}_{-124}$    & -                          & - & 1153$^{+492}_{-375}$ &   -  &                               & 639$^{+104}_{-90}$       \\                                          
MCG $-$01$-$24$-$012                  & 0.0197  & 2   &50/18     & N &616$^{+28}_{-27}$  & 616$^{+28}_{-27}$  & - &  -  &  -      & 381$^{+78}_{-117}$   & 245$^{+17}_{-16}$  \\
                                          &              &     &               & I &  1325$^{+347}_{-251}$& 1325$^{+347}_{-251}$   & - &    -    &  -       & 857$^{+121}_{-273}$ &1325$^{+347}_{-251}$  \\
MCG $-$05$-$23$-$016                  &0.0084   & 2  &123/45    & N&     -                               &            -                           &- &    -  & 409$^{+134}_{-110}$ &          -        & 247$^{+19}_{-18}$         \\
                                         &               &     &               & B &   1223$^{+90}_{-80}$ & 1451$^{+134}_{-149}$   &  - &  -  &   -    &      -          & 1165$^{+27}_{-18}$    \\
                                         &                &       &            & I &   3855$^{+172}_{-158}$  &    -                             & - &   -   &  -     &          -       &3695$^{+335}_{-196}$  \\
Mrk 1210                          & 0.014   & 2   &90/ 24    & N &   564$^{+8}_{-8}$  &   -                                & - & -&   -  &  414$^{+18}_{-18}$&414$^{+18}_{-18}$    \\
                                         &              &       &             & I  &     -                         & 791$^{+122}_{-116}$  & - &670$^{+382}_{-237}$ &760$^{+71}_{-56}$ & 902$^{+114}_{-61}$ & 902$^{+114}_{-61}$    \\
                                        &               &       &             & B & 1374$^{+73}_{-32}$  &     -                          & - &   -   & -     &                       -        & 1937$^{+118}_{-225}$  \\
NGC 612                         & -  &  2   &55/29     & N  & -                               & -                              & - & -    &  -     &    -                           &  -           \\
NGC 788                        &0.0135 &  2   & 69/ 39& N  & 556$^{+16}_{-15}$& 467$^{+153}_{-61}$  & - &  -    & -     &  213$^{+8}_{-7}$& 213$^{+8}_{-7}$ \\
NGC 1052                      & 0.0047     &  2    &67/48  & N  & 792$^{+61}_{-51}$ & -                                & - & -     &  -   & 693$^{+20}_{-20}$    &  -       \\
                                       &             &       &            & B  & 2455$^{+143}_{-128}$ & -                           &- &   -    &   -  &  -  & -      \\                                       
NGC 1142                       &0.0294  &  2   & 61/57   & N  &  -                   & -                           &  - &  -    &  -   &553$^{+49}_{-43}$ &  -         \\
                                       &            &       &             & I    & 934$^{+159}_{-119}$  &             -                  & - &  -      &    -  & -   &-\\     
NGC 1365                      &0.005 & 1.8  &88/35    & N  & -                                 & -                              & - &  - &  - &  -  &787$^{+8}_{-8}$  \\
                                       &            &       &               & B  &1243$^{+100}_{-92}$ &   1363$^{+67}_{-54}$  & 945$^{+38}_{-75}$ &  - & - &  - &1972$^{+85}_{-75}$  \\
NGC 2992                     &0.0077  & 2    & 63/43  & N & 528$^{+56}_{-58}$      &528$^{+56}_{-58}$          & - & - & - & 254$^{+3}_{-8}$&254$^{+3}_{-8}$   \\
                                      &             &       &            & I  &     -                               &           -                            & -  & - & - &869$^{+34}_{-56}$& 869$^{+34}_{-56}$     \\
                                    &                &       &           & B &3186$^{+586}_{-400}$ &  2989$^{+660}_{-488}$  &  - & - & - &   -                    &2056$^{+29}_{-30}$      \\
NGC 3081                   &0.0077   &  2   &39/28  & N  &  546$^{+19}_{-18}$    & 472$^{+212}_{-138}$   & - & - &507$^{+295}_{-177}$& 268$^{+17}_{-16}$  &  183$^{+7}_{-7}$      \\
                                   &                &       &            & I &    -                               &             -                        &  - & - & - &- & 598$^{+70}_{-62}$    \\
NGC 3281                  & 0.0113    &  2   &39/26 & N  & 538$^{+33}_{-31}$     &  534$^{+349}_{-180}$  &  -   & -  &118$^{+23}_{-21}$ & 217$^{+24}_{-23}$ & 123$^{+10}_{-8}$  \\
                                   &                &       &          & I &       -                            &             -                        &  -   & -  &271$^{+78}_{-90}$ &   -  & 300$^{+5}_{-5}$  \\
PKS 0326$-$288           &0.1096       &1.9 & 18/17 &N  &565$^{+53}_{-34}$     &  568$^{+59}_{-51}$     & - & - & - &  -  &    -                \\                                                                                                                                                                                                          
\hline
 \end{tabular}
Notes: (1) Source name; (2) Redshift as reported in Table \ref{tbl:classAGN}; (3) SWIFT70M optical spectral classification; (4) S/N near the Pa$\upbeta$; (5) line components; N: NLR, B: BLR, I: intermediate (see sect \ref{sec:lines} for the classification criteria); (6) to (12) FWHM (\kms) not deconvolved for instrumental resolution.
\noindent
\end{minipage}
\end{table} 

\clearpage
\begin{table}
\begin{minipage}{140mm}
\caption{$EW$ measurements of the emission lines in ISAAC spectra.}
\label{tbl_EWisaacSpettri}
\begin{tabular}{@{}llcccccccccc}
\hline
Object & Redshift & Cl & S/N      & Comp.  & \ion{He}{I}                         & Pa$\upgamma$ &  \ion{O}{II} & [\ion{P}{II}]                & [\ion{S}{IX}]                      & [\ion{Fe}{II}]                     & Pa$\upbeta$  \\
           &               &       & LR/MR       &            &$\lambda$10830\AA\         &           &$\lambda$11287\AA\    &$\lambda$11886\AA\ &$\lambda$12523\AA\ &$\lambda$12570\AA\ &     \\
(1)      & (2)          & (3)  & (4)  &  (5)      &  (6)         &  (7)     &  (8)    &  (9)    &  (10)    &  (11)  & (12)  \\
\hline
2MASX J05054575$-$2351139& 0.035  & 2  &  30/24  & N &  20.5&  5.1 &  - & - & - & - &2.0\\
                                              &              &    &         & I  & -                                 & -                               & - &-  & - &4.8 & 4.8\\
                                              &              &    &         & B &16.7& -                               & -  & - &- &-   & -  \\
3C 105                                  &  0.0884   & 2  & 27/13  & N & -                                 &1.6   & - &- & - & -  & -  \\
                                              &              &    &             & I  & 14.7&   -    & - & - & - & -  & -     \\   
CGCG 420$-$015                       &0.029  &  2    & 64/ 31 & N &75.1  &  - & - &  -   &5.4 &3.1 & 9.4  \\
                                             &            &        &             & I  &18.6 & -   & - &-     & -  & - & 2.28  \\  
ESO 005$-$G004                  &0.006   & 2    & 48/20    &N & -       & -   & - &  -    &   -    & 1.5&  -   \\
ESO 157$-$G023                    & 0.044 &   2   &47/37    & N  & 5.5 & -  & - & - & - &  -  & -  \\
                                           &              &       &              &  I & 35.7 & - & - &- &-  & -   &   -  \\        
  ESO 297$-$G018                &0.0253   & 2     & 47/38     & N  & -     & -  & - &- &     -  & 0.78 &  0.87 \\     
                                          &               &       &                 & I  & 8.1  & -  &- &  - &  -   &  -      & -        \\                                                
 ESO  374$-$G044                & 0.0284  & 2    & 45/18       & N &48.9 & 7.0   & - & 1.9 & 3.2&  - & 12.2 \\
                                           &             &       &                  & I  &      -      &  -       & - &  -  &  -   & 10.2 &  -  \\
ESO 416$-$G002                  & 0.059 & 1.9 &45/17      & N  & 6.2& -   &  - &-    &   -    & -         &   - \\
ESO 417$-$G006                 & 0.0163  & 2   &60/42      & N  &   -       & -      & - &  -    & -    & 6.2 & -      \\
Fairall 272                         & 0.022     & 2    &60/34    & N  & 8.2        & -    &   - &-     & 2.6    &3.5 &0.9   \\
                                          &              &       &             & B & -            & -       &  - &3.3      & -    &  -   & -    \\  
MCG $-$01$-$24$-$012                  &  0.0197 & 2   &50/18     & N &31.9 & 4.0 &  - &-   &  -     & 8.3   & 7.7  \\
                                          &              &     &               & I & 7.9  &2.2 &   - & -    &  -       &3.1 &1.5  \\
                                          &               &       &            & B &   - &   -     &  -&   -     &  -       &  -    &8.9 \\ 
MCG $-$05$-$23$-$016                  & 0.0084 & 2  &123/45    & N&     -  &   -  & - &    -  &  1.3    &  -        & 3.1       \\
                                         &                &     &               & B &  29.8 & 19.5   & -&    -  &   -    &      -          & 27.4    \\
                                         &                &       &            & I & 84.7  &    -    &  - & -   &  -     &          -       &46.1 \\
Mrk 1210                          & 0.014    & 2   &90/ 24    & N & 141.4  &   -                              &  - & -  &   -  &15.1&19.5    \\
                                         &              &       &             & I  &     -                         & 22.0  & - &  6.5  & 11.1   & 21.1 & 26.7    \\
                                        &               &       &             & B & 258.0  &     -                          &  - & -   & -     &                       -        & 21.1  \\
NGC 612                         &   -  &  2   &55/29     & N  & -                               & -                              & - & -    &  -     &    -                           &  -           \\
NGC 788                        & 0.0135  &  2   & 69/ 39   & N  & 26.3  & 2.6   &  -  &-   & -   &   1.7& 5.7  \\
NGC 1052                      & 0.0047    &  2    &67/48  & N  & 10.4   & -                                &  - &-     &  -   & 9.5   &  -       \\
                                       &             &       &            & B  & 40.9   &    -                        & - &  - &   -  &  -  & -      \\  
NGC 1142                       & 0.0294  &  2   & 61/57   & N  &              -         & -                           &  - & -    &  -   & 1.6 &  -         \\
                                       &            &       &             & I    &  6.1         &             -                  & - & -      &    -  & -   &-\\ 
NGC 1365                      & 0.005  & 1.8  &88/35    & N  & -                                 & -                              &  - & - &  - &  -  &15.7\\
                                       &            &       &               & B  &21.3 &  19.3  &12.4 & - & - &  - &15.4 \\
NGC 2992                     & 0.0077  & 2    & 63/43  & N & 22.8     &3.4     & - & - & - & 7.7&5.4   \\
                                      &             &       &            & I  &     -    &           -    & -  & - & -&5.4& 2.1     \\
                                    &                &       &           & B &50.9 & 21.2 &  - & - & - &   -   &22.2    \\
NGC 3081                   & 0.0077  &  2   &39/28  & N  & 44.7   & 5.1  & - & - & 4.3 & 3.4    & 7.8     \\
                                   &                &       &            & I &    -       &    -    &  - & - & - &- & 3.3  \\
NGC 3281                  & 0.0113    &  2   &39/26 & N  & 24.4  &   4.8  &  - & -  & 1.1 & 2.0 & 3.0  \\
                                  &                &       &          & I &       -     &      -   &  -   & -  & 1.2 &   -  & 9.2\\
PKS 0326$-$288           & 0.1096   &1.9 & 18/17 &N  &32.6       &    9.0     & - & - & - &  -  &    -  \\                                                                                                                                                                                                                                                                                                                                                                                                                                                                                     
\hline
 \end{tabular}
Notes: (1) Source name; (2) Redshift as reported in Table \ref{tbl:classAGN}; (3) SWIFT70M optical spectral classification; (4) S/N near the Pa$\upbeta$; (5) line components; N: NLR, B: BLR, I: intermediate (see sect \ref{sec:lines} for the classification criteria); (6) to (12) 
$EW$ (\AA\ ).
\noindent
\end{minipage}
\end{table} 

\clearpage
\begin{table}
\begin{minipage}{140mm}
\caption{Flux measurements of the emission lines in ISAAC spectra.}
\label{tbl_FLUXisaacSpettri}
\begin{tabular}{@{}llcccccccccc}
\hline
Object & Redshift & Cl & S/N & Comp.  & \ion{He}{I}  & Pa$\upgamma$ & \ion{O}{II} & [\ion{P}{II}]                       & [\ion{S}{IX}]                     & [\ion{Fe}{II}]                     & Pa$\upbeta$ \\
           &               &       & LR/MR       &            &$\lambda$10830\AA\         &           &$\lambda$11287\AA\    &$\lambda$11886\AA\ &$\lambda$12523\AA\ &$\lambda$12570\AA\ &     \\
(1)      & (2)          & (3)  & (4)  &  (5)      &  (6)         &  (7)     &  (8)    &  (9)    &  (10)    &  (11) & (12)  \\
\hline
2MASX J05054575$-$2351139& 0.035  & 2  &  30/24  & N & 7.9E-15&1.2E-15& - & - & - & - &0.2E-15\\
                                                 &              &    &         & I  & -                & -                               & - &-  & - &0.5E-15 & 0.4E-15 \\
                                                 &              &    &         & B &6.5E-15& -                               &  - &-  & - &-   & -   \\
3C 105                                  &  0.0884   & 2  & 27/13  & N & -                                 &0.7E-16   & - & - & - & -  & -  \\
                                              &              &    &             & I  &0.6E-15&   -    & - &- & - & -  & -   \\ 
CGCG 420$-$015                       &0.029  &  2    & 64/ 31 & N &21.0E-15 &  - &- &   -   &1.4E-15&0.5E-15 &1.5E-15\\
                                             &            &        &             & I  &5.2E-15& -   &- & -   &   & - & 0.4E-15\\ 
ESO 005$-$G004                  & 0.006 & 2    & 48/20    &N & -                                    & -   & - &  -    &        -                       & 0.4E-15&  -     \\
ESO 157$-$G023                    & 0.044 &   2   &47/37    & N  & 1.6E-15 & -  & - & - & - &  -  & -\\
                                           &              &       &              &  I & 10.3E-15 & - & - &- &-  & -   &   - \\  
 ESO 297$-$G018                & 0.0253 & 2     & 47/38     & N  & -     & -  & - &- &     -  & 0.2E-15 &  0.2E-15\\     
                                          &             &       &                 & I  &1.6E-15  & -  & - & - &  -   &  -      & -         \\                                                                                               
 ESO  374$-$G044                & 0.0284  & 2    & 45/18       & N &11.9E-15 & 7.0   & -&  0.4E-15 &0.6E-15 &  - & 1.8E-15\\
                                           &           &       &                  & I  &      -    &  -     & - &   -  &  -   & 1.6E-15 &  -   \\
                                           &           &       &                 & B   & 4.0E-15 & -    & - &-    &   -   &     -     & 0.6E-15\\                                                                                            
ESO 416$-$G002                  & 0.059  & 1.9 &45/17      & N  & 7.7E-15& -      &- & -    &       & -         &   - \\
ESO 417$-$G006                 & 0.0163    & 2   &60/42      & N  &   -    & -        & - &  -    & -    &10.1E-15& -          \\
Fairall 272                         &  0.022  & 2    &60/34    & N  & 2.1E-15    & -   & - &  -     &  0.7E-15   &0.6E-15 &0.2E-15   \\
                                          &              &       &             & B & -          & -       &- &0.9E-15  & -    &  -   & -    \\      
MCG $-$01$-$24$-$012                  & 0.0197 & 2   &50/18     & N &10.5E-15&1.3E-15 &  - & -    &  -       & 1.3E-15 &1.2E-15 \\
                                          &              &     &               & I & 2.6E-15  &0.7E-15 &  - &  -    &  -       &0.5E-15&0.2E-15 \\
                                         &               &       &            & B &   - &   -     &  -&  -     &  -       &  -    &1.4E-15 \\ 
 MCG $-$05$-$23$-$016                 &   0.0084   & 2  &123/45    & N&     -  &   -  &  - &  -  &2.5E-15  &  -        & 5.5E-15     \\
                                         &               &     &               & B & 59.1E-15& 38.2E-15 & - &   -  &   -    &      -          & 48.8E-15\\
                                         &                &       &            & I &168.0E-15 &    -    &  - & -   &  -     &          -       &82.2E-15 \\     
Mrk 1210                          & 0.014    & 2   &90/ 24    & N & 78.2E-15&   -                              & - &  -  &   -  &4.0E-15&49.0E-15  \\
                                         &              &       &             & I  &     -                         & 11.4E-15& - &2.9E-15 &5.2E-15&5.5E-15& 6.7E-15\\
                                        &               &       &             & B & 142.8E-15 &     -                          &  - & -   & -     &                       -        & 5.3E-15   \\
 NGC 612                         &-  &  2   &55/29     & N  & -                               & -                              & - & -    &  -     &    -                           &  -           \\
 NGC 788                        & 0.0135  &  2   & 69/ 39   & N  & 14.2E-15  & 1.3E-15&  -    & - & -    & 0.9E-15&2.7E-15  \\
 NGC 1052                     & 0.0047  &  2    &67/48  & N  & 16.6E-15  & -                                &  - &-     &  -   &8.5E-15 &  -       \\
                                       &             &       &            & B  & 65.1E-15&    -                        &  - & - &   -  &  -  & -      \\ 
NGC 1142                       & 0.0294   &  2   & 61/57   & N  &              -         & -                           & - &   -    &  -   & 0.5E-15  &  -         \\
                                       &            &       &             & I    & 2.0E-15       &             -                  &  - & -      &    -  & -   &-\\    
NGC 1365                      & 0.005 & 1.8  &88/35    & N  & -              & -            & - &  - &  - &  -  &71.5E-15  \\
                                       &            &       &               & B  &99.6E-15 &90.3E-15  &59.4E-15 & -& - &  - &70.1E-15 \\
 NGC 2992                     & 0.0077  & 2    & 63/43  & N &23.8E-15  &3.4E-15& - & - & - & 8.6E-15&6.4E-15 \\
                                      &             &       &            & I  &     -    &           -    & -  & - & - &6.0E-15&2.5E-15  \\
                                    &                &       &           & B &53.2E-15 &21.5E-15 &  - & - & - &   -   &26.3E-15   \\   
 NGC 3081                   & 0.0077   &  2   &39/28  & N  & 18.1E-15   &2.0E-15 & - & - & 2.8E-15 & 0.9E-15    &2.3E-15     \\
                                   &                &       &            & I &    -       &    -    &  - & - & - &- & 0.9E-15  \\  
NGC 3281                  & 0.0113    &  2   &39/26 & N  & 9.2E-15 &1.7E-15&  - & -  & 0.6E-15& 1.0E-15& 1.4E-15  \\
                                  &                &       &          & I &       -     &      -   &  -   & -  & 0.6E-15&   -  & 4.4E-15\\    
PKS 0326$-$288           & 0.1096       &1.9 & 18/17 &N  &2.0E-15      & 0.5E-15   & - & - & - &  -  &    -  \\
\hline
 \end{tabular}
Notes: (1) Source name; (2) Redshift as reported in Table \ref{tbl:classAGN}; (3) SWIFT70M optical spectral classification; (4) S/N near the Pa$\upbeta$; (5) line components; N: NLR, B: BLR, I: intermediate (see sect \ref{sec:lines} for the classification criteria); (6) to (12) 
Emission line Flux in erg s$^{-1}$ cm$^{-2}$.
\noindent
\end{minipage}
\end{table} 

\clearpage
\begin{landscape}
\begin{table}
\begin{minipage}{140mm}
\caption{FWHM measurements of the emission lines in X-shooter NIR spectra.}
\label{tbl_FWHMxshNIRSpettri}
\resizebox{1.7\textwidth}{!}{
\begin{tabular}{@{}llccccccccccccccc}
\hline
Object & Redshift & Cl & S/N & Comp.  & [\ion{C}{I}]                       & [\ion{S}{VIII}]                & Pa$\updelta$ & \ion{He}{II} & \ion{He}{I}  & Pa$\upgamma$ &  \ion{O}{II} &[\ion{P}{II}]       & [\ion{S}{IX}] & [\ion{Fe}{II}] & Pa$\upbeta$ & [\ion{Fe}{II}] \\
           &               &       &        &             &$\lambda$9853\AA\  &$\lambda$9911\AA\ &    &$\lambda$10122\AA\ &$\lambda$10830\AA\ &  &$\lambda$11287\AA\ &$\lambda$11886\AA\ &$\lambda$12523\AA\  &$\lambda$12570\AA\  & &$\lambda$16436\AA\    \\
(1)      & (2)          & (3)  & (4)  &  (5)      &  (6)         &  (7)     &  (8)    &  (9)    &  (10)    &  (11) &  (12)   & (13)   &  (14)  &  (15)       &  (16)                     &  (17)                           \\
\hline
 ESO 263$-$G013     &  0.0334   &2     &  14  &N  &       -    & -        & -      & 511$^{+288}_{-154}$ & -  &  -  & -  & -  &636$^{+77}_{-68}$  & 636$^{+77}_{-68}$  &  636$^{+77}_{-68}$ & 543$^{+41}_{-34}$  \\
                             &               &       &       & I  &      -      &    -        &  -       &  -                             &  - & -   & -  & 1161$^{+463}_{-498}$& -  &   -                             &     -                          &  -                               \\
 LEDA 093974      & 0.0239    & 2   &57    &N  &500$^{+100}_{-56}$ & -        & -  &356$^{+128}_{-111}$& 319$^{+17}_{-20} $ & -     &    -   & -    &319$^{+17}_{-20} $&  319$^{+17}_{-20} $&319$^{+17}_{-20} $& 319$^{+17}_{-20} $\\
                             &               &       &        &I   &  -                             &   -      & -   &1478$^{+400}_{-379}$ &1107$^{+114}_{-121}$& -    &   -    & -   & -  &840$^{+90}_{-84}$&900$^{+250}_{-286}$&727$^{+55}_{-51}$ \\
MCG $-$05$-$23$-$016     & 0.0084    & 2    &26    & N &     -       & -            &264$^{+55}_{-43}$&264$^{+55}_{-43}$&  233$^{+15}_{-15}$& 233$^{+15}_{-15}$&  - &  - &560$^{+80}_{-77}$ & 560$^{+80}_{-77}$&560$^{+80}_{-77}$& - \\
                             &               &        &       & B &       -     & -           &3468$^{+580}_{-470}$ & -   &2474$^{+67}_{-64}$&1911$^{+105}_{-79}$&  - & - & - &  -             & 2134$^{+93}_{-89}$&- \\    
2MASX J18305065+0928414   & 0.0193   & 2    &37  & N  & -  & - & -        & -             & 216$^{+12}_{-9}$  & -                 &  - & - & - & -              & -                                & -  \\
                                                   &             &       &      &  B  &-   & - &     -    & - &3513$^{+232}_{-213}$&   - & - & - & -                        & -             &-     &-     \\     
ESO 234$-$G050    & 0.0088   & 2    & 40        &  N  & - & - & 124$^{+12}_{-11}$   & - & 122$^{+12}_{-11}$    &122$^{+12}_{-11}$ &  - & - & - &122$^{+12}_{-11}$& 167$^{+11}_{-13}$ &124$^{+12}_{-11}$ \\
                           &              &        &              &  I  & - & - &267$^{+36}_{-31}$   &- &  434$^{+9}_{-9}$    &  272$^{+26}_{-24}$    & - & - & - &490$^{+26}_{-23}$& 343$^{+63}_{-57}$   &498$^{+26}_{-23}$  \\
                           &              &        &              &  B  &- & - &-                              & - &1111$^{+63}_{-59}$  & -                               & -   & - & - & -                           &1305$^{+381}_{-322}$  &-    \\
 NGC 4941         & 0.0038   & 2    &50          & N  & -  &  -  & 123$^{+20}_{-14}$& 123$^{+20}_{-14}$    & 123$^{+20}_{-14}$ & 123$^{+20}_{-14}$ & - &123$^{+20}_{-14}$ &123$^{+20}_{-14}$   &123$^{+20}_{-14}$&123$^{+20}_{-14}$& -\\
                           &              &       &               & I   & -   & -   & -                             & -  &308$^{+8}_{-8}$&246$^{+32}_{-30}$& -   &-&-   &-& 292$^{+24}_{-18}$& - \\  
                           &              &       &               & I   & -   &  -  &-                             &571$^{+40}_{-46}$   & 664$^{+33}_{-31}$&            -    &-&430$^{+142}_{-103}$ &- &479$^{+38}_{-26}$ &-                       &- \\  
NGC 4945          &  0.0017  & 2   &22           & N  & -   &  -   &    -   &   -   &249$^{+30}_{-40}$       &     -    &  -  &  -  & -  &  249$^{+30}_{-40}$ & -                    & 249$^{+30}_{-40}$ \\
                              &           &      &               & I   & -    & -    &    -    &   - & 998$^{+216}_{-172}$   &    -     &  -  &1268$^{+280}_{-272}$&  - &1098$^{+243}_{-146}$& -                   & 1098$^{+243}_{-146}$\\ 
NGC 5643         & 0.0040    & 2   &37           &N  &  -   &  -   &146$^{+21}_{-24}$ &146$^{+21}_{-24}$& 146$^{+21}_{-24}$   &146$^{+21}_{-24}$&  - & 251$^{+31}_{-27}$& - & 146$^{+21}_{-24}$   &146$^{+21}_{-24}$&146$^{+21}_{-24}$ \\
                          &               &       &               & I  &  -    &   -  &330$^{+10}_{-10}$ &560$^{+65}_{-59}$&289$^{+2}_{-5}$       &330$^{+10}_{-10}$& -   &   - & - &376$^{+42}_{-36}$     &330$^{+10}_{-10}$&376$^{+42}_{-36}$ \\
                          &               &       &               & I  &  -    &  -     &-                             & -     & 884$^{+80}_{-84}$  &  -                        &  -  & -   & -  &  -                          & -                     &         -            \\       
NGC 6221         &  0.0045   & 2    & 39          & N &  -   &  -     &141$^{+1}_{-1}$     &    -   & 141$^{+1}_{-1}$      &141$^{+1}_{-1}$   &   -  & 239$^{+31}_{-27}$ & 138$^{+16}_{-18}$  &141$^{+1}_{-1}$       &141$^{+1}_{-1}$ & 141$^{+1}_{-1}$  \\
                          &               &       &               &  I  & -   &  -     & 238$^{+80}_{-28}$  &   -   &343$^{+9}_{-9}$       &343$^{+9}_{-9}$  &  -  &  -   & -  &483$^{+12}_{-12}$    &483$^{+12}_{-12}$ &481$^{+12}_{-8}$ \\
                          &               &       &               & B  &  -  &  -    &   -                              &  -   &2142$^{+110}_{-141}$ & 1433$^{+70}_{-70}$& -  & - & - & -                              & 2257$^{+99}_{-93}$ & -  \\    
 NGC 7314        &  0.0047  & 1.9  & 37          & N  &  -  &    -  &84$^{+2}_{-2}$          & 84$^{+2}_{-2}$         &84$^{+2}_{-2}$          & 84$^{+2}_{-2}$            & -   &132$^{+23}_{-18}$&78$^{+10}_{-10}$    &84$^{+2}_{-2}$       &   84$^{+2}_{-2}$   &143$^{+16}_{-14}$   \\  
                          &               &       &               & I    &  -  &  -    &-                                &  289$^{+60}_{-44}$   &361$^{+10}_{-9}$        & 361$^{+10}_{-9}$      &  -  &   &262$^{+74}_{-40}$ &392$^{+13}_{-13}$ & 361$^{+10}_{-9}$ &-                           \\
                          &               &       &               &B   &  -   &   -   & 1137$^{+94}_{-82}$ &  -                               &1428$^{+46}_{-38}$     &1430$^{+78}_{-99}$ &  -  &    & -  &-                         &1348$^{+15}_{-15}$& -                          \\
                                                                                                                                                                                                                                                                                                                    
\hline
NGC 3783        &  0.0097    & 1    &24          & N   &  -  & 307$^{+53}_{-43}$& 307$^{+53}_{-43}$   &307$^{+53}_{-43}$   & 307$^{+53}_{-43}$        &307$^{+53}_{-43}$  & -  &  - &696$^{+203}_{-150}$&307$^{+53}_{-43}$  & 307$^{+53}_{-43}$  & - \\
                         &                &       &               & I  & -    &   -  &    -                            &  -                                   &1446$^{+98}_{-88}$       & -                              &-  &  -  &   &-                             &    -                                 & - \\
                         &                &       &               & B   &  -  &  -   &3268$^{+318}_{-373}$&2212$^{+617}_{-490}$&5114$^{+205}_{-182}$&3927$^{+252}_{-223}$ &  - & -   &   &-                           & 3500$^{+99}_{-103}$   & -   \\
\hline

 \end{tabular}
 }
Notes: (1) Source name; (2) Redshift as reported in Table \ref{tbl:classAGN}; (3) SWIFT70M optical spectral classification; (4) S/N near the Pa$\upbeta$; (5) line components; N: NLR, B: BLR, I: intermediate (see sect \ref{sec:lines} for the classification criteria); (6) to (17) FWHM (\kms) not deconvolved for instrumental resolution.
\noindent
\end{minipage}
\end{table} 
\end{landscape}

\clearpage
\begin{landscape}
\begin{table}
\begin{minipage}{140mm}
\caption{$EW$ measurements of the emission lines in X-shooter NIR spectra.}
\label{tbl_EWxshNIRSpettri}
\resizebox{1.7\textwidth}{!}{
\begin{tabular}{@{}llccccccccccccccc}
\hline
Object & Redshift & Cl & S/N & Comp.  & [\ion{C}{I}]                       & [\ion{S}{VIII}]                & Pa$\updelta$ & \ion{He}{II} & \ion{He}{I}  & Pa$\upgamma$ &  \ion{O}{II} &[\ion{P}{II}]       & [\ion{S}{IX}] & [\ion{Fe}{II}] & Pa$\upbeta$ & [\ion{Fe}{II}] \\
          &               &       &        &             &$\lambda$9853\AA\  &$\lambda$9911\AA\ &    &$\lambda$10122\AA\ &$\lambda$10830\AA\ &  &$\lambda$11287\AA\ &$\lambda$11886\AA\ &$\lambda$12523\AA\  &$\lambda$12570\AA\  & &$\lambda$16436\AA\ \\
(1)      & (2)          & (3)  & (4)  &  (5)      &  (6)         &  (7)     &  (8)    &  (9)    &  (10)    &  (11) &  (12)   & (13)   &  (14)  &  (15)       &  (16)                     &  (17)                           \\
\hline
 ESO 263$-$G013     & 0.0334   &2     &  14  &N  &       -    &  -        & -      & 3.3 & -  &  -  & -  & -  & 2.9  & 10.5  & 8.1  &10.7  \\
                            &               &       &       & B  &      -    &    -        &  -      &  -   &  - & -   & -  & 5.0& -  &   -   &     -       &  -  \\
LEDA 093974      & 0.0239 & 2   &57    &N  & 3.7     & -            &  -      & 1.1 & 7.5  & -     &    -   & -    & 2.7  &  8.6&2.4&9.3\\
                             &               &       &        &I   &  -     &   -                        & -& 2.9   &12.7  & -    &   -    & -   & -  &9.1 &2.3 &5.9 \\
MCG $-$05$-$23$-$016     & 0.0084    & 2    &26    & N &     -       & -            &1.4&1.6& 8.2& 1.2&  - &  - &1.7 & 0.4 &10.3& -\\
                             &               &        &       & B &       -     & -           &25.0& -   &84.8&31.3&  - & - & - &  -             & 54.9&- \\           
 2MASX J18305065+0928414   & 0.0193  & 2    &37  & N  & -  & - & -    & -     & 1.3 & -   &  - & - & - & -   & -      & -  \\
                                                   &             &       &      &  B  &-   & - &     -    & - &18.6&   - & - & - & -  & -   &-      &-    \\ 
 ESO 234$-$G050    & 0.0088   & 2    & 40        &  N  & - & - & 0.8 & - & 14.8  &3.8&  - & - & - &3.9& 10.4&3.5\\
                           &              &        &              &  I  & - & - &2.4  &1.3& 36.6  & 13.9    & - & - & - &14.8& 10.2&19.7 \\
                           &              &        &              &  B  &- & - &-    & - &13.9  & -                               & -   & - & - & -    &7.1 &-    \\
 NGC 4941         & 0.0038   & 2    &50          & N  & -  &  -  & 0.7& 0.9  &4.6 & 0.8 & - &0.6 &1.0 &0.9&2.5& -\\
                           &              &       &               & I   & -   & -   & -    & -     &13.6&0.9 & -  &-     &-   &-      & 1.4& -\\  
                           &              &       &               & I   & -   &  -  &-      &3.1& 0.9&            -    & -    & 0.7&- &6.0& -   &- \\          
 NGC 4945         &  0.0017 & 2   &22           & N  & -   &  -   &    -   &   -   &4.7 &     -    &  -  &  -  & -  &  6.7 & F  & 6.3  \\
                           &              &      &               & I   & -    & -    &    -    &   - & 8.0  &    -     &  -  &  6.1  &  - &7.4  & -   & 3.7\\       
NGC 5643         & 0.0040   & 2   &37           &N  &  -   &  -   &0.9 & 0.4 & 8.4   &1.9&  - & 1.8& -& 1.9&4.6&2.8 \\
                          &               &       &               & I  &  -    &   -  &1.0 & 3.2&24.7   &1.8& -   &   - & - &5.7 &6.6&6.4 \\
                          &               &       &               & I  &       &       &-      & -   &7.1 &  -       &  -  & -   & -  &  -    & -  &  -    \\     
NGC 6221         &  0.0045    & 2    & 39          & N &  -   &  -     &4.0   &   -  & 10.7  &8.6 &   -  & 2.8  & 1.4  &5.9    &18.6 & 6.1  \\
                          &               &       &               &  I  & -   &  -     &1.1   &   -   &21.7 &2.0 &  -  &  -   & -  &10.3    &9.5 &12.6\\
                          &               &       &               & B  &  -  &  -    &   -    &   -  &29.6  & 9.2 & -  & - & - & -  & 20.5 & -  \\      
 NGC 7314        & 0.0047 & 1.9  & 37          & N  &  -  &    -  &1.9   & 2.3    &11.7   & 3.3 & -   &1.5   & 1.4  &1.4  & 6.4   &2.6  \\  
                          &               &       &               & I    &  -  &  -    &-      &  5.0   &41.0  & 5.6  &  -  &  -     & 2.2  &1.4 & 13.0 &-      \\
                          &               &       &               &B   &  -   &   -  & 24.1&       &114.0 &50.9 &  -  &   -      &  -     &-   &74.9  & -     \\                                                                                                                                          
\hline
NGC 3783        &  0.0097   & 1    &24          & N   &  -  & 3.1 & 2.1   &1.5 & 38.7    & 1.9  & -  &  - &4.4 &2.9 & 4.2 & - \\
                         &                &       &               &  I  &   -  &   -   &    -     &       &39.1  & -       & -  &  -  & -  &-     &    -  & - \\
                         &                &       &               & B   & -   & -     & 28.3  &9.7  &143.4&45.5    & -  & -  &-    &-   & 113.8& -   \\
\hline

 \end{tabular}
 }
Notes: (1) Source name; (2) Redshift as reported in Table \ref{tbl:classAGN}; (3) SWIFT70M optical spectral classification; (4) S/N near the Pa$\upbeta$; (5) line components; N: NLR, B: BLR, I: intermediate (see sect \ref{sec:lines} for the classification criteria); (6) to (17) $EW$ (\AA\ ) .
\noindent
\end{minipage}
\end{table} 
\end{landscape}

\clearpage
\begin{landscape}
\begin{table}
\begin{minipage}{140mm}
\caption{Flux measurements of the emission lines in X-shooter NIR spectra.}
\label{tbl_FLUXxshNIRSpettri}
\resizebox{1.7\textwidth}{!}{
\begin{tabular}{@{}llccccccccccccccc}
\hline
Object & Redshift & Cl & S/N & Comp.  & [\ion{C}{I}]                       & [\ion{S}{VIII}]                & Pa$\updelta$ & \ion{He}{II} & \ion{He}{I}  & Pa$\upgamma$ &  \ion{O}{II} &[\ion{P}{II}]       & [\ion{S}{IX}] & [\ion{Fe}{II}] & Pa$\upbeta$ & [\ion{Fe}{II}] \\
          &               &       &        &             &$\lambda$9853\AA\  &$\lambda$9911\AA\ &    &$\lambda$10122\AA\ &$\lambda$10830\AA\ &  &$\lambda$11287\AA\ &$\lambda$11886\AA\ &$\lambda$12523\AA\  &$\lambda$12570\AA\  & &$\lambda$16436\AA\  \\
(1)      & (2)          & (3)  & (4)  &  (5)      &  (6)         &  (7)     &  (8)    &  (9)    &  (10)    &  (11) &  (12)   & (13)   &  (14)  &  (15)       &  (16)                     &  (17)                           \\
\hline
 ESO 263$-$G013     & 0.0334   &2     &  14  &N  &       -  &  -        & -      & 0.95E-16& -  &  -  & -  & -  & 0.72E-16 &0.3E-15  & 0.3E-15  &0.2E-15  \\
                            &               &       &       & B  & -   &    -        &  -      & -  &  - & -   & -  &1.3E-15& -  &   -   &     -       &  -  \\
LEDA 093974      & 0.0239    & 2   &57    &N  &0.2E-15    & -            &  -      & 0.6E-16 & 0.4E-15  & -     &    -   & -    &0.2E-15  & 0.5E-15&0.1E-15&0.5E-15\\
                             &               &       &      &I   &  -     &   -                        & -& 0.2E-15  &0.7E-15  & -    &   -    & -   & -  &0.5E-15 &0.1E-15 &0.3E-15 \\
MCG $-$05$-$23$-$016   & 0.0084    & 2    &26    & N &     -       & -            &0.4E-15&0.4E-15&2.4E-15&0.3E-15&  - &  - &0.5E-15 & 0.1E-15&3.1E-15& - \\
                             &               &        &       & B &       -     & -           &6.9E-15& -   &24.7E-15&9.1E-15&  - & - & - &  -             & 16.7E-15&- \\       
2MASX J18305065+0928414   & 0.0193  & 2    &37  & N  & -  & - & -        & -             &0.5E-16 & -                 &  - & - & - & -              & -                                & -  \\
                                                   &             &       &      &  B  &-   & - &     -    & - &0.8E-15&   - & - & - & -                        & -             &-          \\   
ESO 234$-$G050    & 0.0088    & 2    & 40        &  N  & - & - & 0.5E-16 & - &0.9E-15   &0.2E-15 &  - & - & - &0.2E-15& 0.5E-15&0.1E-15\\
                           &              &        &              &  I  & - & - &0.2E-15 &0.8E-16& 2.1E-15  &0.8E-15   & - & - & - &0.7E-15& 0.5E-15&0.7E-15 \\   
                           &              &        &              &  B  &- & - &-    & - &0.8E-15  & -                               & -   & - & - & -    &0.4E-15 &-    \\     
 NGC 4941         & 0.0038    & 2    &50          & N  & -  &  -  & 0.6E-16& 0.8E-16  &0.4E-15 & 0.7E-16 & - &0.5E-16 &0.8E-16 &0.7E-16&0.2E-15& -\\
                           &              &       &               & I   & -   & -   & -    & -     &1.2E-15&0.8E-16 & -  &-     &-   &-      & 0.1E-15& -\\  
                           &              &       &               & I   & -   &  -  &-      &0.3E-15& 0.8E-16&            -    & -    & 0.6E-16&- &0.5E-15& -   &- \\   
NGC 4945         &  0.0017 & 2   &22           & N  & -   &  -   &    -   &   -   &0.3E-16&     -    &  -  &  - & -  & 0.4E-16 & -  & 0.3E-16  \\
                          &              &      &               & I   & -    & -    &    -    &   - &0.5E-16  &    -     &  -  &0.4E-16  &  - &0.5E-16  & -   & 0.2E-16\\   
NGC 5643         & 0.0040    & 2   &37           &N  &  -   &  -   &0.9E-16 & 0.5E-16& 0.9E-15   &0.2E-15&  - &0.2E-15 & -& 0.2E-15&0.5E-15&0.3E-15 \\
                          &               &       &               & I  &  -    &   -  &1.0E-16 & 0.3E-15&2.6E-15   &0.2E-15& -   &   - & - &0.6E-15 &0.7E-15&0.6E-15 \\
                          &               &       &               & I  &       &       &-           & -   & 0.7E-15 &  -     &  -  & -   & -  &  -    & -  &  -    \\   
NGC 6221         &  0.0045   & 2    & 39          & N &  -   &  -     &0.3E-15  & -    & 0.7E-15  &0.6E-15 &   -  & 0.2E-15  & 0.1E-16 &0.4E-15    &1.3E-15 & 0.4E-15  \\
                          &               &       &               &  I  & -   &  -     &0.8E-16   &  -    &1.5E-15 &0.1E-15  &  -  &  -              & -            &0.7E-15   &0.7E-15 &0.8E-16\\
                          &               &       &               & B  &  -  &  -    &   -            & -    &2.0E-15  &0.6E-15  & -  & -               & -             & -  & 1.4E-15 & -  \\        
 NGC 7314        & 0.0047  & 1.9  & 37          & N  &  -  &    -  &0.5E-16   & 0.6E-16   &0.3E-15   & 0.9E-16 & -   & 0.4E-16  & 0.4E-16  &0.4E-16  & 0.2E-15   &0.9E-16  \\  
                          &               &       &               & I    &  -  &  -    &-              & 0.1E-15   &1.1E-15  & 0.2E-15 &  -  & -               & 0.7E-16  &0.4E-16 & 0.4E-15 &-      \\
                          &               &       &               &B   &  -   &   -  & 0.7E-15 &                  &3.1E-15 &1.4E-15 &  -  &   -               &                &-             &2.3E-15  & -     \\                                                                                          
\hline
NGC 3783        &  0.0097    & 1    &24          & N   &  -  &0.7E-15  &0.5E-15 &0.3E-15 & 8.0E-15   &0.4E-15   &   -  & - &1.0E-15&0.6E-15& 1.0E-15  & - \\
                         &                &       &               &  I  &   -  &  -           &    -        &  -           &8.1E-15    & -             &   -  &  -  &   &-&    -   & - \\
                         &                &       &               & B   & -   & -           &6.2E-15 & 2.1E-15& 29.8E-15  & 9.6E-15  &  -    &  -  &   &- & 26.0E-15  & -   \\
\hline  
 \end{tabular}
 }
Notes: (1) Source name; (2) Redshift as reported in Table \ref{tbl:classAGN}; (3) SWIFT70M optical spectral classification; (4) S/N near the Pa$\upbeta$; (5) line components; N: NLR, B: BLR, I: intermediate (see sect \ref{sec:lines} for the classification criteria); (6) to (17) Emission line Flux in erg s$^{-1}$ cm$^{-2}$.
\noindent
\end{minipage}
\end{table} 
\end{landscape}

\clearpage
\begin{landscape}
\begin{table}
\begin{minipage}{140mm}
\caption{FWHM measurements of the emission lines in X-shooter VIS spectra.}
\label{tbl_FWHMxshVISSpettri}
\resizebox{1.7\textwidth}{!}{
\begin{tabular}{@{}llccccccccccccccccc}
\hline
Object & Redshift & Cl & S/N & Comp.  & [\ion{Fe}{VII}] & \ion{He}{I} & [\ion{Fe}{VII}] & [\ion{O}{II}] & [\ion{O}{II}] & [\ion{N}{II}] & H$\upalpha$ & [\ion{N}{II}] & [\ion{S}{II}] & [\ion{S}{II}] & [\ion{O}{II}] & [\ion{Fe}{XI}] & [\ion{S}{III}] & [\ion{S}{III}] \\
           &               &      &        &              &$\lambda$5722\AA\ &$\lambda$5877\AA\ &$\lambda$6088\AA\ &$\lambda$6302\AA\ &$\lambda$6366\AA\ &$\lambda$6550\AA\ &   &$\lambda$6585\AA\ &$\lambda$6718\AA\ &$\lambda$6732\AA\ &$\lambda$7322\AA\  &$\lambda$7894\AA\ &$\lambda$9069\AA\ &$\lambda$9531\AA\ \\ 
 (1)         &   (2)           &  (3)    &  (4)      &    (5)          & (6) & (7) &  (8) &   (9)   & (10) &(11)  & (12)  & (13) &  (13)   &  (14)   & (15) & (16)  & (17)   &(18) \\ 
\hline
ESO 263$-$G013   &0.0334  &2     &  13  &N & -      &  -      &  -    &  -      & -     & 56$^{+14}_{-6}$    &56$^{+14}_{-6}$  & 56$^{+14}_{-6}$   &56$^{+14}_{-6}$ & 56$^{+14}_{-6}$   &  -  & -   & 241$^{+103}_{-120}$  &83$^{+9}_{-10}$  \\
                         &              &       &         & I  & -     &   -     &   -     &393$^{+38}_{-30}$& 393$^{+38}_{-30}$     &463$^{8}_{-6}$      & 463$^{8}_{-6}$    & 463$^{8}_{-6}$    &454$^{+29}_{-22}$&454$^{+29}_{-22}$ &  -  & -   &-& 546$^{+54}_{-44}$ \\
LEDA 093974      & 0.0239    & 2   &31 &N  &   -     &  -       &  -        &190$^{+18}_{-9}$ &190$^{+18}_{-9}$ &190$^{+2}_{-2}$ &190$^{+2}_{-2}$ &190$^{+2}_{-2}$   & 197$^{+5}_{-3}$  &197$^{+5}_{-3}$  &208$^{+77}_{-57}$ & -   &223$^{+5}_{-5}$  &223$^{+5}_{-5}$ \\
                             &              &      &     &I  &     -     &   -      &  -        &809$^{+45}_{-38} $&    -   &706$^{+4}_{-4}$&706$^{+4}_{-4}$  &706$^{+4}_{-4}$   &635$^{+8}_{-15}$    &635$^{+8}_{-15}$ &1418$^{+59}_{-80}$&  -   &928$^{+17}_{-20}$ &928$^{+17}_{-20}$\\
MCG $-$05$-$23$-$016     &  0.0084   & 2    &30    & N &384$^{+53}_{-48}$&383$^{+71}_{-55}$&216$^{+34}_{-30}$&195$^{+6}_{-6}$&167$^{+15}_{-15}$   & 215$^{+2}_{-2}$ & 215$^{+2}_{-2}$ & 215$^{+2}_{-2}$ &206$^{+3}_{-5}$  &206$^{+3}_{-5}$  & -   &  - & 215$^{+9}_{-9}$ &209$^{+4}_{-4}$     \\
                             &              &        &        & B & -  & -  &  -                          &         -                &                   -           &        -              &2232$^{+36}_{-38}$      &  -                         &  -                        &  -                    & -  &  -  &-  &  -    \\                             
2MASX J18305065+0928414   & 0.0193  & 2     &37  & N & -   &-   & -  &415$^{+118}_{-112}$&   & 216$^{+12}_{-9}$       & 216$^{+12}_{-9}$               & 216$^{+12}_{-9}$        & 216$^{+12}_{-9}$  & 216$^{+12}_{-9}$& - & -  &226$^{+73}_{-33}$&196$^{+17}_{-13}$\\
                                                  &              &        &       & B  & -  & -  & -  & -  & -  &-                                 &2660$^{+500}_{-460}$     & -                                  & -                             &  -                           & - & -  & - &- \\
                                                  &              &        &       & I  & -  & -  & -  & -  &  - &-                                 &6194$^{+160}_{-160}$    & -                                  & -                             &  -                           & - &  - & - &- \\
 ESO 234$-$G050    & 0.0088    & 2    & 44 &  N  & - &124$^{+33}_{-20}$&-  &175$^{+14}_{-19}$ &175$^{+14}_{-19}$ &117$^{+3}_{-2}$&113$^{+3}_{-4}$             &113$^{+3}_{-4}$     &117$^{+3}_{-2}$&117$^{+3}_{-2}$&288$^{+18}_{-14}$  & - &137$^{+16}_{-26}$&137$^{+16}_{-26}$  \\
                            &              &       &      &  I   & -  &486$^{+79}_{-67}$& - &498$^{+17}_{-44}$&498$^{+17}_{-44}$ &327$^{+7}_{-10}$ &327$^{+7}_{-10}$             &327$^{+7}_{-10}$&366$^{+3}_{-6}$&366$^{+3}_{-6}$&-  & - &391$^{+52}_{-26}$&391$^{+52}_{-26}$   \\
                            &              &       &      &  B  &  - &  -&-  &-  & - &  -                     &972$^{+124}_{-104}$ & -                             &-                        & -                       & -  & - &  -&  -\\
                            
 NGC 4941         & 0.0038   & 2    &24  & N  & - &288$^{+51}_{-36}$&201$^{+68}_{-42}$&107$^{+11}_{-10}$&186$^{+29}_{-19}$& 103$^{+2}_{-1}$&103$^{+2}_{-1}$&103$^{+2}_{-1}$&103$^{+2}_{-1}$&103$^{+2}_{-1}$   & 231$^{+29}_{-22}$& - &106$^{+10}_{-9}$ & 122$^{+5}_{-5}$ \\
                           &              &       &      & I   &  - & -                           & -                            &373$^{+23}_{-34}$& - & 370$^{+4}_{-2}$&370$^{+4}_{-2}$&370$^{+4}_{-2}$&373$^{+2}_{-4}$&373$^{+2}_{-4}$ & - &   -   &359$^{+17}_{-14}$ &372$^{+5}_{-9}$  \\                 
 NGC 4945         &   0.002& 2   &9    & N & - & -  & -  &126$^{+23}_{-20}$  & -    &230$^{+5}_{-6}$     & 230$^{+5}_{-6}$   & 230$^{+5}_{-6}$ &165$^{+22}_{-12}$ &165$^{+22}_{-12}$ & -    &  - &279$^{+31}_{-29}$&234$^{+11}_{-12}$  \\
                           &              &      &       & I  & -  & -  &  -  &583$^{+70}_{-56}$ & -   &794$^{+18}_{-14}$&794$^{+18}_{-14}$&794$^{+18}_{-14}$&527$^{+27}_{-23}$&527$^{+27}_{-23}$ &   -   & -  & -  & -   \\
                           
NGC 5643          & 0.0040    & 2   &25   & N &-                              &-                             & -                            &147$^{+13}_{-8}$  &147$^{+13}_{-8}$  &98$^{+2}_{-2}$   &98$^{+2}_{-2}$ &98$^{+2}_{-2}$  &126$^{+3}_{-6}$    &126$^{+3}_{-6}$    &-                             &-&204$^{+5}_{-4}$&109$^{+1}_{-1}$\\
                           &              &     &        &  I  & 402$^{+48}_{-42}$&342$^{+19}_{-17}$ &332$^{+43}_{-31}$&402$^{+11}_{-11}$& 402$^{+11}_{-11}$&273$^{+2}_{-2}$&273$^{+2}_{-2}$&273$^{+2}_{-2}$&472$^{+34}_{-60}$&472$^{+34}_{-60}$&310$^{+14}_{-12}$&-&600$^{+5}_{-4}$&346$^{+1}_{-1}$\\  
                           &              &     &        & I  &       -                       &     -                         &      -                       &        -                    &            -                 &401$^{+2}_{-3}$    &401$^{+2}_{-3}$&401$^{+2}_{-3}$  &306$^{+8}_{-10}$  &306$^{+8}_{-10}$&-                            &-                         &-&-\\  
                           &             &      &       &   I &       -                       &          -                    &         -                    &           -                 &               -               &249$^{+23}_{-11}$&-                        &249$^{+23}_{-11}$&-                            & -                           & -                           &-                         &-&-\\    
NGC 6221          & 0.0045  & 2   & 44  & N  &386$^{+61}_{-54}$&142$^{+13}_{-9}$  &200$^{+55}_{-28}$&134$^{+13}_{-10}$&147$^{+40}_{-25}$&90$^{+4}_{-3}$&90$^{+4}_{-3}$&90$^{+4}_{-3}$&90$^{+4}_{-3}$ &90$^{+4}_{-3}$&125$^{+31}_{-13}$ & -&241$^{+103}_{-120}$&83$^{+9}_{-10}$  \\
                           &             &      &       &  I  & -                            &-&- &-&-&192$^{+8}_{-16}$&192$^{+8}_{-16}$&192$^{+8}_{-16}$&192$^{+8}_{-16}$ & 192$^{+8}_{-16}$&- &- &- & - \\
                           &             &      &       &  I & - &442$^{+34}_{-22}$&558$^{+66}_{-114}$ &456$^{+17}_{-23}$&519$^{+102}_{-108}$&500$^{+12}_{-7}$&500$^{+12}_{-7}$&500$^{+12}_{-7}$&500$^{+12}_{-7}$ &500$^{+12}_{-7}$ &765$^{+90}_{-50}$&- &- &546$^{+54}_{-44}$  \\
                           &             &      &       &  I & - & -&- &- & -&77$^{+16}_{-8}$  &77$^{+16}_{-8}$  &77$^{+16}_{-8}$  &77$^{+16}_{-8}$  &77$^{+16}_{-8}$  &- &- &- &  -\\
                            &             &      &      &  B & - &- &- &- &- &-  &1630$^{+12}_{-11}$  &- &- & -&- &- & -& - \\
NGC 7314           & 0.0047 & 1.9  &    45     & N  &191$^{+11}_{-10}$&130$^{+8}_{-8}$      &246$^{+8}_{-8}$&70$^{+2}_{-4}$ &62$^{+2}_{-4}$& 89$^{+1}_{-1}$ &  89$^{+1}_{-1}$ & 89$^{+1}_{-1}$ &  87$^{+1}_{-1}$&  87$^{+1}_{-1}$     &94$^{+8}_{-11}$ &- &86$^{+1}_{-2}$&90$^{+1}_{-1}$  \\
                              &          &        &              & I   &   -                         & -                              &  -                      &180$^{+5}_{-5}$&180$^{+5}_{-5}$& 187$^{+2}_{-2}$&270$^{+2}_{-2}$  & 187$^{+2}_{-2}$ &179$^{+4}_{-4}$& 190$^{+3}_{-3}$&236$^{+19}_{-16}$ &-  &219$^{+2}_{-2}$ &207$^{+1}_{-2}$ \\
                              &           &        &              &B   &   -                        &1561$^{+50}_{-59}$&   -                     & -                       &  -                      & -                        & 1330$^{+5}_{-4}$& -                      &          -                &      -                   &  -& -& -&- \\
\hline
 
NGC 3783        &  0.0097  & 1    & 53          & N   &-                                 &524$^{+45}_{-30}$       &575$^{+17}_{-17}$         &153$^{+7}_{-7}$    &158$^{+22}_{-13}$ &212$^{+4}_{-2}$    &212$^{+4}_{-2}$        &212$^{+4}_{-2}$    &212$^{+4}_{-2}$ &212$^{+4}_{-2}$&316$^{+17}_{-13}$& -                            &128$^{+23}_{-18}$& 164$^{+6}_{-4}$\\ 
                          &                &      &               & I    &830$^{+28}_{-23}$    & -                                   & 1796$^{+146}_{-125}$  &566$^{+40}_{-43}$ &757$^{+44}_{-54}$ &1167$^{+9}_{-11}$&1167$^{+9}_{-11}$     &1167$^{+9}_{-11}$& -                         & -                       &-                             &746$^{+56}_{-52}$ &369$^{+8}_{-17}$  &456$^{+14}_{-12}$\\                          
                          &               &       &              & B   & -                                &8533$^{+124}_{-115}$  &-                                      &-                             &-                              &-                            & 5290$^{+24}_{-23}$ &   -                          &   -                      &       -                  & -                            & -                            &-                             \\                                                             
\hline 
\end{tabular}
}
Notes: (1) Source name; (2) Redshift as reported in Table \ref{tbl:classAGN}; (3) SWIFT70M optical spectral classification; (4) S/N near the H$\upalpha$; (5) line components; N: NLR, B: BLR, I: intermediate (see sect \ref{sec:lines} for the classification criteria); (6) to (18) FWHM (\kms) not deconvolved for instrumental resolution.
\noindent
\end{minipage}
\end{table} 
\end{landscape}

\clearpage
\begin{landscape}
\begin{table}
\begin{minipage}{140mm}
\caption{$EW$ measurements of the emission lines in X-shooter VIS spectra.}
\label{tbl_EWxshVISSpettri}
\resizebox{1.7\textwidth}{!}{
\begin{tabular}{@{}llccccccccccccccccc}
\hline
Object & Redshift & Cl & S/N & Comp.  & [\ion{Fe}{VII}] & \ion{He}{I} & [\ion{Fe}{VII}] & [\ion{O}{II}] & [\ion{O}{II}] & [\ion{N}{II}] & H$\upalpha$ & [\ion{N}{II}] & [\ion{S}{II}] & [\ion{S}{II}] & [\ion{O}{II}] & [\ion{Fe}{XI}] & [\ion{S}{III}] & [\ion{S}{III}] \\
           &               &      &        &              &$\lambda$5721\AA\ &$\lambda$5876\AA\ &$\lambda$6087\AA\ &$\lambda$6300\AA\ &$\lambda$6363\AA\ &$\lambda$6548\AA\ &   &$\lambda$6583\AA\ &$\lambda$6716\AA\ &$\lambda$6731\AA\ &$\lambda$7323.0\AA\  &$\lambda$7892\AA\ &$\lambda$9069\AA\ &$\lambda$9531\AA\ \\ 
 (1)         &   (2)           &  (3)    &  (4)      &    (5)          & (6) & (7) &  (8) &   (9)   & (10) &(11)  & (12)  & (13) &  (13)   &  (14)   & (15) & (16)  & (17)   &(18) \\ 
\hline
ESO 263$-$G013   & 0.0334  &2     &  14  &N &    -   &  -      & -     &  -   &  -    & 2.1    & 4.7 & 6.4   & 2.0 & 1.6   & -   &  -  & 13.0   &12.1 \\
                         &              &       &         & I  &  -    &   -     &   -     &  12.6  & 4.9 &23.3  & 52.5&70.1 &19.7&21.4 & -   &  -  & -   &34.6 \\
 LEDA 093974  & 0.0239    & 2   &57    &N  &   -     &   -      & -         & 2.0       &  0.8     &5.0   &8.9 &15.4  & 5.5  & 4.1   & 0.3    & -   &3.3  &9.6\\
                         &              &      &          &I  &    -      &   -      & -         &4.3        &   -    &11.2&14.5 &34.1 &8.9   &10.7 & 3.5    &   -  &9.5 &13.6\\      
MCG $-$05$-$23$-$016  & 0.0084    & 2    &26    & N &1.3  &1.3  & 1.1  &3.7 &0.9 & 5.4& 19.4 & 16.0 &4.3  &4.6  &  -  & -   & 3.2   & 7.5\\
                          &              &        &        & B & -  & -  &  -  &  -  & -  &   -  &34.2  &  -      &  -                        &  -                         &  -  &   - & -   & -  \\   
2MASX J18305065+0928414   & 0.0193    & 2     &37  & N & -  & -  &-  &  2.3  &  - &2.6       & 3.3 & 7.8   &1.6  & 2.0  &  -  & -  & 1.7&4.0\\
                                                  &              &        &       & B  & - &  - & -  & -  & -  &-         &9.8            & -    & -        &  - & - & -  & - &- \\
                                                  &              &        &       & I  &  - & -  & -  & -  & -  &-        &90.9          & -    & -        &  - & - & -  & - & -\\
ESO 234$-$G050    &  0.0088 & 2    & 40 &  N  & - & 1.2   & - &5.3&1.7 &1.5 &29.7&4.5   &5.4  &4.9  &4.7  &-  & 5.5 & 13.9 \\
                            &              &       &      &  I   & -  & 2.4 & - &9.9&3.6 &3.8 &56.2 &11.5 &15.3&15.4& - & - & 5.7 &18.3   \\
                            &              &       &      &  B  & -  & -    & - & -   & -   &  -    &20.7 & -     &-       & -   & -  & - & - & - \\   
NGC 4941           & 0.0038  & 2    &50  & N   & - &2.6&1.0 &4.2&3.3&4.8&10.4&14.3&4.5&5.3& 2.6&- &5.3 & 16.7\\
                           &              &       &      & I   &  - & -    & - &8.6& -  &19.4&32.5&57.9&14.9&19.1& - &   -   &17.1&37.6  \\ 
NGC 4945          & 0.0017  & 2   &22   & N & - & -  & -  &2.0  & -  &11.7&8.8   & 36.4 &4.0 &3.1 & -    &  - &1.9&3.5  \\
                           &              &      &       & I  & -  & -  &  - &7.5 & -   &25.7&21.9&80.1 &13.0&12.2 &   -   & -  & -  & -   \\
NGC 5643          & 0.0040   & 2   &37   & N &-    &-     & -   &3.6 &1.1 &2.6 &6.5 &7.7    &3.8 &4.2  &-     &-&15.7&11.3\\
                           &              &     &        &  I  &1.4&2.9 &1.6&7.6&2.4  &14.8&45.6&44.1&5.8&4.1    &2.8 &-&10.9&48.9\\ 
                           &              &     &        & I  &   - & -    &  -  & -   &-      &6.2 &15.6 &18.6 &9.9  & 1.4 &-    &-&-&-\\  
                           &             &      &       &   I &  -  & -    &  -  & -   &  -    &0.6&-        &1.9   &-     & -      & -  &-&-&-\\    
NGC 6221          &0.0045  & 2   & 44  & N  &0.7&2.0&0.5&1.3&0.5&9.9&48.0&29.9&4.4&5.3&2.6 & -&13.0&12.0  \\
                           &             &      &       &  I  & -  &-&- &-&-&8.8&50.9&26.6&5.5 & 5.3&- &- &- & - \\
                           &             &      &       &  I & - &1.9&0.6&3.3&1.1&11.3&45.7&33.9&9.1&10.6 &4.7&- &- &34.6  \\
                           &             &      &       &  I & - & -&- &- & -&0.4&1.1 &1.1  &0.3&0.3&- &- &- &  -\\
                            &             &      &      &  B & - &- &- &- &- &-  &59.0&- &- & -&- &- & -& - \\     
 NGC 7314           & 0.0047& 1.9  &    45     & N  &4.0 &5.3  &7.7&9.5   & 2.5  & 13.5& 45.2&40.6 &15.4 & 13.6  &1.3 & -&16.0 &50.0  \\
                              &          &        &              & I   &  -   &-     &-    &6.2   & 3.8  &8.5&66.8 & 25.5  &8.2    & 11.6   &1.6 & - &19.4 &56.9 \\
                              &           &        &              &B   &  -  &2.3 & -   & -     &  -     & -   & 223.0& -      &  -      &  -       & -    & - & -&- \\                                                  
\hline
 
NGC 3783        &  0.0097    & 1    & 53          & N   &-       &2.7    &7.2 &2.7 &1.0  &5.3  &22.0     &19.5    &5.7& 6.2 &1.9& -     &2.7 & 11.0\\ 
                          &                &      &               & I    &7.5   & -        &5.1 &3.6 &4.1 &40.1&147.4     &78.5   & -  & -      &-    &3.4 &8.5  &19.0\\                          
                          &               &       &              & B   & -       &57.2  &-     &-      &-              &-            & 446.0 &-   &   -    & -   & -   & -     &-      \\                                                                                                                                                                 
\hline                         
 \end{tabular}
 }
Notes: (1) Source name; (2) Redshift as reported in Table \ref{tbl:classAGN}; (3) SWIFT70M optical spectral classification; (4) S/N near the H$\upalpha$; (5) line components; N: NLR, B: BLR, I: intermediate (see sect \ref{sec:lines} for the classification criteria); (6) to (18)  $EW$ (\AA\ ) .
\noindent
\end{minipage}
\end{table} 
\end{landscape}

\clearpage
\begin{landscape}
\begin{table}
\begin{minipage}{140mm}
\caption{Line flux measurements of the emission lines in X-shooter VIS spectra.}
\label{tbl_FLUXxshVISSpettri}
\resizebox{1.7\textwidth}{!}{
\begin{tabular}{@{}llccccccccccccccccc}
\hline
Object & Redshift & Cl & S/N & Comp.  & [\ion{Fe}{VII}] & \ion{He}{I} & [\ion{Fe}{VII}] & [\ion{O}{II}] & [\ion{O}{II}] & [\ion{N}{II}] & H$\upalpha$ & [\ion{N}{II}] & [\ion{S}{II}] & [\ion{S}{II}] & [\ion{O}{II}] & [\ion{Fe}{XI}] & [\ion{S}{III}] & [\ion{S}{III}] \\
           &               &      &        &              &$\lambda$5721\AA\ &$\lambda$5876\AA\ &$\lambda$6087\AA\ &$\lambda$6300\AA\ &$\lambda$6363\AA\ &$\lambda$6548\AA\ &   &$\lambda$6583\AA\ &$\lambda$6716\AA\ &$\lambda$6731\AA\ &$\lambda$7323.0\AA\  &$\lambda$7892\AA\ &$\lambda$9069\AA\ &$\lambda$9531\AA\ \\ 
 (1)         &   (2)           &  (3)    &  (4)      &    (5)          & (6) & (7) &  (8) &   (9)   & (10) &(11)  & (12)  & (13) &  (13)   &  (14)   & (15) & (16)  & (17)   &(18) \\ 
\hline
ESO 263$-$G013   & 0.0334   &2     &  14  &N &   -    &  -      &  -     &    -    &   -   & 0.6E-16   & 1.4E-16 & 1.9E-16   &0.5E-16 & 0.4E-16 & -    &  -  & 2.2E-16&2.2E-16    \\
                         &              &       &         & I  &  -    &   -     &   -     & 3.7E-16       &   1.4E-16   &7.0E-16  & 1.5E-15  &2.1E-15    &5.2E-16&5.7E-16& -   &  -  & -& 6.4E-16   \\
  LEDA 093974  & 0.0239    & 2   &57    &N  &   -     &  -       &   -       &   1.2E-16     &  0.5E-16     &3.3E-16   &5.8E-16&9.9E-16  & 3.5E-16   &2.6E-16  & 0.2E-16    &  -  & 2.2E-16 &6.3E-16\\
                         &              &      &          &I  &      -    &   -      &    -      &  2.7E-16      &   -    &7.3E-16   &9.4E-16 &2.2E-15 &5.7E-16&6.9E-16 & 2.3E-16    &  -   &6.4E-16  &8.9E-16 \\  
 MCG $-$05$-$23$-$016 & 0.0084   & 2    &26    & N &2.3E-16 &2.4E-16&2.0E-16 & 6.9E-16&1.7E-16 &1.1E-15& 4.0E-15 & 3.3E-15 &3.3E-15  &0.9E-15 & 0.9E-15&  -  & 7.1E-16   &1.7E-15 \\
                          &              &        &        & B & -  & -  & -  & -   &  - &   -  &7.1E-15  &  -      &  -                        &  -                         & -   &  -  &  -  & -  \\       
 2MASX J18305065+0928414   & 0.0193     & 2     &37  & N & -  & -  & -  & 1.1E-16 & -  &1.4E-16 &1.8E-16 &4.2E-16 &0.9E-16  &1.1E-16  &  -  &  - &  1.2E-16 &2.5E-16\\
                                                  &              &        &       & B  & - &  - &  - & -  & -  &-         &5.3E-16     & -    & -        &  - & - & -  & - &- \\
                                                  &              &        &       & I  & -  & -  & -  & -  & -  &-        &4.9E-15      & -    & -        &  - & - & -  & - &- \\        
 ESO 234$-$G050    &0.0088  & 2    & 40 &  N  &-   &1.6E-16  & - & 7.1E-16&2.3E-16&2.1E-16&4.2E-15&6.3E-16&7.4E-16&6.8E-16    &6.1E-16  & - &5.9E-16  &1.3E-15  \\
                            &              &       &      &  I   &-   &3.2E-16  & - &1.3E-15  &4.8E-16  &5.4E-16&7.8E-15 &1.6E-15 &2.1E-15&2.1E-15& - & - &6.0E-16  &1.7E-15   \\
                            &              &       &      &  B  & -  &     -        & - &-  &  -&  -   &3.4E-15& -&- & -  & -  & - & - & - \\         
NGC 4941           & 0.0038    & 2    &50  & N   & - &3.5E-16&1.5E-16&6.3E-16 &4.9E-16&9.3E-16&2.0E-15&2.8E-15&7.3E-16&8.5E-16&4.4E-16&- &9.3E-16&3.0E-15\\
                           &              &       &       & I   &  - & -           & -          &1.3E-15& -            &3.8E-15&6.3E-15&1.1E-14&2.4E-15&3.1E-15& - &   -   &3.0E-15&6.8E-15  \\          
NGC 4945          & 0.0017  & 2   &22   & N & - & -  & -  &5.5E-18 & -   &4.2E-17 &3.1E-17&1.3E-16 &1.8E-17 &1.4E-17& F    &  - &2.1E-17&4.2E-17  \\
                           &              &      &       & I  & -  & -  &  - &2.1E-17& -   &9.2E-17&7.7E-17&2.8E-16&5.7E-17&5.3E-17&   -   & -  & -  & -   \\

NGC 5643          & 0.0040 & 2   &37   & N &-    &-     & -           &4.7E-16 &1.5E-16 &3.9E-16 &9.8E-16 &1.2E-15   &5.8E-16 &6.3E-16  &-     &-&2.6E-15&2.0E-15\\
                           &              &     &        &  I  &1.6E-16&3.3E-16 &2.1E-16 &9.9E-16 &3.2E-16  &2.2E-15&6.9E-15&6.6E-15 &8.7E-16&6.2E-16    &4.5E-16 &-&1.9E-15&8.7E-15\\ 
                           &              &     &        & I  &   - & -    &  -  & -   &-      &9.4E-16 &2.4E-15 &2.8E-15 &1.5E-15  & 2.1E-15 &-    &-&-&-\\  
                           &             &      &       &   I &  -  & -    &  -  & -   &  -    &9.3E-17&-        &2.8E-16   &-     & -      & -  &-&-&-\\              

NGC 6221          & 0.0045  & 2   & 44  & N  &4.0E-17&1.1E-16&2.8E-17&7.5E-17&2.6E-17&6.1E-16&2.9E-15&1.8E-15&2.7E-16&3.2E-16&2.8E-17& -&2.2E-16&2.2E-16  \\
                           &             &      &       &  I  & -  &-&- &-&-&5.2E-16&3.1E-15&1.6E-15&3.3E-16& 4.0E-16&- &- &- & - \\
                           &             &      &       &  I & - &1.1E-16&3.4E-17&1.9E-16&6.3E-17&7.0E-16&2.8E-15&2.1E-15&5.5E-16&6.5E-16 &1.6E-16&- &- &6.4E-16  \\
                           &             &      &       &  I & - & -&- &- & -&2.8E-17&6.9E-17 &8.3E-17&1.6E-17&1.8E-17&- &- &- &  -\\
                            &             &      &      &  B & - &- &- &- &- &-  &3.6E-15&- &- & -&- &- & -& - \\      
 NGC 7314           &0.0047 & 1.9  &    45     & N  &2.5E-17 &3.4E-17  &5.8E-17&9.1E-17 & 2.4E-17  &1.9E-16&6.5E-16&5.8E-16  &1.7E-16 &1.9E-16 & 2.0E-17   &-    &3.3E-16&9.0E-16\\
                              &          &        &              & I   &  -          &-             &-           &5.9E-17  & 3.6E-17  &1.2E-16&9.5E-16&3.6E-16  &1.4E-16  &1.0E-16 & 2.4E-17   &-    & 4.0E-16  &1.0E-15\\
                              &           &        &              &B   &  -         &1.5E-16& -           & -             &  -            & -          &3.2E-15& -             &  -            & -          &  -              & -    & -  & -\\  
                              
 \hline
 
NGC 3783        &  0.0097     & 1    & 53          & N   &-            &8.9E-16  &2.4E-15 &9.3E-16 &3.4E-16 &1.8E-15 &7.8E-15     &6.8E-15   &2.0E-15& 2.2E-15 &6.1E-16& -     &7.1E-16 & 2.8E-15\\ 
                          &                &      &               & I    &2.4E-15 & -             &1.7E-15 &1.2E-15 &1.4E-15 &1.4E-14&5.2E-14     &2.7E-14   & -           & -             &-    &1.0E-15 &2.2E-15  &4.8E-15\\                          
                          &               &       &              & B   & -             &1.9E-14  &-            &-            &-             &-            &1.6E-13 &-   &   -    & -   & -   & -     &-      \\    
                                                                                                                                                                                                                                                                                                                                  
\hline                         
 \end{tabular}
 }
Notes: (1) Source name; (2) Redshift as reported in Table \ref{tbl:classAGN}; (3) SWIFT70M optical spectral classification; (4) S/N near the H$\upalpha$; (5) line components; N: NLR, B: BLR, I: intermediate (see sect \ref{sec:lines} for the classification criteria); (6) to (18) Emission line Flux in erg s$^{-1}$ cm$^{-2}$.
\noindent
\end{minipage}
\end{table} 
\end{landscape}

\clearpage
\begin{landscape}
\begin{table}
\begin{minipage}{140mm}
\caption{FWHM measurements of the emission lines in X-shooter UVB spectra.}
\label{tbl_FWHMxshUVBSpettri}
\resizebox{1.7\textwidth}{!}{
\begin{tabular}{@{}llcccccccccccccccc}
\hline
Object & Redshift & Cl & S/N & Comp.  & [\ion{Ne}{V}] & [\ion{Ne}{V}]& [\ion{O}{II}]  & [\ion{O}{II}]& [\ion{Fe}{VII}] & [\ion{Ne}{III}] & [\ion{Fe}{V}] & [\ion{Ne}{III}] & H$\upgamma$& \ion{He}{II} & H$\upbeta$& [\ion{O}{III}] & [\ion{O}{III}]  \\
           &               &      &        &              &$\lambda$3346\AA\ &$\lambda$3426\AA\ &$\lambda$3728\AA\  & $\lambda$3730\AA\ &$\lambda$3760\AA\ &$\lambda$3869\AA\ &$\lambda$3893\AA\ &$\lambda$ 3968\AA\  &   &$\lambda$4686\AA\ &  &$\lambda$4959\AA\  &$\lambda$5007\AA\ \\ 
 (1)         &   (2)           &  (3)    &  (4)      &    (5)          & (6) & (7) &  (8) &   (9)   & (10) &(11)  & (12)  & (13) &  (13)   &  (14)   & (15) & (16)&(17)   \\ 
\hline
ESO 263$-$G013   &0.0334  &2     &  14  &N   &293$^{+61}_{-73}$  &293$^{+61}_{-73}$  & -  &-  &  - & 390$^{+25}_{-29}$ &390$^{+25}_{-29}$  &349$^{+53}_{-62}$  &330$^{+117}_{-78}$ &384$^{+203}_{-110}$  &  98$^{+2}_{-2}$       & 98$^{+2}_{-2}$   & 98$^{+2}_{-2}$    \\
                          &              &       &     &I   &    -                            &  -                               &508$^{+20}_{-17}$&-  &-           &-      &  - & -  & -  & - &560$^{+274}_{-137} $&  440$^{+19}_{-14}$ &440$^{+19}_{-14}$ \\
 LEDA 093974  &0.0239  & 2   &57 &N  &  -    &  -    & 99$^{+11}_{-6}$& 99$^{+11}_{-6}$ & - &149$^{+10}_{-10}$ & -  & -  & -  & - &157$^{+2}_{-2}$&157$^{+2}_{-2}$&157$^{+2}_{-2}$ \\
                         &               &      &      &I  &    -    &  -    &596$^{+15}_{-18}$ &596$^{+15}_{-18}$ &  -&673$^{+59}_{-59}$& - &  - &  - & - &-& 732$^{+8}_{-8}$&732$^{+8}_{-8}$ \\   
 MCG $-$05$-$23$-$016 & 0.0084    & 2    &26    & N &255$^{+56}_{-48}$ &246$^{+16}_{-4}$& 210$^{+8}_{-12}$ &210$^{+8}_{-12}$ & -  &207$^{+7}_{-10}$  &310$^{+97}_{-94}$&185$^{+28}_{-33}$ &190$^{+28}_{-27}$&262$^{+47}_{-39}$  &221$^{+4}_{-3}$ &221$^{+4}_{-3}$ &221$^{+4}_{-3}$   \\ 
                          &              &       &       & I   &        -                       &       -                    &       -                      &      -                     &  - &     -                       &        -                     &       -                      &     -                        &     -                         &        -                   & -                        &      -                    \\                         
 2MASX J18305065+0928414   & 0.0193   & 2     &37  & N & -  & - & 415$^{+124}_{-96}$&- &449$^{+281}_{-209}$&  -  &  - &  - &  -  &  -  &176$^{+7}_{-8}$& 176$^{+7}_{-8}$  & 176$^{+7}_{-8}$     \\
                                                   &              &        &      &  I  &  - & - &-                               &- & -   &  -  &   - &  - & -  &  -  & -   & 663$^{+74}_{-75}$&663$^{+74}_{-75}$  \\                                                                     
 ESO 234$-$G050    &0.0088   & 2    & 40 &  N  & - & - &132$^{+4}_{-2}$    &132$^{+4}_{-2}$    &-  &166$^{+12}_{-12}$ &166$^{+12}_{-12}$&308$^{+35}_{-25}$&161$^{+12}_{-14}$& - &172$^{+2}_{-2}$  &172$^{+2}_{-2}$  &172$^{+2}_{-2}$    \\
                            &              &       &       &  I  &  - & - &425$^{+13}_{-6}$  &425$^{+13}_{-6}$ &-   &481$^{+2}_{-2}$&  -                          &  -                           &371$^{+59}_{-44}$&  - &481$^{+2}_{-2}$ &481$^{+2}_{-2}$ &481$^{+2}_{-2}$   \\  
 
 NGC 4941           & 0.0038 & 2    &50  & N  &133$^{+4}_{-6}$&133$^{+4}_{-6}$&138$^{+4}_{-4}$&138$^{+4}_{-4}$  & - &129$^{+6}_{-6}$&129$^{+6}_{-6}$  &169$^{+9}_{-15}$&195$^{+13}_{-6}$&241$^{+14}_{-9}$&109$^{+2}_{-2}$&109$^{+2}_{-2}$&109$^{+2}_{-2}$   \\
                             &             &       &      & I   &549$^{+68}_{-56}$&549$^{+68}_{-56}$&386$^{+9}_{-6}$&386$^{+9}_{-6}$ & - &422$^{+16}_{-20}$&422$^{+16}_{-20}$&-  & - & - &359$^{+10}_{-11}$&299$^{+7}_{-8}$ &299$^{+7}_{-8}$  \\  
                             &             &       &      & I   &  -                        &  -  &       -                 & -                         &-  & -  & - & -  & - & - &  -                 &691$^{+28}_{-62}$&691$^{+28}_{-62}$ \\             
NGC 4945          & 0.0017   & 2   &22   & N  & - & -  & -  &- & -   &-&-&-&- &-& -    &  -&- \\       

NGC 5643          & 0.0040 & 2   &37   & N &-                           & -                                    &-                               &  -                           &  -                          & -                            &  -  & -                             &  -                         & -   &93$^{+5}_{-5}$   &-                           &-                         \\
                           &              &      &       & N &284$^{17}_{-10}$ &284$^{17}_{-10}$         &187$^{+4}_{-7}$       &  187$^{+4}_{-7}$  &158$^{+18}_{-18}$ &245$^{+8}_{-4}$    &  -  &319$^{+18}_{-18}$ & 202$^{+5}_{-7}$  &290$^{+10}_{-12}$    &251$^{+2}_{-2}$    &251$^{+2}_{-2}$  &251$^{+2}_{-2}$ \\
                           &              &      &       & I &  -                            & -                                  &428$^{+32}_{-19}$  &        -                      &   -                          & 603$^{+24}_{10}$ &  -  &  -                           &473$^{+46}_{-32}$ &568$^{+81}_{-72}$    &398$^{+41}_{-26}$&291$^{+5}_{-4}$ &291$^{+5}_{-4}$ \\
                           &              &      &      & I  & -                             &1023$^{+960}_{-498}$ & 750$^{+96}_{-76}$ &   -                          & -                            &   -                          &  -  &  -                           & -                             &    -                            &880$^{+130}_{-116}$& 697$^{+11}_{-12}$  &  697$^{+11}_{-12}$    \\
 NGC 6221         &0.0045   & 2   & 50  & N  & - & -&126$^{+2}_{-4}$&126$^{+2}_{-4}$& -&- &125$^{+2}_{-6}$&- &77$^{+2}_{-2}$&- & 79$^{+1}_{-1}$&177$^{+4}_{-2}$ &177$^{+4}_{-2}$  \\  
                           &              &      &       & I  & - &- &569$^{+10}_{-8}$&- &- & -&- &- & 144$^{+1}_{-2}$&- &146$^{+1}_{-4}$&125$^{+5}_{-5}$  & 125$^{+5}_{-5}$\\
                            &             &     &       & I  & - & -&551$^{+60}_{-53}$& -& -&636$^{+28}_{-38}$&- &- &518$^{+19}_{-19}$&- &541$^{+13}_{-13}$ &212$^{+3}_{-4}$ &212$^{+3}_{-4}$ \\  
                            &             &    &       & I  & - & -& -&- &- &- &- &- &- &- &   -                           &851$^{+16}_{-10}$ &851$^{+16}_{-10}$ \\  

 NGC 7314        &  0.0047  & 1.9  & 11 & N  &68$^{+1}_{-1}$   & 68$^{+1}_{-1}$      & 56$^{+6}_{-4}$ &56$^{+6}_{-4}$   &- &56$^{+1}_{-1}$ &75$^{+9}_{-4}$        &91$^{+5}_{-6}$ &67$^{+2}_{-2}$ &78$^{+2}_{-7}$     &56$^{+1}_{-1}$ &56$^{+1}_{-1}$  &56$^{+1}_{-1}$  \\
                          &              &        &      & I   &178$^{+4}_{-3}$  &178$^{+4}_{-3}$   & 121$^{+6}_{-5}$&121$^{+6}_{-5}$&- &171$^{+4}_{-3}$ &224$^{+28}_{-27}$ &-                        &222$^{+8}_{-8}$ &203$^{+8}_{-5}$ &68$^{+1}_{-1}$  &68$^{+1}_{-1}$  &68$^{+1}_{-1}$   \\
                          &              &        &      & I   &-                           &-                           &135$^{+6}_{-4}$ &135$^{+6}_{-4}$ &- &-                        &-                               &-                        & -                        &-                          &213$^{+1}_{-1}$   &213$^{+1}_{-1}$    &213$^{+1}_{-1}$    \\ 
                          &              &        &      & B  &-                           &-                           &-                         &-                          & -& -                      & -                               &-                       &  -                       &-                           &1097$^{+57}_{-63}$ & -& -\\                                                                                                                                                                                    
 \hline             
 NGC 3783        &  0.0097 &1     & 21 & N  &133$^{+5}_{-6}$   & 133$^{+5}_{-6}$      &146$^{+3}_{-2}$&146$^{+3}_{-2}$         &-                                  &234$^{+7}_{-4}$     &      -                           &114$^{+21}_{-20}$      &241$^{+10}_{-11}$     &226$^{+60}_{-52}$& 322$^{+1}_{-1}$        & 114.0$^{+2}_{-2}$& 114.0$^{+2}_{-2}$\\ 
                          &              &       &      & I  & 529$^{+7}_{-5}$    &529$^{+7}_{-5}$       &548$^{+17}_{-10}$&548$^{+17}_{-10}$ & 566$^{+30}_{-37}$    &668$^{+7}_{-10}$ & 473$^{+20}_{-10}$   &437$^{+17}_{-27}$       &     -                            &  -                        & -                                 & 322$^{+1}_{-1}$   & 322$^{+1}_{-1}$   \\
                          &              &       &      & I   &495$^{+82}_{-78}$&1556$^{+38}_{-38}$ &      -                   &             -                     &918$^{+168}_{-170}$ &            -                 &       -                          &1216$^{+158}_{-107}$& 1012$^{+14}_{-22}$  & 705$^{+59}_{-77}$ & 2102$^{+14}_{-13}$   & 685$^{+2}_{-2}$    & 685$^{+2}_{-2}$   \\
                          &              &       &      & B  &         -                   &    -                            &       -                  &              -                    &-                                  &           -                  &        -                         &             -                     &   5186$^{+37}_{-44}$ &          -                 & 5549$^{+21}_{-18}$   & -                             & -\\
 \hline                 
 \end{tabular}
 }
Notes: (1) Source name; (2) Redshift as reported in Table \ref{tbl:classAGN}; (3) SWIFT70M optical spectral classification; (4) S/N near the H$\upbeta$; (5) line components; N: NLR, B: BLR, I: intermediate (see sect \ref{sec:lines} for the classification criteria); (6) to (17) FWHM (\kms) not deconvolved for instrumental resolution.
\noindent
\end{minipage}
\end{table} 
\end{landscape}

\clearpage
\begin{landscape}
\begin{table}
\begin{minipage}{140mm}
\caption{$EW$ measurements of the emission lines in X-shooter UVB spectra.}
\label{tbl_EWxshUVBSpettri}
\resizebox{1.7\textwidth}{!}{
\begin{tabular}{@{}llcccccccccccccccc}
\hline
Object & Redshift & Cl & S/N & Comp.  & [\ion{Ne}{V}] & [\ion{Ne}{V}]& [\ion{O}{II}]  & [\ion{O}{II}]& [\ion{Fe}{VII}] & [\ion{Ne}{III}] & [\ion{Fe}{V}] & [\ion{Ne}{III}] & H$\upgamma$& \ion{He}{II} & H$\upbeta$& [\ion{O}{III}] & [\ion{O}{III}]  \\
           &               &      &        &              &$\lambda$3346\AA\ &$\lambda$3426\AA\ &$\lambda$3728\AA\ & $\lambda$3730\AA\ &$\lambda$3760\AA\ &$\lambda$3869\AA\ &$\lambda$3893\AA\ &$\lambda$ 3968\AA\  &   &$\lambda$4686\AA\ &  &$\lambda$4959\AA\  &$\lambda$5007\AA\ \\ 
 (1)         &   (2)           &  (3)    &  (4)      &    (5)          & (6) & (7) &  (8) &   (9)   & (10) &(11)  & (12)  & (13) &  (13)   &  (14)   & (15) & (16)& (17)   \\ 
\hline
ESO 263$-$G013   &0.0088  &2     &  14  &N   &26.6 &39.8   & -      &- & - &51.6  & 19.7  & 17.2  & 9.9  & 5.3 & 5.2   &22.7   & 69.0    \\
                          &              &       &        &I   &    -    &  -      &111.2&-  &  -& -   & -  & -  &  - &-  &9.2&  37.9 &108.0 \\
LEDA 093974  & 0.0239    & 2   &57 &N  &   -   &  -    & 1.8& 2.8 & - & 3.0  & -  & -  &  - & - &1.6&3.9&11.7 \\
                         &               &      &      &I  &   -     &  -    &5.9&26.5 & -  &5.1   & -  & -  & -  & - &-& 5.7&16.9 \\     
MCG $-$05$-$23$-$016 & 0.0084      & 2    &26  & N & 6.3 & 22.3   & 13.1 &12.9 & -  &18.1  & 3.1 & 2.9 & 2.0 &2.0  &3.3 &14.1 &42.2   \\     
                         &              &       &       & I   &  -  & -          &    -   &   -    &  - &   -     &  -    &  -    &  -    & -    & -    & -      &   -     \\       
  2MASX J18305065+0928414   & 0.0193  & 2     &37  & N & -  & - &10.7&- & 9.0  &  -  &-   & -  & -   &  -  &0.8& 4.4& 13.3    \\
                                                   &              &        &      &  I  & -  & - &-      & -&  -  & -   &   - & -  & -  &  -  & -     & 2.3&7.0  \\   
   ESO 234$-$G050    & 0.0088   & 2    & 40 &  N  & - & - &11.5    &12.9 &-  &2.5&2.0&2.0&4.0& - &10.7 &8.9  &26.8    \\
                              &              &       &       &  I  &  - & - &29.2    & 23.1 &-   &5.5&  -  &  -  &3.0&  - &10.0 &15.4 &46.3  \\                                           
   NGC 4941           & 0.0038   & 2    &50  & N  &3.9&13.2 &9.7&6.1 & - &14.7&1.4 &5.4&3.1&3.6&2.3&11.7&35.3\\
                             &             &       &      & I   &2.7&10.6&12.1&34.8 & - &24.7&5.4&-  & - & - &9.0&29.9&90.2  \\  
                             &             &       &      & I   &  -  &  -  & -       & -   &-  & -  & - & -  & - & - &  -   &12.7&38.3 \\   
 NGC 4945          & 0.0017  & 2   &22   & N  & - & -  & -  &- & -   &-&-&-&- &-& -    &  -&- \\    
 NGC 5643          &0.0040    & 2   &37   & N &-    & -    &-     &  -      &  -  & -     &  -  & -     &  -   & -   &1.5  &-        &-    \\
                           &              &      &       & N &4.5 &15.0&19.9&16.1 &1.0&12.5 &  -  &5.2 &4.1 &3.7 &12.0 &47.0  &144.6\\
                           &              &      &       & I &  -    & -     &27.4&  -     &-    &10.0&  -  &  -    &2.0 &1.3 &4.5  &7.1 &22.3 \\
                           &              &      &      & I  & -     &12.7&12.6&   -    & -   &   -   &  -  &  -    & -    &    - &2.7 & 12.1& 33.7  \\
 NGC 6221         &0.0045   & 2   & 50  & N  &-  &- &4.2  &3.6 &- &-    &1.3 &- &1.4 &- & 4.7&2.1&6.4 \\  
                            &              &      &       & I  & - & -&10.8 & -   &- &-    &  -   & -&5.2 &- &12.9&0.6  &2.0\\
                            &             &     &       & I    & - &- &1.1  &-     & -&2.0 & -   & -&2.5 &- &7.8&1.4&4.3 \\  
                            &             &    &       & I     &  -& -&-     &-     & - &-   &   -  & -& -    &- &   -  &3.0&9.7 \\       
NGC 7314        &  0.0047  & 1.9  & 11 & N  &6.3  &20.0&16.2 &23.5 &25.4 &3.6 &- &5.7 &8.5   &8.2 &17.9&83.6  &250.4 \\
                          &              &        &      & I   &9.1 &29.4&2.4 &10.5   &31.2 &4.7 &- &-     &13.8 &14.7 &5.5 &23.1  &69.1 \\
                          &              &        &      & I   & -    & -     &1.9 &13.4   &-       &   - &- &-     &-       &- &29.5  &120.5    &361.0 \\ 
                          &              &        &      & B  & -    &-      & -    &-        &-       &   - &- &-     &-       &- &31.2    & -& -\\                                                                      
\hline             
 NGC 3783        & 0.0097  &1     & 21 & N  &0.2  & 0.5  &1.0&1.0&-         &3.4 &   -  &0.3  &1.8   &1.0  & 7.1 & 3.4 &10.1\\ 
                          &              &       &      & I  &1.2   &5.0    &0.8&1.2& 1.7   &5.2 & 1.7 &2.7   &   -   &  -    & -    &20.8   & 63.7  \\
                          &              &       &      & I   &0.3   &3.5    &   -&-    &0.9   &  -   &  -     &11.7& 5.8  & 2.7 & 27.4 & 14.9& 43.2\\
                          &              &       &      & B  & -      &    -   &  -  &  -  &-      &  -   &  -     &   -   &43.6 & -     & 88.9   & -    & -\\
                                                                                                                                                                                                                             
 \hline                         
 \end{tabular}
 }
Notes: (1) Source name; (2) Redshift as reported in Table \ref{tbl:classAGN}; (3) SWIFT70M optical spectral classification; (4) S/N near the H$\upbeta$; (5) line components; N: NLR, B: BLR, I: intermediate (see sect \ref{sec:lines} for the classification criteria); (6) to (17) $EW$ (\AA\ ) .
\noindent
\end{minipage}
\end{table} 
\end{landscape}

\clearpage
\begin{landscape}
\begin{table}
\begin{minipage}{140mm}
\caption{Line flux measurements of the emission lines in X-shooter UVB spectra.}
\label{tbl_FLUXxshUVBSpettri}
\resizebox{1.7\textwidth}{!}{
\begin{tabular}{@{}llcccccccccccccccc}
\hline
Object & Redshift & Cl & S/N & Comp.  & [\ion{Ne}{V}] & [\ion{Ne}{V}]& [\ion{O}{II}]  & [\ion{O}{II}]& [\ion{Fe}{VII}] & [\ion{Ne}{III}] & [\ion{Fe}{V}] & [\ion{Ne}{III}] & H$\upgamma$& \ion{He}{II} & H$\upbeta$& [\ion{O}{III}] & [\ion{O}{III}]  \\
           &               &      &        &              &$\lambda$3346\AA\ &$\lambda$3426\AA\ &$\lambda$3728\AA\ & $\lambda$3730\AA\ &$\lambda$3760\AA\ &$\lambda$3869\AA\ &$\lambda$3893\AA\ &$\lambda$ 3968\AA\  &   &$\lambda$4686\AA\ &  &$\lambda$4959\AA\  &$\lambda$5007\AA\ \\ 
 (1)         &   (2)           &  (3)    &  (4)      &    (5)          & (6) & (7) &  (8) &   (9)   & (10) &(11)  & (12)  & (13) &  (13)   &  (14)   & (15) & (16)& (17)   \\ 
\hline
ESO 263$-$G013   &0.0334   &2     &  14  &N   &0.2E-15  &0.3E-15 & -           &- & - & 0.5E-15 & 0.2E-15  &0.2E-15   &1.5E-16   &0.1E-16  &  0.8E-16   & 3.7E-16   & 1.1E-15    \\
                          &              &       &        &I   &    -          &  -            &1.2E-15&-  & -  &  - & -  & -  &  - & - &1.4E-16&  6.2E-16 &1.7E-15 \\
 LEDA 093974  & 0.0239     & 2   &57 &N  & -     &  -    & 0.3E-16& 0.5E-16&-  & 0.6E-16  & -  & -  & -  & - &0.8E-16&1.8E-16&5.4E-16 \\
                         &               &      &      &I  &   -     &   -   &1.0E-16 &4.7E-16 & -  &1.0E-16   &  - &  - & -  & - &-& 2.6E-16&7.9E-16\\     
  MCG $-$05$-$23$-$016 &0.0084    & 2    &26  & N & 2.5E-16  &9.4E-16 & 6.8E-16 &6.6E-16 &  - & 9.6E-16  & 1.6E-16 & 1.9E-16 &2.2E-16&2.5E-16  &4.4E-16 &1.9E-15 &5.6E-15  \\     
                            &              &       &       & I   &  -  & -          &    -   &   -    &  - &   -     &  -    &  -    &  -    & -    & -    & -      &   -     \\   
  2MASX J18305065+0928414   & 0.0193  & 2     &37  & N &   &  &0.4E-16 & -& 0.4E-16  &  -  &  - & -  &  -  &    &0.2E-16&0.9E-16 & 2.6E-16    \\
                                                   &              &        &      &  I  &   &  &-             &- & -   &  -  &  -  &  - & -  &  -  & -     & 0.5E-16&1.4E-16  \\    
 ESO 234$-$G050                           &0.0088   & 2    & 40 &  N  & - & - &7.5E-16&8.4E-16 &-  &1.8E-16&1.4E-16&1.4E-16&3.5E-16& - &8.9E-16 &8.1E-16  &2.4E-15    \\
                                                   &              &       &       &  I  &  - & - &1.9E-15&1.5E-15 &-   &4.0E-16&  -         &  -         &2.6E-16&  - &8.4E-16&1.4E-15 &4.2E-15  \\ 
  NGC 4941                               & 0.0038 & 2    &50  & N  &5.9E-17&2.1E-16 &3.2E-16&2.0E-16 & - &4.6E-16&4.3E-17 &2.4E-16&3.1E-16&3.8E-16&2.6E-16&1.4E-15&4.1E-15\\
                                                  &             &       &      & I   &4.1E-17&1.7E-16&4.0E-16&1.1E-15& - &7.7E-16&1.7E-16&-  & - & - &1.0E-15&3.5E-15&1.0E-14  \\  
                                                  &             &       &      & I   &  -  &  -  & -       & -   &-  & -  & - & -  & - & - &  -   &1.5E-15&4.5E-15\\             
 NGC 4945          & 0.0017   & 2   &22   & N  & - & -  & -  &- & -   &-&-&-&- &-& -    &  -&- \\ 
NGC 5643          & 0.0040    & 2   &37   & N &-            & -          &-            &  -          &  -  & -     &  -  & -     &  -   & -   &8.5E-17 &-        &-    \\
                           &              &      &       & N &3.5E-17 &1.1E-16&2.9E-16&2.3E-16 &2.7E-17&3.6E-16 &  -  &1.9E-16 &2.0E-16 &2.5E-16 &6.9E-16 &3.4E-15  &1.1E-14\\
                           &              &      &       & I &  -            & -           &3.9E-16&  -         &-             &2.9E-16&  -  &  -           &1.0E-16 &8.8E-17&2.6E-16 &5.2E-16&1.6E-15\\
                           &              &      &      & I  & -             &9.6E-17&1.8E-16&   -        & -             &   -        &  -  &  -           & -           &    -        &1.6E-16 &8.8E-16&2.5E-15  \\       
NGC 6221          &0.0045   & 2   & 50  & N   &-  &- &1.1E-16  &9.5E-17 &-  &-            &4.5E-17 &- &5.9E-17 &- & 2.2E-16&1.1E-16&3.2E-16\\  
                            &             &      &       & I     & - & -&2.9E-16 & -            &-  &-            &  -          & -&2.2E-16 &- &6.2E-16 &3.2E-17  &9.9E-17\\
                            &             &      &       & I     & - &- &2.9E-17 &-             & - &7.0E-17& -           & -&1.1E-16 &- &3.8E-16&7.2E-17&2.1E-16 \\  
                            &             &      &       & I     &  -& -&-            &-              & - &-           &   -         & -& -           &- &   -        &1.4E-16&4.8E-16 \\         
 NGC 7314         & 0.0047  & 1.9  & 11 & N  &1.7E-17  &5.3E-17&5.5E-17&8.0E-17  &6.7E-17 &9.4E-18 &- &1.5E-17 &3.2E-17  &3.5E-17&7.5E-17  &4.0E-16  &1.2E-15 \\
                           &              &        &      & I   &2.4E-17 &7.8E-17&8.3E-18 &3.6E-17   &8.2E-17 &1.2E-17 &- &-            &5.1E-17 &6.2E-17 &2.3E-17  &1.1E-16  &3.3E-16 \\
                           &              &        &      & I   & -            & -           &6.4E-18&4.6E-17   &-            &   -          &- &-            &-            &-             &1.2E-16  &5.8E-16    &1.7E-15\\ 
                           &              &        &      & B  & -            &-            & -          &-               &-            &   -         &- &-             &-           &-             &1.3E-16  & -                & -\\          
\hline                              
NGC 3783        & 0.0097  &1     & 21 & N  &2.0E-16  & 6.4E-16  &8.9E-16&8.9E-16&-                &2.6E-15 &   -          &1.7E-16  &8.9E-16    &5.1E-16   & 3.4E-15 & 1.6E-15&4.9E-15\\ 
                          &              &       &      & I  &2.2E-15  &5.9E-15    &7.4E-16&1.2E-15& 1.5E-16   &4.1E-15 & 1.3E-15 &1.8E-15   &   -            &  -            & -            &1.0E-14   & 3.1E-14  \\
                          &              &       &      & I   &3.5E-16  &4.1E-15    &   -       &-            &7.8E-=16   &  -         &  -            &1.2E-15   &2.8E-15   & 1.5E-15  & 1.3E-14 & 7.2E-15& 2.1E-14\\
                          &              &       &      & B  & -            &    -           &  -         &  -         &-                 &  -         &  -           &   -            &2.1E-14   & -             & 4.3E-14   & -    & -\\                                                                                                                                                                                                                                                                                                                                                                                                                                                                                                                                                                                                  
 \hline                              
 \end{tabular}
 }
Notes: (1) Source name; (2) Redshift as reported in Table \ref{tbl:classAGN}; (3) SWIFT70M optical spectral classification; (4) S/N near the H$\upbeta$; (5) line components; N: NLR, B: BLR, I: intermediate (see sect \ref{sec:lines} for the classification criteria); (6) to (17) Emission line Flux in erg s$^{-1}$ cm$^{-2}$.
\noindent
\end{minipage}
\end{table} 
\end{landscape}

\bsp	
\label{lastpage}
\end{document}